\documentclass{dissertation}
\begin{document}
\includepdf{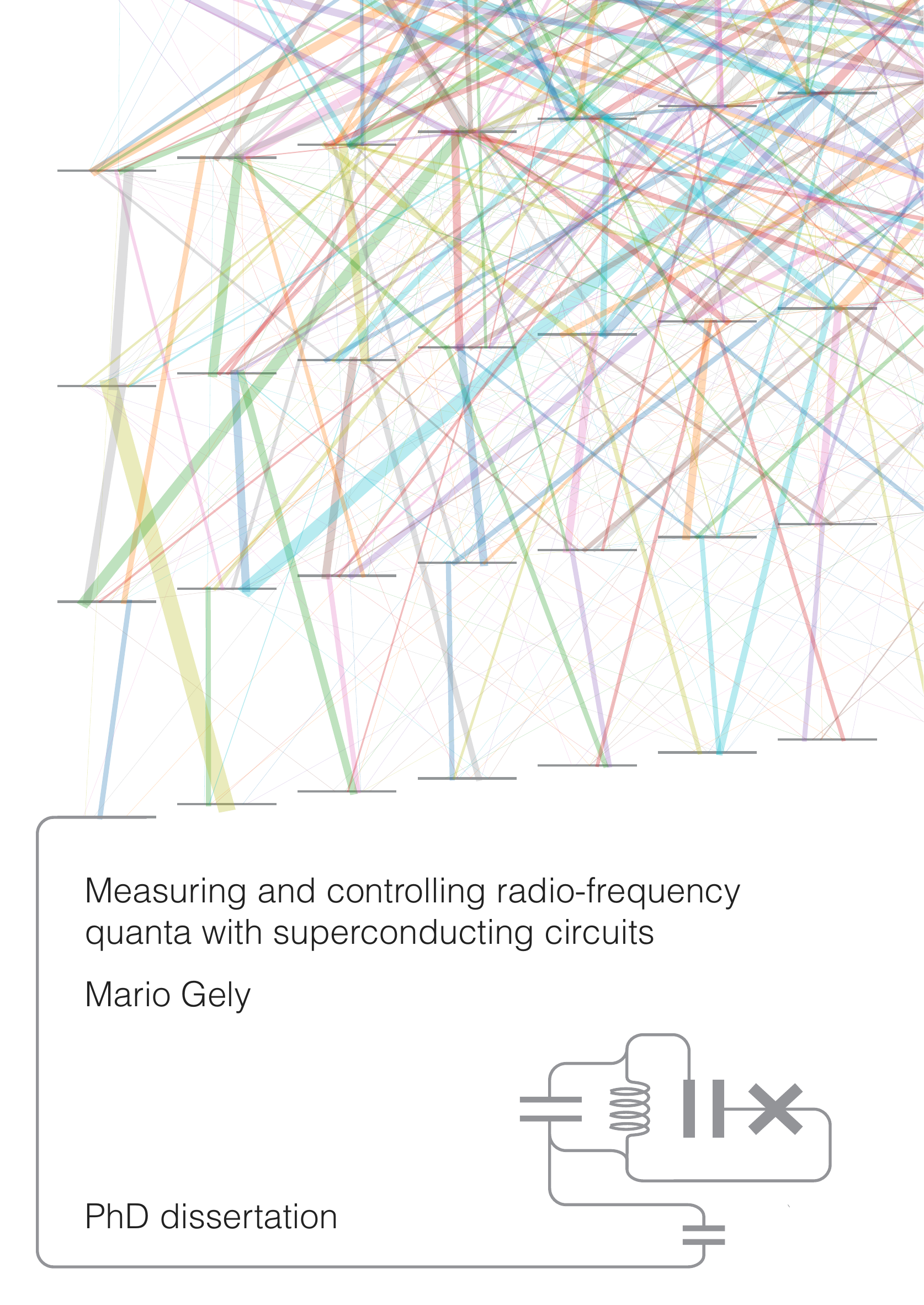}

\title[]{Measuring and controlling radio-frequency quanta with superconducting circuits}
\author{Mario Florentin}{Gely}

\frontmatter

\begin{titlepage}







\begin{center}


\vspace*{2\bigskipamount}

{\makeatletter
\titlestyle\bfseries\Large\@title
\makeatother}

{\makeatletter
\ifx\@subtitle\undefined\else
    \bigskip
    \titlefont\titleshape\Large\@subtitle
\fi
\makeatother}

\vfill

\bigskip
\bigskip
\bigskip
\bigskip

{\titlefont\bfseries\large  Dissertation}
\bigskip
\bigskip

for the purpose of obtaining the degree of doctor
at Delft University of Technology
by the authority of the Rector Magnificus 
Prof.dr.ir. T.H.J.J. van der Hagen,
chair of the Board for Doctorates
to be defended publicly on
Thursday 9 April 2020, at 10:00 o’clock

\bigskip
\bigskip

by

\bigskip
\bigskip

{\makeatletter
\titlefont\bfseries\large Mario Florentin GELY
\makeatother}\\
\bigskip
Master of Science in Applied Physics, Delft University of Technology, the Netherlands\\
Master of Science in Engineering, Ecole Centrale de Nantes, France\\
born in Montpellier, France


\bigskip
\bigskip

\vspace*{2\bigskipamount}

\end{center}

\clearpage
\thispagestyle{empty}


\noindent This dissertation has been approved by the promotors.\\

\noindent Composition of the doctoral committee:

\medskip\noindent
\begin{tabular}{p{4cm}l}
    Rector Magnificus, & chairperson \\
    Prof.\ dr.\ G.\ A.\ Steele, & Delft University of Technology, promotor \\
    Dr. T. van der Sar, & Delft University of Technology, copromotor \\

    \medskip
    \mbox{\emph{Independent members:}} & \\
    Prof.\ dr.\ D.\ DiVincenzo, & Delft University of Technology \\
    Prof.\ dr.\ K.\ M{\o}lmer, & Aarhus University, Denmark \\
    Prof.\ dr.\ H.\ Ulbricht, & University of Southampton, United Kingdom \\
    Dr.\ Z.\ Leghtas, & Mines ParisTech, France \\
    Dr.\ A.\ R.\ Akhmerov, & Delft University of Technology \\
    Prof.\ dr.\ S.\ Otte, & Delft University of Technology, Reserve member \\



\end{tabular}


\vfill
\begin{center}
    \centering
\includegraphics[height=0.55in]{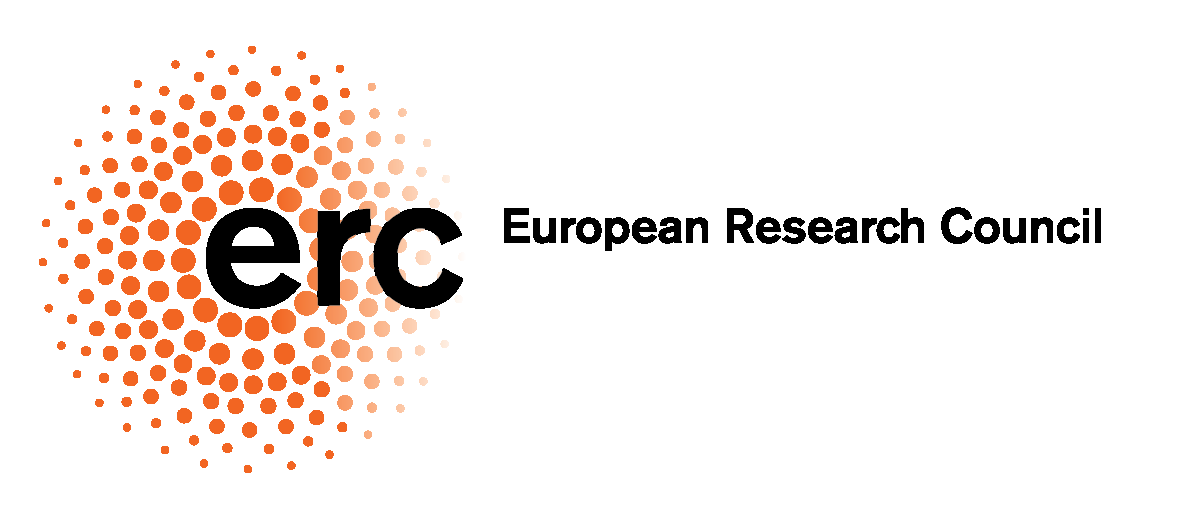}
    \hspace{2em}
    \centering
\includegraphics[height=0.5in]{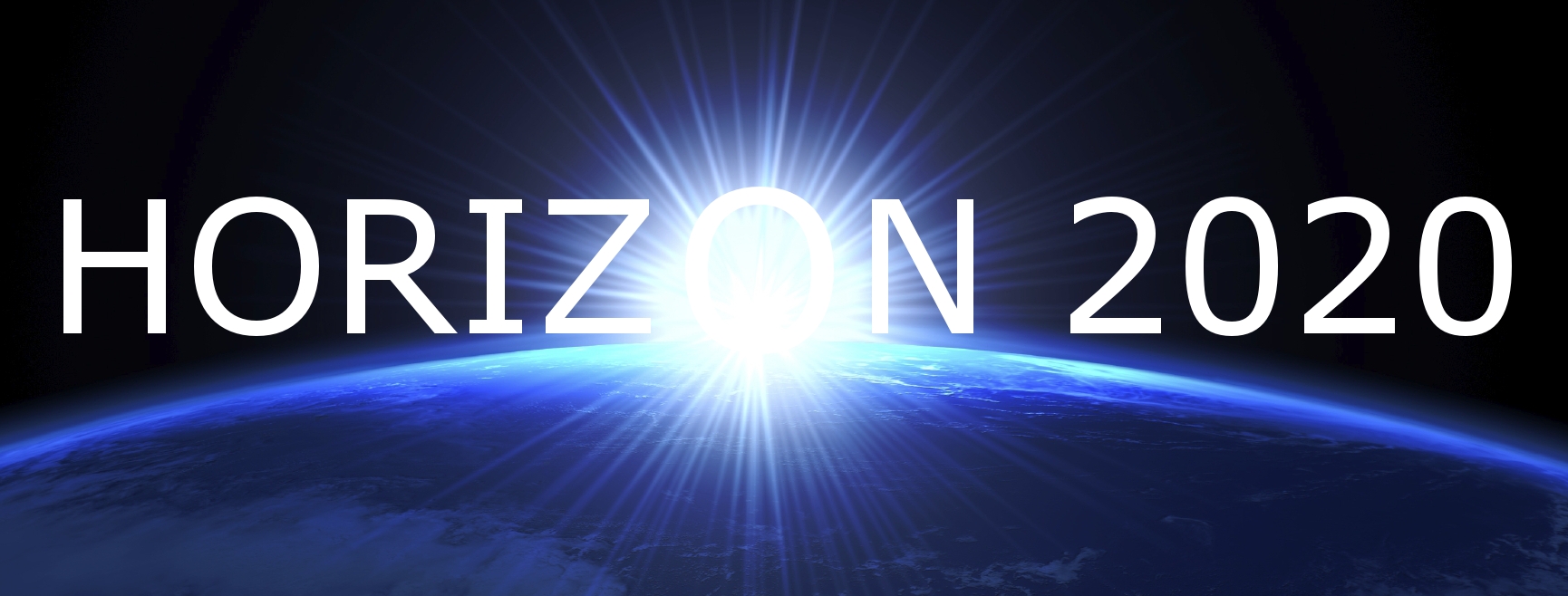} \\
    \vspace{2em}
    \centering
\includegraphics[height=0.45in]{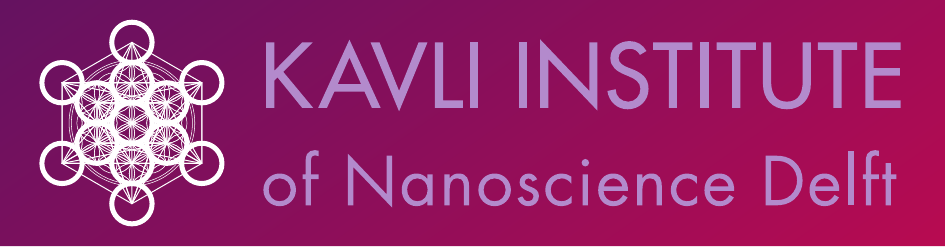}
    \hspace{2em}
    \centering
\includegraphics[height=0.5in]{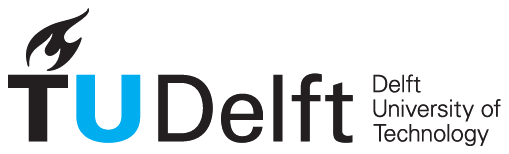}
\end{center}
\vfill

\noindent
\begin{tabular}{@{}p{0.2\textwidth}@{}p{0.8\textwidth}}
    \textit{Printed by:} & Gildeprint, Enschede \\[\medskipamount]
\end{tabular}

\vspace{4\bigskipamount}

\noindent Copyright \textcopyright\ 2020 by M.~Gely

\medskip
\noindent Casimir PhD Series, Delft-Leiden 2020-09

\medskip
\noindent ISBN: 978-90-8593-436-3

\medskip
\noindent An electronic version of this dissertation is available at \\
\url{http://repository.tudelft.nl/}.

\end{titlepage}

\dedication{To my parents. For endless love and support.\\
A mes parents. Pour leur inépuisable amour et soutien.\\
\vspace{10cm}
\epigraph{Slow down and remember this:\\ most things make no difference.\\Being busy is a form of laziness\\ — lazy thinking and indiscriminate action.}{Timothy Ferriss}}
\tableofcontents

\setlength{\parskip}{0.5em}


\chapter*{Summary}
\addcontentsline{toc}{chapter}{Summary}
\setheader{Summary}

In this thesis, we will present the theoretical and experimental work that led to the realization of Radio-Frequency Circuit Quantum Electro-Dynamics (RFcQED).
In chapter \textbf{1}, I will introduce the field of circuit quantum electrodynamics (QED), and the motivations for extending this field to radio frequencies.
In chapter \textbf{2}, we provide a detailed derivation of the Hamiltonian of circuit QED formulated in the context of the Rabi model, and extract expressions for the cross-Kerr interaction. 
The resulting requirements for the coupling rate in RFcQED are discussed, one of them being the need to dramatically increase the coupling rate compared to typical circuit QED device.
In chapter \textbf{3} we cover two experimental approaches to increasing the coupling in a circuit QED system, one making use of a high impedance resonator, the second utilizing a large coupling capacitor.
In chapter \textbf{4}, we combine these two approaches to implement RFcQED.
Through strong dispersive coupling, we could measure individual photons in a megahertz resonator, demonstrate quantum control by cooling the resonator to the ground state or preparing Fock states, and finally observe with nanosecond resolution the re-thermalization of these states.
In chapter \textbf{5} we present QuCAT or Quantum Circuit Analyzer Tool in Python, a software package that can be used for the design of circuit QED systems such as the one presented here in this thesis.
In chapter \textbf{6} we discuss how certain interplays between general relativity and quantum mechanics cannot be described using our current laws of physics.
In particular, we show how radio-frequency mechanical oscillators are perfect candidates to perform experiments in this regime.
In chapter \textbf{7} we present the prospects for coupling such mechanical oscillator to weakly anharmonic superconducting circuits such as the transmon qubits or RFcQED systems.
In chapter \textbf{8}, we provide an outlook.

\chapter*{Samenvatting}
\addcontentsline{toc}{chapter}{Samenvatting}
\setheader{Samenvatting}

{\selectlanguage{dutch}

In dit proefschrift zullen we het theoretische en experimentele werk presenteren dat heeft geleid tot de realisatie van Radiofrequentie Circuit Quantum Electro-Dynamics  (RFcQED). In hoofdstuk 1 zal ik het veld van circuit Quantum Electro-Dynamics (cQED) introduceren en de motieven presenteren om dit vervolgens uit te breiden naar radiofrequenties. In hoofdstuk 2 geven we een gedetailleerde afleiding van de Hamiltoniaan van het Rabi-model geformuleerd in de context van cQED en deduceren we expressies voor de cross-Kerr-interactie. De resulterende voorwaarden voor de koppelingen tussen systemen in RFcQED worden besproken, waaronder een daarvan is om deze drastisch te verhogen in vergelijking met een cQED systeem. In hoofdstuk 3 behandelen we twee experimentele benaderingen voor het verhogen van de koppeling in een RFcQED-systeem: een met behulp van een resonator met hoge impedantie en de tweede met een grote koppelcondensator. In hoofdstuk 4 combineren we deze twee technieken om RFcQED te implementeren. Doormiddel van een sterke dispersieve koppeling konden we individuele fotonen in een megahertz-resonator meten en demonstreerden we controle over de toestand van de resonator door deze vervolgens naar de grondtoestand te brengen. Ook konden we Fock-toestanden preparen en ten slotte met nanoseconde-resolutie de re-thermalisatie van deze toestanden observeren. In hoofdstuk 5 presenteren we QuCAT of Quantum Circuit Analyzer Tool in Python, een softwarepakket dat kan worden gebruikt voor het ontwerpen van cQED-systemen zoals die hier in dit proefschrift worden gepresenteerd. In hoofdstuk 6 bespreken we hoe bepaalde fenomenen in de kwantummechanica niet kunnen worden beschreven in het kader van de algemene relativiteitstheorie met behulp van onze huidige natuurwetten. In het bijzonder laten we zien hoe mechanische radiofrequentie-oscillatoren ingezet kunnen worden in experimenten om de relatie tussen deze twee kaders beter in kaart te brengen. In hoofdstuk 7 presenteren we de vooruitzichten voor het koppelen van dergelijke mechanische oscillator aan zwak anharmonische supergeleidende circuits zoals de transmon qubits of RFcQED-systemen. In hoofdstuk 8 geven we een vooruitblik.

}
\chapter*{Résumé}
\addcontentsline{toc}{chapter}{Résumé}
\setheader{Résumé}

{
Dans cette thèse, nous présentons les travaux théoriques et expérimentaux qui ont conduit à l'étude d'électrodynamique quantique de circuits à fréquences radio (RFcQED).
Dans le chapitre \textbf{1}, je présenterai le domaine de l'électrodynamique quantique de circuits (cQED), et les motivations pour étendre ce domaine aux radiofréquences. 
Dans le chapitre \textbf{2}, nous fournissons une dérivation détaillée de l'Hamiltonien de cQED formulé dans le contexte du modèle Rabi, et extrayons des expressions pour le couplage cross-Kerr. 
Les exigences résultantes pour le taux de couplage en RFcQED sont discutées, l'une d'entre elles étant la nécessité d'augmenter considérablement ce taux par rapport à ce qui est typiquement utilisé en cQED. 
Dans le chapitre \textbf{3}, nous couvrons deux approches expérimentales pour augmenter le couplage en cQED, l'une utilisant un résonateur à haute impédance, la seconde utilisant un grand condensateur de couplage. 
Dans le chapitre \textbf{4}, nous combinons ces deux approches pour implémenter RFcQED. 
Grâce à un fort couplage dispersif, nous avons pu mesurer des photons, individuellement, dans un résonateur mégahertz, démontrer un contrôle quantique en refroidissant le résonateur à l'état fondamental ou préparer des états de Fock du résonateur, et enfin observer avec une résolution en nanosecondes la re-thermalisation de ces états. 
Dans le chapitre \textbf{5}, nous présentons QuCAT ou "Quantum Circuit Analyzer Tool in Python", un logiciel qui peut être utilisé pour la conception de systèmes cQED tels que ceux présenté dans cette thèse. 
Dans le chapitre \textbf{6}, nous discutons comment certaines interactions entre la relativité générale et la mécanique quantique ne peuvent pas être décrites en utilisant nos lois connues de physique. 
En particulier, nous montrons comment les oscillateurs mécaniques radiofréquences sont des candidats parfaits pour réaliser des expériences dans ce régime.
Dans le chapitre \textbf{7}, nous présentons les perspectives de couplage d'un tel oscillateur mécanique à des circuits supraconducteurs faiblement anharmoniques tels que les qubits de type transmon ou le système RFcQED.
Dans le chapitre \textbf{8}, nous concluons et recommandons des pistes de recherches futures.
 
}

\mainmatter
\thumbtrue

\FloatBarrier\chapter{Introduction}
\label{chapter_1}

\begin{abstract}
\end{abstract}

\newpage

\noindent 

\section{Circuit quantum electrodynamics}

\subsection{Historical context}

\subsubsection{Artificial atoms and the birth of circuit QED}
\begin{figure}
\centering
\includegraphics[width=0.8\textwidth]{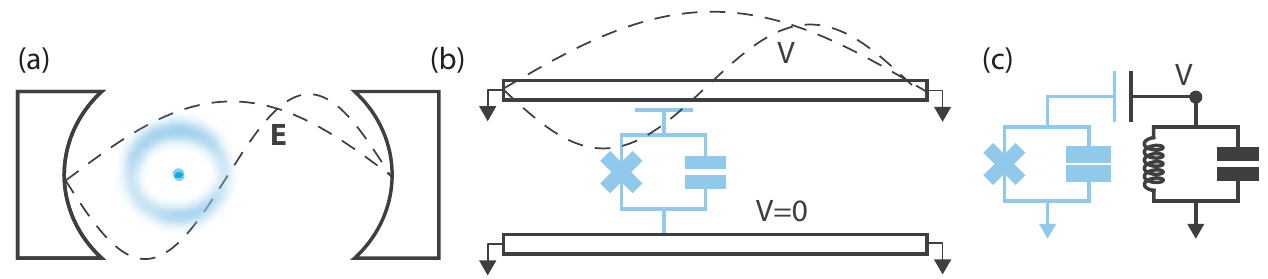}
                                     
\caption{(a) Schematic representation of cavity QED: an atom (in blue) coupled to the electric field of the cavity modes (dashed black lines). (b) Circuit QED example: an artificial atom circuit, of the transmon type, coupled capacitively to the voltage of the modes of a waveguide resonator. (c) Alternative circuit QED system or lumped-element, single-mode equivalent of (b): an artificial atom capacitively coupled to an LC oscillator.}
\label{fig:atom_photon}
\end{figure}

Circuit quantum electro-dynamics (circuit QED) is a circuit implementation of quantum electrodynamics (QED) -- the study of photons (light) interacting with atoms (matter).
The prototypical QED setting is of an atom in a cavity (two face-to-face mirrors) as shown in Fig.~\ref{fig:atom_photon}(a).
The electric field of the different resonance modes of the cavity then interacts with the dipole formed by the negatively charged electrons and positively charged nucleus of the atom.
Constructing a circuit implementation of the cavity is an easy task.
The analogue of the empty space of the cavity is a waveguide, two conductors between which a voltage $V$ can build up.
And the analogue of the mirrors are short circuits to ground which impose $V=0$ (just as mirrors impose a boundary condition $\textbf{E}=0$ for the electric field of the cavity), as shown in Fig.~\ref{fig:atom_photon}(b).
Such systems are called waveguide resonators.
But whereas QED studies atoms coming from your regular periodic table, in circuit QED atoms are actually circuits, made by interconnecting superconducting electrical components.
The surprising fact that superconducting circuits, sometimes large enough to be seen with the naked eye, could show quantum behavior, was demonstrated by a few pioneers in the end of the 1990s and beginning of 2000s.
These feats include the demonstration of energy quantization~\cite{martinis1985energy}, the quantum superposition of a charge being present -- and not -- on a piece of metal~\cite{bouchiat1998quantum}, or of current flowing simultaneously in two different directions~\cite{friedman2000quantum,van2000quantum}.
One important experiment~\cite{nakamura2001rabi} demonstrated for the first time the ability to manipulate a circuit from one quantum state (let's call it $\ket{0}$) to another ($\ket{1}$) or to superpositions of these states, for example $(\ket{0}+\ket{1})/\sqrt{2}$.
Such a system is called a quantum bit or qubit analogously to bits in computers.
The two states $\ket{0}$, $\ket{1}$ are then the possible carriers of quantum information for a quantum computer.
The prospect of building a quantum computer has been a major fuel for the field of superconducting circuits ever since.
Another name for the qubit, which we will adopt from here on, is that of artificial atom.
A critical addition to these artificial atoms came in 2004, with the theoretical proposal~\cite{blais2004cavity}, and then the experimental realization~\cite{wallraff2004strong} of coupling these atoms to waveguide resonators.
This work demonstrated a viable method to control and measure these atoms whilst keeping their fragile quantum states isolated from the environment.
This has been crucial to the development of quantum computing with these systems.
Further, it broadened the scope of these atoms by enabling the exploration of QED topics with engineerable atoms, which can notably have much larger coupling rates than those achievable with natural atoms.

\subsubsection{Exploring the new periodic table}

A central topic from the beginning of this field has been extending the coherence time of these atoms.
That is the average time after which quantum information, stored in the atom, tends to be lost through the interaction of the atom with its environment.
In quantum computing, this sets the amount of computation time that one has available.
When coupling the atom to an auxiliary system, such as a cavity, it sets the maximum time these two systems have to interact.
As a general rule, the more coherence the better, as it is directly correlated to ones experimental prospects.
One way to improve coherence times was by changing the circuit components used, their inter-connection, and the different relations between their component values, leading to atoms with vastly different properties.
The first measure of coherence in a so-called charge qubit~\cite{nakamura2001rabi} preserved the quantum information of a superposition state for a nanosecond.
After the flux~\cite{mooij1999josephson,friedman2000quantum,van2000quantum}, phase~\cite{martinis2002rabi,yu2002coherent}, and quantronium qubit~\cite{vion2002manipulating,cottet2002implementation}, the values of a qubit nick-named the transmon were discovered~\cite{koch_charge-insensitive_2007,schreier2008suppressing}.
The transmon was an extension of the charge qubit, which promised protection against the charge qubits most prevalent decoherence channel, charge noise.
It also turns out to be relatively easy to fabricate.
These two reasons have made the transmon the most prevalent of these circuits in modern experiments, the qubit of choice for scaling up quantum computing efforts, and the artificial atom we will mostly talk about in this dissertation.
Together with considerable progress in fabrication methods and shielding from radiation, standard coherence times for the transmon are now in the tens of micro-seconds~\cite{kjaergaard2019superconducting}.
Longer coherence times could be enabled by even more advanced atoms such as the fluxonium~~\cite{manucharyan2009fluxonium,nguyen2019high} or 0-pi~\cite{brooks2013protected,gyenis2019experimental} qubits which promise even higher coherence times.

\subsubsection{The quest for a quantum computer}

These improvements of artificial atoms have been greatly motivated by the prospect of building a quantum computer.
One class of quantum computers are quantum annealers.
These tackle very specific optimization tasks.
Physical implementations of quantum annealers currently boast thousands of superconducting-circuit-based qubits, and are already commercialized~\cite{mcgeoch2019practical}.
Analog quantum computers are thought of as the next level of complexity in quantum computation.
The idea is to study the dynamics of a fully-controllable quantum system (the computer), which is simulating a problem of interest.
First evoked by Richard Feynman as a way to study quantum many-body problems~\cite{feynman1982simulating}, a few first simulations have been implemented in circuit QED~\cite{barends2016digitized,barends2015digital,salathe2015digital}.
An even greater challenge of the field is to implement gate-based or digital quantum computation that is not problem specific.
In a recent experiment, a digital quantum processor has outperformed a classical computer, a result heralded as quantum supremacy~\cite{arute2019quantum}.
The task used to compare the quantum and classical computers was however specifically designed for the purpose of demonstrating quantum supremacy.
There are still a harsh limitation in the amount of time one has available to run a quantum algorithm~\cite{nielsen2002quantum}, which is imposed by the decoherence time of the qubits.
This has hindered the implementation of a "useful" quantum algorithm.
The current challenge to be addressed in order to run such "useful" algorithms is an implementation of quantum error correction (QEC).
QEC is the quantum analogue of classical error correction schemes.
These schemes protect the information stored, for example in every hard-drive or compact disk, from interactions with its environment.
An implementation of QEC, and its associated challenges: high coherence qubits, high fidelity control of their state, reproducibility in fabrication, increase in the number of qubits and hardware, forms the main goal of most modern quantum computing efforts.
One promising method of QEC is to store quantum information in protected "cat states" of harmonic oscillators~\cite{mirrahimi2014dynamically}.
Experimentally pursued in Yale, it has already been demonstrated that QEC can extend the lifetime of quantum information in a qubit~\cite{ofek2016extending}.
Another method is surface code~\cite{fowler2012surface}, which uses redundancy in the number of qubits used to store information.
This approach is pursued by multiple research groups~\cite{riste_detecting_2015,versluis2017scalable,barends2014superconducting,corcoles2015demonstration}.

\subsubsection{Pushing QED to new heights}

Quantum computing has been a been a central topic in circuit QED.
But the flexibility offered by engineerable atoms has also enabled researchers to push the results achieved with natural atoms in the field of QED to another level.
The level of control over quantum states possible in QED culminated in a Nobel prize awarded in 2012 to Serge Haroche and David Wineland~\cite{haroche2013nobel,raimond2001manipulating}.
This prize celebrates results such as measuring the decoherence that occurs during measurement~\cite{brune1996observing}, enabled by strong-coupling between Rydberg atoms and the electromagnetic cavity field.
Strong coupling~\cite{brune1996quantum}, the "holy grail" that the Haroche group chased since the early 1980s~\cite{haroche2013nobel}, did not present a big challenge in circuits.
It was in fact achieved in the first circuit QED experimental paper~\cite{wallraff2004strong}.
The ability to engineer such strong couplings enabled many fundamental QED experiments.
These experiments demonstrate the interesting ways in which atoms and light can interact on a quantum level~\cite{gu2017microwave}.
There are some other advantages of working with circuits: for example it requires a lot less man-power.
A PhD student with access to a clean-room, a cryostat, some microwave equipment, and a few experienced colleagues to whom ask questions can build a basic circuit QED experiments on the time-scale of a few months.
One downside is the lack of coherence of artificial atoms compared to their natural counter-parts.
But the experimental prospects of an engineerable QED set-up seem plentiful as more and more research groups are joining this exciting field.

\subsubsection{Hybrid systems}

Finally, another application of interest is that of building hybrid systems.
That is a system which contains circuit QED components coupled to objects typically belonging to another field of physics~\cite{xiang2013hybrid}.
These could be for example spins~\cite{kubo2010strong} or mechanical oscillators~\cite{armour2002entanglement,cleland2004superconducting,lahaye2004approaching,lahaye2009nanomechanical,o2010quantum,gustafsson2014propagating,lecocq2015resolving,reed2017faithful,chu2017quantum,Viennot2018}.
One could exploit their large coherence times and use them as a quantum memory for a circuit QED based quantum computer, or conversely use the circuits to measure and control the auxiliary system. 
A particularly attractive prospect is to create quantum states of mechanical oscillators which have a very large mass (by quantum standards).
The relativistic effect of time dilation associated with mass is currently incompatible with the theory of quantum mechanics~\cite{penrose1996gravity}.
One reason is the lack of experimental data -- quantum experiments with heavy objects is technically very challenging.
Circuit QED coupled to mechanical oscillators could be a promising platform in that regard (see Chapters~\ref{chapter_gravity},\ref{chapter_phonon_res}).

\subsection{Basic theoretical concepts}

In order to go further and place the work of this thesis in the context of circuit QED, we first have to introduce some technical details of the most common circuit, that of an artificial atom coupled to a cavity.
Specifically, we will focus on the most popular artificial atom, the transmon, coupled to an LC-oscillator through a capacitor as shown in Fig.~\ref{fig:atom_photon}(c).
An LC oscillator can represent a mode of a waveguide resonator as shown in Fig.~\ref{fig:atom_photon}(b), or can be built using lumped element circuit components.
We will present only an overview of some basic physical effects and quantities, whereas a detailed derivation of the circuit Hamiltonian and of the dependence of the anharmonicity $A$ and the linear and cross-Kerr coupling rates $g$ and $\chi$, as a function of circuit parameters, will be given in Chapter~\ref{chapter_2}.

\subsubsection{Harmonic oscillator}

An LC oscillator is a harmonic oscillator and as such posses discrete quantum eigenstates $\ket{0},\ \ket{1},\ \ket{2},\ ...$ with energies $\hbar\omega/2$, $3\hbar\omega/2$, $5\hbar\omega/2$, ... where $\hbar$ designates Planck's reduced constant and $\omega = 1/\sqrt{LC}$ designates the oscillators angular resonance frequency determined by the capacitance $C$ and inductance $L$ of the circuit.
One important feature of the eigenstates $\ket{n}$, used to understand the physics of the transmon, is that the quantum fluctuations $\langle n| \hat i^2 |n\rangle$, of the current $\hat i$ traversing the oscillators inductor, increase with $n$.

\subsubsection{The transmon}

The transmon is an extension of this oscillator, where the inductor is replaced by a Josephson junction, an inductance $L_J(i)$ which varies with the current going through it.
For a transmon, this inductance increases with the fluctuations of current traversing it.
The frequency of this oscillator $\omega_n = 1/\sqrt{L_J(i)C}$ will thus acquire a dependence on $n$, the state of the circuit, such that $\omega_n = \omega_{n-1}+(\omega_0-nA/\hbar)$, where $A$ is the anharmonicity of the transmon.
It is this property which warrants the name of artificial atom: since the frequency of transitions between levels are all different, by tuning an AC source coupled to the circuit to a specific frequency $\omega_n-\omega_{n-1}$, one can transfer the population of state $\ket{n-1}$ to the state $\ket{n}$ and back.
If one knows the initial state of the circuit, any superposition of eigenstates is possible.
These states are analogous to the eigenstates of a natural atom, and as such we will tend to call the states of the artificial atom $\ket{g}$, $\ket{e}$, $\ket{f}$, ... (the two first referring to ground and excited states) rather than using numbers which will be reserved for the photons of the harmonic LC oscillator.

\subsubsection{Strong coupling}

By connecting an LC oscillator to a transmon with a capacitor, we allow oscillations of charge on one side of the capacitor to induce oscillations of charge on the other side.
This is referred to as "coupling".
One way of quantifying the coupling is through the "linear coupling" $g$.
To understand $g$, let us assume resonance: that the frequency of the LC oscillator $\omega$ matches the transition frequency of the first two states of the transmon $\omega_{ge} = \omega_{e}-\omega_g$.
If we populate the oscillator with a single photon, such that the state of the total system is $\ket{1g}$, and let the system evolve, this quanta will oscillate back and forth between the oscillator and transmon at a rate $g$, realizing Rabi oscillations.
The system is in the "strong coupling" regime if $g$ is sufficiently large such that a Rabi oscillation can occur before the interaction with the environment affects the quantum state of the system.
The latter timescales will be called $1/\kappa$ and $1/\gamma$ for the oscillator and transmon respectively.
Note that the larger the coupling capacitor, the larger $g$ will be.

\subsubsection{Strong dispersive coupling}

Away from resonance, when $\omega-\omega_{ge}\gg g$, the coupling can be interpreted in a different manner, and is better quantified by $\chi\propto g^2$, called the cross-Kerr interaction strength.
In this regime, oscillation of charge on one side of the coupling capacitor will induce a small current oscillations on the other side.
This has two consequences.
First, the junction inductance $L_J(i)$ will have a small contribution to the total inductance of the harmonic oscillator, such that when the atom changes state, changing the inductance $L_J(i)$, the frequency of the oscillator also changes.
The transition energies between levels $\ket{n,g}\leftrightarrow\ket{n+1,g}$ then differs from that of $\ket{n,e}\leftrightarrow\ket{n+1,e}$ by $\chi$, called the cross-Kerr interaction strength.
Secondly, exciting the oscillator will lead to a change in the current fluctuations in the junction, changing the inductance $L_J(i)$ and thus the transition frequencies of the transmon.
The transition energies between levels $\ket{n,g}\leftrightarrow\ket{n,e}$ then differs from that of $\ket{n+1,g}\leftrightarrow\ket{n+1,e}$ by $\chi$ again.
If $\chi/\hbar \gg\kappa$, by measuring the resonance frequency of the oscillator, one can thus infer the state of the transmon, and conversely with $\chi/\hbar \gg\gamma$ the frequency of the transmon will reflect the state of the oscillator.
In the latter case, one can also exert control over the oscillator in addition to measure its state.
Certain transitions of the system, for example $\ket{0,g}\leftrightarrow\ket{1,e}$, equal to the transition frequency of $\ket{0,g}\leftrightarrow\ket{1,g}$ plus that of $\ket{1,g}\leftrightarrow\ket{1,e}$, will have a unique frequency and the oscillator can be driven to its first excited state, which is impossible without its coupling to the atom.
A circuit QED system where $\chi$ is sufficiently large for such measurements or control of the system is said to be in the strong dispersive coupling regime.

\section{Beyond the microwave band}
Note: this section is based on the excellent Ref.~\cite{chris_blog}.

\subsection{Why microwaves}

\begin{figure}
\centering
\includegraphics[width=0.8\textwidth]{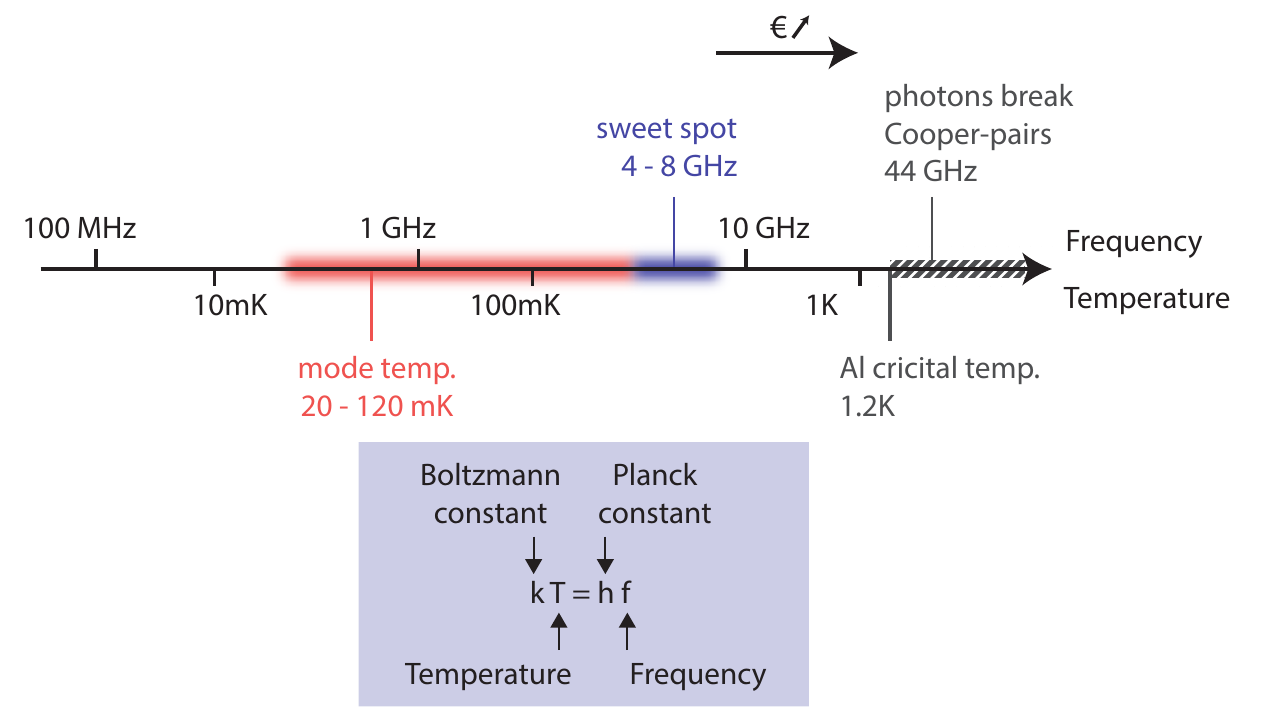}
                                     
\caption{The relevant energy scales of a circuit QED system are plotted on a logarithmic scales. 
The 20-110 mK range is the temperature to which a circuit QED system typically thermalizes in a dilution refrigerator.
Above it the sweet spot for circuit QED, the 4 to 8 GHz band.
There the thermal energy of the environment is not sufficient to disturb the system, the cost of electronic equipment remains relatively low, and both the critical temperature and superconducting gap are far away.}
\end{figure}

\subsubsection{Upper limit in frequency: superconductivity and equipment price}

Let's start by discussing what an upper limit in the operation frequency of superconducting circuits could be.
To have this discussion, let's pick a superconductor to use for our circuit: aluminum.
The Josephson junctions typically used in artificial atoms require constructing a small ($\sim 100 \text{nm}^2$) overlap between two superconducting leads, separated by a tunnel barrier made from an insulating material a few nanometers thick.
Aluminum is very common material used in such nanofabrication and has a native oxide layer of the right thickness to serve as a tunnel barrier.
As such, it is the most practical superconductor to make Josephson junctions~\cite{dolan1977offset}.
The temperature at which aluminum becomes superconducting is $T_c=1.2$ K~\cite{cochran1958superconducting}\footnote{Value for bulk aluminum, this value increases for thinner films such as those used in Josephson junctions~\cite{ferguson2007energy}}. 
Conversely, the superconducting gap, \textit{i.e.} the energy necessary for an incoming photon to break Cooper-pairs, is given by $1.76 k_BT_c$~\cite{tinkham2004introduction}, which translates to 44 GHz.
These form hard upper limits to the temperatures and frequencies we may choose for our artificial atom.
Using different superconducting metals for our junction could increase these values, but a very practical reason makes this undesirable: the dramatic increase in price of electrical equipment with operation frequency.
Even with aluminum as a superconductor, it is desirable to move down in frequency from the 44 GHz mark as much as possible, because of the equipment cost and because the quality factor of oscillators tends to decrease and increase respectively as the frequency decreases~\cite{bruno2015reducing}.

\subsubsection{Lower limit: thermal population}

\label{sec:intro_thermal_population}
Let's now explore what the lower limit in frequency is.
To do so, we need to discuss two different energy scales.
Through Boltzmann's constant $k_B$ and Planck's constant $h$, energy $E$ can be interpreted as temperature $T$ and frequency $f$
\begin{equation}
 	E = k_B T = hf\ .
\end{equation}
To clarify the equivelance between these two measures of energy, we take the example of a harmonic oscillator, with oscillation frequency $f$, in equilibrium with a thermal bath of temperature $T$.
The average number of quanta in the oscillator $\bar n$ is  
\begin{equation}
	\bar n = \frac{1}{\exp\left[hf/k_BT\right]-1}\ .
\end{equation}
This provides a interpretation of the equivalence of these energy scales which will be relevant for this discussion: the temperature $T = hf/k_B$ is the temperature necessary for a coupled thermal bath to excite an an oscillator with frequency $f$ to an average population of $\bar n = 1/(e-1) \simeq 0.6$.
If $T \ll hf/k_B$, the oscillator has an average population close to zero, and the system is in its ground state.
The condition $T \ll hf_{ge}/k_B$, where $f_{ge}$ is the transition frequency between the ground state and first excited state of an atom is critical to manipulating an atom, natural or artificial, at a quantum level.
In the opposite case, the environment has enough thermal energy to significantly raise the average number of quanta.
Meaning that the initial state of the system at the start of any experiment will be unknown, and that the environment will have the ability to influence our experiment in unpredictable ways (it will be a source of decoherence).
%

%
For natural atoms, such as the hydrogen atom, $f_{ge} = 2466$ THz.
The low temperature condition is then satisfied at room temperature, with the temperature $T = 118349$ K necessary to thermally excite the atom being much higher.
However for circuits operating in the GHz range, room temperature would lead to average populations in the thousands of quanta.
We thus need a method to cool the circuit, together with the substrate on which it sits and the cables connected to it, to lower temperatures.
The lowest achievable temperatures are a few miliKelvin, through the use of a dilution refrigerator. 
In practice however, the relevant degrees of freedom of the circuit QED system in a dilution refrigerator are hotter than the temperature measured by the fridge~\cite{serniak2018hot}.
In our experiments, we measured circuit temperatures ranging from 20 to 110 mK, whereas the temperature sensor of our fridge was out of range, indicating a temperature below 7 mK.
These temperatures correspond to frequencies $k_BT/h = 416$ MHz and $2.3$ GHz for 20 and 110 mK respectively.

\subsubsection{The sweet spot: 4-8 GHz}

Considering all these requirements, the C-band of microwave frequencies spanning 4 to 8 GHz is a sweet spot for circuit QED.
It is a popular frequency for telecommunication or radar, translating to a lot of commercially available equipment at a relatively affordable price.
This band is sufficiently below the superconducting gap frequency, and sufficiently high in frequency to ensure populations of the resonators and artificial atoms stay well below 1.

\subsection{Radio-frequency circuit QED}

In this section we will discuss the motivations for building a circuit QED setup operating in the MHz frequency range.
We are specifically referring to a circuit QED setup with a resonator frequency low enough for the miliKelvin temperatures of a dilution refrigerator to excite photons.
This corresponds to a few hundred MHz or less.
The artificial atom of the system would remain in the usual 4-8 GHz band.
Since the resonator is at the frequency at which FM radio operates, we refer to such a system as radio-frequency circuit QED (RFcQED).
This system should also be in the strong dispersive coupling regime, enabling the measurement and control of radio-frequency photons through the artificial atom (a transmon), the latter being in its ground state.

\subsubsection{What lies below the GHz range}

\begin{figure}
\centering
\includegraphics[width=0.8\textwidth]{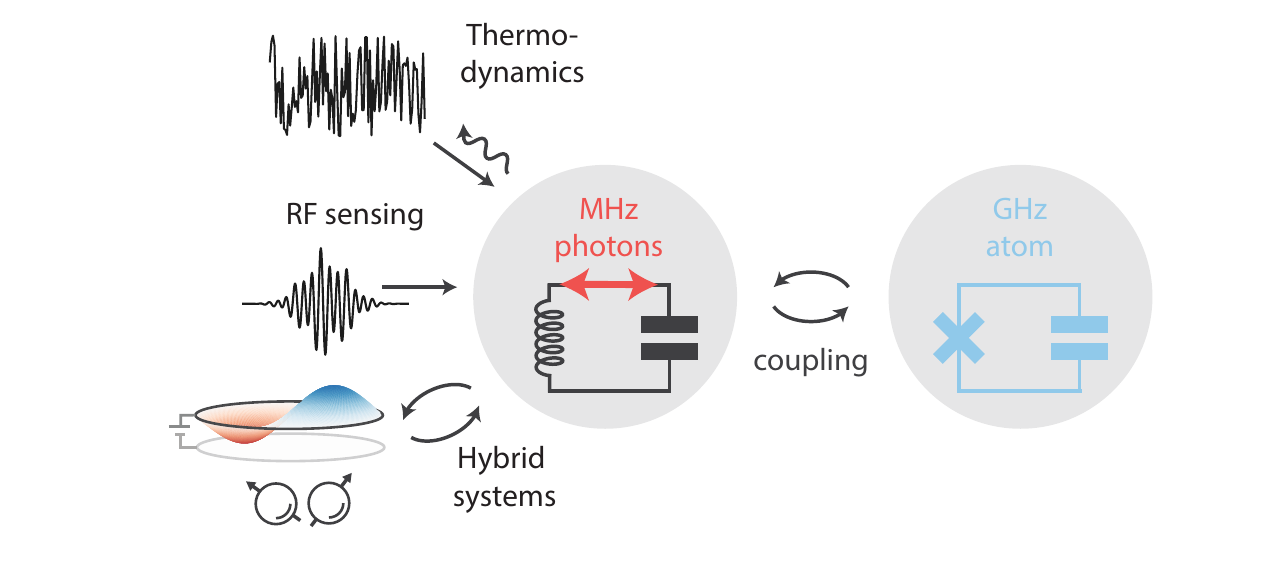}
                                     
\caption{Schematic description of the radio-frequency circuit QED (RFcQED) system and its applications.
In RFcQED a microwave frequency artificial atom, in its ground state, interacts with the megahertz photons of a resonator in thermal equilibrium with a hot bath.
Applications could include studying quantum thermodynamics, detecting radio-frequency signals or coupling to other megahertz frequency systems such as mechanical oscillators or spins.
}
\end{figure}

\textbf{Thermodynamics -- }
Working with a resonator which is not in its ground state is a technical challenge, but also of fundamental interest.
The thermodynamics of the resonator interacting with its environment will be an omnipresent feature to tame but also to study.
Combined with quantum measurement and control of the resonator photons, we enter the realm of quantum thermodynamics~\cite{merali2017new}, and are in measure to explore overlap between the two fields of physics.
With a number of open questions~\cite{alicki2018introduction}, quantum thermodynamics is a topic of growing interest.

\noindent 
\textbf{Detection -- }
The use of circuit QED as detectors of single \textit{microwave} photons has been actively pursued in the recent years~\cite{murch2018single,chen2011microwave,inomata2016single,besse2018single,kono2018quantum}.
However many fields critically rely on the detection of very weak signals at much lower \textit{megahertz} frequencies, for example in radio-astronomy or nuclear magnetic resonance.
The photon resolution achievable in the strong dispersive regime of RFcQED could thus be highly relevant for such detection applications.
\noindent 
\textbf{Controlling and measuring MHz systems -- }
By extending circuit QED to the megahertz regime, we also allow resonantly interfacing quantum circuits with megahertz systems such as spin systems in low magnetic fields~\cite{ares2016sensitive} or mechanical oscillators.
The keyword here is "resonant interaction".
Concerning mechanics, this has only been achieved for GHz acoustic resonators, typically coupled to a circuit QED system through piezoelectricity, resulting in spectacular results such as the observation or preparation of quantum states of motion~\cite{o2010quantum,chu2018creation,satzinger2018quantum,arrangoiz2019resolving}.
However there is fundamental interest in extending that level of control to systems featuring larger displacements and heavier mass, which inevitably translates to lower frequencies.
Controlling such macroscopic mechanical oscillators is the main personal motivation behind this project and merits further explanation in a chapter of its own (Chapter~\ref{chapter_gravity}).

\subsubsection{Challenges to building RF circuit QED, and possible solutions}

\noindent 
\textbf{Temperature -- }
One challenge is the effect of the hot environment on the resonator, which potentially manifests in a few ways. 
First is the mixed thermal state the resonator will relax to, with an average population superior to one, which imposes an inevitable uncertainty in the initial state of the resonator at the beginning of any experiment.
This uncertainty in photon number, combined with the photon-number-dependence of the frequency of the resonator caused by the Josephson junction also causes dephasing in the system.
Taken to the extreme, the current fluctuations through the junction excited by the thermal bath could potentially exceed the critical current of the junction, possibly incapacitating the artificial atom.
A solution at least to the first issue is making use of reservoir engineering to cool, but also manipulate the resonator.
This is a well explored technique in circuit QED.
With the prospect of building a quantum computer, or to demonstrate fundamental phenomena, experiments have shown the cooling or reset of qubits to their ground-state~\cite{Valenzuela2006,Geerlings2013,Magnard2018}, also in the megahertz regime~\cite{Vool2018}, quantum state stabilization~\cite{Murch2012,Shankar2013,Leghtas2015}, and quantum error correction~\cite{ofek2016extending}.
Using superconducting circuits, reservoir engineering is commonplace in electromechanical systems~\cite{Teufel2011}, but with weak non-linearity, such schemes have only limited quantum control~\cite{ockeloen2018stabilized,Viennot2018} compared to typical cQED systems.
Despite the many applications of quantum state engineering in cQED, obtaining control over the quantum state of a hot resonator, where the environment temperature is a dominant energy scale, remains a largely unexplored and challenging task.
\noindent 
\textbf{Large frequency detuning -- }
Measuring the state of the resonator, and manipulating it through reservoir engineering requires a strong dispersive coupling $\chi\hbar\gg\gamma$.
However, as derived in detail in chapter~\ref{chapter_2} (see Eq.~(\ref{eq:shifts_beyond})),
\begin{equation}
 	\chi = 4A g^2\frac{\omega^2}{\omega_{ge}^4}\ .
\end{equation} 
In RFcQED, we have $\omega\ll\omega_{ge}$.
And we want to use a transmon as an artificial atom, we have to satisfy the condition $A\ll\hbar\omega_{ge}$.
The two restrictions mean that extremely large coupling rates $g$ are required for this concept to work.
Exploring how large this coupling can be is a topic that is thoroughly covered in this work in chapters~\ref{chapter_2} and~\ref{chapter_3}.

\section{Outline}

In this thesis, we will present the theoretical and experimental work that led to the realization of RFcQED.
In \textbf{chapter~\ref{chapter_2}}, we provide a detailed derivation of the Hamiltonian of circuit QED formulated in the context of the Rabi model, and extract expressions for the cross-Kerr interaction. 
The resulting requirements for the coupling rate in RFcQED are discussed, one of them being the need to dramatically increase the coupling rate compared to typical circuit QED device.
In \textbf{chapter~\ref{chapter_3}} we cover two experimental approaches to increasing the coupling in a circuit QED system, one making use of a high impedance resonator, the second utilizing a large coupling capacitor.
In \textbf{chapter~\ref{chapter_4}}, we combine these two approaches to implement RFcQED.
Through strong dispersive coupling, we could measure individual photons in a megahertz resonator, demonstrate quantum control by cooling the resonator to the ground state or preparing Fock states, and finally observe with nanosecond resolution the re-thermalization of these states.
In \textbf{chapter~\ref{chapter_qucat}} we present QuCAT or Quantum Circuit Analyzer Tool in Python, a software package that can be used for the design of circuit QED systems such as the one presented here in this thesis.
In \textbf{chapter~\ref{chapter_gravity}} we discuss how certain interplays between general relativity and quantum mechanics cannot be described using our current laws of physics.
In particular, we show how radio-frequency mechanical oscillators are perfect candidates to perform experiments in this regime.
In \textbf{chapter~\ref{chapter_phonon_res}} we present the prospects for coupling such mechanical oscillator to weakly anharmonic superconducting circuits such as the transmon qubits or RFcQED systems.

\FloatBarrier\chapter{How large should the coupling be for RFcQED?}
\label{chapter_2}

\begin{abstract}
We provide a detailed derivation of the Hamiltonian of circuit QED formulated in the context of the Rabi model, and extract expressions for the cross-Kerr interaction. 
The resulting requirements for the coupling rate in radio-frequency circuit QED are discussed.
\end{abstract}

\newpage

\noindent 

In this chapter, we derive the Hamiltonian of circuit QED, formulated as the Rabi model.
We first consider a transmon coupled to multiple modes of a waveguide resonator, then to only a single mode.
We then focus on the single mode case, and present the physics of the dispersive and resonant regime.
The basis change underlying this analysis will be derived in detail.
Finally, we will extract conditions on the coupling rate and other circuit parameters to achieve RFcQED.

\section{Derivation of the Hamiltonian of circuit QED}\label{sec:chapter-2-Hamiltonian_derivation}

In this section, we present a detailed derivation of the Hamiltonian of circuit QED.
This Hamiltonian will describe the physics of a transmon interacting with the modes of a waveguide resonator.
We start by determining the lumped element equivalent of a waveguide resonator.
We then perform circuit quantization, and focus on the different parts of the resulting Hamiltonian.
These include the resonator, transmon, and interaction Hamiltonian.
Finally, we simplify the Hamiltonian to the case of a transmon interacting with a single mode of the resonator.
This will be the system of interest in most of this thesis.

This derivation forms a part of the publication M. F. Gely*, A. Parra-Rodriguez*, D. Bothner, Y. M. Blanter, S. J. Bosman, E. Solano and G. A. Steele, \textit{Convergence of the multimode quantum Rabi model of circuit quantum electrodynamics}, {\href{https://journals.aps.org/prb/abstract/10.1103/PhysRevB.95.245115}{Phys. Rev. B \textbf{95}, 245115  (2017)}}. 
A. P-R and E. S. thank Enrique Rico and \'I\~nigo Egusquiza for useful discussions. M. G. and G. S. thank  Yuli V. Nazarov for useful discussions. The authors acknowledge funding from UPV/EHU UFI 11/55, Spanish MINECO/FEDER FIS2015-69983-P, Basque Government IT986-16 and PhD grant PRE2016-1-0284, the Netherlands Organization for Scientific Research (NWO) in the Innovational Research Incentives Scheme – VIDI, project 680-47-526, the Dutch Foundation for Fundamental Research on Matter (FOM), and the European Research Council (ERC) under the European Union’s Horizon 2020 research and innovation program (grant agreement No 681476 - QOMD).
M.F.G. and A.P.-R. contributed equally to this work.   

\subsection{Lumped element equivalent of waveguide resonators}

\begin{figure}[h!]
\centering
\centering
\includegraphics[width=0.8\textwidth]{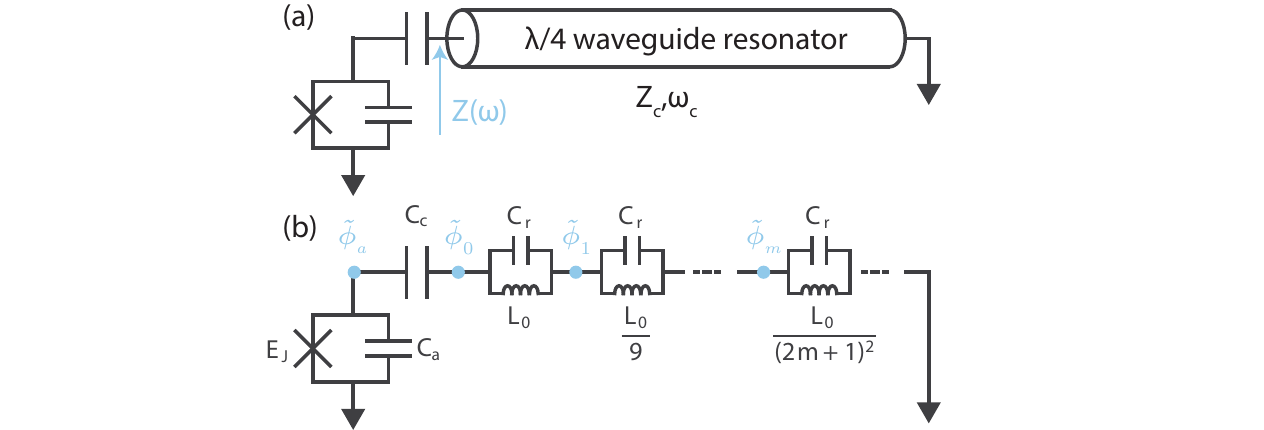}
\caption{\textbf{(a)} Transmon qubit on the left coupled capacitively to a waveguide resonator on the right.
A relation between the voltage $v$ and current $i$ at the left boundary of the waveguide $Z(\omega) = v/i$ is imposed by the shorting of the waveguide on the right boundary~\cite{pozar2009microwave}.
\textbf{(b)} Lumped element equivalent of the circuit in (a) where the series assembly of LC oscillators implements the same impedance $Z(\omega)$ seen by the transmon.
Note that if the waveguide resonator had an open circuit instead of a short on the right boundary, the equivalent circuit would be identical.
}
\label{fig:chapter-2_circuit_notations}
\end{figure}

In this section we will present two implementations of a waveguide resonator and their lumped element circuit equivalents.
There are multiple ways of constructing a waveguide resonator, corresponding to different boundary conditions imposed on a section of waveguide.
We will focus on resonators where one end is an open circuit, creating a voltage anti-node, and the other end is either shorted or opened, corresponding to $\lambda/4$ or $\lambda/2$ resonators respectively.
We will consider the case where the transmon is situated at the voltage anti-node.
The resulting system is shown in Fig.~\ref{fig:chapter-2_circuit_notations}(a) for a $\lambda/4$ resonator.

We start by treating the $\lambda/4$ resonator.

The input impedance of a shorted transmission line, at a distance $\lambda_0/4$ from the short, \textit{i.e.} the impedance seen by the transmon, is given by~\cite{pozar2009microwave}
\begin{equation}
  Z(\omega) = i Z_c \tan\bigg(\frac{\pi}{2}\frac{\omega}{\omega_c}\bigg)\ ,
\end{equation}
where $Z_c$ is the characteristic impedance of the waveguide, $\omega_c/2\pi$ is the resonance frequency and $\lambda_0$ the wavelength of the fundamental mode of the quarter wave resonator when the transmon is replaced by an open termination.
%
%
Note that the wavelength $\lambda_0$ is related to the frequency $\omega_c$ through $\lambda_0 = 2\pi v_\varphi /\omega_c$, where $v_\varphi$ is the phase velocity of the waveguide~\cite{pozar2009microwave}.
The partial fraction expansion of the tangent
\begin{equation}
  \tan(z) = \sum_{m=0}^\infty \frac{-2z}{z^2-(m+\frac{1}{2})^2\pi^2}
\end{equation}
leads to an expression for the resonators imput impedance which is equal to that of an infinite number of parallel LC resonators. Each of them corresponds to a resonance mode
\begin{equation}
  Z(\omega) = \sum_{m=0}^\infty \frac{1}{iC_r\omega+\frac{1}{iL_m\omega}}\ ,
\end{equation}
\begin{equation}
  C_m = C_r = \frac{\pi}{4\omega_c Z_c}\ ,
\end{equation}
\begin{equation}
  L_m = \frac{1}{(2m+1)^2}\frac{4Z_c}{\pi \omega_c}\ .
\end{equation}
A similar expression can be derived for a $\lambda/2$ resonator.
The impedance at a distance $\lambda_0/2$ of an open termination of a transmission line is $Z(\omega) = -iZ_c\cot(\pi\omega/\omega_c)$.
A partial fraction expansion leads to the same equivalent circuit, but with apacitances and inductances given by
\begin{equation}
	C_m = C_r=\frac{\pi}{2\omega_c Z_c}      \ ,
\end{equation}
\begin{equation}
	L_m  = \frac{1}{(2m+1)^2}\frac{2Z_c}{\pi \omega_c}    \ .
\end{equation}
We consider the first $M$ resonators by constructing a new circuit where the waveguide resonator is replaced by $M$ LC oscillators as shown in Fig.~\ref{fig:chapter-2_circuit_notations}.

\subsection{Derivation of the total Hamiltonian}
In this section we will apply the tools of circuit quantization to obtain the Hamiltonian of the previously derived lumped-element circuit. 
Following the methodology given in Refs.~\cite{devoret1995quantum,vool2017introduction}, we start by defining a set of nodes of the circuit and their corresponding fluxes. We define the flux $\tilde{\phi}$ from the voltage $v$ of that node to ground as
\begin{equation}
  \tilde{\phi}(t) = \int_{-\infty}^t v(t')dt'\ .
\end{equation}
 As described in Fig. \ref{fig:chapter-2_circuit_notations}, the node corresponding to the superconducting island of the transmon is denoted by the subscript $a$, and we number from $0$ to $M-1$ the nodes corresponding to the fluxes from the $m$-th LC oscillator to the coupling capacitor. The Lagrangian of the system is given by
 \begin{equation}
 \begin{split}
   \mathcal{L} &= C_a\frac{\dot{\tilde{\phi}}_a^2}{2} + C_c\frac{(\dot{\tilde{\phi}}_a-\dot{\tilde{\phi}}_0)^2}{2} + \sum_{m=0}^{m<M-1} C_r\frac{(\dot{\tilde{\phi}}_m-\dot{\tilde{\phi}}_{m+1})^2}{2} + C_r\frac{(\dot{\tilde{\phi}}_{M-1})^2}{2} \\
   &+E_J\cos\bigg(\frac{\tilde{\phi}_a}{\Phi_0}\bigg) - \sum_{m=0}^{m<M-1} (2m+1)^2\frac{(\tilde{\phi}_m-\tilde{\phi}_{m+1})^2}{2L_0} - (2M-1)^2\frac{(\tilde{\phi}_{M-1})^2}{2L_0}\ ,
   \end{split}
 \end{equation}
where $\Phi_0 = \hbar/2e$ corresponds to the reduced flux quantum and is not to be confused with $\tilde{\phi}_0$. We now make the change of variables $\phi_m = \tilde{\phi}_{m} - \tilde{\phi}_{m+1}$ for $0\le m<M-1$, leaving the remaining two variables unchanged $\phi_{M-1} = \tilde{\phi}_{M-1}$ and $\phi_a = \tilde{\phi}_a$. The Lagrangian then reads
 \begin{equation}
 \begin{split}
   \mathcal{L} &= C_a\frac{\dot{\phi}_a^2}{2} + C_c\frac{\big(\dot{\phi}_a-\sum_{m=0}^{m<M}\dot{\phi}_m\big)^2}{2} + \sum_{m=0}^{m<M} C_r\frac{(\dot{\phi}_m)^2}{2}\\
   &+E_J\cos\bigg(\frac{\phi_a}{\Phi_0}\bigg) - \sum_{m=0}^{m<M} (2m+1)^2\frac{(\phi_m)^2}{2L_0}\ .
   \end{split}
 \end{equation}
Now the variables $\dot{\phi}_m$ correspond directly to the voltage difference across the capacitance of the $m$-th LC oscillator. With the objective of writing a Hamiltonian, it is useful to express the capacitive part of the Lagrangian in matrix notation
\begin{equation}
  \mathcal{L} = \frac{1}{2}\boldsymbol{\dot{\phi}}^T \boldsymbol{C} \boldsymbol{\dot{\phi}} +E_J\cos\bigg(\frac{\phi_a}{\Phi_0}\bigg) - \sum_{m=0}^{m<M} (2m+1)^2\frac{(\phi_m)^2}{2L_0}\ ,
\end{equation}
\begin{equation}
  \boldsymbol{\dot{\phi}}^T = 
  \begin{bmatrix}
 \dot{\phi}_a    & \dot{\phi}_0     &  \dot{\phi}_1  &  \dot{\phi}_2    & \cdots & \dot{\phi}_{M-1}
\end{bmatrix}\ ,
\end{equation}

\begin{equation}
  \boldsymbol{C} = 
\begin{bmatrix}
 C_a+C_c     & -C_c    &  -C_c     & -C_c      & \cdots & -C_c \\
 -C_c      &  C_r+C_c     & C_c & C_c &  \cdots & C_c  \\
-C_c      &  C_c     & C_r+C_c & C_c &  & \\
 -C_c      &  C_c     & C_c & C_r+C_c &  & \\
 \vdots & \vdots &  &        &     \ddots &\\
 -C_c      &  C_c     &  &  &  & C_r+C_c\\
\end{bmatrix}\ .
\end{equation}
The canonical momenta (dimensionally charges) are equal to
\begin{equation}
  q_i = \frac{\partial \mathcal{L} }{\partial \dot{\phi}_i} = \sum_j C_{ij}\dot{\phi}_j\ .
\end{equation}
The Hamiltonian $H = \sum_i q_i\dot{\phi}_i - \mathcal{L}$ is then given by
\begin{equation}
  H = \frac{1}{2}\boldsymbol{q}^T\boldsymbol{C}^{-1}\boldsymbol{q}  -E_J\cos(\frac{\phi_a}{\Phi_0}) + \sum_{m=0}^{m<M} (2m+1)^2\frac{(\phi_m)^2}{2L_0}\ ,
\end{equation}
\begin{equation}
  \boldsymbol{q}^T = 
  \begin{bmatrix}
 q_B    & q_0     &  q_1  &  q_2    & \cdots & q_{M-1}
\end{bmatrix}
\end{equation}
and the inverse of the capacitance matrix is
\begin{equation}
\begin{split}
  \boldsymbol{C}^{-1} &= \frac{1}{C_r(MC_cC_a+C_r(C_c+C_a))}\\
  &\times
\begin{bmatrix}
 C_r^2+MC_rC_c    & C_rC_c     &  C_rC_c       & \cdots  \\
C_rC_c    &  C_r(C_c+C_a)+(M-1)C_cC_a    & -C_aC_c  &   \cdots   \\
C_rC_c    &  -C_aC_c     & C_r(C_c+C_a)+(M-1)C_cC_a &   \\
 \vdots & \vdots &  &           \ddots \\
\end{bmatrix}\ .
\end{split}
\end{equation}
It is easy to check this result in a very general way by veryfing that $\boldsymbol{CC}^{-1} = \boldsymbol{C}^{-1}\boldsymbol{C} = \boldsymbol{I}$. We now quantize the canonical variables $q_i\rightarrow\hat{q_i}$, $\phi_i\rightarrow\hat{\phi_i}$, postulating the commutation relation

\begin{equation}
  [\hat{\phi}_i,\hat{q}_j] = i\hbar\delta_{ij}\ .
\end{equation}
This results in the Hamiltonian
\begin{equation}
  \hat{H}^{(M)} = \hat{H}^{(M)}_a + \hat{H}^{(M)}_{\text{cav}} + \hat{H}^{(M)}_{\text{int}}\ ,
\end{equation}
where we have divided the Hamiltonian into the atomic term, the resonator, or cavity term, and the interaction term.
The Hamiltonian, and many of its constitutive quantities will depend on the number of modes $M$ considered in the model.
However, to avoid cumbersome notation, we will no longer explicitly point this out as it is done above.

\subsection{The resonator Hamiltonian}
The resonator Hamiltonian is
\begin{equation}
   \hat{H}_{\text{cav}} = \sum_{m =0}^{m<M}\frac{1}{2\tilde C_r}\hat{q}_m^2  + \sum_{m=0}^{m<M} (2m+1)^2\frac{(\phi_m)^2}{2L_0}\ ,
\end{equation}
where the effective capacitance of each oscillator is given by
\begin{equation}
  \tilde C_r = \frac{C_r\left(MC_cC_a+C_r(C_c+C_a)\right)}{(M-1)C_cC_a+C_r(C_c+C_a)}\ .
\end{equation}
We define the creation and annihilation operators
\begin{equation}
\begin{split}
  \hat{\phi}_m &= \phi_{\text{zpf},m}(\hat{b}_m+\hat{b}_m^\dagger),\ \phi_{\text{zpf},m} = \sqrt{\frac{\hbar}{2}Z_m},\ Z_m=\sqrt{\frac{L_m}{\tilde C_r}},\\
  \hat{q}_m &= -iq_{\text{zpf},m}(\hat{b}_m-\hat{b}_m^\dagger),\ q_{\text{zpf},m} = \sqrt{\frac{\hbar}{2}\frac{1}{Z_m}}
\end{split}
\end{equation}

reducing the cavity Hamiltonian to 
\begin{equation}
   \hat{H}_{\text{cav}} = \sum_{m =0}^{m<M}\hbar\omega_m\hat{b}^\dagger\hat{b}\ ,
\end{equation}
\begin{equation}
   \omega_m = \frac{2m+1}{\sqrt{L_0\tilde C_r}}\ ,
\end{equation}
where we have dropped the constant energy contributions $\hbar\omega_m/2$.

\subsection{The transmon Hamiltonian}
The transmon Hamiltonian is
\begin{equation}
  \hat{H}_a = \frac{1}{2\tilde C_a}\hat{q}_B^2 -E_J\cos\bigg(\frac{\hat\phi_a}{\Phi_0}\bigg)\ ,
\end{equation}
where the atoms capacitance is given by
\begin{equation}
  \tilde C_a = \frac{C_r(MC_cC_a+C_r(C_c+C_a))}{C_r^2+MC_rC_c}\ .
\end{equation}
\subsubsection{Harmonic oscillator basis}
We introduce creation and annihilation operators for the transmon
\begin{equation}
  \hat{\phi}_a = \phi_\text{zpf,a}(\hat{a}+\hat{a}^\dagger),\ \phi_\text{zpf,a}=\sqrt{\frac{\hbar}{2}Z_a},\ Z_a=\sqrt{\frac{L_a}{\tilde C_{\text{tr}}}},
\end{equation}
\begin{equation}
  \hat{q}_a = -iq_\text{zpf,a}(\hat{a}-\hat{a}^\dagger),\ q_\text{zpf,a} = \sqrt{\frac{\hbar}{2}\frac{1}{Z_a}}\ .
\end{equation}
By imposing the transmon approximation 
\begin{equation}
	\phi_\text{zpf,a}\ll\Phi_0\ ,
\end{equation}
such that when the number of excitations in the transmon is small (\text{i.e.} the expectation value of powers of $\hat{b}_m+\hat{b}_m^\dagger$ does not make the argument of the cosine large) we can Taylor expand the cosine potential of the junction to obtain
\begin{equation}
  \hat{H}_a = \frac{1}{2\tilde C_a}\hat{q}_B^2 +\frac{E_J}{2\Phi_0^2}\phi_a^2 - \frac{E_J}{24}\frac{\phi_a^4}{\Phi_0^4}\ .
  \label{eq:original_transmonCPBHamiltonian}
\end{equation}
where the Josephson inductance is given by 
\begin{equation}
	L_J = \Phi_0^2/E_J\ .
\end{equation}
The transmon Hamiltonian finally reduces to 
\begin{equation}
  \hat{H}_a = \hbar\bar\omega_a\hat{a}^\dagger\hat{a}- \frac{E_J}{24}\left(\varphi_\text{zpf,a}\left(\hat{a}+\hat{a}^\dagger\right)\right)^4\ ,
\end{equation}
where the transmon frequency is $\bar\omega_a = 1/\sqrt{L_J\tilde C_a}$, and the zero-point fluctuations in phase across the junction is $\varphi_\text{zpf,a} = \varphi_\text{zpf,a}/\Phi_0$.
%

\subsubsection{Charge basis}
Alternatively, we can write the Hamiltonian in the charge basis.
Usually, the charge is expressed in number of Cooper pairs $\hat{q}_B = 2e\hat{N}$ and the charging energy $E_C = e^2/2\tilde C_{\text{tr}}$ is introduced.
In the basis where $\ket{N}$ is an eigenstate of $\hat{N}$, the Josephson energy has a practical form~\cite{schuster2007circuit}, yielding the Hamiltonian
\begin{equation}
  \hat{H}_a = 4E_C\hat{N}^2 -\frac{E_J}{2}\sum_{N=-\infty}^{+\infty} \ket{N}\bra{N+1} + \ket{N+1}\bra{N}\ .
\end{equation}
The transmon approximation of small flux has not been made here, and in regimes of large flux through the junction, the physics of the system critically depends on the equilibrium charge $N_\text{env}$ on the capacitor imposed by the electric environment of the system.
Taking this into account leads to the Cooper-pair box Hamiltonian
\begin{equation}
  \hat{H}_a = 4E_C(\hat{N}-N_\text{env})^2 -\frac{E_J}{2}\sum_{N=-\infty}^{+\infty} \ket{N}\bra{N+1} + \ket{N+1}\bra{N}\ .
  \label{eq:CPB_Hamiltonian}
\end{equation}

\subsection{Interaction}
The interaction term $\hat{H}_{\text{int}}$ is given by
\begin{equation}
  \hat{H}_{\text{int}} = -\sum_{m=0}^{m<M}\sum_{m'=m+1}^{m'<M}\hbar G_{m,m'}(\hat{b}_m-\hat{b}_m^\dagger)(\hat{b}_{m'}-\hat{b}_{m'}^\dagger)
  - \sum_{m=0}^{m<M}\hbar g_{m}(\hat{a}-\hat{a}^\dagger)(\hat{b}_m-\hat{b}_m^\dagger)\ ,
\end{equation}
Here $G_{m,m'}$ quantifies the coupling between the $m$-th and $m'$-th modes of the resonator through the presence of the capacitances introduced by the transmon
\begin{equation}
  \hbar G_{m,m'} = -\frac{C_cC_a}{MC_rC_cC_a+C_r^2(C_c+C_a)}q_{\text{zpf},m}q_{\text{zpf},m'}\ ,
\end{equation}
and 
\begin{equation}
  \hbar\bar{g}_{m}= q_{\text{zpf},m} q_{\text{zpf},a}\tilde C_c /C_aC_r
\end{equation} 
quantifies the coupling between the $m$-th mode of the resonator and the transmon. 
The effective coupling capacitance is given by
\begin{equation}
  \tilde C_c = \frac{C_aC_rC_c}{MC_cC_a+C_r(C_c+C_a)}\ .
\end{equation}
We can also write the Hamiltonian in the basis of eigenstates of the transmon Hamiltonian. Defining the eigenstates $\{\ket{i}\}$ and eigenvalues $E_i$ by $\hat{H}_a\ket{i} = E_i\ket{i}$ and making the transformation $\hat{H} \rightarrow \sum_i\ket{i}\bra{i}\hat{H}\sum_j\ket{j}\bra{j}$ we obtain

\begin{equation}
\begin{split}
  \hat{H} &= \sum_{m=0}^{m<M}\hbar\omega_m\hat{b}_m^\dagger \hat{b}_m \\
  &+ \sum_i E_i \ket{i}\bra{i} \\
  &+  \sum_{i,j}\sum_{m=0}^{m<M}\hbar g_{m,i,j}\ket{i}\bra{j}(\hat{b}_m+\hat{b}_m^\dagger)
   \\
  &+ \sum_{m=0}^{m<M}\sum_{m'=m+1}^{m<M}\hbar G_{m,m'}(\hat{b}_m+\hat{b}_m^\dagger)(a_{m'}+a_{m'}^\dagger)
  \end{split}
  \label{eq:circuit_Hamiltonian_eigenstate_basis}
\end{equation}
where the coupling $g_{m,i,j}$ is given by
\begin{equation}
\begin{split}
  \hbar g_{m,i,j} &= q_{\text{zpf},m}2e\tilde C_c /C_aC_r\bra{i}\hat{N}\ket{j}\ \text{or}\\
  &= q_{\text{zpf},m}q_{\text{zpf},a}\tilde C_c /C_aC_r\bra{i}(\hat{a}-\hat{a}^\dagger)\ket{j}
\end{split}
\end{equation}

\subsection{Single resonator mode}

In this section, we simplify the circuit QED Hamiltonian to the case with only a single resonator mode ($M=1$).
This is the limit we will work in until the end of the chapter and in most of the experimental settings discussed in following chapters.
Note that the multi-mode case is however essential to the correct modeling of the experiment of Sec.~\ref{sec:MMUSC}.
Additionally the dependence of circuit parameters on the number of modes $M$ has some important consequences discussed further in Ref.~\cite{gely2017convergence} as well as in Sections.~\ref{sec:outlook_renormalization} and \ref{sec:outlook_atom_renormalization}.
When we only consider a single mode of the resonator, we will be replacing the mode indexing $m=0$ with $r$, for example with the mode frequency $\omega_{m=0}\rightarrow\omega_r$.
The only interaction that remains is that of the transmon and single resonator mode
\begin{align}
\begin{split}
	\hat{H} &= (\hbar\omega_a+E_C) \hat{a}^\dagger\hat{a}-\frac{E_C}{12} \left(\hat{a}+\hat{a}^\dagger\right)^4\\
	&+\hbar\omega_r \hat{b}^\dagger\hat{b}+\hbar g \left(\hat{a}-\hat{a}^\dagger\right)\left(\hat{b}-\hat{b}^\dagger\right)\ .
	\label{eq:Hamiltonian_a_b_int_anh}
\end{split}
\end{align}
Here $\hat{b}$ is the annihilation operator for photons in the resonator, $\omega_r$ its frequency and $g$ the coupling strength.
Compared to the Hamiltonian of Eq.~(\ref{eq:Hamiltonian_a}), we replaced the frequency $\bar \omega_a$ with $\omega_a+E_C/\hbar$.
Doing so will ensure that $\omega_a = \bar \omega_a-E_C/\hbar$ corresponds to the frequency of the first atomic transition, independent of the anharmonicity $E_C$.
In this single-resonator case, the following form for the coupling is the most useful:
\begin{equation}
\begin{split}
	g&=\frac{1}{2}\sqrt{\frac{\bar \omega_a\omega_r}{(1+\frac{C_a}{C_c})(1+\frac{C_r}{C_c})}}\ \text{for }C_c\gg C_a,C_r\ ,\\
  &<\frac{1}{2}\sqrt{\bar \omega_a\omega_r}\ .
  \label{eq:chapter_2_coupling_large_Cc}
\end{split}
\end{equation}

\section{Physics of the dispersive regime}

In this section, we will present the physics emerging from the Hamiltonian of a transmon interacting with a single resonator mode in the dispersive regime.
The dispersive regime occurs when the transmon and resonator frequencies are far apart relative to the coupling rate.
In order to attain an understanding of the physics at play, we will first focus on an intuitive explanation of how anharmonicity emerges in the transmon circuit.
Using these concepts, we will then explore the physics of cross-Kerr interaction occurring when the transmon is dispersively coupled to a resonator.
The formulae derived here will be key to determining the requirements to implement RFcQED.

The derivations presented here form a part of the publication
M. F. Gely, G. A. Steele and D. Bothner, \textit{Nature of the Lamb shift in weakly anharmonic atoms: From normal-mode splitting to quantum fluctuations}, {\href{https://journals.aps.org/pra/abstract/10.1103/PhysRevA.98.053808}{Phys. Rev. A \textbf{98}, 053808 (2018)}}.
This project has received funding from the European Research Council (ERC) under the European Union’s Horizon 2020 research and innovation program (Grant Agreement No. 681476 - QOMD), and from the Netherlands Organisation for Scientific Research (NWO) in the Innovational Research Incentives Scheme – VIDI, project 680-47-526.

\subsection{Origin of anharmonicity for the transmon}

In this section, we provide an intuitive explanation of how anharmonicity arises in the transmon circuit.

\subsubsection{Intuition behind transmon physics with currents}

\begin{figure*}[h!]
\centering
\includegraphics[width=0.8\textwidth]{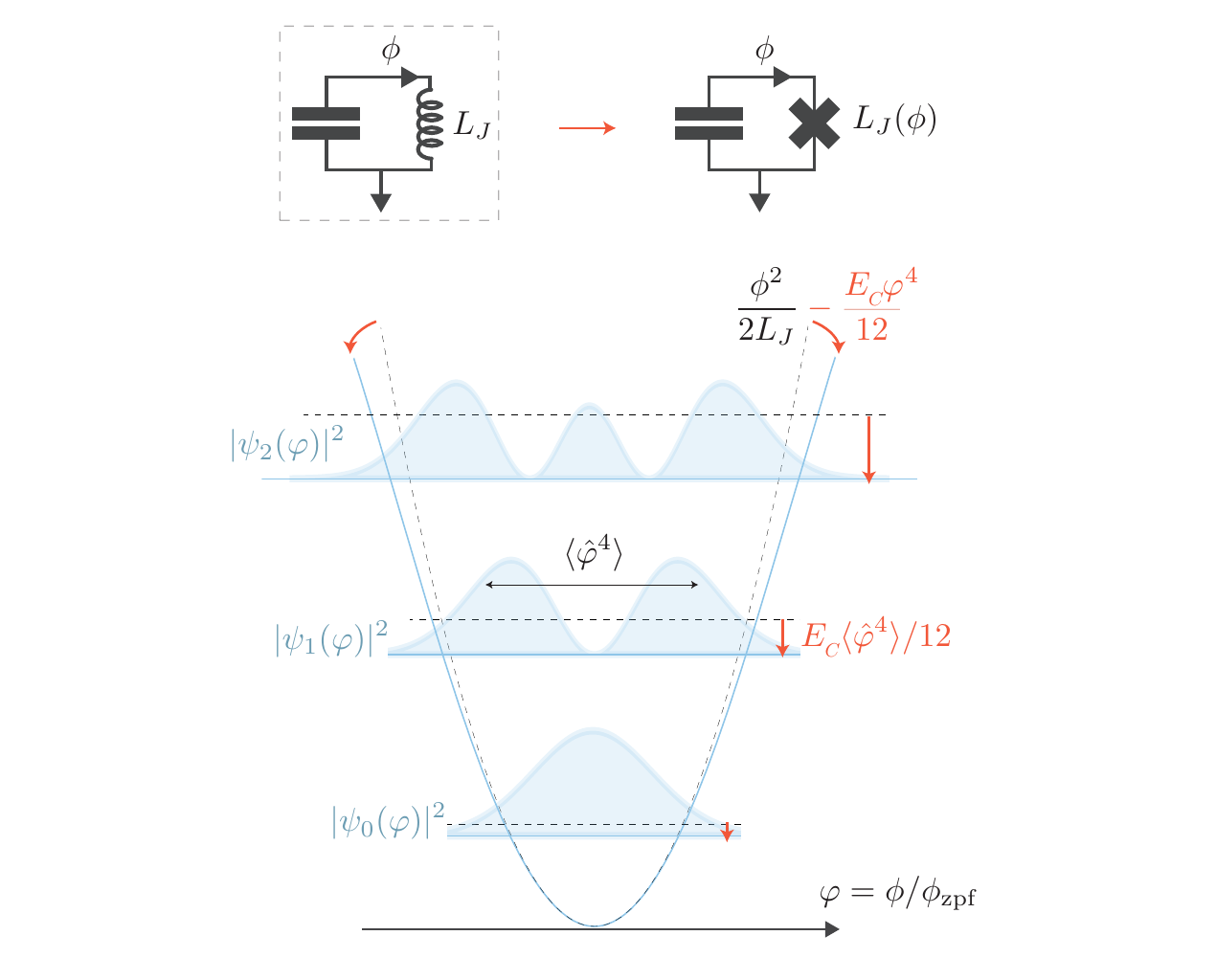}
\caption{
Replacing the linear inductance of an $LC$ oscillator with a Josephson junction (JJ) results in a weakly-anharmonic artificial atom.
To first order, the energy level $n$ is shifted proportionally to $\bra{n}\hat{\phi}^4\ket{n}$, where $\ket{n}$ is a Fock state of the harmonic system.
}
\label{fig:fig1}
\end{figure*}

The intuitive picture behind the anharmonicity of the transmon is to think of it as an $LC$ oscillator where the inductor is replaced by the non-linear inductance $L_J\left(I\right)$ of a Josephson junction (JJ).
The transmon is weakly-anharmonic if its zero-point fluctuations in current are much smaller than the junctions critical current $I_c$.
The current $I$ traversing the JJ when only a few excitations populate the circuit is then much smaller than $I_c$ and $L_J\left(I\right)\simeq L_J\left(1+I^2/2I_c^2\right)$.
Intuitively, the expectation value of the current squared $\langle I^2 \rangle$, on which the inductance depends, will increase with the number of excitations in the circuit.
So with increasing number of excitations $n$ in the circuit, the effective inductance of the circuit increases and the energy of each photon number state $E_n$ will tend to decrease with respect to the harmonic case.

\subsubsection{Intuition behind transmon physics with flux}

For a rigorous quantum description of the system, the flux $\phi(t) = \int^t_{-\infty}V(t')dt'$, where $V$ is the voltage across the JJ, is a more practical variable to use than current as we saw in the previous derivations.
The transmon Hamiltonian from Eq.~(~\ref{eq:Hamiltonian_a_b_int_anh}) is
\begin{equation}
  \hat{H}_a = \underbrace{(\hbar\omega_a+E_C) (\hat{a}^\dagger\hat{a}+\frac{1}{2})}_{\hat{H}_\text{HO}}\underbrace{-\frac{E_C}{12} \left(\hat{a}+\hat{a}^\dagger\right)^4}_{\hat{H}_\text{anh}}\ .
  \label{eq:Hamiltonian_a}
\end{equation}
Recall that the flux relates to the annihilation operator through $\hat{\phi} = \phi_\text{zpf}(\hat{a}+\hat{a}^\dagger)$.
We can recover the intuition gained by describing the system with currents, by plotting the eigen-states in the normalized flux basis $\hat{\varphi} = \hat{\phi}/\phi_\text{zpf}$ of the harmonic oscillator in Fig.~\ref{fig:fig1}. 
The fluctuations in flux increases with the excitation number, hence the expectation value of the fourth-power of the flux $\langle \hat{\phi}^4\rangle\propto\langle \hat{H}_\text{anh}\rangle$ will increase.
The energy of each eigen-state will then decrease, deviating from a harmonic level structure.

\subsubsection{Perturbative treatment of the transmon non-linearity}

The deviation from a harmonic energy level structure in the transmon can be derived from perturbation theory.
Corrections to the eigen-energies of $\hat{H}_\text{HO}$ due to anharmonicity are to first order equal to $-(E_C/12) \bra{n}(\hat{a}+\hat{a}^\dagger)^4\ket{n}$, with $\ket{n}$ a number state.
We can expand $(\hat{a}+\hat{a}^\dagger)^4$ and only consider terms that preserve the number of excitations $n$, since only they will give a non-zero contribution to the first-order correction
\begin{equation}
\hat{H}_\text{anh} \simeq -\frac{E_C}{2}\left(\left(\hat{a}^\dagger\hat{a}\right)^2 +\hat{a}^\dagger\hat{a}+ \frac{1}{2}\right)\ ,
\label{eq:levels}
\end{equation}
leading to energy levels
\begin{equation}
E_n \simeq \hbar\omega_a\left(n+\frac{1}{2}\right)-E_C\left(\frac{n^2}{2}-\frac{n}{2}-\frac{1}{4}\right)\ .
\label{eq:levels}
\end{equation}
If we write the transition frequencies of the atom $E_{n+1}-E_n=\hbar\omega_a-n E_C$, the weakly-anharmonic level structure shown in Fig.~\ref{fig:fig1} becomes apparent.

\subsection{Approximate formulae for the cross-Kerr coupling}

We now study the effect of coupling a LC oscillator to the transmon. 
As derived in the previous section (see Eq.~(\ref{eq:Hamiltonian_a_b_int_anh})), the relevant Hamiltonian is
\begin{equation}
\begin{split}
  \hat{H} &= (\hbar\omega_a+ E_C) \hat{a}^\dagger\hat{a}-\frac{ E_C}{12} \left(\hat{a}+\hat{a}^\dagger\right)^4\\
  &+\hbar\omega_r \hat{b}^\dagger\hat{b}-\hbar g \left(\hat{a}-\hat{a}^\dagger\right)\left(\hat{b}-\hat{b}^\dagger\right)\ .
\end{split}
\end{equation}
Here $\hat{b}$ is the annihilation operator for photons in the resonator, $\omega_r$ its frequency and $g$ the coupling strength.
We omitted the ground-state energies $\hbar\omega_r/2$ and $(\hbar\omega_a+ E_C)/2$ in this Hamiltonian; their presence plays no role in the results derived here.

\subsubsection{Normal-mode basis}

To describe the dispersive regime $g\ll|\Delta|$, $\Delta = \omega_r-\omega_a$, of this interaction, we first move to the normal-mode basis.
The technical details of this change of basis are described in Sec.~\ref{sec:normalmodebasisLS2}, we simply use the obtained results below.
We introduce normal-mode frequencies $\tilde{\omega}_{r}$, $\tilde{\omega}_{a} = \omega_{a}-\delta_\text{NM}$ and operators $\hat{\alpha},\hat{\beta}$ which eliminate the coupling term in Eq.~(\ref{eq:Hamiltonian_a_b_int_anh}) whilst preserving canonical commutation relations
\begin{align}
\begin{split}
  \hat{H}&=(\hbar\tilde{\omega}_a+ E_C) \hat{\alpha}^\dagger\hat{\alpha}+\hbar\tilde{\omega}_r \hat{\beta}^\dagger\hat{\beta}\\ 
  &\underbrace{- \frac{1}{12}\left( A_a^{1/4}\left(\hat{\alpha} + \hat{\alpha}^\dagger\right)+ A_r^{1/4}\left(\hat{\beta}+\hat{\beta}^\dagger\right)\right)^4}_{\hat{H}_\text{anh}}\ .
  \label{eq:Hamiltonian_a_b_anh}
\end{split}
\end{align}
The operators $\hat{\alpha},\hat{\beta}$ have a linear relation to $\hat{a},\hat{b}$, which determines the value of $ A_a$ and $ A_r$ (see Sec.~\ref{sec:normalmodebasisLS2}). 
Expanding the anharmonicity leads to
\begin{align}
\begin{split}
  \hat{H}_\text{anh}=&-\frac{ A_{a}}{2}\left(\left(\hat{\alpha}^\dagger\hat{\alpha}\right)^2 +\hat{\alpha}^\dagger\hat{\alpha}+ \frac{1}{2}\right)\\
  &-\frac{ A_{r}}{2}\left(\left(\hat{\beta}^\dagger\hat{\beta}\right)^2 +\hat{\beta}^\dagger\hat{\beta}+ \frac{1}{2}\right)\\
  &-\chi\left(\hat{\alpha}^\dagger\hat{\alpha}+\frac{1}{2}\right)\left(\hat{\beta}^\dagger\hat{\beta}+\frac{1}{2}\right)\ ,\\
  \label{eq:shifts}
\end{split}
\end{align}
if we neglect terms which do not preserve excitation number, irrelevant to first order in $ E_C$. This approximation is valid for $ E_C/\hbar \ll |\Delta|,|3\omega_a-\omega_r|,|\omega_a-3\omega_r|$. 
The anharmonicity (or self-Kerr) of the normal-mode-splitted atom and resonator $ A_a$ and $ A_r$ is related to the cross-Kerr interaction (or AC Stark shift) $\chi$ through
\begin{equation}
  \chi=2\sqrt{ A_{a} A_{r}}\ .
\end{equation}
The cross-Kerr interaction is the change in frequency one mode acquires as a function of the number of excitations in the other.

\subsubsection{Intuitive explanation}

\begin{figure*}[h!]
\centering
\includegraphics[width=0.8\textwidth]{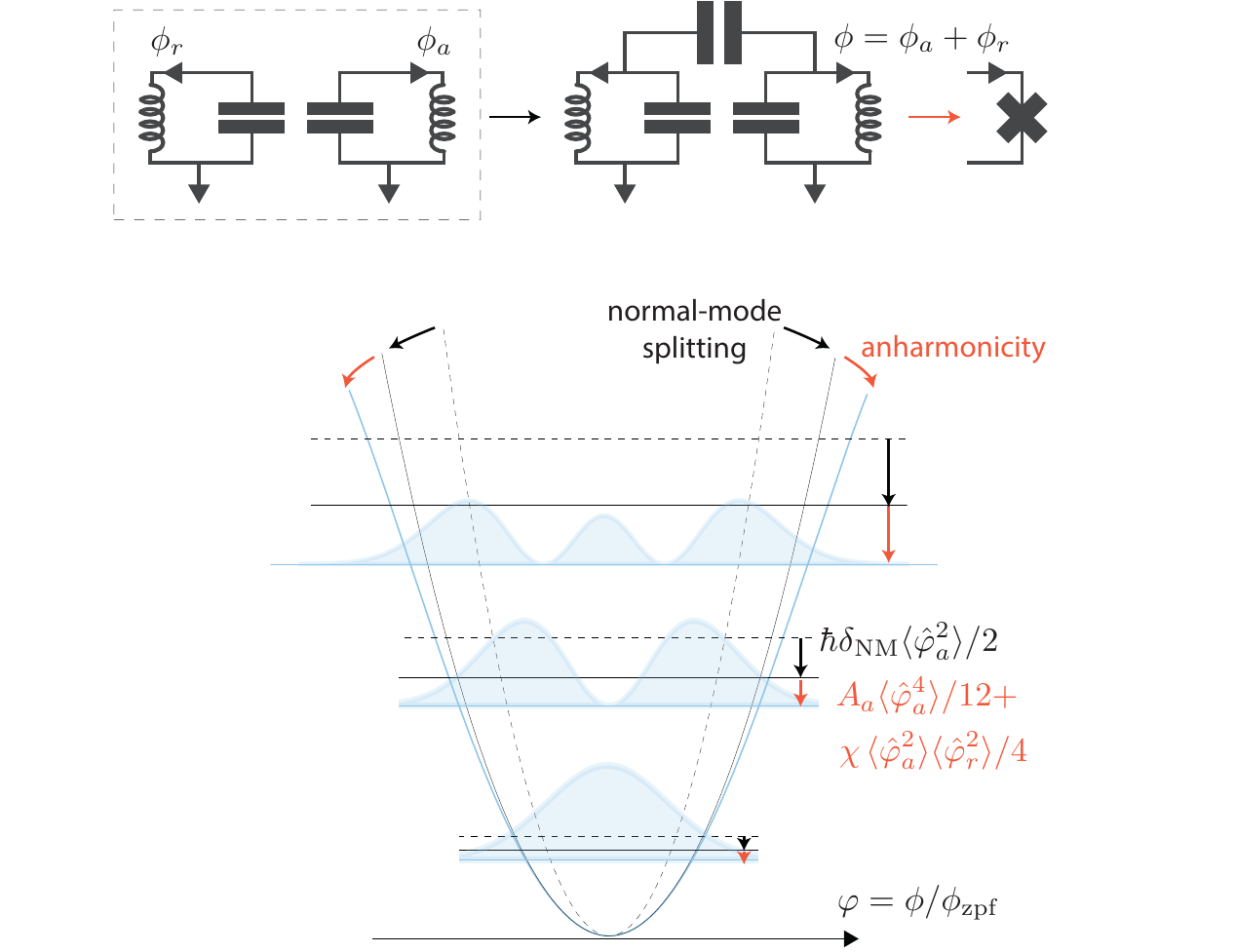}
\caption{Two coupled harmonic oscillators undergo normal-mode splitting, resulting in a frequency shift $\delta_\text{NM}$.
In the normal mode basis, the flux traversing one of the inductances $\phi$, is composed of the flux from both normal mode oscillations $\phi = \phi_a+\phi_r$.
Replacing an inductor with a JJ leads to the same shift as in the isolated atom $ A_a\langle\hat{\phi}_a^4\rangle$, but also to a shift due to quantum fluctuations of the coupled oscillator $\chi\langle\hat{\phi}_a^2\rangle \langle\hat{\phi}_r^2\rangle/4$.}
\label{fig:fig1_2}
\end{figure*}

The appearance of a cross-Kerr interaction and the resonators anharmonicity can be understood from the mechanism of normal-mode splitting. 
When the transmon and $LC$ oscillator dispersively couple, the normal-mode corresponding to the $LC$ oscillator will be composed of currents oscillating through its inductor but also partly through the JJ.
We can decompose the current $I$ traversing the JJ into the current corresponding to atomic excitations $I_a$ and resonator excitations $I_r$.
In Eq.~(\ref{eq:Hamiltonian_a_b_anh}), this appears in the terms of flux as $\phi=\phi_a+\phi_r\propto  A_a^{1/4}(\hat{\alpha} + \hat{\alpha}^\dagger)+ A_r^{1/4}(\hat{\beta}+\hat{\beta}^\dagger)$.
Consequently the value of the JJ inductance is not only dependent on the number of excitations in the atom but also in the resonator.
Since the frequency of the normal-mode-splitted transmon and resonator depends on the value of this inductance, the atomic frequency is a function of the number of excitations in the resonator (cross-Kerr interaction), and the resonator frequency changes as it is excited (the resonator acquires some anharmonicity).
Even when the resonator mode is in its ground state, vacuum current fluctuations shift the atomic frequency.
This can be verified by the presence of $1/2$ in the cross-Kerr term of Eq.~(\ref{eq:shifts}) which arise from commutation relations $[\hat{\alpha},\hat{\alpha}^\dagger]=[\hat{\beta},\hat{\beta}^\dagger]=1$, mathematically at the origin of vacuum fluctuations.

To summarize, compared to an isolated harmonic oscillator the energy levels of the coupled atom are shifted by: \textit{(1)} normal-mode splitting $\delta_\text{NM}$, \textit{(2)} its anharmonicity $ A_a$ which arises from the quantum fluctuations of its eigen-states, and \textit{(3)} the shift proportional to $\chi$ arising from the quantum fluctuations of the resonator it is coupled to.
These different effects are depicted in Fig.~\ref{fig:fig1_2}.

\subsubsection{Approximate formulae}
In the RWA $g\ll|\Delta|\ll\Sigma$, where $\Sigma = \omega_a +\omega_r$, and considering the relations relating $\hat a, \hat b$ to $\hat \alpha, \hat \beta$ derived in Sec.~\ref{sec:normalmodebasisLS2}, the following approximations hold
\begin{align}
\begin{split}
  \tilde{\omega}_a &= \omega_a- \delta_{NM} \simeq \omega_a- \frac{g^2}{\Delta} -  \frac{E_C}{\hbar}\frac{g^2}{\Delta^2}\ ,\\
  \tilde{\omega}_r &\simeq \omega_r+ \frac{g^2}{\Delta} +  \frac{E_C}{\hbar}\frac{g^2}{\Delta^2}\ ,\\
  A_a &\simeq  E_C\left(1-2\frac{g^2}{\Delta^2}\right)\ ,\\
  A_r &= E_C\frac{g^4 }{\Delta^4}\ ,\\
  \chi &\simeq  2E_C\frac{g^2 }{\Delta^2} ,
  \label{eq:shifts_rwa}
\end{split}
\end{align}
valid to leading order in $g/\Delta$.
The expression for the cross-Kerr interaction was also derived by Koch \textit{et al.}~\cite{koch_charge-insensitive_2007} from perturbation theory, given in the form $ E_C g^2/\Delta(\Delta- E_C)$. 
Applying perturbation theory to the Hamiltonian of Eq.~(\ref{eq:Hamiltonian_a_b_int_anh}), however, fails to predict the correct shift beyond the RWA and does not make the distinction between the physical origin of the different shifts.
%

\begin{figure}[h!]
\centering
\includegraphics[width=0.8\textwidth]{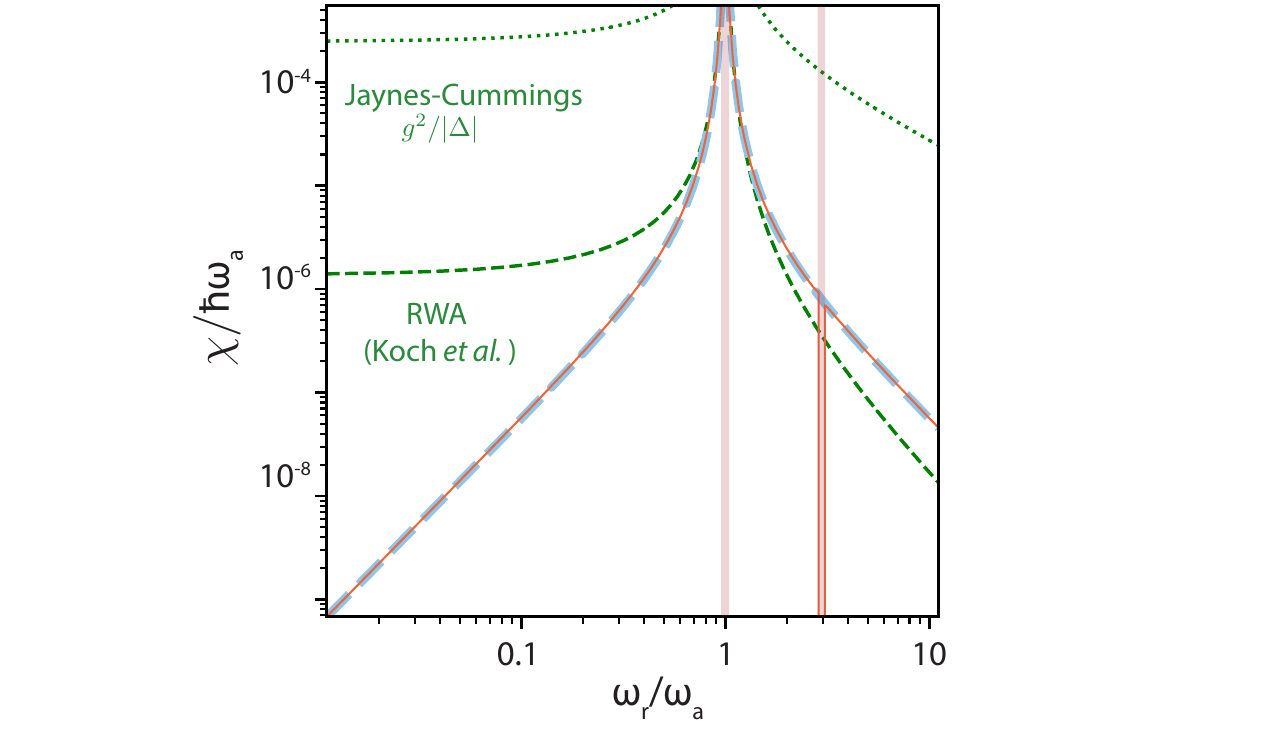}
\caption{Vacuum-fluctuations-induced shift $\chi$ beyond the RWA fixing $ E_C/\hbar\omega_a=0.01$ and $g/\omega_a=0.02$.
Numerical calculation (full red line), are compared to the analytical expression of Eq.~(\ref{eq:shifts_beyond}) (dashed blue line) and Eq.~(\ref{eq:shifts}) (dashed green).
Resonances invalidating our approximations are denoted by red bars.}
\label{fig:fig3}
\end{figure}

Beyond the RWA to regimes of large detuning $g\ll|\Delta|\sim\Sigma$ (and $g\ll\omega_a,\omega_r$) the approximate expressions of the different shifts are given by

\begin{align}
\begin{split}
  \tilde{\omega}_a &\simeq \omega_a- g^2\frac{2\omega_r}{\Delta\Sigma}- 4 \frac{E_C}{\hbar} g^2\frac{\omega_r\omega_a }{\Delta^2\Sigma^2}\ ,\\
  \tilde{\omega}_r &\simeq \omega_r+ g^2\frac{2\omega_a}{\Delta\Sigma}+ 4 \frac{E_C}{\hbar} g^2\frac{\omega_a^2 }{\Delta^2\Sigma^2}\ ,\\
  A_a &\simeq  E_C\left(1-4g^2\frac{\omega_r \left(\omega_a^2 + \omega_r^2\right)}{\omega_a \Delta^2 \Sigma^2}\right)\ ,\\
  A_r &= \mathcal{O}(\frac{g^4 }{\Delta^4})\ ,\\
  \chi &\simeq 8 E_C g^2\frac{ \omega_r^2}{\Delta^2\Sigma^2} .
  \label{eq:shifts_beyond}
\end{split}
\end{align}
An important difference with the RWA is that the cross-Kerr interaction $\chi$ scales with $\omega_r^2$, decreasing with the frequency of a coupled resonator as shown in Fig.~\ref{fig:fig3}.
This notably explains why the transmon is insensitive to low frequency charge fluctuations as compared to the highly anharmonic Cooper pair box.
It also explains why implementing radio-frequency circuit QED is so hard.
Contrary to the cross-Kerr interaction in the RWA, this expression cannot be derived by applying perturbation theory to Eq.~(\ref{eq:Hamiltonian_a_b_int_anh}).
\section{Physics of the resonant regime}

In this section, we cover the physics resulting from the coupling of a single resonator mode and a transmon when they have the same frequency.
Whilst this topic is not essential to answering the question raised in the title of this chapter (How large should the coupling be for RFcQED?), the theoretical results derived here will be used at various points in this thesis, and easily follow from the previous derivations.
The main feature of the resonant regime (when transmon and LC oscillator have the same frequency) is the hybridization of the transmon and resonator, such that the first two excited levels of the combined system become $(\ket{g,1}+\ket{e,0})/\sqrt{2}$ and $(\ket{g,1}-\ket{e,0})/\sqrt{2}$, separated in energy by $2\hbar g$.
In this case there is no longer a clear distinction between atomic and harmonic degrees of freedom.
The normal-mode basis can also be used to analyze the resonant regime in the large coupling regime $g\gg E_C$, where the non-linearity can still be treated perturbative.
In the opposite regime $g\ll E_C$, we will see why this approach no longer works, and cover a method to obtain similar physics non the less.
\subsection{Large coupling}\label{sec:chapter-2_resonant_large_coupling}
Here we study the regime $E_C\ll g\ll\omega$.
Here $\omega$ is the common frequency of both the resonator and the linear part of the transmon Hamiltonian $\omega_a+ E_C/\hbar = \omega_r = \omega$.
The anharmonicity here is assumed to be the smallest part of the Hamiltonian, and we will thus treat it as such by performing the normal-mode transformation as done in the dispersive regime.
The two normal modes, in the limit $g\ll \omega$, are characterized by
\begin{align}
\begin{split}
\hat{a} &\simeq \frac{\hat{\alpha}_-+\hat{\alpha}_+}{\sqrt{2}} - \frac{g}{2\omega}\frac{\hat{\alpha}_-^\dagger-\hat{\alpha}_+^\dagger}{\sqrt{2}}\\
\hat{b} &\simeq \frac{\hat{\alpha}_--\hat{\alpha}_+}{\sqrt{2}} - \frac{g}{2\omega}\frac{\hat{\alpha}_-^\dagger+\hat{\alpha}_+^\dagger}{\sqrt{2}}\\
\tilde{\omega}_{+} &\simeq\omega+g-\frac{g^2}{2\omega}\ ,\\
\tilde{\omega}_{-} &\simeq\omega-g-\frac{g^2}{2\omega}\ .\\
\label{eq:chapter-2_normal_mode_resonant}
\end{split}
\end{align}
These approximations are valid to first order in $g/\omega$.
We index the two normal modes by $\pm$ as no distinction can now be made as to the atomic or harmonic nature of the mode.
The Hamiltonian in the normal-mode basis is then to 0-th order in $g/\omega$ given by
\begin{equation}
  \hat H = \hbar\omega_+\hat{\alpha}_+^\dagger\hat{\alpha}_++\hbar\omega_-\hat{\alpha}_-^\dagger\hat{\alpha}_--\frac{E_c/4}{12}(\hat \alpha_+ +\hat \alpha_+^\dagger+\hat \alpha_- +\hat \alpha_-^\dagger)^4
\end{equation}
Expanding the quartic term reveals and keeping terms which will be relevant in first order perturbation theory leads to
\begin{align}
\begin{split}
  \hat{H}\simeq&\hbar\omega_+\hat{\alpha}_+^\dagger\hat{\alpha}_++\hbar\omega_-\hat{\alpha}_-^\dagger\hat{\alpha}_-\\
  &-\frac{ A_{+}}{2}\left(\left(\hat{\alpha}_+^\dagger\hat{\alpha}_+\right)^2 +\hat{\alpha}_+^\dagger\hat{\alpha}_++ \frac{1}{2}\right)\\
  &-\frac{ A_{-}}{2}\left(\left(\hat{\alpha}_-^\dagger\hat{\alpha}_-\right)^2 +\hat{\alpha}_-^\dagger\hat{\alpha}_-+ \frac{1}{2}\right)\\
  &-\chi\left(\hat{\alpha}_+^\dagger\hat{\alpha}_++\frac{1}{2}\right)\left(\hat{\alpha}_-^\dagger\hat{\alpha}_-+\frac{1}{2}\right)\ ,\\
  \label{eq:chapter-2_resonant_Hamiltonian_large_g}
\end{split}
\end{align}
with anharmonicities $A_\pm = E_C/4$ and a cross-Kerr interaction $\chi = E_C/2$.
In this approximation we neglected terms proportional to $(\hat{\alpha}_+\hat{\alpha}_-^\dagger +  \hat{\alpha}_+^\dagger\hat{\alpha}_-)$.
Note that in the RWA-applicable dispersive regime such terms were neglected as the largest contribution to the energy scales they may have is $\propto E_C^2g^2/\Delta^3$ following second-order perturbation theory.
In the resonant regime however, following second-order perturbation theory, they have a contribution $\propto E_C^2/(\omega_+-\omega_-) = E_C^2/2g$.
When the coupling becomes comparable to the anharmonicity, the approximations we made after expanding the quartic term are no longer valid, this case is treated in the following section.
\subsection{Small coupling}\label{sec:chapter-2_resonant_small_coupling}
Here we study the regime $g\ll E_C\ll\omega$.
Since the anharmonicity dominates over the coupling, the latter should be treated as a perturbation, we will see why a better definition of the resonant condition is then $\omega_a = \omega_r = \omega$.
We first expand the quartic non-linearity of the transmon, and neglect terms which vanish following first order perturbation theory, to obtain the Hamiltonian
\begin{equation}
\begin{split}
  \hat{H} &= \hbar\omega \hat{a}^\dagger\hat{a}-\frac{E_C}{2} \hat{a}^{\dagger2}\hat{a}^2\\
  &+\hbar\omega \hat{b}^\dagger\hat{b}+\hbar g \left(\hat{a}-\hat{a}^\dagger\right)\left(\hat{b}-\hat{b}^\dagger\right)\ .
\end{split}
\end{equation}
Let's now analyze which coupling terms are relevant here using first order perturbation theory.
The impact of the terms $\hbar g(\hat{a}\hat{b} + \hat{a}^\dagger\hat{b}^\dagger)$ is $\propto g^2/2\omega$ which is assumed to be very small here.
The other terms $\hbar g(\hat{a}\hat{b}^\dagger + \hat{a}^\dagger\hat{b})$ are divided into two categories by writing $\hat{a}$ in terms of transmon excitations $\ket{i}$.
These have relative energies $E_{i+1}-E_i=\hbar\omega-i E_C$ given by the transmon part of the Hamiltonian $\hbar\omega \hat{a}^\dagger\hat{a}-\frac{E_C}{2} \hat{a}^{\dagger2}\hat{a}^2$.
The terms $\sum_{i\le 1}\hbar \sqrt{i+1}g(\ket{i}\bra{i+1}\hat{b}^\dagger + \ket{i+1}\bra{i}\hat{b})$ couple first and higher excited states of the transmon $\ket{i}$, $i\ge 1$.
The impact of these coupling terms is $\propto g^2/(\omega - (E_{i+1}-E_i)/\hbar) \propto g^2/ E_C$ which is also assumed to be small here.
The remaining terms which couple the ground and first excited level of the transmon to the resonator $\hbar g(\ket{g}\bra{e}\hat{b}^\dagger + \ket{e}\bra{g}\hat{b})$ cannot be treated perturbatively as they couple resonant levels.
Since this is the only relevant interaction, it makes sense to truncate the Hilbert space of the transmon to its two first levels, resulting in the Hamiltonian
\begin{equation}
  \hat{H} = \hbar\omega \ket{e}\bra{e} +\hbar\omega \hat{b}^\dagger\hat{b}+ \hbar g(\ket{g}\bra{e}\hat{b}^\dagger + \ket{e}\bra{g}\hat{b})\ 
\label{eq:chapter-2_jaynes_cummings}
\end{equation}
Also known as the Jaynes-Cummings Hamiltonian.
This Hamiltonian is exactly solvable~\cite{jaynes_comparison_1963,bishop2010circuit} and in this resonant condition has eigenstates
\begin{equation}
  \ket{n,\pm} = \frac{\ket{g,n}\pm\ket{e,n-1}}{\sqrt{2}}
\end{equation}
with energies
\begin{equation}
  \hbar\omega_{n,\pm} =  n\hbar\omega\pm\sqrt{n}\hbar g\ .
\end{equation}
As in the large coupling limit, the two first excited states of the system are separated in energy by $2\hbar g$.

\FloatBarrier
\section{Details of the basis change (Bogoliubov transformation)}\label{sec:normalmodebasisLS2}

In this section we cover the details of the change of basis, or Bogoliubov transformation, that is used in analyzing the physics of both the dispersive and resonant regime.
Considering the Hamiltonian of two coupled harmonic modes, we derive the change of variables, which allow us to write the Hamiltonian as two independent normal-modes.

The Hamiltonian 
\begin{equation}
    \hat{H}=(\hbar\omega_a+ E_C) \hat{a}^\dagger\hat{a}+\hbar\omega_r \hat{b}^\dagger\hat{b}+\hbar g (\hat{a}-\hat{a}^\dagger)(\hat{b}-\hat{b}^\dagger)
    \label{eq:ham_no_bogo}
\end{equation}
describes two harmonic oscillators with a linear interaction between them.
It corresponds to the Hamiltonian of Eq.~(\ref{eq:Hamiltonian_a_b_int_anh}), stripped of the anharmonicity term. 
It can be compactly written as
\begin{align}
\begin{split}
  \hat{H}&= \boldsymbol{v} ^T \boldsymbol{H} \boldsymbol{v}\ ,\\
\boldsymbol{v} ^T&= [\hat{a},\hat{b},\hat{a}^\dagger,\hat{b}^\dagger]\ ,\\
  \boldsymbol{H}= \frac{1}{2}&\begin{bmatrix}
 0 & \hbar g  & (\hbar\omega_a+ E_C) &-\hbar g     \\
 \hbar g & 0 &  -\hbar g &   \hbar\omega_r \\
 (\hbar\omega_a+ E_C) & -\hbar g  & 0 &\hbar g     \\
 -\hbar g & \hbar\omega_r & \hbar g &   0 
\end{bmatrix}\ ,
    \label{eq:ham_no_bogo_matrix}
\end{split}
\end{align}
omitting constant contributions. Using this notation, the canonical commutation relations read
\begin{align}
\begin{split}
    [\boldsymbol{v},\boldsymbol{v} ^T]=\boldsymbol{v}\boldsymbol{v}^T-(\boldsymbol{v}\boldsymbol{v} ^T)^T=\boldsymbol{J}= \begin{bmatrix}
 0 & \boldsymbol{I}_2    \\
 -\boldsymbol{I}_2  & 0 \\\end{bmatrix}\ ,
\label{eq:commutation}
\end{split}
\end{align}
where $\boldsymbol{I}_2$ is the $2\times2$ identity matrix.
The objective of this section is to rewrite (\ref{eq:ham_no_bogo}) as the Hamiltonian of two independent harmonic oscillators, or normal-modes
\begin{equation}
    \hat{H}=(\hbar\tilde{\omega}_a+ E_C) \hat{\alpha}^\dagger\hat{\alpha}+\hbar\tilde{\omega}_r \hat{\beta}^\dagger\hat{\beta}\ ,
    \label{eq:ham_after_bogo}
\end{equation}
which we write in compact notation as
\begin{align}
\begin{split}
    \hat{H}&=\boldsymbol{\eta} ^T \boldsymbol{\Lambda} \boldsymbol{\eta}\ ,\\
    \boldsymbol{\eta} ^T&= [\hat{\alpha},\hat{\beta},\hat{\alpha}^\dagger,\hat{\beta}^\dagger]\\
    \boldsymbol{\Lambda}&=\frac{1}{2}\begin{bmatrix}
 0 & 0&(\hbar\tilde{\omega}_a+ E_C)&0   \\
 0&0&0  & \hbar\tilde{\omega}_r  \\
(\hbar\tilde{\omega}_a+ E_C)&0&0  & 0 \\
 0&\hbar\tilde{\omega}_r &0  & 0\\\end{bmatrix}\ .
    \label{eq:ham_bogo_matrix}
\end{split}
\end{align}
To do so, we need to find a matrix which maps $\boldsymbol{v}$ to a new set of annihilation and creation operators of the normal-modes $\boldsymbol{\eta}$ which should also satisfy the commutation relations (\ref{eq:commutation}). 

We start by noticing that the matrix $\boldsymbol{\Lambda}\boldsymbol{J}$ is diagonal
\begin{equation}
    \boldsymbol{\Lambda}\boldsymbol{J} = \frac{1}{2}\begin{bmatrix}
 -(\hbar\tilde{\omega}_a+ E_C) & 0&0&0   \\
 0&-\hbar\tilde{\omega}_r&0  & 0 \\
 0&0&(\hbar\tilde{\omega}_a+ E_C)  & 0 \\
 0&0&0  & \hbar\tilde{\omega}_r \\\end{bmatrix}\ ,
\end{equation}
and we define it as the diagonal form of the matrix $\boldsymbol{H}\boldsymbol{J}$. 
In other words, we can determine the value of $\tilde{\omega}_a$ and $\tilde{\omega}_r$ by diagonalizing $\boldsymbol{H}\boldsymbol{J}$.
An exact expression for these normal-mode frequencies is given by
\begin{align}
\begin{split}
&\tilde{\omega}_{ar} = \frac{1}{\sqrt{2}}\bigg((\omega_a+ E_C/\hbar)^2 + \omega_r^2 \\
  &\pm \sqrt{ ((\omega_a+ E_C/\hbar)^2-\omega_r^2)^2 + 16 g^2 (\omega_a+ E_C/\hbar) \omega_r }\bigg)^{\frac{1}{2}}\ .
\end{split}
\label{eq:LS2_exact_mode_frequencies}
\end{align}
As we will now demonstrate, defining $\boldsymbol{\Lambda}$ in this way will lead to operators with the correct commutation relations. 
We define the matrix of eigen-vectors that diagonalizes $\boldsymbol{H}\boldsymbol{J}$ as $\boldsymbol{F} = [\boldsymbol{w}_0,\boldsymbol{w}_1,\boldsymbol{w}_2,\boldsymbol{w}_3]$, such that
\begin{equation}
    \boldsymbol{H}\boldsymbol{J} = \boldsymbol{F}\boldsymbol{\Lambda}\boldsymbol{J}\boldsymbol{F}^{-1}
    \label{eq:eigen_vectors}
\end{equation}
The matrix $\boldsymbol{F}$ can be normalized in such a way that it satisfies an important condition, it can be made symplectic
\begin{equation}
    \boldsymbol{F}^T\boldsymbol{J}\boldsymbol{F}=\boldsymbol{F}\boldsymbol{J}\boldsymbol{F}^T=\boldsymbol{J}\ .
    \label{eq:sympletic}
 \end{equation} 
If the eigenvectors are normalized such that $\boldsymbol{w}_i^T \boldsymbol{w}_i=1$, the operation that leads to symplecticity is 
 \begin{align}
\begin{split}
\boldsymbol{w}_0' &= \pm\boldsymbol{w}_0/\sqrt{|\boldsymbol{w}_0^T\boldsymbol{J}\boldsymbol{w}_2|}\ ,\\
\boldsymbol{w}_1' &= \pm\boldsymbol{w}_1/\sqrt{|\boldsymbol{w}_1^T\boldsymbol{J}\boldsymbol{w}_3|}\ ,\\
\boldsymbol{w}_2' &= \pm\boldsymbol{w}_2/\sqrt{|\boldsymbol{w}_0^T\boldsymbol{J}\boldsymbol{w}_2|}\ ,\\
\boldsymbol{w}_3' &= \pm\boldsymbol{w}_3/\sqrt{|\boldsymbol{w}_1^T\boldsymbol{J}\boldsymbol{w}_3|}\ ,
\label{eq:normalization}
\end{split}
\end{align}
where the $+$ or $-$ sign is chosen such that if we redefine $\boldsymbol{F}= [\boldsymbol{w}_0',\boldsymbol{w}_1',\boldsymbol{w}_2',\boldsymbol{w}_3']$ it is of the form 
\begin{equation}
  \boldsymbol{F}= 
\begin{bmatrix}
 \boldsymbol{A}     & \boldsymbol{B}     \\
 \boldsymbol{B}     &   \boldsymbol{A}     \\
\end{bmatrix}\ ,
  \end{equation}
and such that $\boldsymbol{F}=\boldsymbol{I}_4$ in the limit $g=0$.
With $\boldsymbol{F}$ a symplectic matrix, we can define $\boldsymbol{\eta}$ as
\begin{equation}
    \boldsymbol{\eta} = \boldsymbol{F}^T \boldsymbol{v}\ 
    \label{eq:transformation}
\end{equation} 
and (\textbf{Proposition 1}) $\boldsymbol{\eta}$ will respect the commutation relations~(\ref{eq:commutation}) whilst ensuring that (\textbf{Proposition 2}) the two Hamiltonians (\ref{eq:ham_no_bogo_matrix}) and (\ref{eq:ham_bogo_matrix}) are equivalent. Proof of these proposition is provided at the end of this section. With the relation~(\ref{eq:identity}), we can invert~(\ref{eq:transformation}) to obtain
\begin{equation}
    \boldsymbol{v} = -\boldsymbol{J}\boldsymbol{F}\boldsymbol{J} \boldsymbol{\eta}\ .
    \label{eq:transformation2}
\end{equation} 
Using the software Mathematica, we diagonalize $\boldsymbol{H}\boldsymbol{J}$ symbolically and perform the normalizations of Eqs.~(\ref{eq:normalization}) to obtain $\boldsymbol{F}$. 
As written in Eq.~\ref{eq:transformation2}, $\boldsymbol{F}$ leads to the transformation between the operators $\hat a$,$\hat b$ and $\hat\alpha$,$\hat\beta$. 
By Taylor expanding the resulting expressions for small values of $g$ ($g\ll|\Delta|,\Sigma,\omega_a,\omega_r$), we obtain
\begin{align}
\begin{split}
\hat{a} &\simeq \left( 1- g^2\frac{2  (\omega_a+ E_C/\hbar) \omega_r}{\Delta'^2 \Sigma'^2}\right) \hat{\alpha}-\frac{g}{\Delta'}\hat{\beta} \\
&-g^2\frac{\omega_r}{(\omega_a+ E_C/\hbar)}\frac{1}{\Sigma'\Delta'} \hat{\alpha}^\dagger- \frac{g}{\Sigma'}\hat{\beta}^\dagger\ ,  \\
\hat{b} &\simeq \frac{g}{\Delta'}  \hat{\alpha}+ \left(1- g^2\frac{2  (\omega_a+ E_C/\hbar) \omega_r}{\Delta'^2 \Sigma'^2}\right) \hat{\beta} \\
&- \frac{g}{\Sigma'}  \hat{\alpha}^\dagger +g^2\frac{(\omega_a+ E_C/\hbar)}{\omega_r}\frac{1}{\Sigma'\Delta'} \hat{\beta}^\dagger\ .\\
\end{split}
\end{align}
These approximations are valid to second order in $g$ and we define $\Delta'=\Delta- E_C/\hbar$ and $\Sigma'=\Sigma+ E_C/\hbar$.
Using these relations, we can express the anharmonicity $ E_C (\hat{a}+\hat{a}^\dagger)/12$ as a function of $\hat{\alpha}$ and $\hat \beta$, leading to expressions for $A_a$ and $A_r$.
In the same approximation, the eigen-frequencies write
\begin{align}
\begin{split}
\tilde{\omega}_{a} &\simeq\omega_a-\frac{2g^2\omega_r}{\Sigma'\Delta'}\ ,\\
\tilde{\omega}_{r} &\simeq\omega_r+\frac{2g^2\omega_a}{\Sigma'\Delta'}\ .\\
\end{split}
\end{align}
leading to the expression for the normal mode splitting $\delta_\text{NM}$.
Finally, we provide proofs for the two propositions used above.

\textbf{Proposition 1:} 
The vector $\boldsymbol{\eta} = \boldsymbol{F}^T \boldsymbol{v}$ satisfies the canonical commutation relations written in compact form in Eq.~\ref{eq:commutation} if these commutation relations are also satisfied by $\boldsymbol{v}$ and if $\boldsymbol{F}$ is symplectic (Eq.~(\ref{eq:sympletic})).

Proof:
\begin{align}
\begin{split}
    [\boldsymbol{\eta},\boldsymbol{\eta} ^T]&=\boldsymbol{\eta}\boldsymbol{\eta}^T-(\boldsymbol{\eta}\boldsymbol{\eta} ^T)^T\\
        &\stackrel{\text{(\ref{eq:transformation})}}{=}\boldsymbol{F}^T (\boldsymbol{v}\boldsymbol{v}^T)\boldsymbol{F} - \boldsymbol{F}^T (\boldsymbol{v}\boldsymbol{v}^T)^T\boldsymbol{F}\\
        &=\boldsymbol{F}^T [\boldsymbol{v},\boldsymbol{v} ^T]\boldsymbol{F}\\
        &\stackrel{\text{(\ref{eq:commutation})}}{=}\boldsymbol{F}^T \boldsymbol{J}\boldsymbol{F} \\
        &\stackrel{\text{(\ref{eq:sympletic})}}{=}\boldsymbol{J}\ ,
\end{split}
\end{align}

\textbf{Proposition 2: } 
The two Hamiltonians (\ref{eq:ham_no_bogo_matrix}) and (\ref{eq:ham_bogo_matrix}) are equivalent if we define $\boldsymbol{\eta} = \boldsymbol{F}^T \boldsymbol{v}$ and $\boldsymbol{F}$ is symplectic (Eq.~(\ref{eq:sympletic})).

Proof:
Multiplying Eq.~(\ref{eq:sympletic}) with $\boldsymbol{J}$, we find
 \begin{equation}
    -\boldsymbol{F}\boldsymbol{J}\boldsymbol{F}^T\boldsymbol{J}=-\boldsymbol{J}\boldsymbol{F}\boldsymbol{J}\boldsymbol{F}^T=-\boldsymbol{J}^2=\boldsymbol{I}_4\ ,
    \label{eq:identity}
 \end{equation}
 where $\boldsymbol{I}_4$ is the $4\times4$ identity matrix.
 This relation allows us to introduce the matrix $\boldsymbol{F}$ into Eq.~(\ref{eq:ham_no_bogo_matrix})
 \begin{align}
 \begin{split}
 \hat{H}/\hbar&= \boldsymbol{v} ^T \boldsymbol{H} \boldsymbol{v}\\
        &\stackrel{\text{(\ref{eq:identity})}}{=}-\boldsymbol{v} ^T  \boldsymbol{H}\boldsymbol{J}\boldsymbol{F}\boldsymbol{J}\boldsymbol{F}^T \boldsymbol{v}\\
        &\stackrel{\text{(\ref{eq:eigen_vectors})}}{=}-\boldsymbol{v} ^T \boldsymbol{F}\boldsymbol{\Lambda}\boldsymbol{J}\underbrace{\boldsymbol{F}^{-1}\boldsymbol{F}}_{=\boldsymbol{I}_4}\boldsymbol{J}\boldsymbol{F}^T \boldsymbol{v}\\
        &\stackrel{\text{(\ref{eq:eigen_vectors})}}{=}-\boldsymbol{v} ^T \boldsymbol{F}\boldsymbol{\Lambda}\underbrace{\boldsymbol{J}\boldsymbol{J}}_{=-\boldsymbol{I}_4}\boldsymbol{F}^T \boldsymbol{v}\\
        &=(\boldsymbol{F}^T \boldsymbol{v})^T\boldsymbol{\Lambda}(\boldsymbol{F}^T \boldsymbol{v})\ ,
 \end{split}
 \end{align}

\section{Requirements for strong dispersive coupling in RFcQED}

In this section we will make use of the previously derived formulae to determine requirements for radio-frequency circuit QED.
The most relevant formula is the cross-Kerr interaction, Eq.~(\ref{eq:shifts_beyond}), which in the limit $\omega_r\ll\omega_a$ simplifies to
\begin{equation}
	\chi\simeq8 E_C \frac{ g^2\omega_r^2}{\omega_a^4}\ ,
\end{equation}
and the coupling, (see Eq.~(\ref{eq:chapter_2_coupling_large_Cc})) is
\begin{equation}
	g\simeq\frac{1}{2}\sqrt{\frac{ \omega_a\omega_r}{(1+\frac{C_a}{C_c})(1+\frac{C_r}{C_c})}}
  \label{eq:chapter-2-summary_coupling_rate}
\end{equation}
where the frequencies for $C_c\gg C_a,C_r$, are given by
\begin{equation}
\begin{split}
  \omega_a &= \frac{1}{\sqrt{(C_a+C_r)L_J}}\\
  \omega_r &= \frac{1}{\sqrt{(C_a+C_r)L_r}}\\
\end{split}
\end{equation}

We shall now fill in parameters in the equations above to determine what is necessary to achieve RFcQED.
One objective is to obtain strong dispersive coupling, requiring $\chi/\hbar \gg \gamma$ where $\gamma$ is the decoherence rate of the transmon.
We fix the transmon frequency to $\omega_a \simeq 2\pi\times 6$ GHz, in order to be in the ground state (Sec.~\ref{sec:intro_thermal_population}).
We would like to have $\omega_r$ in the radio-frequency regime, and to be conservative, let's pick $\omega_r=2\pi\times 200$ MHz.
The anharmonicity should be sufficiently large that the transmon behaves as an atom, yet small enough to be in the transmon limit $\hbar \omega_a\gg E_C$, a typical value is $E_C=h\times 200$ MHz.
Now we should choose a coupling $g$. 
For a typical circuit QED system with $g/\sqrt{\omega_a\omega_d} = 0.01$, the cross-Kerr interaction is extremely small $\chi=h\times 5.9$ Hz.
For this coupling, the normal-mode frequencies, calculated from Eq.~(\ref{eq:LS2_exact_mode_frequencies}), are approximatively equal to the uncoupled frequencies.
This first calculation is summarized in the table below

\medskip
\begin{center}
\begin{tabular}{ll}
$\omega_a = 2\pi\times 6$ GHz & $\tilde\omega_a = 2\pi\times 6.00$ GHz \\
$\omega_r = 2\pi\times 0.2$ GHz & $\tilde\omega_r = 2\pi\times 0.20$ GHz \\
$g/\sqrt{\omega_a\omega_d} = 0.01$ & $\chi =h\times5.9$ Hz \\
\end{tabular}
\end{center}
\medskip

\noindent
Note that the tilde frequencies ($\tilde\omega_a,\tilde\omega_r$) correspond to the mode frequencies shifted by the normal mode splitting induced by the coupling $g$.
Increasing $\chi$ can only come from an increase in $g$ at this point, we try $g/\sqrt{\omega_a\omega_d} = 0.2$

\medskip
\begin{center}
\begin{tabular}{ll}
$\omega_a = 2\pi\times 6.00$ GHz & $\tilde\omega_a = 2\pi\times 6.00$ GHz \\
$\omega_r = 2\pi\times 0.20$ GHz & $\tilde\omega_r = 2\pi\times 0.183$ GHz \\
$g/\sqrt{\omega_a\omega_d} = 0.2$ & $\chi = h\times 2.38$ kHz \\
\end{tabular}
\end{center}
\medskip

\noindent
which still yields a small $\chi = h\times 2$ kHz, since the typical values for $\gamma$ measured in our lab are $\gamma\sim 2\pi\times 1$ MHz.
Pushing the coupling even further does increase the cross-Kerr interaction

\medskip
\begin{center}
\begin{tabular}{ll}
$\omega_a = 2\pi\times 6.00$ GHz & $\tilde\omega_a = 2\pi\times 6.00$ GHz \\
$\omega_r = 2\pi\times 0.20$ GHz & $\tilde\omega_r = 2\pi\times 0.087$ GHz \\
$g/\sqrt{\omega_a\omega_d} = 0.45$ & $\chi = h\times12.0$ kHz \\
\end{tabular}
\end{center}
\medskip

\noindent
however we notice that in this coupling regime, the normal-mode frequency of the resonator mode $\tilde\omega_r$ is getting pushed far below our target frequency.
By correcting for this through a higher bare resonator frequency $\omega_r$, we obtain cross-Kerr interaction rates which start to make the experiment seem feasible

\medskip
\begin{center}
\begin{tabular}{ll}
$\omega_a = 2\pi\times 6.00$ GHz & $\tilde\omega_a = 2\pi\times 6.02$ GHz \\
$\omega_r = 2\pi\times 0.50$ GHz & $\tilde\omega_r = 2\pi\times 0.217$ GHz \\
$g/\sqrt{\omega_a\omega_d} = 0.45$ & $\chi = h\times 190.1$ kHz \\
\end{tabular}
\end{center}
\medskip

\begin{center}
\begin{tabular}{ll}
$\omega_a = 2\pi\times 6.00$ GHz & $\tilde\omega_a = 2\pi\times 6.08$ GHz \\
$\omega_r = 2\pi\times 1.00$ GHz & $\tilde\omega_r = 2\pi\times 0.196$ GHz \\
$g/\sqrt{\omega_a\omega_d} = 0.49$ & $\chi = h\times1.9$ MHz \\
\end{tabular}
\end{center}
\medskip

\begin{center}
\begin{tabular}{ll}
$\omega_a = 2\pi\times 4.76$ GHz & $\tilde\omega_a = 2\pi\times 6.17$ GHz \\
$\omega_r = 2\pi\times 3.93$ GHz & $\tilde\omega_r = 2\pi\times 0.182$ GHz \\
$g/\sqrt{\omega_a\omega_d} = 0.4991$ & $\chi > h\times 2$ MHz \\
\end{tabular}
\end{center}
\medskip

\noindent
In the last case, notice that the un-coupled frequencies of the atom and resonator, $\omega_a$ and $\omega_r$, are nearly identical, and the coupling $g =2\pi\times 2.16$ GHz is so large that $g\sim\Delta =\omega_a-\omega_r $ and we are no longer in the dispersive regime.
The approximations of the two previous sections break down, and a new method to determine $\chi$ is necessary.
All we can say is that is may be larger than the previous scenario $\chi > h\times 2$ MHz, potentially larger than our typical $\gamma\sim2\pi\times 1$ MHz.

Following Eq.~(\ref{eq:chapter-2-summary_coupling_rate}), the required coupling rates should be reached by maximizing $C_c/C_r$ and $C_c/C_a$, this is explored experimentally by using a high impedance resonator and a vacuum-gap capacitor respectively in the following chapter.
After going over the experimental details, we will be in measure to envision a chip combining the two approaches, which will implement RFcQED.
The corresponding experiments will be details in chapter~\ref{chapter_4}.
The parameters in that experiment correspond to those of the last table, and so a different approach to the circuit quantization will also be presented in~\ref{chapter_4}.


\FloatBarrier\chapter{Large coupling in circuit QED: towards RFcQED}
\label{chapter_3}

\begin{abstract}
In this chapter we experimentally explore two avenues to increasing the coupling in circuit QED.
Specifically, we consider coupling a superconducting transmon to a high-impedance microwave resonator or increasing the coupling capacitance through a vacuum-gap capacitor. 
The two different approaches lead to coupling rates, measured on resonance $\omega_a=\omega_r$, of $g\simeq 0.07\sqrt{\omega_a\omega_r}$ and $g\simeq 0.13\sqrt{\omega_a\omega_r}$ respectively.
Combining the two approaches shows the potential of reaching $g\sim0.5\sqrt{\omega_a\omega_r}$.
\end{abstract}

\newpage

In this section we will present the experimental results from two circuit QED devices, where we strove to increase the coupling through two different means:
first through increasing the resonator impedance, and secondly through increasing the coupling capacitance.
An easy way to measure the coupling in these systems is through the normal mode splitting which occurs when the transmon and resonator are resonant.
As explored in the previous chapter, the two first transitions of the system should be separated by $\sim 2g$.
This estimate will systematically be supported by a fit of a diagonalization of the Hamiltonian to the available spectrum.

\section{High impedance resonator}\label{sec:VSC}

A modified version of this section was published as S. J. Bosman*,M. F. Gely*, V. Singh, D. Bothner, A. Castellanos-Gomez and G. A. Steele, \textit{Approaching ultrastrong coupling in transmon circuit QED using a high-impedance resonator}, {\href{https://journals.aps.org/prb/abstract/10.1103/PhysRevB.95.224515}{Phys. Rev. B \textbf{95}, 224515 (2017)}}.
M. F. G. led the theoretical modeling of the device, the data analysis, the fitting and the writing of the published manuscript with input from D. B., S. J. B. and G. A. S.
The design of the device, fabrication and measurements were carried out by S. J. B., V. S., A. C-G. and G. A. S.
All authors provided comments to the published manuscript.
The authors thank Alessandro Bruno, Leo DiCarlo, Nathan Langford, Adrian Parra-Rodriguez, and Marios Kounalakis for useful discussions. 
This project has received funding from the Dutch Foundation for Fundamental Research on Matter (FOM), the European Research Council (ERC) under the European Union’s Horizon 2020 research and innovation program (grant agreement No 681476 - QOMD) and from the Netherlands Organisation for Scientific Research (NWO) in the Innovational Research Incentives Scheme – VIDI, project 680-47-526.
S.J.B. and M.F.G. contributed equally to this manuscript.  

\begin{figure}[h!]
\centering
\includegraphics[width=0.8\textwidth]{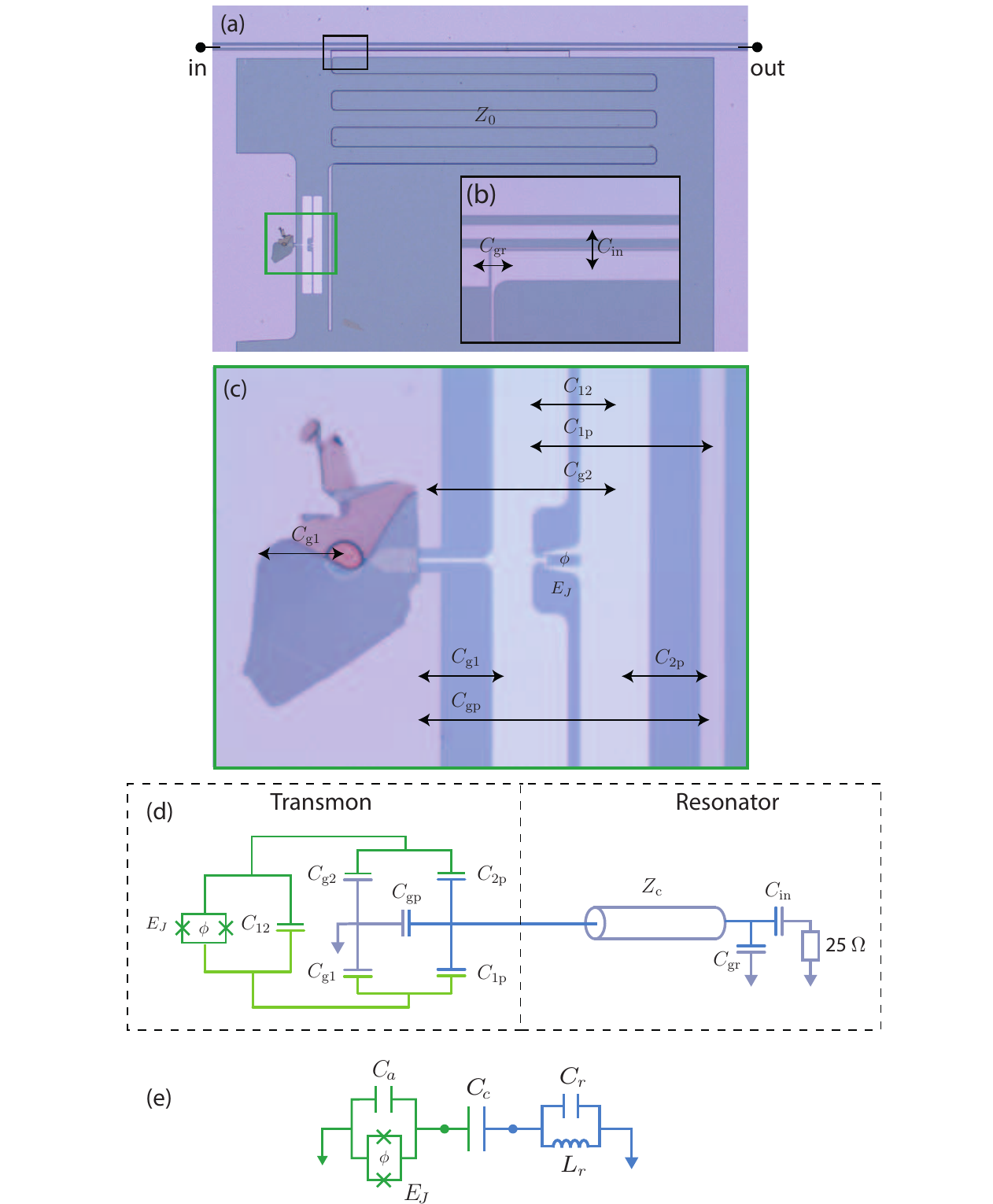}
\caption{(a) Optical image of the device, showing the $\lambda/2$-microwave resonator formed by a meandering $1 \ \mu$m wide stripline capacitively coupled to a $50\ \Omega$ feedline on one end and a transmon on the other end. (b) Zoom in on the capacitive coupler between the resonator and the feedline. (c) Zoom in on the transmon, showing the SQUID loop formed by two Josephson junctions and the parallel plate vacuum-gap capacitor formed by an aluminum transmon as bottom electrode and graphite as top electrode. (d) Full schematic representation of the devices electrical circuit. (e) Equivalent lumped element circuit of the device.}
\label{fig:device}
\end{figure}

\subsection{Device}
Our device, depicted in Fig.~\ref{fig:device}, consists of a high-impedance 645 $\Omega$ superconducting $\lambda/2$ microwave resonator capacitively over-coupled to a $50\ \Omega$ feedline on one end and coupled to a transmon on the other. 
The resonator is a $1\ \mu$m wide, $180$ nm thick and $\sim 6.5$ mm long meandering conductor. It is capacitively connected to a back ground plane through the $275\ \mu$m Silicon substrate as well as through vacuum/Silicon to the side ground planes.
The transmon is in part coupled to ground through a vacuum gap capacitor, see Fig.~\ref{fig:device}(b). 
Its bottom electrode constitutes one island of the transmon, the other plate is a suspended $50$ nm thick graphite flake. 
The diameter of this capacitor is $15\ \mu$m with a gap of $150$ nm. 
This device was initially designed to couple the mechanical motion of the suspended graphite to the transmon, where the coupling is mediated by a DC voltage offset~\cite{lahaye2009nanomechanical,pirkkalainen_hybrid_2013}. 
In this experiment however, no DC voltage was applied, rendering the coupling of the transmon to the motion negligible. 
To enable tunability of the transmon frequency, a SQUID loop is incorporated such that the Josephson energy $E_J$ can be modified using an external magnetic field. $E_J(\phi)$ is a function of the flux through the SQUID loop $\phi$ following $E_J(\phi)=E_{J,\text{max}} \cos(\pi\phi /\phi_0)$ where $\phi_0=h/2e$ is the superconducting flux quantum.

\subsection{Fabrication}\label{sec:VSC_fabrication}
We fabricate our devices in a three-step process. 
First we define our microwave resonators on a $275\ \mu$m Silicon substrate using reactive ion etching of molybdenum-rhenium (MoRe) alloy.
Subsequently, $\mbox{Al/AlO}_x \mbox{/Al}$ Josephson junctions are fabricated using aluminum shadow evaporation~\cite{dolan1977offset}. 
Finally, we stamp a graphite flake on the $15\mu$m diameter opening in the ground plane using deterministic dry viscoelastic stamping technique~\cite{castellanos2014deterministic}. 
From room temperature resistance measurements, optical and SEM images we observe that the flake is suspended, though folded, and that it does not short the transmon to ground.

\subsection{Equivalent circuit and Hamiltonian}
\subsubsection{Transmon parameters}
The capacitance network of the transmon and its coupling to the resonator shown in Fig.~\ref{fig:device}(d) can be reduced to using two capacitances $C_a,C_c$ as shown in Fig.~\ref{fig:device}(e) following the methodology in appendix A of Ref.~\cite{koch_charge-insensitive_2007}. Using the finite element solver ANSYS Maxwell, we compute the value of each of the physical capacitances. This results in the values $C_a = 51$ fF, $C_c=9$ fF.\\

\subsubsection{Circuit parameters}
The waveguide resonator has a characteristic impedance $Z_c=645\Omega$, terminated by a capacitive connection $C_s=C_{\text{in}}+C_{\text{gr}}$ to ground (we ignore the $25\ \Omega$ load that constitutes the feedline along with all other sources of dissipation). If this capacitance is small (\textit{i.e.} $\omega Z_c C_s \ll 1$ with $\omega$ in the GHz range), the resonator is a $\lambda/2$ cavity  where $\omega_c$ would be the resonance frequency of the fundamental mode, if the transmon were replaced with an open circuit.
As established using QUCS~\cite{brinson2009qucs}, the second mode of the resonator shifts the spectrum by $~5$ MHz and the third by less than $1$ MHz. These shifts being smaller than the measured linewidths, we can only consider a singe mode of the resonator.
As derived in chapter~\ref{chapter_2}, the fundamental mode of the circuit can then be approximated by an LC oscillator characterized by
\begin{equation}
Z_r  = \frac{2Z_c}{\pi},\ 
C_r =\frac{\pi}{2\omega_c Z_c},\ 
L_r = \frac{2Z_c}{\pi \omega_c}    \ .
\end{equation}

\subsubsection{Validity of the waveguide assumption}
Although the capacitance of the resonator on the coupler end is substantial, warranting the name of over-coupled $\lambda/2$ resonator, the decision to approximate it as an ideal $\lambda/2$ can be verified by calculating the phase velocity $v_\varphi=\lambda_1 /2\pi\omega_r$ in the line. The wavelength of the fundamental mode $\lambda_1$ is given by twice the total length of the line $\lambda_1=2\times 7.4$ mm (including the length of the coupling pads), yielding a phase velocity of $\sim1\times10^8$ m/s. Approximating the distance from the conductor to side ground plane to a constant $\sim1$ mm (the dominating capacitance coming from the back plane), we calculate an effective relative dielectric constant of $\epsilon_r\simeq6.9$ \cite{wadell1991transmission} corresponding to a theoretical phase velocity $c/\sqrt{\epsilon_r}\simeq1.14\times10^8\ \text{m.s}^{-1}$. This value is a bit larger than the measured speed as expected from over-coupling our resonator. The impedance extracted from our fits is then a slight over-estimate.\\

\subsubsection{Hamiltonian}
The Hamiltonian of the circuit in Fig.~\ref{fig:device}(e) was derived and extensively discussed in chapter~\ref{chapter_2}.
Note that all numerical calculations below are based on the diagonalization in the basis of eigenstates of the transmon Eq.~(\ref{eq:circuit_Hamiltonian_eigenstate_basis}), where the eigenstates are computed from the Cooper-pair box Hamiltonian Eq.~(\ref{eq:CPB_Hamiltonian}). 
This topic is expanded on in Sec.~\ref{sec:MMUSC_fitting}.

\subsection{Measurement methods}

\begin{figure}[h!]
\centering
\includegraphics[width=0.8\textwidth]{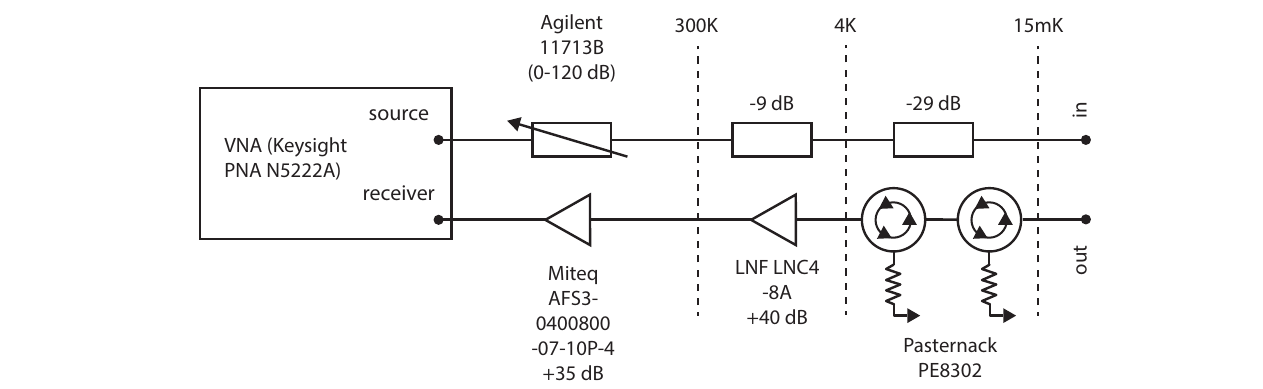}
\caption{Schematic representation of the experimental setup described in the main text. The input microwave line is heavily attenuated down to the 15 mK stage where the sample is placed. On the return line the sample is isolated from the noise of the amplification stage through the use of two circulators. A HEMT amplifier is placed at the 4 K stage and a room temperature amplifier completes the return line.}
\label{fig:VSC_setup}
\end{figure}

We characterize our device at a temperature of 15 mK, mounted in a radiation-tight box. From a vector network analyzer we send a microwave tone that is heavily attenuated before being launched on the feedline of the chip. 
The transmitted signal is sent back to the vector network analyzer through a circulator and a low-noise HEMT amplifier. 
A schematic representation of our experimental setup is presented in Fig~\ref{fig:VSC_setup}.
It allows us to probe the absorption of our device and thus the energy spectrum of the Hamiltonian.
At high driving power we measure the bare cavity resonance~\cite{bishop2010response} to have a total line-width of $\kappa=2\pi \times 29.3$ MHz and a coupling coefficient of $\eta=\kappa_\text{c}/\kappa=0.96$, giving the ratio between the coupling rate $\kappa_\text{c}$ and total dissipation rate $\kappa=\kappa_\text{c}+\kappa_\text{i}$, where $\kappa_\text{i}$ is the internal dissipation rate.
\subsection{Results}
\begin{figure}[h!]
\centering
\includegraphics[width=0.8\textwidth]{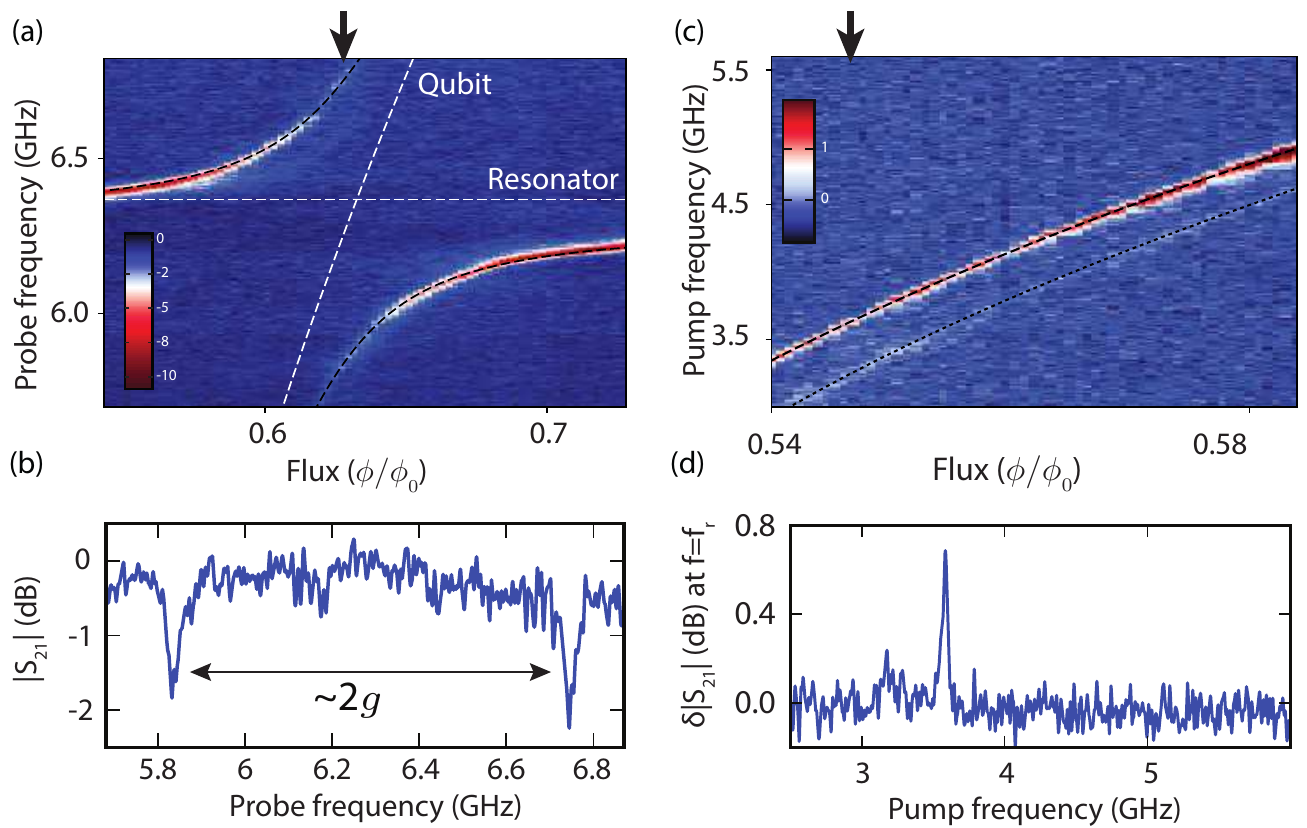}
\caption{ (a) Color plot of the feedline transmission $|S_{21}|$ as a function of magnetic field and frequency, showing the coupling between the resonator and transmon. The magnetic flux penetrating the SQUID loop incorporated in the transmon changes the Josephson energy, which results in the tunability of the $\ket{g}-\ket{e}$ transition energy of the transmon. Close to resonance the transmon and resonator show an avoided crossing centered at $6.23$ GHz. Black dashed lines correspond to fits to this data. The horizontal (oblique) white dashed line corresponds to the bare resonator (transmon) frequency, $\omega_r/2\pi$ ($\omega_a(\phi)/2\pi$). The lack of symmetry around the crossing point is due to the flux dependence of the coupling strength $g$. (b) Trace of the microwave response where the transmon and cavity are close to resonance $\phi=0.63\phi_0$ showing an anti-crossing of $2g\simeq2\pi \times 910$ MHz, and a line-width of $28$ MHz.
(c) Color plot of the change in transmission ($\delta|S_{21}|$) of a microwave tone at resonance with the cavity at single-photon power as a function of frequency of a secondary transmon-drive tone and magnetic field. Due to the transmon-state dependent frequency shift of the cavity, the transmission changes as the drive tone excites the transmon, tracing the transmon $\ket{g}-\ket{e}$ transition frequency as a function of magnetic field. The secondary faint resonance corresponds to the $\ket{e}-\ket{f}$ transition of the thermally excited population in state $\ket{e}$. Power broadening is visible when the transmon transition frequency is close to the cavity. The power of the drive tone delivered to the feedline is constant, but close to the cavity a larger portion of the power is delivered to the transmon. Dashed lines correspond to fits to the data. (d) Line-cut of the color plot where the transmon frequency is at $\omega_a=2\pi \times 3.586$ GHz with a line-width of $38$ MHz. The second peak corresponds to the first to second exited state transition, which corresponds to the (dressed) anharmonicity, a rough approximation of $E_c/h= 370$ MHz.}
\label{fig:single_tone}
\end{figure}
\subsubsection{Avoided crossing}
With a current biased coil, we can control the magnetic field and tune the effective $E_J$ to bring the transmon in resonance with the cavity. 
Where the transmon and resonator frequencies cross, we measure an avoided crossing which gives an estimate of the coupling rate $2g \simeq 910$~MHz as shown in Fig.~\ref{fig:single_tone}.
This corresponds to $g \simeq 0.07 \sqrt{\omega_a\omega_r}$.

\subsubsection{Two-tone spectroscopy}
In Fig.~\ref{fig:single_tone}, we show the result of performing two-tone spectroscopy to probe the transmon frequency~\cite{blais2004cavity,wallraff2005approaching}, necessary to accurately fit the circuit parameters of our model to the data. 
When the transmon is detuned from the cavity, the resonator acquires a frequency shift which is dependent on the state of the transmon through cross-Kerr interaction. 
Hence, probing the transmission of the feedline at the cavity resonance (shifted by the transmon in the ground state), while exciting the transmon with another microwave tone, will cause the transmission to change by a value $\delta|S_{21}|$ due to the transmon-state dependent shift. 
In Fig.~\ref{fig:single_tone}(c) we measure the spectral response of the transmon for different magnetic fields. 
As the magnetic flux through the SQUID loop tunes the transmon frequency we track the ground to first exited state transition as a function of magnetic field. 
Since the probe power is kept constant during this experiment a clear power broadening of the transmon is visible, because more of the power is delivered to the transmon as it is closer to the cavity in frequency. 
The secondary faint resonance corresponds to the spectral response of the first to second exited state transition of the transmon due to some residual occupation of the first excited state. 
The difference in frequency between both transitions provides an estimate of the charging energy (or equivalently the anharmonicity of the transmon), $E_c/h \sim 370$ MHz.

\subsubsection{Transmon linewidth}
The $38$ MHz line-width of this resonance translates to very short coherence times ($T_1\sim40$~ns) compared to typical implementations \cite{houck2008controlling}. 
Purcell losses contribute less than $2$ MHz to this line-width. The Ohmic losses in the graphite can also not explain this high dissipation. 
From the bulk conductivity of graphite, we find that the resistance of our $50$ nm thick flake is $7\ \Omega/\square$~\cite{primak1954electrical}. 
Simulating the effect of a $7\ \Omega$ resistance at the appropriate position in the transmons capacitance network using QUCS~\cite{brinson2009qucs} reveals that this has a negligible contribution of 200kHz to the total transmon line-width. 
Other possibilities include dielectric losses in the interface between silicon and MoRe (see fabrication details in Sec.~\ref{sec:VSC_fabrication}). 
Indeed, the high quality factors previously obtained with this superconductor~\cite{singh2014molybdenum} were very sensitive to a surface preparation which was incompatible with the complexity of the device presented here. 
Another relevant source of dissipation could be the resistance associated with the conversion of a supercurrent into a normal current at the interface between MoRe and the graphite~\cite{boogaard2004resistance}.
Finally, a likely source of dissipation is probably the absence of shielding against magnetic fields, as well as many forms of radiation, which is today common practice in the field.

\subsection{Fitting}

We fit a numerical diagonalization of the Hamiltonian to the acquired data, obtaining the fits shown as dashed lines in Figs.~\ref{fig:single_tone}. 
We thereby obtain the Hamiltonian parameters $E_c/h = 300$ MHz, $g/2\pi=455$ MHz (on resonance), $\omega_r/2\pi=6.367$ GHz and $E_{J,\text{max}}/h=46$ GHz. 
Combined with our knowledge of the capacitances $C_a$ and $C_c$, we extract the following values for the circuit elements of the resonator: $C_r=57.1$ fF and $L_r=9.65$ nH. 
If we assume that the parallel LC oscillator corresponds to the fundamental mode of a $\lambda/2$ resonator, then the resonators effective impedance $Z_r=\sqrt{L_r/C_r}=411\ \Omega$ is related to the characteristic impedance of the transmission line through $Z_c=\pi Z_r/2$, yielding a value $Z_c=645\ \Omega$. \\
We describe a similar fitting routine in greater detail in Sec.~\ref{sec:MMUSC_fitting}.

\section{Increasing the coupling capacitance}\label{sec:MMUSC}
A modified version of this section was published as S. J. Bosman,M. F. Gely, V. Singh, A. Bruno, D. Bothner and G. A. Steele, \textit{Multi-mode ultra-strong coupling in circuit quantum electrodynamics}, {\href{https://www.nature.com/articles/s41534-017-0046-y}{npj Quantum Information \textbf{3}, 46 (2017)}}.
S.\@ J.\@ B.\@ and G.\@ A.\@ S.\@ conceived the experiment. S.\@ J.\@ B.\@ designed and fabricated the devices. V.\@ S.\@ , A.\@ B.\@ and G.\@ A.\@ S.\@ provided input for the fabrication. S.\@ J.\@ B.\@ and M.\@ F.\@ G.\@ did the measurements with input of D.\@ B.\@ and G.\@ A.\@ S.\@. M.\@ F.\@ G. and D.\@ B.\@ performed data analysis with input of S.\@ J.\@ B.\@ and G.\@ A.\@ S.\@. Manuscript was written by S.\@ J.\@ B.\@, M.\@ F.\@ G.\@ and G.\@ A.\@ S.\@, and all authors provided comments to the manuscript. G.\@ A.\@ S.\@ supervised the work.
We acknowledge Enrique Solano, Adrian Parra-Rodriguez and Enrique Rico Ortega for valuable input and discussions.
This project has received funding from the Dutch Foundation for Fundamental Research on Matter (FOM), the European Research Council (ERC) under the European Union’s Horizon 2020 research and innovation program (grant agreement No 681476 - QOMD) and from the Netherlands Organisation for Scientific Research (NWO) in the Innovational Research Incentives Scheme – VIDI, project 680-47-526.

\begin{figure*}[h!]
\centering
\includegraphics[width=0.8\textwidth]{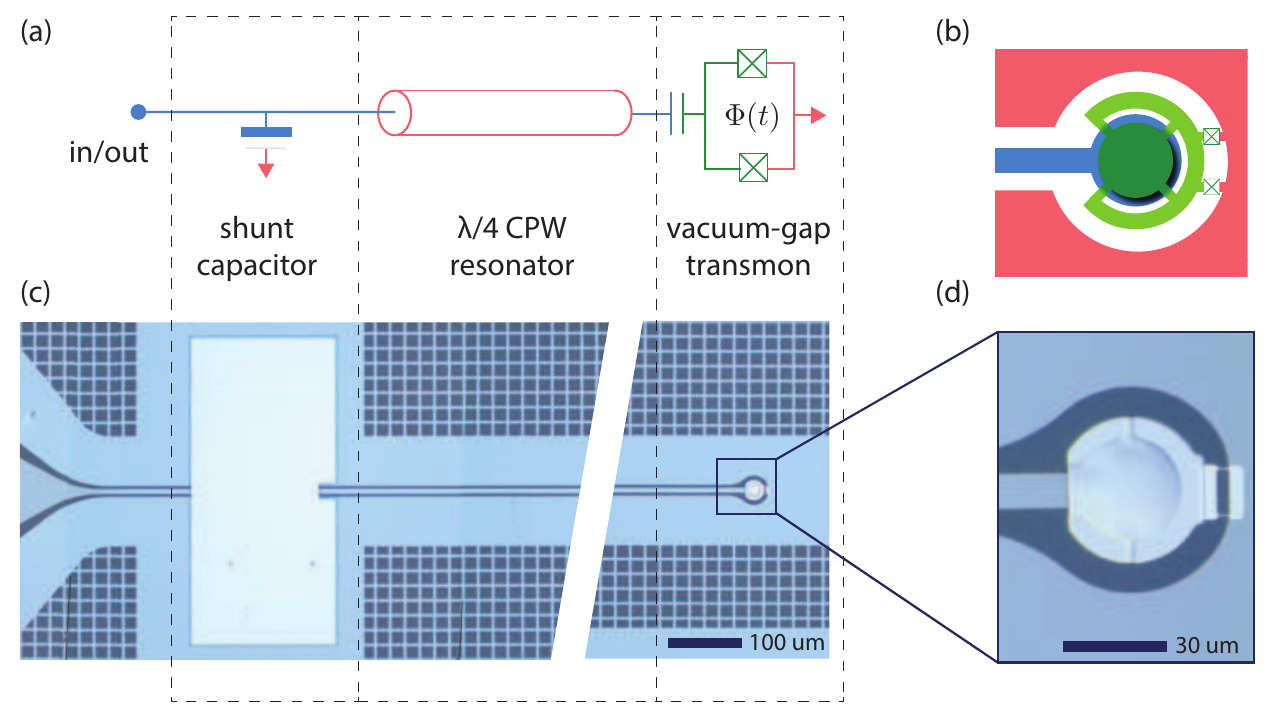}
\caption{\textbf{Vacuum-gap transmon circuit architecture. (a) }
  Schematic diagram of the equivalent circuit containing a $\lambda/4$ microwave cavity, which on the left is coupled through a shunt capacitor~\cite{bosman2015broadband} to a 50 $\Omega$ port for reflection measurements. On the right, at the voltage anti-node of the resonator, it is capacitively coupled to a transmon. 
  \textbf{(b)} 
  Detailed schematic of the transmon showing the vacuum-gap capacitor between the center conductor of  the resonator and a suspended superconducting island, which is connected to ground with two Josephson junctions in SQUID geometry (note the matched colors with (a)). 
  \textbf{(c)}
  Optical image of the device implementing the circuit. 
  \textbf{(d)}
  Zoom-in on the transmon showing the suspended capacitor plate above the end of the resonator connected to ground by the junctions.
 }
 \label{fig:device}
\end{figure*}

\subsection{Device}
Here, we realize a circuit QED system where the coupling capacitance dominates the total capacitance of the transmon. The transmon consists of a superconducting island shorted to ground by two Josephson junctions in parallel (or SQUID), which is suspended above the voltage anti-node of a quarter wavelength ($\lambda/4$) coplanar waveguide microwave cavity as shown in Figs.~\ref{fig:device}(a,b). 

\subsection{Fabrication}
Our sample, depicted in Fig.~\ref{fig:device}(c,d), is fabricated on a sapphire substrate and uses as superconductor an alloy of molybdenum-rhenium (MoRe)~\cite{singh2014molybdenum} (with Josephson junctions fabricated out of aluminum).
In a five step electron beam lithography process we pattern the microwave resonator, shunt capacitor dielectric, vacuum-gap sacrificial layer and lift-off mask for the MoRe suspension (see methods for more details). 
In the first step, we define the bottom metalization layer of the cavity, including the bottom layers of the shunt-capacitor and the vacuum-gap capacitor, on top of a sapphire substrate.
We use magnetron sputtering to deposit a $45\,$nm thick layer of $60-40$ molybdenum-rhenium (MoRe) alloy and pattern it by means of electron-beam lithography (EBL) and SF$_6$/He reactive ion etching (RIE).
For the definition of the shunt-capacitor dielectric, we deposit a $100\,$nm thick layer of silicon nitride by means of plasma-enhanced chemical vapor deposition and perform the patterning by EBL and wet etching in buffered hydrofluoric acid.
In a third EBL step we pattern the sacrificial layer for the vacuum-gap capacitor, which in our samples consists of a $\sim 160\,$nm thick layer of the electron-beam resist PMGI SF7 diluted $2:1$ with cyclopentanone. 
After the development of the sacrificial layer in L-ethyl-lactate, stopped by rinsing with isopropanol, we reflow the patterned PMGI for $180\,$s at $250$ $^{\circ}$C in order to slightly smooth the stepped edge, facilitating the sidewall metalization in the next step.
The shunt-capacitor and vacuum-gap capacitor top electrodes are fabricated subsequently by means of lift-off technique.
First, we perform EBL to pattern the corresponding PMMA resist layer and secondly, we sputter deposit a $120\,$nm thick layer of MoRe on top.
We do the lift-off in hot xylene, while the sacrificial layer of the vacuum-gap capacitor is not attacked in this process and thus remains unchanged.
In the last step, we fabricate the Josephson junctions using a PMGI/PMMA bilayer lift-off mask, EBL and aluminum shadow-evaporation.
We perform a simultaneous aluminum lift-off and the drum release in the resist stripper PRS3000. Finally, we dry the sample in an IPA filled beaker on a hot plate.

\subsection{Measurement methods}
\begin{figure}[h!]
\centering
\includegraphics[width=0.8\textwidth]{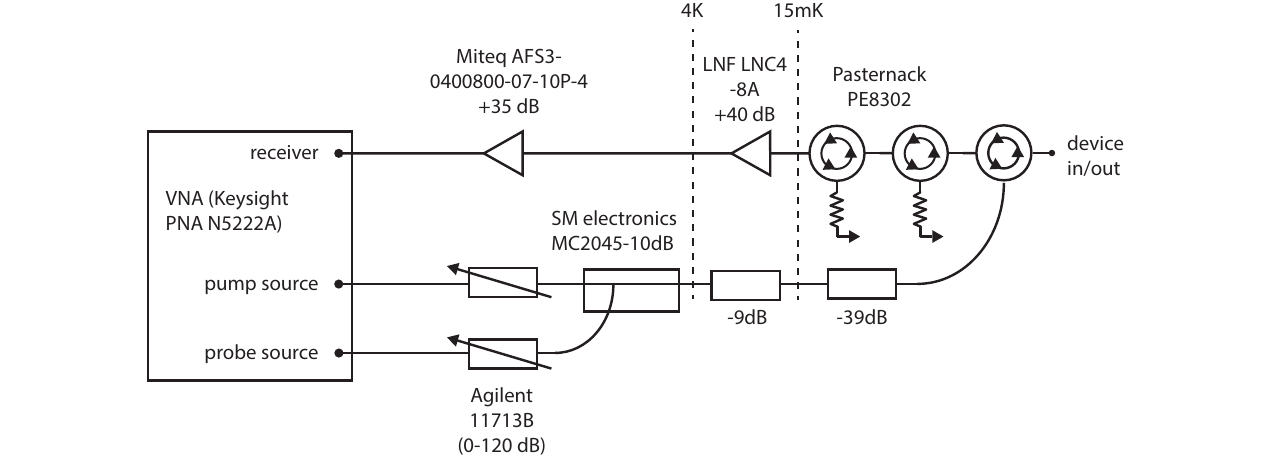}
\caption{ Schematic of the measurement setup.  
}
\label{fig:MMUSC_setup}
\end{figure}

Fig.~\ref{fig:MMUSC_setup} shows schematically the measurement setup used for the device characterization. The vector network analyzer (VNA) outputs one or two continuous wave (CW) signals that are sent through a variable attenuator (0-120 dB) and combined with a directional coupler. From there the signal is sent into the dilution fridge, where it is attenuated (48 dB) before reaching the sample through a circulator. The reflected signal from the device is sent back to the VNA using two isolators and amplifiers. 

\subsection{Results}
\begin{figure*}[h!]
\centering
\includegraphics[width=0.8\textwidth]{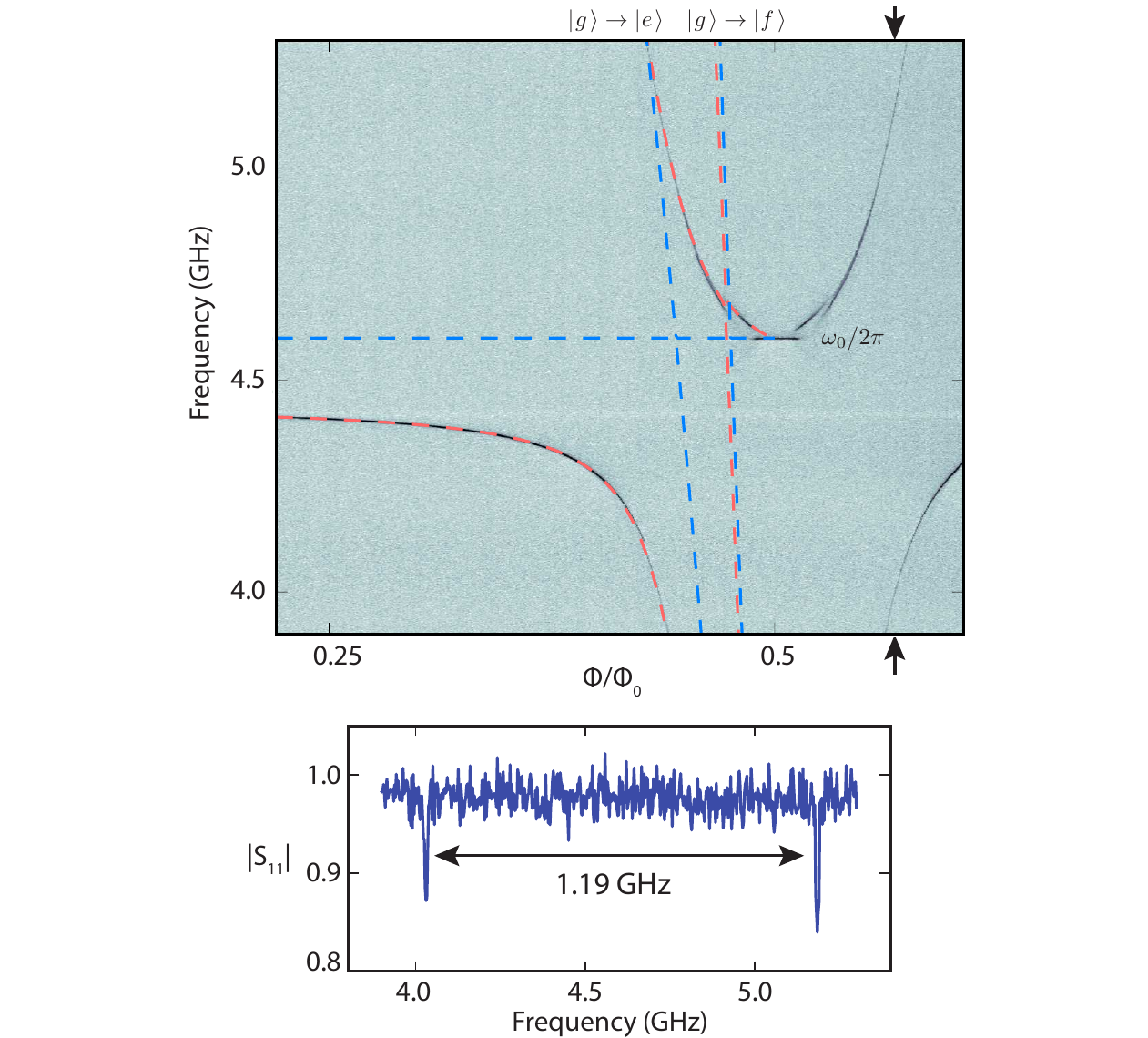}
\caption{\textbf{Avoided crossing. (a) }
The spectral response is shown as a function of flux in a single-tone reflection measurement plotted as $|S_{11}|$. 
The blue dashed lines indicate the bare (uncoupled) frequency of the fundamental cavity mode, $\omega_0$, and the transition frequencies of the transmon from the ground state $\ket{g}$, to its first and second excited state, $\ket{e}$ and $\ket{f}$ respectively. 
The red lines show the hybridized state transitions of the coupled system as fitted from the full spectrum.
\textbf{(b)}
Line-cut showing the avoided crossing of the transmon transition with $\omega_0$, resulting in a separation of $2g\simeq 1.19$ GHz.
}
 \label{fig:MMUSC_single_tone}
\end{figure*}

\begin{figure*}[h!]
\centering
\includegraphics[width=0.8\textwidth]{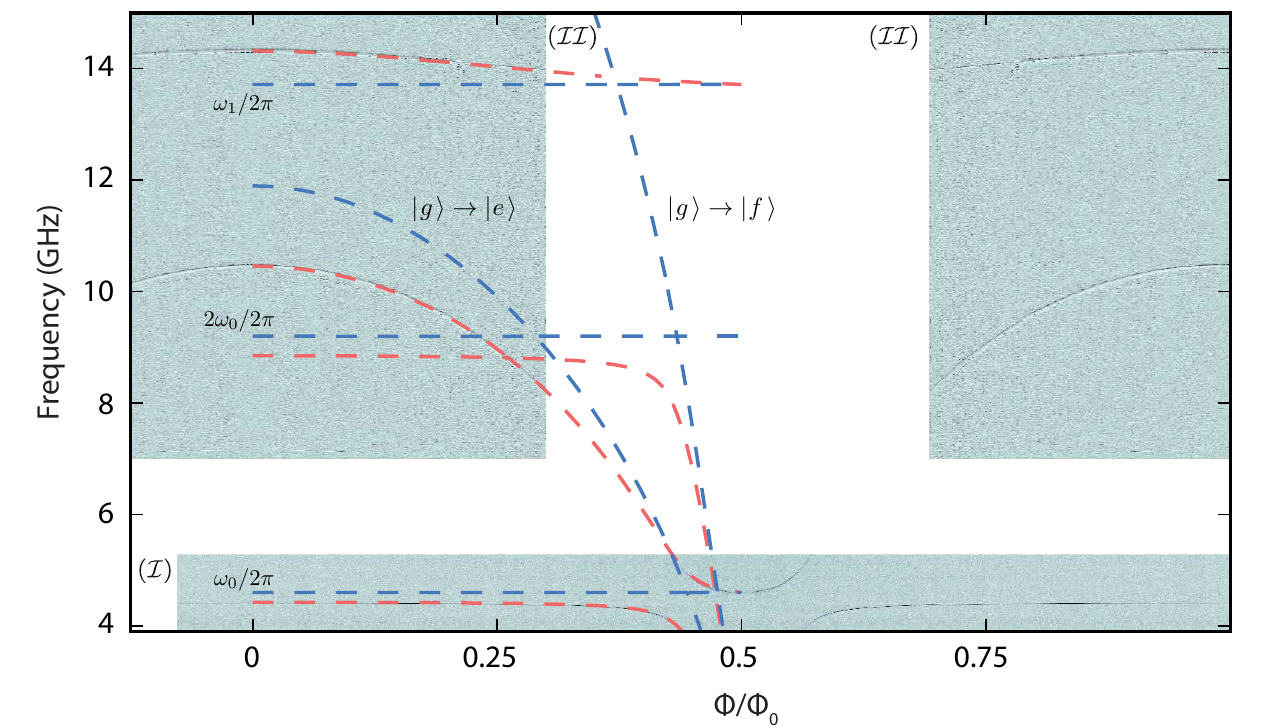}
\caption{\textbf{Multi-mode spectrum. (a) }
Color plot composed of two spectroscopic measurements. 
The bottom frame, $(\mathcal{I})$, shows data obtained as in Fig.~\ref{fig:MMUSC_single_tone}. 
The top frame, $(\mathcal{II})$, shows data from a two-tone spectroscopy measurement from the change in the reflection phase of a weak probe tone at the cavity frequency ($\omega_0$), as a function of a secondary drive tone. 
We plot the derivative of this phase $\delta\theta_{11}=\partial \varphi[\tilde{S}_{11}(\omega)] /\partial\omega$. 
Frame $(\mathcal{II}^*)$ is a mirrored copy of the same data.  
The blue dashed lines show the transition frequencies of the uncoupled cavities ($\omega_0, \omega_1$) and the first transition ($\ket{g}$ to $\ket{e}$) of the transmon. 
The red (dashed lines show the hybridized spectrum after diagonalization of the Hamiltonian. 
 }
\label{fig:MMUSC_multimode}
\end{figure*}

\textbf{Avoided crossing --}
Figure~\ref{fig:MMUSC_single_tone} shows the spectral response using single-tone microwave reflectometry at $\sim$14 mK. 
By measuring the complex scattering parameter $S_{11}(\omega)$ of the circuit as a function of an external magnetic field, we can probe the absorption of the circuit at a given frequency within the circulator and amplifier bandwidth of 4-8 GHz.
The transitions of the circuit appear as a dip in the magnitude of the scattering parameter, $|S_{11}|$, thereby mapping the spectrum of the circuit. 
At high measurement drive powers, we measure the bare fundamental mode of the resonator to have a frequency of $4.60$ GHz, with internal and external quality factors of $Q_\text{int}=3800$ and $Q_\text{ext}=2600$ respectively.
We measured the avoided crossing to be $2g\simeq 1.19$ GHz, corresponding to a ratio $g/\sqrt{\omega_a\omega_0}\simeq 0.13$.
The linewidth of the resonator and transmon is measured to be $\kappa=2\pi\times 3$ MHz and $\gamma\sim2\pi\times 20$ MHz respectively

\textbf{At half a flux quantum --}
Close to half a flux quantum ($\Phi/\Phi_0 \simeq 0.5$), $E_J$ becomes small, such that the transmon frequency goes towards zero for very symmetric junctions, and negligible loop inductance, and the transmon becomes more like a Cooper-pair box (CPB) as the ratio of $E_J/E_c$ drops~\cite{koch_charge-insensitive_2007,}.
In this flux region, we observe two notable features.
The first is an anti-crossing at $\Phi/\Phi_0 \sim 0.471$, which we attribute to an avoided crossing with the $\ket{g}$ to $\ket{f}$ transition of the CPB (indicated with the blue dashed line). 
This shows that in this flux region the transmon behaves like a CPB as such transitions are exponentially suppressed in the transmon regime~\cite{koch_charge-insensitive_2007}. 
The second feature is a jump of the dressed cavity to the frequency of the bare cavity at $\Phi/\Phi_0 \sim 0.485$.
Although simulations indicate that this jump is due to the CPB becoming thermal populated as its frequency decreases, we have yet to obtain an intuitive picture of the physics in this regime.

\textbf{Coupling to higher modes --}
Fig.~\ref{fig:MMUSC_multimode} is a composition of the single tone measurement (Fig.~\ref{fig:MMUSC_single_tone}(a)), combined with a two-tone spectroscopy measurement~\cite{wallraff2004strong}. 
In such a measurement, the change in cavity response $\tilde{S}_{11}(\omega)$ is monitored using a weak probe tone as a function of a second drive tone at the transmon. 
Due to the transmon-state dependent dispersive shift of the cavity at $\omega_0$, the reflection of the weak probe tone changes as the drive tone excites the transmon. 
The measured transmon linewidth is $\gamma\sim2\pi\times 20$ MHz. 
From the loss tangent of previous aluminum vacuum-gap capacitor measurements~{\cite{cicak2010low}}, we would expect a $100$ kHz linewidth. 
Our experiment differs by the use of MoRe with PMGI as a sacrificial layer; differences in the superconductor to vacuum interface could thus explain the large losses. 
Another explanation could be the absence of shielding agains many source of radiation and against noise in the magnetic field.
The maximum frequency of the transmon is $\sim 12$ GHz, and the first harmonic of the cavity is at $\sim 14$ GHz, so this mode and the transmon do not cross. 
The second energy level of the fundamental mode of the resonator is indicated in the blue dashed lines (labeled $2\omega_0/2\pi$), and the corresponding state is found to hybridize with the $\ket{g} \rightarrow \ket{f}$ transition.

\subsection{Fitting} \label{sec:MMUSC_fitting}
\subsubsection{Hamiltonian}

In the GHz regime we are working in ($\omega/2\pi>3.5$ GHz), the impedance of the $C_s=30$ pF shunt capacitor is small ($|1/i\omega C_{s}|\simeq2\ \Omega\ll Z_c=50\ \Omega$) such that it can effectively be considered as a short to ground~\cite{bosman2015broadband}. 
We can then use the model derived in chapter~\ref{chapter_2} (see Eqs.~(\ref{eq:circuit_Hamiltonian_eigenstate_basis},\ref{eq:CPB_Hamiltonian}), adding however two elements. 
First, the flux dependent Josephson term (see Ref. \cite{koch_charge-insensitive_2007})
\begin{equation}
- E_J \cos (\hat{\delta})\rightarrow- E_J \cos  (\hat{\delta})\cos \bigg(\pi\frac{\Phi}{\Phi_0}\bigg) - dE_J \sin (\hat{\delta})\sin \bigg(\pi\frac{\Phi}{\Phi_0}\bigg)\ ,
\label{eq:Ej_flux_dep}
\end{equation}
where $\hat{\delta}$ is the superconducting phase difference across the SQUID. 
We defined the total Josephson energy by $E_J = E_{J,1} + E_{J,2}$ and the asymmetry by $d =( E_{J,1} - E_{J,2})/(E_{J,1} + E_{J,2})$, where $E_{J,1}$ and $E_{J,2}$ are the two different Josephson energies of the SQUID junctions. 
Secondly, we offset the quantum number of Cooper pairs on the transmon island by a constant value $N_\text{env}$ to model an environmental offset charge present in any realistic system in the Cooper pair box regime~\cite{koch_charge-insensitive_2007}.

\subsubsection{Hamiltonian diagonalization}
\begin{figure}[h!]
  \centering
  \centering
\includegraphics[width=0.8\textwidth]{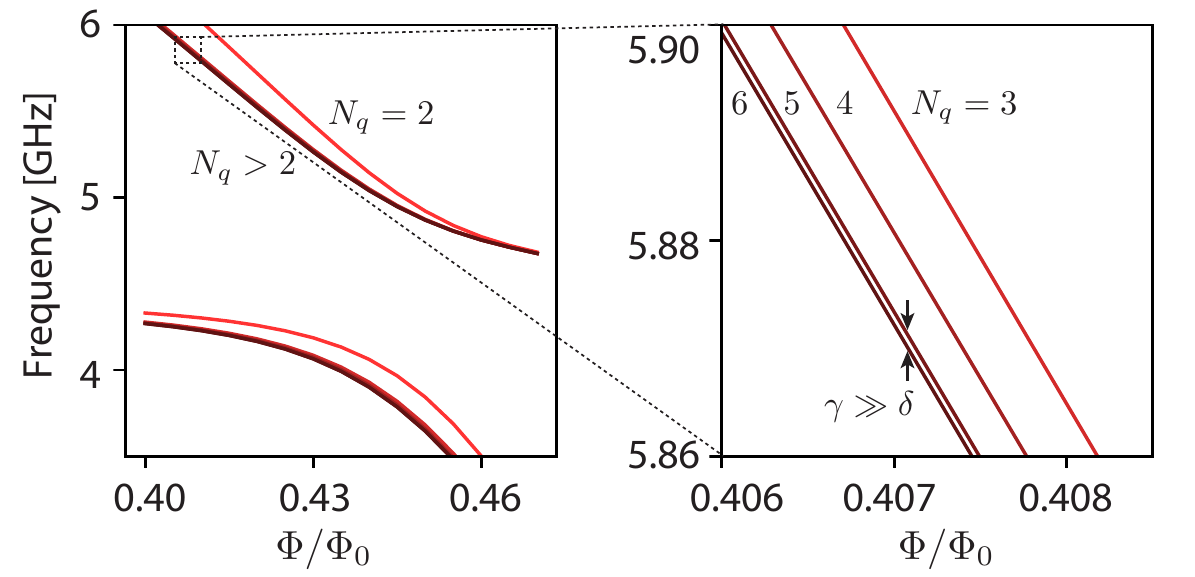}
  \caption{Dressed spectrum shown in Fig.~\ref{fig:MMUSC_single_tone}(a) illustrating the importance of higher transmon levels in obtaining the correct spectrum. The spectrum is plotted for different values of the number of included transmon levels $N_q=2,...,6$. Including a 6-th level leads to a difference in the spectrum $\delta$ which is more than 10 times smaller than the measured linewidth. In the right panel, the relevant linewidth is that of the transmon $\gamma\sim 2\pi\times 20$ MHz and $\delta \sim 2\pi\times 1.2$ MHz.}
  \label{fig:s5}
 
\end{figure}

The Hamiltonian diagonalization is performed in two steps, first a diagonalization of the Cooper pair box Hamiltonian in the charge basis Eq.~(\ref{eq:CPB_Hamiltonian}), secondly of the multi-mode circuit QED Hamiltonian in the Cooper-pair box eigenstate basis Eq.~(\ref{eq:circuit_Hamiltonian_eigenstate_basis}).
In order to perform numerics, four simulation parameters should be fixed: (i) a maximum number of Cooper pairs on the transmons charge island $N_\text{max}$, (ii) the number of transmon levels $N_q$, (iii) the number of resonator modes $N_\text{modes}$ (iv) each with a certain number of photons $\{N_{m}\}_{m = 0,..,N_\text{modes}-1}$.
The size of the Hilbert space used for the diagonalization of the Cooper pair box Hamiltonian scales with $2N_\text{max}+1$ and for the diagonalization of the multi-mode circuit QED Hamiltonian it scales with $N_q\prod_m N_{m}$. 
We should therefore have a small enough Hilbert space such that the diagonalizations are feasible with the computer resources at our disposal, whilst ensuring that the neglected degrees of freedom do not significantly change the spectrum if we would have included them.
We consider that a degree of freedom which changes the computed spectrum by less than a tenth of the measured linewidth can be neglected. 
This condition leads to the following simulation parameters $N_\text{max} = 20,N_\text{modes}=4,N_q=5$.
The number of photon levels to include depends on the strength of the coupling and we have used 6,4,3 and 3 photon levels in the fundamental, first, second and third modes of the resonator. 
We show the impact of higher levels in Fig.~\ref{fig:s5}, where the spectrum of Fig.~\ref{fig:MMUSC_single_tone}(a) is plotted for different values of $N_q$.

\subsubsection{Fitting routine}\label{sec:MMUSC_fitting}

We first experimentally perform a broad flux-dependent single tone measurement of the dressed cavity in order to extract the current periodicity and the point of maximum transmon frequency.
This allows us to convert current (that we apply to a coil to bias the SQUID) to the flux through the SQUID (in units of flux quantum).
Nine free parameters are to be determined: $\omega_0, C_c, C_a, N_\text{env}, E_J, d$. 
Due for example to the effect of the shunt capacitor (neglected in the model) we also correct the frequency of the higher modes leading to three more parameters: $\alpha_n$, such that $\omega_m = \alpha_m(2m-1)\omega_0$, $m\ge1$. 
We expect $\alpha_m\simeq1$ and shall maintain this hypothesis until the last stage of the fit. 
Two other parameters can be fixed easily:
\begin{enumerate}
  \item Through a high power single-tone measurement, we can drive the (dressed) fundamental mode of the resonator to its harmonic regime \cite{bishop2010response} and determine the bare resonator frequency $\omega_0$.
  \item At a flux of $\Phi_0/2$, the dressed cavity frequency resumes its bare frequency which indicates that the coupling is negligible at that flux point. 
  In simulation, this can only be achieved for very low asymmetry $d\simeq0.01$.
\end{enumerate}
We then perform a least-square minimization routine, fitting the extracted data -- the dressed first and second cavity mode, and dressed first transmon transition frequency, as a function of flux -- to a diagonalization of the Hamiltonian. 
The free parameters in this routine are $C_c$, $C_a$, $N_\text{env}$, $E_J$. 
Due to the considerable time necessary to perform the diagonalization (approximatively ten seconds per flux point on a regular laptop computer), we parallelize the diagonalization over the different flux points using a high performance computing cluster. 
Finally we adjust the frequencies of the higher modes through the parameters $\alpha_m$.
The results are given in Table~\ref{tab:MMUSC_fitresults}.
\begin{table}[h!]
    \begin{tabular} {llll}
        \hline
        Quantity & Symbol  & Value  & unit \\ \hline
        \textbf{resonator parameters} &  & \\ 
        bare fundamental frequency & $\omega_0/2\pi$  & 4.603  & GHz \\
        internal quality factor & $Q_\text{int}$  & 3800  &  \\
        external quality factor & $Q_\text{ext}$  & 2600 &  \\
        resonator impedance & $Z_c$  & 50  & $\Omega$\\
        harmonics deviation & $[\alpha_m]_m$& [1.0,0.994,1.0,1.0] \\ 
        \textbf{transmon parameters} &  & \\
        charging energy & $E_c/h$  & 426 &  MHz \\
        Josephson energy at 0 flux & $E_J(0)/h$  \hspace{5pt} & 36.3 &  GHz \\
        maximum transmon frequency at 0 flux &$ \omega_{a}/2\pi $  & 10.67 &  GHz \\ 
        transmon capacitance to ground  &$C_a$  & 5.13 & fF\\
        vacuum-gap capacitor &$C_c$ & 40.3 & fF\\
        
    \end{tabular} \\ \vskip .5cm

    \caption{\textbf{Fit results}, reulting from the fitting routine described in Sec.~\ref{sec:MMUSC_fitting}}
    \label{tab:MMUSC_fitresults}   
\end{table}

\section{Combining the two approaches for RFcQED}

Let us recall the coupling rate
\begin{equation}
	g\simeq\frac{1}{2}\sqrt{\frac{ \omega_a\omega_r}{(1+\frac{C_a}{C_c})(1+\frac{C_r}{C_c})}}
\end{equation}
where the un-coupled frequencies for $C_c\gg C_a,C_r$, are given by
\begin{equation}
\begin{split}
  \omega_a &= \frac{1}{\sqrt{(C_a+C_r)L_J}}\ ,\\
  \omega_r &= \frac{1}{\sqrt{(C_a+C_r)L_r}}\ .\\
\end{split}
\end{equation}

In the previous section, we derived a possible combination of coupling rate and uncoupled frequencies which could allow us to exceed a MHz of cross-Kerr coupling
\begin{center}
\begin{tabular}{ll}
$\omega_a = 2\pi\times 4.76$ GHz & $\tilde\omega_a = 2\pi\times 6.17$ GHz \\
$\omega_r = 2\pi\times 3.93$ GHz & $\tilde\omega_r = 2\pi\times 0.182$ GHz \\
$g/\sqrt{\omega_a\omega_d} = 0.4991$ & $\chi > h\times 2$ MHz \\
\end{tabular}
\end{center}
\medskip

\noindent
In the spirit of the vacuum-gap transmon device, we propose to increase the coupling capacitance to even higher values by increasing the area covered by the capacitor and filling it with a dielectric.
By simply tuning both the area spanned by the capacitor and the thickness of the dielectric layer, we can tune $g/\sqrt{\omega_a\omega_d}$ to arbitrary values.
The values for the other capacitances are fixed by the transmon regime.
The highest ratio of $E_c/E_J$ for which we remain in the transmon regime is $E_c/E_J > 1/20$~\cite{koch_charge-insensitive_2007}. 
With a transmon frequency $\omega_a=2\pi\times 4.76$ GHz, we obtain a value of the transmons total capacitance $\tilde C_a \simeq C_a+C_r > 50$ fF and and a Josephson inductance $L_J<20$ nH.

Fixing the resonator frequency to $\omega_r = 2\pi\times 3.93$ GHz, combined with the capacitance $\tilde C_r \simeq \tilde C_a > 50$ fF, leads to a value for resonators characteristic impedance $\sqrt{L_r/\tilde C_r}>786\ \Omega$.
An increase in resonator impedance compared to the usual $50\Omega$, as was presented in the first section of this chapter, is thus inevitable.
An additional, practical requirement would be to remove the higher modes of the resonator, which complicate the system by hybridizing with the degrees of freedom of interest, as we learned in the vacuum-gap transmon device.
A way to address these two criteria is to make use of a lumped element inductor, for example a spiral inductor.
As we will see in the next chapter, these can be constructed with a very high characteristic impedance.

 
\FloatBarrier\chapter{Radio frequency circuit QED}
\label{chapter_RFcQED}
\label{chapter_4}

\begin{abstract}
Here we present our implementation of radio-frequency circuit QED.
Using a gigahertz transmon, we observe the quantization of a megahertz radio-frequency resonator, cool it to the ground-state and stabilize Fock states.
Releasing the resonator from our control, we observe its re-thermalization with nanosecond resolution.
\end{abstract}

\newpage
A modified version of this chapter was published as M. F. Gely, M. Kounalakis, C. Dickel, J. Dalle, R. Vatré, B. Baker, M. D. Jenkins and G. A. Steele, \textit{Observation and stabilization of photonic Fock states in a hot radio-frequency resonator}, {\href{https://science.sciencemag.org/content/363/6431/1072}{Science \textbf{363}, 1072 (2019)}}.
MFG and RV developed the theoretical description of the experiment. 
MFG designed the device. 
MFG fabricated the device with help from JD and MK. 
MFG, MK, CD, JD and MJ participated in the measurements.
MFG and CD analyzed the data.
BB provided the software and input for the adaptive rotating-wave-approximation simulation.
MFG wrote the manuscript with input from MK, CD and GAS.
All co-authors reviewed the manuscript and provided feedback.
GAS supervised the project.
The authors acknowledge Ya. M. Blanter, S. M. Girvin, J. D. P. Machado for useful discussions.
This work was supported by the European Research Council under the European Union’s H2020 program under grant agreements 681476 - QOM3D and 785219 - GrapheneCore2, by the Dutch Foundation for Scientific Research (NWO) through the Casimir Research School, and by the Army Research Office through Grant No.\ W911NF-15-1-0421.
Raw data as well as all measurement, data-analysis and simulation code used in the generation of main and supplementary figures is available in Zenodo with the identifier 10.5281/zenodo.2551258

\noindent 
\section{Experimental highlights}

\subsection{Strong dispersive coupling}
\FloatBarrier

\subsubsection{Circuit and Hamiltonian}
In Fig.~\ref{fig:LFT_1} we present the circuit enabling radio-frequency circuit QED.
The circuit (Fig.~\ref{fig:LFT_1}(a)) comprises of a Josephson junction ($L_\text{J}=41$ nH) connected in series to a capacitor ($C_L=11$~pF) and a spiral inductor ($L=28$ nH).
At low frequencies, the parasitic capacitance of the spiral inductor is negligible, and the equivalent circuit (Fig.~\ref{fig:LFT_1}(b)) has a first transition frequency $\omega_r = 2\pi\times173\ \text{MHz}$.
At gigahertz frequencies, $C_L$ behaves as a short, and the capacitance of the spiral inductor $C_H=40$~fF becomes relevant instead.
The resulting parallel connection of $L_\text{J}$, $L$ and $C_H$ (Fig.~\ref{fig:LFT_1}(c)) has a first transition frequency $\omega_a=2\pi\times5.91\ \text{GHz}$.
The two modes share the Josephson junction.
The junction has an inductance that varies with the current fluctuations traversing it, and consequently the resonance frequency of the transmon shifts as a function of the number of excitations in the resonator and vice versa.
This cross-Kerr interaction is quantified by the shift per photon $\chi = 2\sqrt{A_aA_r}$, where the anharmonicity of the resonator and transmon $A_r= h\times495\ \text{kHz}$ and $A_a=h\times192\ \text{MHz}$ are given by (see~Sec.~\ref{sec:black_box})
\begin{equation}
  A_r = -\frac{e^2}{2C_L}\left(\frac{L_\text{J}}{L+L_\text{J}}\right)^3,\ A_a = -\frac{e^2}{2C_H}\left(\frac{L}{L+L_\text{J}}\right).
  \label{eq:anharmonicities}
\end{equation}

The system is described by the Hamiltonian (see~Sec.~\ref{sec:black_box})
\begin{equation}
    \hat{H} = \hbar\omega_a\hat{a}^\dagger\hat{a}+ \hbar\omega_r\hat{b}^\dagger\hat{b}
      -\frac{A_a}{2}\hat{a}^\dagger\hat{a}^\dagger\hat{a}\hat{a}-\frac{A_r}{2}\hat{b}^\dagger\hat{b}^\dagger\hat{b}\hat{b}
    -\chi\hat{a}^\dagger\hat{a}\hat{b}^\dagger\hat{b}\ ,\\
    \label{eq:main_hamiltonian}
\end{equation}
where $\hat a$ ($\hat b$) is the annihilation operator for photons in the transmon (resonator).
The second line describes the anharmonicity or Kerr non-linearity of each mode.
The last term describes the cross-Kerr interaction.
By combining it with the first term as $(\hbar\omega_a-\chi\hat{b}^\dagger\hat{b})\hat{a}^\dagger\hat{a}$, the dependence of the transmon resonance on the number of photons in the resonator becomes apparent.

\subsubsection{Photon number resolution}
\begin{figure}[]
\centering
\includegraphics[width=0.8\textwidth]{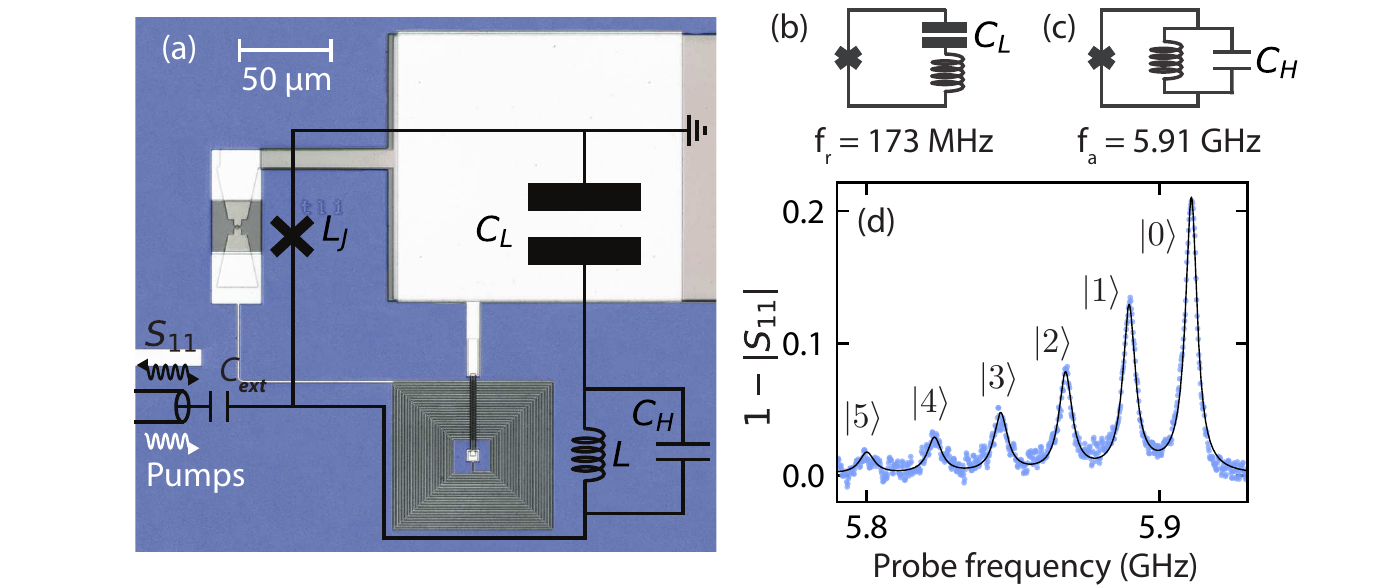}
\caption[]{
\textbf{Cross-Kerr coupling between a transmon and radio-frequency resonator. (a) }
False-colored optical micrograph of the device overlaid with the equivalent lumped element circuit. 
\textbf{(b-c) }
Effective circuit at low and high frequencies.
At low (high) frequencies, the femtofarad (picofarad) capacitances of the circuit are equivalent to open (short) circuits, and the device is equivalent to a series (parallel) JJ-inductor-capacitor combination.
The circuit has thus two modes, a near-harmonic mode at $173$~MHz and an anharmonic or transmon mode at $5.9$~GHz.
\textbf{(d) }
Microwave response $|S_{11}|$.
Through cross-Kerr coupling, quantum fluctuations of a photon number state $|n=0,1,..\rangle$ in the resonator shift the transmon transition frequency.
Peak heights are proportional to the occupation of state $|n\rangle$, and we extract a thermal occupation $n_\text{th} = 1.6$ in the resonator corresponding to a temperature of 17~mK. 
}
\label{fig:LFT_1}
\end{figure}

The cross-Kerr interaction manifests as photon-number splitting~\cite{schuster_resolving_2007} in the measured microwave reflection $S_{11}$ (Fig.~\ref{fig:LFT_1}(d)).
Distinct peaks correspond to the first transition frequency of the transmon $|g,n\rangle\leftrightarrow|e,n\rangle$, with frequencies $\omega_a-n\chi/\hbar$ where $\chi/h =21$~MHz.
We label the eigenstates of the system $|j,n\rangle$, with $j=g,e,f, ...$ ($n=0,1,2, ...$ ) corresponding to excitations of the transmon (resonator).
The amplitude of peak $n$ is proportional to
\begin{equation}
  P_n\gamma_\text{ext}/\gamma_n\ ,
  \label{eq:peak_amplitude}
\end{equation}
where $P_n$ is the occupation of photon-number level $|n\rangle$ in the resonator and $\gamma_\text{ext}/\gamma_n$ is the ratio of external coupling $\gamma_\text{ext}/2\pi=1.6\cdot 10^6s^{-1}$ to the total line-width $\gamma_n$ of peak $n$.
From the Bose-Einstein distribution of peak heights $P_n$, we extract the average photon occupation $n_\text{th} = 1.6$ corresponding to a mode temperature of $17$~mK.
The resolution of individual photon peaks is due to the strong dispersive coupling condition $\gamma_n\ll\chi/\hbar$.
The peak line-widths increase with $n$ following $\gamma_n = \gamma(1+4 n_\text{th}^{(a)})+2\kappa(n+(1+2n)n_\text{th})$, where $\gamma/2\pi=3.7\cdot 10^6s^{-1}$ is the dissipation rate of the transmon, $n_\text{th}^{(a)}\simeq0.09$ its thermal occupation (see Fig.~\ref{fig:S_temperature}), and $\kappa/2\pi=23\cdot 10^3s^{-1}$ is the dissipation rate of the resonator (obtained through time-domain measurement Fig.~\ref{fig:LFT_4}).
The condition $\gamma_n\ll A_a/\hbar$ ensures that the transmon is an artificial atom, making it possible to selectively drive the $|g,n\rangle\leftrightarrow|e,n\rangle$ and $|e,n\rangle\leftrightarrow|f,n\rangle$ transitions.
Despite its low dissipation rate $\kappa$, the resonator has a line-width of a few~MHz (measured with two-tone spectroscopy, Fig.~\ref{fig:S_two_tone}) which originates in thermal processes such as $|g,n\rangle\rightarrow|e,n\rangle$ occurring at rates $\sim\gamma n_\text{th}^{(a)}$ larger than $\kappa$ (see~Sec.~\ref{sec:LFT_supp_low}).
The resonator line-width is then an order of magnitude larger than $A_r$, making it essentially a harmonic oscillator.

\subsection{Ground state cooling}
\subsubsection{Four-wave mixing}
The junction non-linearity enables transfer of population between states by coherently pumping the circuit at a frequency $\omega_\text{p}$.
%
%
The cosine potential of the junction imposes four-wave mixing selection rules, only allowing interactions that involve 4 photons.
One such interaction is
\begin{equation}
  \begin{split}
    \hat H_\text{int}=&-\hbar J\sqrt{n+1}|f,n\rangle\langle g,n+1|+h.c.\ ,\\
  \end{split}
  \label{eq:cooling_int}
\end{equation}
activated when driving at the energy difference between the two coupled states $\omega_\text{p}=2\omega_a-\omega_r-\left(2n\chi+A_a\right)/\hbar$.
This process, enabled by a pump photon, annihilates a photon in the resonator and creates two in the transmon. 
The number of photons involved in the interaction is four, making it an allowed four-wave mixing process.
The induced coupling rate is $J=A_a^{\frac{3}{4}}A_r^{\frac{1}{4}}\xi_\text{p}$, where $|\xi_\text{p}|^2$ is the amplitude of the coherent pump tone measured in number of photons (see~Sec.~\ref{sec:LFT_fourwave}).

\begin{figure}[]
\centering
\includegraphics[width=0.8\textwidth]{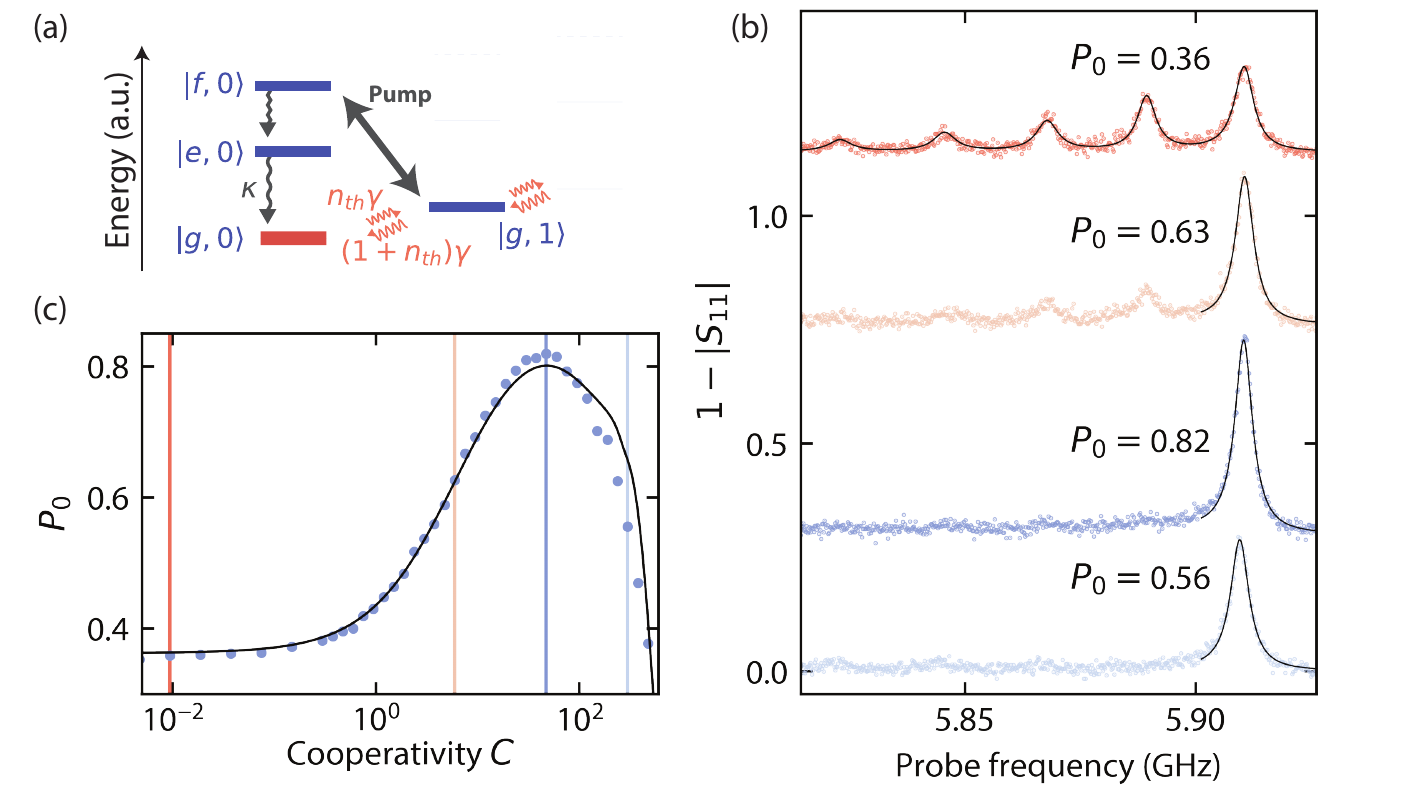}
\caption[]{
\textbf{Ground-state cooling of the radio-frequency resonator. (a) }
Energy ladder of the coupled transmon and resonator.
Meandering arrows indicate relaxation and thermal processes. 
The resonator is cooled by driving a transition (black arrow) that transfers excitations from the resonator to the transmon, where they are quickly dissipated.
\textbf{(b) }
Photon-number spectroscopy of the resonator for different cooperativities $C$ (proportional to cooling-pump power).
$C = 0.01,\ 6,\ 47,\ 300$ from top to bottom.
Ground-state occupations $P_0$ are extracted from Lorentzian fits (black curves). 
\textbf{(c) }
Vertical lines indicate the the datasets of panel (b).
A simulation (curve) predicts the measured (dots) high-$C$ decrease of $P_0$ through the off-resonant driving of other sideband transitions.
}
\label{fig:LFT_2}
\end{figure}

\subsubsection{Cooling mechanism}
We use this pump tone in combination with the large difference in mode relaxation rates to cool the megahertz resonator to its ground-state (Fig.~\ref{fig:LFT_2}(a)).
The pump drives transitions between $|g,1\rangle$ and $|f,0\rangle$ at a rate $g$.
The population of $|g,1\rangle$, transfered to $|f,0\rangle$, subsequently decays at a rate $2\gamma$ to the ground-state $|g,0\rangle$.
Cooling occurs when the thermalization rate of the resonator $n_\text{th}\kappa$ is slower than the rate $C\kappa$ at which excitations are transfered from $|g,1\rangle$ to $|g,0\rangle$, where $C = 2J^2/\gamma\kappa$ is the cooperativity (proportional to cooling-pump power (see~Sec.~\ref{sec:LFT_fourwave})).
For different cooling pump strengths, we measure $S_{11}$ (Fig.~\ref{fig:LFT_2}(b)).
The pump frequency is adapted at each power since the AC-stark effect increasingly shifts the transmon frequency as a function of power (see Fig.~\ref{fig:S_cooling}).
The data is fitted to a sum of complex Lorentzians, with amplitudes given by Eq.~(\ref{eq:peak_amplitude}) and line-widths $\gamma_n$, from which $P_n$ is extracted.
Thermal effects lead to the ratio $P_{n+1}/P_n = n_\text{th}/(1+n_\text{th})$ between neighboring photon-number states for $n\ge 1$, and the cooling pump changes the ratio of occupation of the first two states
\begin{equation}
  \frac{P_1}{P_0} \simeq \frac{n_\text{th}}{1+n_\text{th}+C}\ .
\label{eq:main_P1_over_P0}
\end{equation}
The ground-state occupation hence increases with cooperativity and we attain a maximum $P_0=0.82$.
At higher cooperativity, $P_0$ diminishes due to the off-resonant driving of other four-wave mixing processes such as $|f,n+1\rangle\langle g,n|+h.c.$ which tend to raise the photon number of the resonator.
This effect is simulated using an adaptive rotating-wave approximation~\cite{baker2018adaptive} (Fig.~\ref{fig:LFT_2}(c) and \ref{fig:S_tree}).

\subsection{Fock-state stabilization}

\begin{figure}[]
\centering
\includegraphics[width=0.8\textwidth]{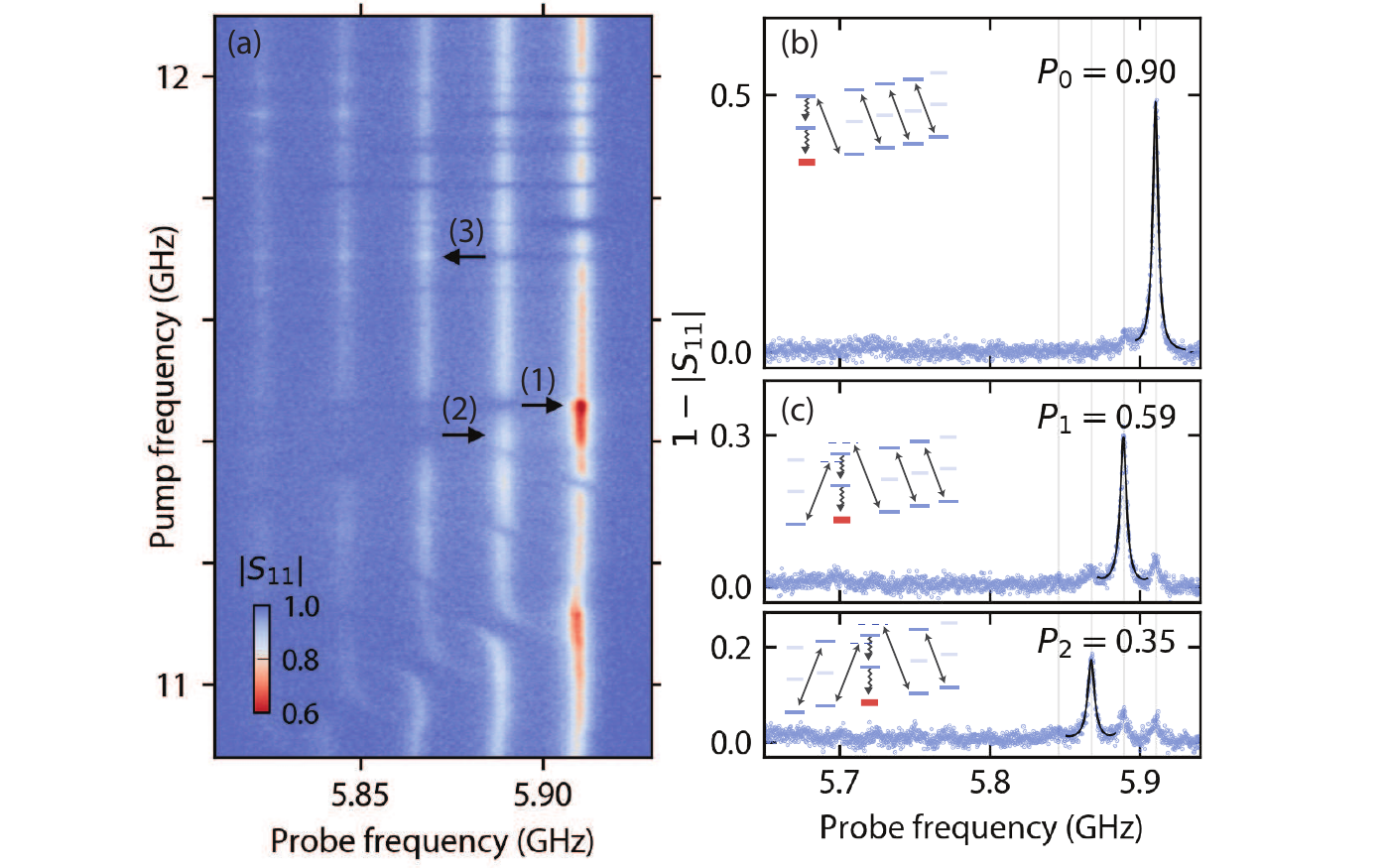}
\caption[]{
\textbf{Enhanced cooling and Fock-state stabilization using multiple tones. }
\textbf{(a)} $|S_{11}|$ as a function of pump and probe frequency.
Vertical lines correspond to the photon-number splitted transmon frequencies. 
Horizontal and diagonal features appear at pump frequencies enabling the transfer of population between Fock states of the resonator. 
Arrows indicate three example transitions:
\textbf{(1)} the cooling transition of Fig.~\ref{fig:LFT_2},
\textbf{(2)} the transition $|g,2\rangle\leftrightarrow|f,1\rangle$ transferring $|2\rangle$ to $|1\rangle$, and
\textbf{(3)} the transition $|g,1\rangle\leftrightarrow|f,2\rangle$ which raises $|1\rangle$ to $|2\rangle$.
\textbf{(b) }
By simultaneously driving four cooling transitions ($|g,n+1\rangle\leftrightarrow|f,n\rangle$), cooling is enhanced to $P_0 = 0.9$. 
\textbf{(c) } 
Using these transitions in conjunction with raising transitions $|g,n\rangle\leftrightarrow|f,n+1\rangle$, we stabilize Fock states 1 and 2 with $59\%$ and $35\%$ fidelity.
We fit a sum of complex Lorentzians to the spectrum, showing only the relevant Lorentzian (black curve) whose amplitude provides $P_n$.
Off-resonant driving results in population transfer to higher energy states visible as features in the lower frequencies of the spectrum.
}
\label{fig:LFT_3}
\end{figure}

\subsubsection{Four-wave mixing spectrum and enhanced cooling}
Neighboring four-wave mixing processes are measured by sweeping the pump frequency whilst monitoring the spectrum (Fig.~\ref{fig:LFT_3}(a)).
When cooling with a single pump they eventually limit performance, but can be resonantly driven to our advantage.
By driving multiple cooling interactions $|g,n\rangle\leftrightarrow|f,n-1\rangle$, less total pump power is required to reach a given ground-state occupation, hence minimizing off-resonant driving.
By maximizing the ground-state peak amplitude as a function of the power and frequency of four cooling tones, we achieve $P_0 = 0.90$ (Fig.~\ref{fig:LFT_3}(b)).
An identical ground-state occupation would be achieved by thermalizing the low frequency mode to a $T_\text{cool} = 3.5$ mK bath.

\subsubsection{Fock state stabilization}
By combining cooling $|g,n\rangle\leftrightarrow|f,n-1\rangle$ and raising $|g,n\rangle\leftrightarrow|f,n+1\rangle$ tones (inset of Fig.~\ref{fig:LFT_3}(c)), we demonstrate stabilization of higher Fock states, non-Gaussian states commonly considered as non-classical phenomena.
The optimum frequencies for the raising and cooling tones adjacent to the stabilized state were detuned by a few~MHz from the transition frequency (see dashed lines in the inset of Fig.~\ref{fig:LFT_3}(c)), otherwise one pump tone would populate the $|f\rangle$ level, diminishing the effectiveness of the other.

\subsection{Fock-state-resolved thermalization dynamics}
\begin{figure}[]
\centering
\includegraphics[width=0.4\textwidth]{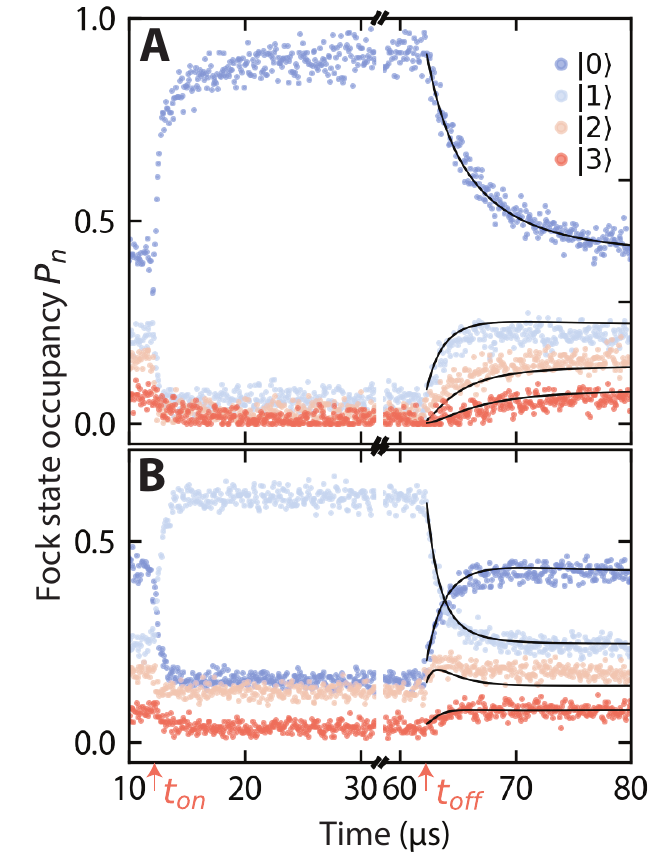}
\caption[]{\textbf{Fock-state-resolved thermalization-dynamics of the resonator.}
At $t_\text{on}$, pumps are turned on and the resonator evolves into the ground-state (\textbf{(a)}) or a single photon state (\textbf{(b)}). 
At $t_\text{off}$, control is released and we observe photon-number resolved thermalization of the resonator. 
The extracted Fock-state occupation (dots) is fitted to Eq.~\ref{eq:main_rate_equation} (black curve). 
}
\label{fig:LFT_4}
\end{figure}
Finally we investigate dynamics in a photon resolved manner (Fig.~\ref{fig:LFT_4}).
Whilst probing $S_{11}$ at a given frequency, we switch the cooling or single photon stabilization pumps on and off for intervals of $50\ \mu$s.
We perform this for a sequence of probe frequencies, resulting in $S_{11}$ as a function of both frequency and time (see full spectrum in Fig.~\ref{fig:S_temperature}).
The spectrum is fitted at each time to extract $P_n$ as a function of time.
After reaching the steady state, the pumps are turned off and we observe the thermalization process which follows the semi-classical master equation
\begin{equation}
\begin{split}
  \dot P_n &= -n \kappa (n_\text{th}+1) P_n + n\kappa n_\text{th}P_{n-1}\\
  & -(n+1)P_n\kappa n_\text{th} + (n+1)P_{n+1}\kappa (n_\text{th}+1)\ .
\end{split}
\label{eq:main_rate_equation}
\end{equation}

\section{Fabrication}
The circuit is fabricated on a high resistivity silicon substrate using aluminum as a superconductor.
Using shadow evaporation, we first pattern Al/AlOx/Al Josephson junctions, the bottom plate of the capacitor and an underpass (a line connecting the center of the spiral inductor to the capacitor).
We then deposit $260$ nm of hydrogenated amorphous silicon (a-Si:H) as a dielectric layer, motivated by its expected low dielectric loss~\cite{o2008microwave}, using PECVD (plasma enhanced chemical vapor deposition)
The a-Si:H is patterned to form a dielectric layer for the parallel plate capacitor, a bridge over the spirals underpass, and a protection layer above the junctions.
Finally we sputter-deposit and pattern aluminum to form the rest of the circuitry, after an argon-milling step to ensure a galvanic connection to the first aluminum layer.
The resulting circuit is shown in detail in Fig.~\ref{fig:S_setup}.

\section{Experimental setup}
\begin{figure}[]
\centering
\includegraphics[width=0.8\textwidth]{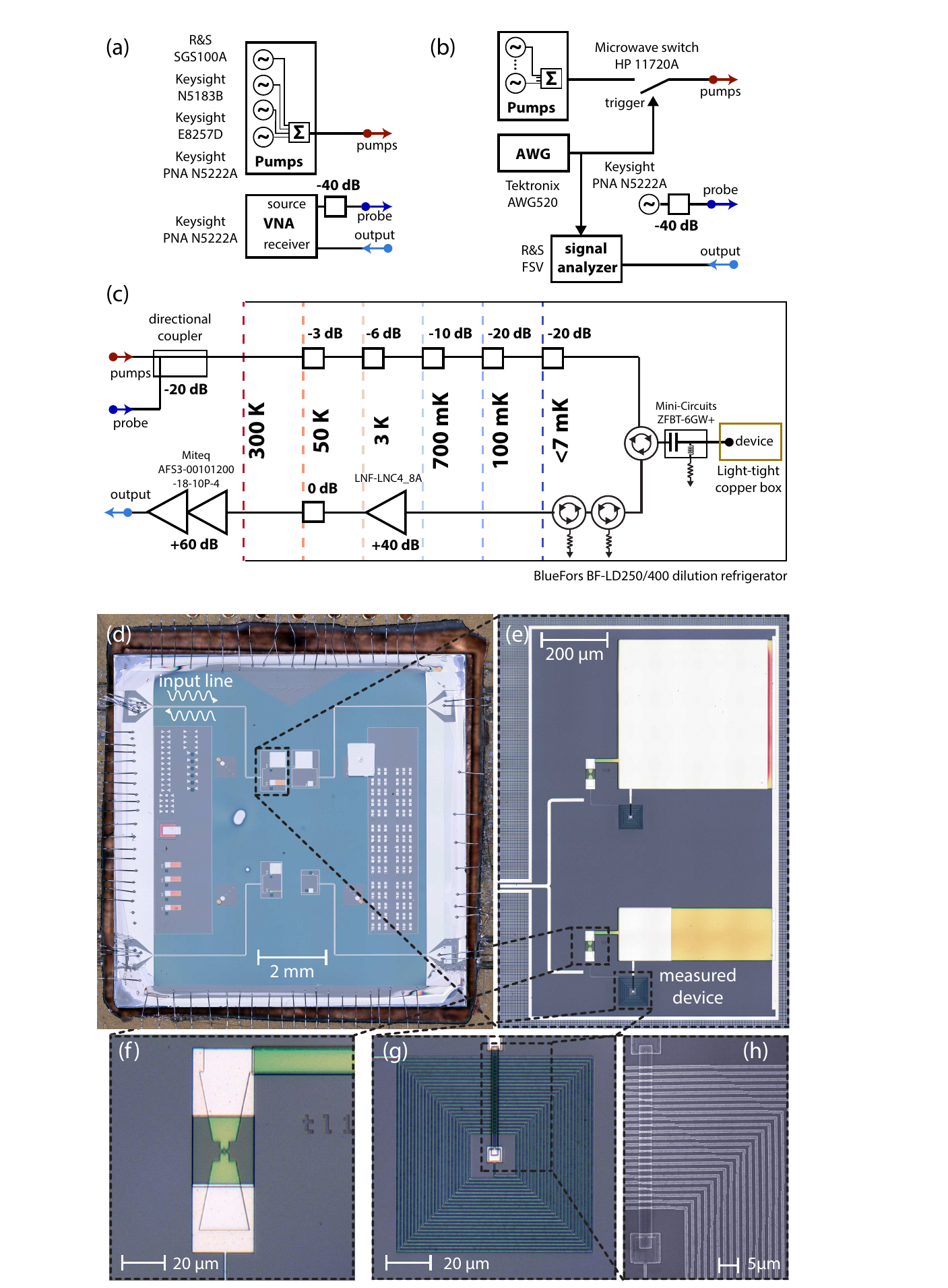}
\caption[Experimental setup and device]{\textbf{Experimental setup and device. (a) }
room temperature setup for spectroscopy experiments, \textbf{(b) }room temperature setup for time-domain experiments.
These setups are connected to the fixed setup shown in \textbf{C}.
\textbf{(c) }Cryogenic setup.
\textbf{(d) }Optical image of the chip, wire-bonded to a surrounding printed circuit board (PCB).
The PCB is mounted in a copper box which is cooled below $7$~mK (\textit{i.e.} under the range of our fridge thermometry) in our dilution refrigerator.
\textbf{(e) }Optical image of the two circuits connected to the measured feedline.
Due a small cross-Kerr to line-width ratio, photon-number splitting was not achieved in the top device, where the low (high) mode was designed to resonate at $\sim50$~MHz ($\sim7.2$ GHz).
\textbf{(f) }Optical image of the SQUID, under a protective a-Si:H layer to avoid damage from Ar milling in the last step.
\textbf{(g-h) }Optical and SEM image of the 23-turn spiral inductor which has a $1.5\ \mu$m pitch and a $500\ $nm wire width.
}
\label{fig:S_setup}
\end{figure}

A complete description of the experimental setup is provided in figure~\ref{fig:S_setup}.

\section{Data filtering}
In Figs~1(a), 3A, \ref{fig:S_caveats}B, \ref{fig:S_temperature}B,C, \ref{fig:S_all_transitions}, \ref{fig:S_time_domain}A,B,C we applied a Gaussian filter with a standard deviation of one increment in the x-axis (and y-axis when applicable).
The filtering was used in the construction of the figure for clarity.
No filtering was applied before fitting the data.

%

\section{Theory}

\setlength{\parskip}{1em}

\begin{figure}[]
\centering
\includegraphics[width=0.8\textwidth]{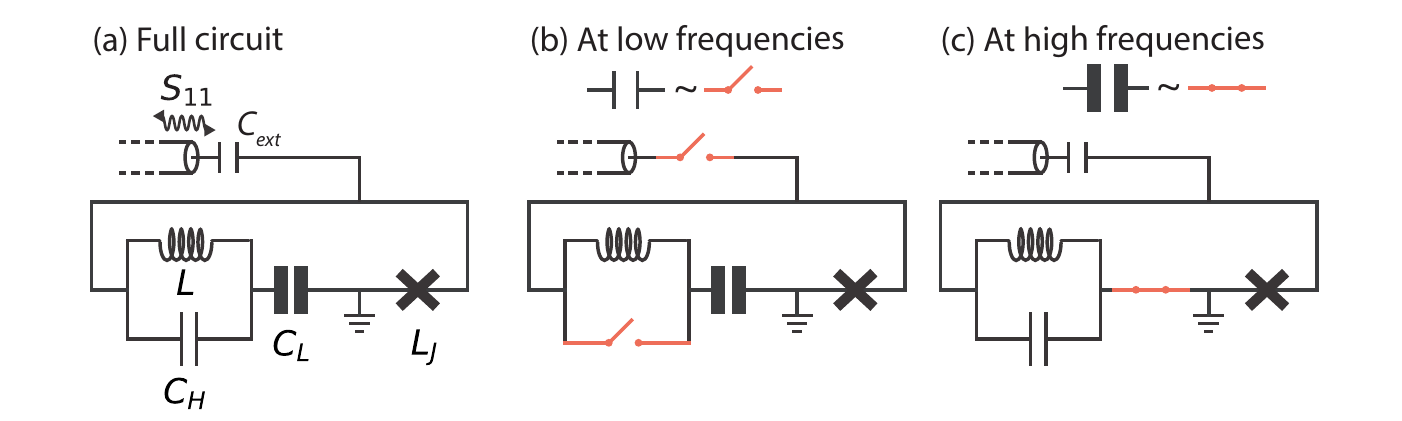}
\caption[Detailed circuit diagram]{The circuit studied in this work (a) and approximate circuits for the low-frequency (b) and high-frequency regime (c).
}
\label{fig:S_circuit}
\end{figure}

\subsection{Circuit Hamiltonian}
\label{sec:black_box}

\begin{figure}[]
\centering
\includegraphics[width=0.8\textwidth]{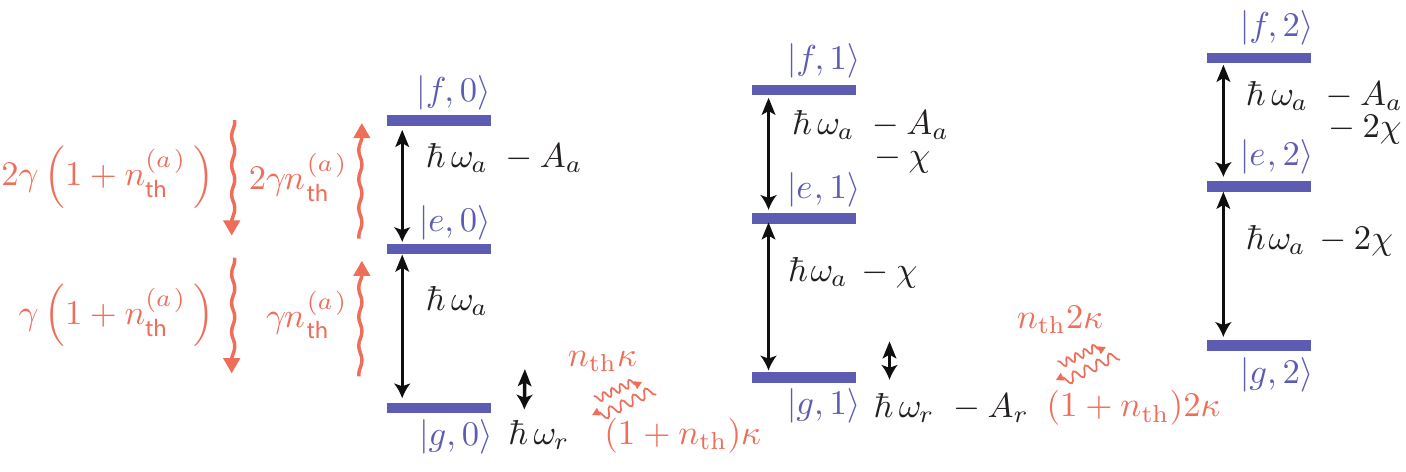}
\caption[Detailed energy diagram]{\textbf{Detailed energy diagram of the system}. 
We depict the first three levels of both high and resonator along with their dissipation and thermalization rates.
Transition energies are written with $\hbar=1$.
}
\label{fig:energy_diagram}
\end{figure}

In this section, we derive the Hamiltonian for the circuit shown in Fig.~\ref{fig:S_circuit}(a) using the black-box quantization method~\cite{nigg_black-box_2012}.
This method allows the systematic derivation of the resonance frequency $\bar\omega_m$ and anharmonicity $A_m$ of the different modes $m$ of a circuit from the admittance $Y(\omega) = 1/Z(\omega)$ across the Josephson junction if we replace the latter by a linear inductor $L_\text{J} = \hbar^2/4 e^2 E_\text{J}$.
The resonance frequencies $\bar\omega_m$ are the zeros of the admittance $Y(\bar\omega_m) = 0$, and the anharmonicities are given by 
\begin{equation}
	A_m = -\frac{2e^2}{L_\text{J}\bar\omega_m^2(\text{Im}Y'(\bar\omega_m))^2}\ .
	\label{eq:anharmonicity}
\end{equation}
The idea is to quantify through $A_m$ the amount of current traversing the Josephson junction for an excitation in mode $m$.
The Hamiltonian of the circuit is then
\begin{equation}
	\begin{split}
	\hat{H} &= \sum_m\hbar\bar\omega_m\hat{a}_m^\dagger\hat{a}_m + \underbrace{E_\text{J}[1-\cos{\hat{\varphi}}]-E_\text{J}\frac{\hat{\varphi}^2}{2}}_\text{junction non-linearity}\ ,\\
	\text{where }\hat{\varphi} &= \sum_m\left(2 A_m/E_\text{J}\right)^{1/4}(\hat{a}_m^\dagger+\hat{a}_m)\ .
	\label{eq:Hamiltonian_8th_order}
\end{split}
\end{equation}
In the circuit of Fig.~\ref{fig:S_circuit}(a) , there are two modes, a high-frequency one and a low-frequency one.
By comparing to a black-box quantization of the full circuit, we find that taking the approximation of $C_H\simeq0$, $C_\text{ext}\simeq0$ for the low-frequency mode and $C_L\simeq\infty$ for the Transmon results in corrections of only $0.2$, $1.2$, $0.3$ and $2.1$ \% in the value of $\omega_r$, $\omega_a$, $A_r$ and $A_a$ respectively.
It is therefore a good approximation, which has the additional advantage of producing simple analytical equations for the frequencies and anharmonicities of the circuit.
Starting with the low-frequency mode shown in Fig.~\ref{fig:S_circuit}(b), we find the (imaginary part of the) admittance across the linearized junction to be
\begin{equation}
	\text{Im}Y(\omega) = \frac{1}{\omega L_J}\frac{\left(\frac{\omega}{\omega_r}\right)^2-1}{1-\left(\omega\sqrt{LC_L}\right)^2}\ ,
\end{equation}
yielding the resonance frequency
\begin{equation}
	\omega_r = \frac{1}{\sqrt{(L+L_J)C_L}}\ .
	\label{eq:wl}
\end{equation}
Taking the derivative of the imaginary part of the admittance at $\omega = \omega_r$ yields:
\begin{equation}
	\text{Im}\frac{\partial Y}{\partial \omega}(\omega_r) = 2C_L\left(\frac{L+L_J}{L_J}\right)^2
\end{equation}
Substituting this into Eq.~(\ref{eq:anharmonicity}) yields 
\begin{equation}
	A_r = -\frac{e^2}{2C_L}\left(\frac{L_J}{L+L_J}\right)^3\ .
	\label{eq:Al}
\end{equation}
Turning to the Transmon shown in Fig.~\ref{fig:S_circuit}(c), we find the (imaginary part of the) admittance across the linearized junction to be 
\begin{equation}
	\text{Im}Y(\omega) = C_H\omega\left(1-\frac{\omega_a^2}{\omega^2}\right)\ ,
\end{equation}
yielding the resonance frequency
\begin{equation}
	\omega_a = \sqrt{\frac{L+L_J}{LL_JC_H}}\ .
	\label{eq:wh}
\end{equation}
Taking the derivative of the imaginary part of the admittance at $\omega = \omega_a$ yields:
\begin{equation}
	\text{Im}\frac{\partial Y}{\partial \omega}(\omega_a) = 2C_H
\end{equation}
Substituting this into Eq.~(\ref{eq:anharmonicity}) yields 
\begin{equation}
	A_a = -\frac{e^2}{2C_H}\left(\frac{L}{L+L_J}\right)\ .
	\label{eq:Ah}
\end{equation}
A Taylor expansion of the junctions cosine potential is justified if the anharmonicities are weak and only a few photons populate the circuit.
Whilst numerical calculations in this work consider the 8-th order expansion, much understanding can be gleaned by stopping the expansion at the fourth-order
\begin{equation}
\begin{split}
		\hat{H}_{4,\text{diag}} =& \hbar\omega_a\hat{a}^\dagger\hat{a} -\frac{A_a}{2}\hat{a}^\dagger\hat{a}^\dagger\hat{a}\hat{a}\\
		&+ \hbar\omega_r\hat{b}^\dagger\hat{b} -\frac{A_r}{2}\hat{b}^\dagger\hat{b}^\dagger\hat{b}\hat{b}\\
		&-\chi\hat{a}^\dagger\hat{a}\hat{b}^\dagger\hat{b}\ ,\\
		\label{eq:H4_diag}
\end{split}
\end{equation}
where $\chi$ is the cross-Kerr coupling: the amount by which the transmon transition shifts as a result of adding an excitation in the resonator and vice versa.
We defined the first transition frequencies of both modes
\begin{equation}
\begin{split}
	\hbar\omega_a=\hbar\bar\omega_a-A_a-\frac{\chi}{2}\ ,&\\
	\hbar\omega_r=\hbar\bar\omega_r-A_r-\frac{\chi}{2}\ .&\\
\end{split}
\end{equation}
In Eq.~(\ref{eq:H4_diag}), we have neglected terms in the expansion which are off-diagonal in the Fock basis and do not modify the eigenenergies to leading order perturbation theory.
The eigenfrequencies of the system are summarized in the energy diagram of Fig.~\ref{fig:energy_diagram}

\subsection{Translating the measured $S_{11}$ to a quantum operator}

We now introduce a driving term in the Hamiltonian and consider losses to both the environment and the measurement port.
Following input-output theory~\cite{vool2017introduction,clerk2010introduction}, the quantum Langevin equation for $\hat{a}(t)$ is
\begin{equation}
	\frac{\text{d}}{\text{d}t}\hat{a}(t) = \frac{i}{\hbar}[\hat{H}_\text{undr},\hat{a}(t)] - \frac{\gamma}{2}\hat{a}(t)+\sqrt{\gamma_\text{ext}}\tilde{a}_\text{in}(t)\ .
\end{equation}
Where the undriven Hamiltonian $\hat{H}_\text{undr}$ corresponds to that of Eq.~(\ref{eq:Hamiltonian_8th_order}), where the degree of expansion of the non-linearity is yet unspecified.
The microwave reflection measured in spectroscopy (here in the time-domain) is given by
\begin{equation}
	S_\text{11}(t) = \frac{\tilde{a}_\text{out}(t)}{\tilde{a}_\text{in}(t)} = 1-\sqrt{\gamma_\text{ext}}\frac{\hat{a}(t)}{\tilde{a}_\text{in}(t)}\ ,
	\label{eq:in_out}
\end{equation}
where $\tilde{a}_\text{in}(t)$ ($\tilde{a}_\text{out}(t)$) is the incoming (outgoing) field amplitude, $\gamma_\text{ext}$ ($\gamma$) is the external (total) coupling rate of the Transmon.
The coupling of the resonator to the feedline $\kappa_\text{ext}/2\pi=2s^{-1}$ is much smaller than coupling of the transmon to the feedline $\gamma_\text{ext}/2\pi=1.63\cdot10^6s^{-1}$, we therefore assume that a drive tone only affects the Transmon.
For a coherent drive, characterized by a drive frequency $\omega_\text{d}$ and an incoming power $P_\text{in}$ (equal to the average power $\langle P(t) \rangle$ of the oscillating input signal), the wave amplitude is
\begin{equation}
	\tilde{a}_\text{in}(t)=\sqrt{\frac{P_\text{in}}{\hbar\omega_\text{H}}}  e^{-i\omega_\text{d} t}\ ,
\end{equation}
and the drive term can be incorporated in the Hamiltonian of the system
\begin{align}
\begin{split}
	\frac{\text{d}}{\text{d}t}\hat{a}(t) &= \frac{i}{\hbar}[\hat{H}_\text{undr}+\hat{H}_\text{drive},\hat{a}(t)] - \frac{\gamma_\text{ext}}{2}\hat{a}(t)\ ,\\
	\text{where }\hat{H}_\text{drive} &= i\hbar\epsilon_\text{d}\left(e^{-i\omega_\text{d} t}\hat{a}^\dagger(t) - e^{i\omega_\text{d} t}\hat{a}(t)\right)\ ,\\
	\epsilon_\text{d} &= \sqrt{\frac{\gamma_\text{ext}P_{in}}{\hbar\omega_\text{H}}}\ .
\end{split}
\end{align}
%
Additionally, we also remove the time-dependence in the drive Hamiltonian by moving to a frame rotating at $\omega_\text{d}$ with the unitary transformation $U_\text{probe} = e^{i\omega_\text{d} t \hat{a}^\dagger\hat{a}}$, 
\begin{equation}
	\frac{\text{d}}{\text{d}t}\hat{a} = \frac{i}{\hbar}[U_\text{probe}^\dagger\hat{H}_\text{undr}U_\text{probe}+\tilde{H}_\text{drive},\hat{a}] - \frac{\gamma_\text{ext}}{2}\hat{a}\ ,
	\label{eq:langevin}
\end{equation}
where $\hat a e^{i\omega_d t} = \hat a(t)$ and 
\begin{equation}
	\tilde{H}_\text{drive} =  -\hbar\omega_\text{d} \hat{a}^\dagger  \hat{a}\\
	+i\hbar\epsilon_\text{d}\left(\hat{a}^\dagger -\hat{a}\right)\ .
	\label{eq:drive_hamiltonian}
\end{equation}
In this rotating frame, the reflection coefficient becomes
\begin{equation}
	\hat{S}_\text{11}(\omega_\text{d}) = 1-\frac{\gamma_\text{ext}}{\epsilon_\text{d}}\hat{a}\ ,
	\label{eq:in_out_freq}
\end{equation}
of which we measure the expectation value when probing the system.
From now on, and in the main text we use the shorthand $S_{11}(\omega_\text{d}) = \langle\hat{S}_\text{11}(\omega_\text{d})\rangle$.
Note that by casting the quantum Langevin Eq.~(\ref{eq:langevin}) in the form
\begin{equation}
\begin{split}
	\frac{\text{d}}{\text{d}t}\hat{a} =& \frac{i}{\hbar}[U_\text{probe}^\dagger\hat{H}_\text{undr}U_\text{probe}+\hat{H}_\text{drive},\hat{a}]\\ 
	+&\left(L^{\dagger}\hat{a}L - \frac{1}{2}\left(\hat{a}L^{\dagger}L + L^{\dagger}L\hat{a}\right)\right)\ ,\\
	\text{where }L=&\gamma_\text{ext} \hat{a}\ ,
\end{split}
\end{equation}
it can be readily transformed to a Lindblad equation
\begin{equation}
\begin{split}
	\frac{\text{d}}{\text{d}t}\rho =& -\frac{i}{\hbar}[U_\text{probe}^\dagger\hat{H}_\text{undr}U_\text{probe}+\hat{H}_\text{drive},\rho] \\
	+& \left(L \rho L^{\dagger}-\frac{1}{2}\left(\rho L^{\dagger}L + L^{\dagger}L\rho\right)\right)\ \\
	=& -\frac{i}{\hbar}[U_\text{probe}^\dagger\hat{H}_\text{undr}U_\text{probe}+\hat{H}_\text{drive},\rho] + \gamma_\text{ext}\mathcal{L}[\hat{a}]\ ,
	\label{eq:probed_lindbald_early}
\end{split}
\end{equation}
better suited to numerical calculations using QuTiP~\cite{johansson2012qutip}.

\subsection{Derivation of reflection coefficient}
\begin{figure}[]
\centering
\includegraphics[width=0.8\textwidth]{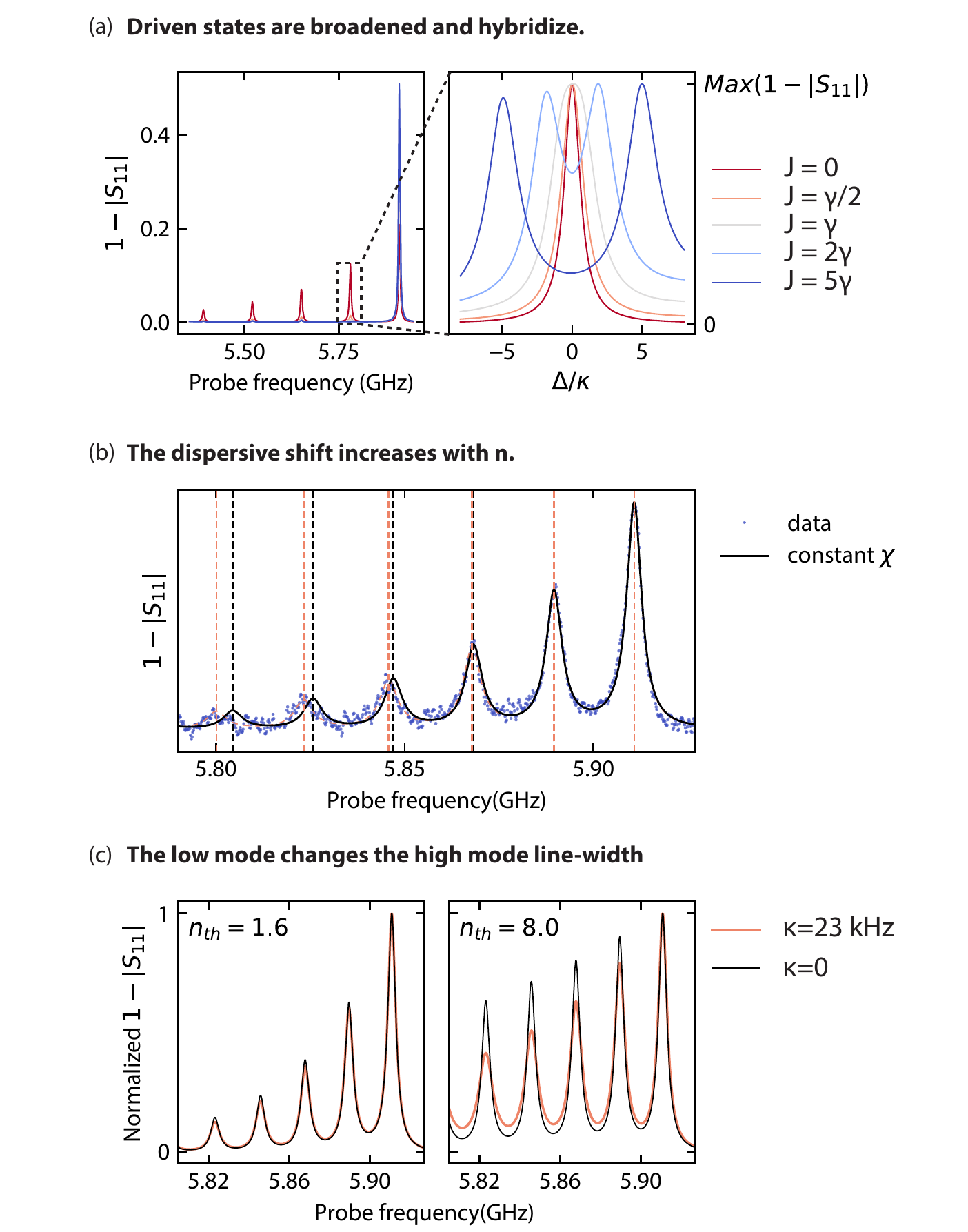}
\caption[Possible caveats in fitting a sum of Lorentzians to $S_{11}$]{\textbf{Possible caveats in fitting a sum of Lorentzians to $S_{11}$}.
\\\hspace{0.8\textwidth}
\textbf{(a) driven states are broadened then hybridize.}
As we increase the coupling $J$ induced by a cooling pump resonant with $\ket{g,1}\leftrightarrow\ket{f,0}$,the resonator is cooled as shown in the left panel.
In the right panel, we zoom in to the normalized $n=1$ peak.
As a consequence of the coupling between levels $\ket{g,1}$ and $\ket{f,0}$, this peak first broadens then splits into two distinct peaks,
The slight asymmetry arises from the tail of the $n=0$ peak.
We used the device parameters with an increased $\chi= h\times130$~MHz and $A_a= h\times600$~MHz in order to minimize the visibility of the tail of the other peaks.
\\\hspace{0.8\textwidth}
\textbf{(b) the dispersive shift increases with $n$.}
The Hamiltonian presented in Eq.~(2), which only considers the diagonal contributions of the quartic term of the JJ non-linearity, results in a constant shift of the transmon frequency $\omega_a-n\chi/\hbar$.
As shown in black, overlaid on the blue dots of the same data as in panel (a), this results in a slight misalignment of the peaks.
By diagonalizing the Hamiltonian of Eq.~(\ref{eq:Hamiltonian_8th_order}), with the JJ non-linearity Taylor expanded to the 8-th order, we achieve a more realistic prediction of the system frequencies, and find that the shift increases with the number of photons in the resonator, as shown with red lines.
\\\hspace{0.8\textwidth}
\textbf{(c) $\kappa$ and $n_\text{th}$ modify the transmon line-width $\gamma_{j,n}$.}
As shown in Eq.~(\ref{eq:sum_of_lorentzians}), the transmon line-width not only depends on transmon dissipation rate $\gamma$, but also on the dissipation $\kappa$ and thermal occupation $n_\text{th}$ of the resonator.
As $\kappa\ll\gamma$, this effect is subtle for low thermal occupations, but if neglected, can lead to an underestimation of the resonator occupation at higher temperatures.
}
\label{fig:S_caveats}
\end{figure}

In this section we derive the spectrum of the transmon for arbitrary states of the resonator.
We append the Lindblad equation of Eq.~(\ref{eq:probed_lindbald_early}) to take into account additional interactions of the system with the environment.
Internal dissipation of the transmon $\gamma_\text{int}$, is added to the external dissipation rate to constitute its total dissipation rate $\gamma = \gamma_\text{int}+\gamma_\text{ext}$.
The resonator is attributed a dissipation rate $\kappa$.
The average thermal occupation of the two modes are denoted by $n_\text{th}^{(a)}$ and $n_\text{th}$ for the high and resonator respectively.
We can estimate the response function $S_{11}(\omega_\text{d})$ analytically using the Hamiltonian of Eq.~(\ref{eq:H4_diag}).
The unitary $U_\text{probe}$ leaves this Hamiltonian unchanged and the complete Lindblad equation is then
\begin{equation}
\begin{split}
	\frac{\text{d}}{\text{d}t} \rho = &-\frac{i}{\hbar}[H_\text{4,diag}+\hat{H}_\text{drive},\rho] \\
	&+\gamma (n_\text{th}^{(a)}+1) \mathcal{L}[\hat{a}]
	+\gamma n_\text{th}^{(a)} \mathcal{L}[\hat{a}^\dagger]\\
	&+\kappa (n_\text{th}+1) \mathcal{L}[\hat{b}]
	+\kappa n_\text{th} \mathcal{L}[\hat{b}^\dagger]\ .
	\label{eq:probed_lindbald}
\end{split}
\end{equation}
In the un-driven case $\epsilon_\text{d}=0$, we assume the steady-state solution to be a diagonal density matrix $\rho_\text{ss}$ as a consequence of thermal effects
\begin{equation}
	\rho_\text{ss} = 
\begin{bmatrix}
    P_g & 0&0 &    \\
    0 & P_e & 0 &   \\
    0 & 0 & P_f &   \\
     &  &  & \ddots \\
\end{bmatrix}_H
\otimes
\begin{bmatrix}
    P_0 & 0&0 &    \\
    0 & P_1 & 0 &   \\
    0 & 0 & P_2 &   \\
     &  &  & \ddots \\
\end{bmatrix}_L\ ,
\label{eq:thermal_steady_state}
\end{equation}
where $P_g,P_e,P_f$ ($P_0,P_1,P_2$) corresponds to the occupation of high (low) mode levels.
Note that when we pump the system, effectively coupling levels of the high and resonator, this approximation breaks down and that particular limit is discussed below.
We now look for a perturbative correction to this matrix at a small driving rate $\epsilon_\text{d}$
\begin{equation}
	\rho = \rho_\text{ss} + \epsilon_\text{d} \rho_\text{pert} \ ,
\end{equation}
where $\rho_\text{pert}$ has the unit time. 
The objective is to determine the expectation value of the reflection coefficient 
\begin{equation}
	S_\text{11}(\omega_\text{d}) = \text{Tr}\left[\rho \left(1-\frac{\gamma_\text{ext}\hat{a}}{\epsilon_\text{d}}\right)\right] \ .
\end{equation}
We substitute the perturbative expansion of $\rho$ into Eq.~(\ref{eq:probed_lindbald}) and keep only terms to first order in $\epsilon_\text{d}$.
This equation is solved analytically in reduced Hilbert-space sizes using the software Wolfram Mathematica.
The largest Hilbert-space sizes for which Mathematica could provide an analytical solution in a reasonable amount of time were: (4,0), (3,2), (2,5) where the first (second) number designates the number of levels included in the high (low) mode.
We extrapolate the obtained results to construct the reflection coefficient
\begin{equation}
\begin{split}
	S_\text{11}(\omega_\text{d}) &= 1-(P_g-P_e)\sum_n P_n\frac{\gamma_\text{ext}}{i\Delta_{g,n}+\gamma_{g,n}}\\
	&\ \ \ \ \ -(P_e-P_f)\sum_n P_n\frac{2\gamma_\text{ext}}{i\Delta_{e,n}+\gamma_{e,n}}\ ,\\
	\text{where }\gamma_{g,n} &= \gamma(1+4 n_\text{th}^{(a)})+2\kappa(n+(1+2n)n_\text{th})\ ,\\
	\gamma_{e,n} &= \gamma(3+8 n_\text{th}^{(a)})+2\kappa(n+(1+2n)n_\text{th})\ .\\
\end{split}
\label{eq:sum_of_lorentzians}
\end{equation}
which corresponds to a sum of Lorentzian functions, each associated to transmon level $i$ and a resonator level $n$, with line-width $\gamma_{i,n}$ centered around $\Delta_{i,n} = 0$, where
\begin{equation}
 	\begin{split}
 		&\Delta_{g,n} = \omega_a-n\chi/\hbar-\omega_\text{d}\ ,\\
		&\Delta_{e,n} = \omega_a-(n\chi-A_a)/\hbar-\omega_\text{d}\ .
 	\end{split}
 	\label{eq:deltas}
 \end{equation} 
Note that in the main text we use the notation $\gamma_{g,n}=\gamma_n$.

%
By numerically computing $S_{11}$ as described in Sec.~\ref{sec:numerics_S11}, we find that the expression for the line-widths $\gamma_{i,n}$ remains accurate, whilst the center of the Lorentzians $\Delta_{i,n}$ will slightly shift from Eq.~(\ref{eq:deltas}), as shown in Fig.~\ref{fig:S_caveats}(b).
When fitting data, we hence use the Eqs.~(\ref{eq:sum_of_lorentzians}) whilst fixing $\Delta_{i,n}$ with a diagonalization of the Hamiltonian Eq.~(\ref{eq:Hamiltonian_8th_order}) Taylor expanded to the 8-th order.
In Fig.~\ref{fig:S_temperature}(c), we show that Eq.~(\ref{eq:sum_of_lorentzians}) is in excellent agreement with both data and numerics.

\subsubsection{The impact of a pump tone on $S_{11}(\omega)$}
Pump tones can invalidate Eq.~(\ref{eq:sum_of_lorentzians}) in different ways.
As an example let us take the cooling scheme where a pump tone couples the levels $\ket{g,1}$ and $\ket{f,0}$ at a rate $J$.
This is simulated by numerically finding the steady state of the Hamiltonian
\begin{equation}
\begin{split}
		\hat{H} =& \hbar\Delta\hat{a}^\dagger\hat{a} -\frac{A_a}{2}\hat{a}^\dagger\hat{a}^\dagger\hat{a}\hat{a}\\
		&+ (2\hbar\Delta-A_a-A_r)\hat{b}^\dagger\hat{b} -\frac{A_r}{2}\hat{b}^\dagger\hat{b}^\dagger\hat{b}\hat{b}\\
		&-\chi\hat{a}^\dagger\hat{a}\hat{b}^\dagger\hat{b}\\
		&+J(\ket{g,1}\bra{f,0}+\ket{f,0}\bra{g,1})\\
		&+i\hbar\epsilon_\text{d}(\hat{a}^\dagger-\hat a) ,\\
\end{split}
\end{equation}
written in a frame rotating at the probe frequency $\Delta = \omega_a-\omega_d$ and where the levels $\ket{g,1}$ and $\ket{f,0}$ are made resonant.
As shown in Fig.~\ref{fig:S_caveats}(a), a peak corresponding to a transition to or from a level which is being pumped will be broadened in line-width and eventually will split into two peaks with increasing $J$.
This is not an issue in the cooling scheme since we do not use the driven $n=1$ peak to extract Fock-state fidelity, only the $n=0$ peak.

We do however off-resonantly pump $\ket{g,0}\leftrightarrow\ket{f,1}$ for example, along with many other transitions involving either state $\ket{g,0}$ or $\ket{e,0}$.
Off-resonant pumping should also lead to line-width broadening, this time of a peak used in extracting a Fock state fidelity.
To mitigate this issue we extract $P_n$ -- when stabilizing the $n$-th Fock state -- by using a fixed line-width $\gamma_n$ defined in Eq.~(\ref{eq:sum_of_lorentzians}).
This means that we always give a lower bound to $P_n$.
By comparing the pumped and un-pumped line-width of $n=0$ peak (see Fig.~\ref{fig:S_cooling}(b)), we notice no change in line-width with increasing pump power, indicating that our underestimation is certainly not drastic.
Finally, pump tones could drive the steady-state away from our assumption of a purely diagonal density matrix Eq.~(\ref{eq:thermal_steady_state}).
However we find that in the cooling experiment of Fig.~\ref{fig:LFT_2}, the adaptive rotating-wave simulation suggests that at maximum $P_0$, all off-diagonal terms of the density matrix are below $2.3\times10^{-3}$. 
This issue can safely be disregarded.

\subsection{Four wave mixing}\label{sec:LFT_fourwave}
\subsubsection{Analytical derivation of the pump-induced coupling rates}

In this section we will consider the probe tone to be very weak and hence negligible.
Following Refs.~\cite{Leghtas2015}, we add a pump tone driving the transmon with frequency $\omega_\text{p}$ and strength $\epsilon_\text{p}$ to the system Hamiltonian
\begin{equation}
\begin{split}
	\hat{H}_\text{4,dr} &= \hbar\bar\omega_a\hat{a}^\dagger\hat{a}+\hbar\bar\omega_r\hat{b}^\dagger\hat{b} +E_\text{J}[1-\cos{\hat{\varphi}}]-\frac{E_\text{J}}{2}\hat{\varphi}^2 \\
	&+ \hbar \left(\epsilon_\text{p}e^{-i\omega_\text{p} t}+\epsilon_\text{p}^*e^{i\omega_\text{p} t}\right)\left(\hat{a}^\dagger +\hat{a}\right)\ ,\\
	\text{where }\hat{\varphi} &= \left(2 A_a/E_\text{J}\right)^{1/4}(\hat{a}^\dagger+\hat{a})+\left(2 A_r/E_\text{J}\right)^{1/4}(\hat{b}^\dagger+\hat{b})\ .
\end{split}
\end{equation}
We move to the displaced frame of the pump through the unitary transformation
\begin{equation}
	U_\text{pump} = e^{-\tilde\xi_\text{p}\hat{a}^\dagger+\tilde\xi_\text{p}^*\hat{a}}\ ,
\end{equation}
Where $\tilde\xi_\text{p}$ is defined by the differential equation
\begin{equation}
	\frac{d\tilde\xi_\text{p}}{dt}=-i\bar\omega_a\tilde\xi_\text{p}-i\left(\epsilon_\text{p}e^{-i\omega_\text{p} t}+\epsilon_\text{p}^*e^{i\omega_\text{p} t}\right)-\frac{\gamma}{2}\tilde\xi_\text{p}\ .
\end{equation}
For $t\gg1/\gamma$, and for far detuned drives $|\omega_a-\omega_\text{p}|\gg\gamma$, this equation is solved by
\begin{equation}
	\begin{split}
		\tilde\xi_\text{p}&\simeq \frac{\epsilon_\text{p}e^{-i\omega_\text{p} t}}{\omega_\text{p}-\bar\omega_a} -\frac{\epsilon^*_\text{p}e^{i\omega_\text{p} t}}{\omega_\text{p}+\bar\omega_a} \ .
	\end{split}
\end{equation}
In this frame, the Hamiltonian becomes
\begin{equation}
\begin{split}
	\hat{H}_\text{4,dr} =& \hbar\bar\omega_a\hat{a}^\dagger\hat{a}+\hbar\bar\omega_r\hat{b}^\dagger\hat{b} +E_\text{J}[1-\cos{\tilde{\varphi}}]-\frac{E_\text{J}}{2}\tilde{\varphi}^2 \\
	\text{where }\tilde{\varphi} =& \left(2 A_a/E_\text{J}\right)^{1/4}(\hat{a}^\dagger+\hat{a})+\left(2 A_r/E_\text{J}\right)^{1/4}(\hat{b}^\dagger+\hat{b})\\
	& +\left(2 A_a/E_\text{J}\right)^{1/4}(\xi_\text{p}^*+\xi_\text{p})
\end{split}
\end{equation}
where $\xi$ is defined such that $\xi_\text{p}^*+\xi_\text{p} = \tilde\xi_\text{p}^*+\tilde\xi_\text{p}$ but $\xi$ only oscillates at on frequency, $\xi = |\xi|e^{-i\omega_\text{p} t}$
We now Taylor expand the cosine non-linearity to fourth-order, neglecting terms which are off-diagonal in the Fock basis except when they depend on $\xi_\text{p}$ .
The latter can be made relevant depending on our choice of $\omega_\text{p}$.
\begin{equation}
\begin{split}
		\hat{H}_{4,\text{pumped}} &= \hat{H}_{4,\text{diag}} + \hat{H}_\text{p}\ ,\\
\end{split}
\label{eq:hamiltonian_pumped}
\end{equation}
Where $\hat{H}_{4,\text{diag}}$ was given in Eq.~(\ref{eq:H4_diag}).
The terms dependent on the pump power and frequency are assembled in the term $\hat{H}_\text{p}$ and written in Table \ref{tab:pump_terms}, along with the approximate pumping frequency $\omega_\text{p}$ necessary to eliminate their time-dependence.
As shown in the next paragraph, this occurs when the pump frequency matches the transition frequency between the two states coupled by the interaction term.

\begin{table}[]
\centering
\caption[Four-wave mixing terms]{\textbf{Four-wave mixing terms}
Only half of terms are shown, the other half can be obtained by taking the hermitian conjugate of all these terms.
Terms become approximately time-independent around the frequency $\omega_\text{p}$ given in the left column.
}
\label{tab:pump_terms}
\begin{tabular}{c c c}
\\
$\omega_\text{p}  \simeq $       & prefactor  & interaction \\ \hline \\
\multicolumn{3}{c}{ \textbf{Stark shift}}\\
					&$-2A_a    |\xi_\text{p}|^2$&									$\hat a^\dagger\hat a$\\
					&$-\chi    |\xi_\text{p}|^2$&									$\hat b^\dagger\hat b$\\
\multicolumn{3}{c}{\textbf{Heating interactions}}\\
$(\omega_a+\omega_r)/2$					&$-A_a^{\frac{3}{4}}A_r^{\frac{1}{4}}   (\xi_\text{p}^*)^2$&		$\hat{a}\hat{b}$\\
$\omega_a+2\omega_r$						&$-\chi\xi_\text{p}^*/2$&										$\hat{a}\hat b^2 $\\
$2\omega_a+\omega_r$					&$-A_a^{\frac{3}{4}}A_r^{\frac{1}{4}}\xi_\text{p}^*$&			$\hat a^2\hat{b}$\\
\multicolumn{3}{c}{\textbf{Cooling interactions}}\\
$(\omega_a-\omega_r)/2$					&$-A_a^{\frac{3}{4}}A_r^{\frac{1}{4}}   (\xi_\text{p}^*)^2$&		$\hat{a}\hat{b}^\dagger$\\
$\omega_a-2\omega_r$						&$-\chi\xi_\text{p}^*/2$&										$\hat{a}(\hat b^\dagger)^2$\\
$2\omega_a-\omega_r$					&$-A_a^{\frac{3}{4}}A_r^{\frac{1}{4}}\xi_\text{p}^*$&			$\hat a^2 \hat{b}^\dagger$\\
\multicolumn{3}{c}{\textbf{Unused interactions}}\\
$3\omega_a$					&$- A_a\xi_\text{p}/3$&										$\hat a^3$\\
$\omega_a/3$		 			&$-A_a (\xi_\text{p}^*)^3/3$&									$\hat a$\\
$\omega_a$					&$-A_a   (\xi_\text{p}^*)^2/2$&									$\hat a^2 $\\
$\omega_a$					&$-\chi\xi_\text{p}$&										$\hat{a}\hat b^\dagger\hat b $\\
$\omega_a$					&$-A_a\xi_\text{p}$&											$\hat a^\dagger\hat a^2$\\
$\omega_a$					&$- A_a(\xi_\text{p}^*)^3$&										$\hat{a}$\\
$\omega_a$					&$-A_a\xi_\text{p}$&											$\hat a$\\
$\omega_a$					&$-\chi\xi_\text{p}/2$&										$\hat{a}$\\
$3\omega_r$					&$-A_a^{\frac{1}{4}}A_r^{\frac{3}{4}}\xi_\text{p}/3$&		$\hat b^3$\\
$\omega_r/3$					&$-A_a^{\frac{3}{4}}A_r^{\frac{1}{4}}(\xi_\text{p}^*)^3/3$&		$\hat{b}$\\
$\omega_r$					&$-\chi    \xi_\text{p}^2/4$&								$\hat b^2 $\\
$\omega_r$					&$-2A_a^{\frac{3}{4}}A_r^{\frac{1}{4}}\xi_\text{p}$&			$\hat a^\dagger\hat a\hat{b}$\\
$\omega_r$					&$-A_a^{\frac{1}{4}}A_r^{\frac{3}{4}}\xi_\text{p}$&			$\hat b^\dagger\hat b^2$\\
$\omega_r$					&$-A_a^{\frac{3}{4}}A_r^{\frac{1}{4}}(\xi_\text{p}^*)^3$&		$\hat{b}$\\
$\omega_r$					&$-A_a^{\frac{1}{4}}A_r^{\frac{3}{4}}\xi_\text{p}$&			$\hat b$\\
$\omega_r$					&$-A_a^{\frac{3}{4}}A_r^{\frac{1}{4}}\xi_\text{p}$&			$\hat{b}$\\
\end{tabular}
\end{table}

We now move to the interaction picture through the unitary transformation 
\begin{equation}
\begin{split}
	U_\text{int}&=e^{i \hat{H}_{4,\text{diag}}t/\hbar}\ ,
\end{split}
\end{equation}
$\hat{H}_{4,\text{diag}}$ is diagonal in the Fock state basis $\left\{\ket{j,n}\right\}_{\substack{n=0,1,2,.. \\ j=g,e,f,..}}$
\begin{equation}
\begin{split}
	\hat{H}_0&=\sum_{\substack{n=0,1,2,.. \\ j=g,e,f,..}}\hbar\epsilon_{j,n}\ket{j,n}\bra{j,n}\ ,\\
	\text{where }\epsilon_{j,n}&= n\omega_r - \frac{A_r}{2\hbar}\left(n^2-n\right)\\
	&+j\omega_a - \frac{A_a}{2\hbar}\left(j^2-j\right)\\
	&-nj\chi/\hbar\ .
\end{split}
\end{equation}
To determine $\hat H_\text{p}$ in this frame, it suffices to know the expression of annihilation operators in this frame.
We will take as an example the term we use for cooling, which reads in the interaction picture
\begin{equation}
\begin{split}
	&U_\text{int}\left(-A_a^{\frac{3}{4}}A_r^{\frac{1}{4}}   (\xi_\text{p}^*)\hat{a}^2\hat b^\dagger\right) U_\text{int}^\dagger +h.c. \\
	= -A_a^{\frac{3}{4}}&A_r^{\frac{1}{4}}   (\xi_\text{p}^*)(U\hat{a}U^\dagger)^2 (U\hat bU^\dagger)^\dagger +h.c.\ .
\end{split}
	\label{eq:cooling_out_of_interaction_picture}
\end{equation}
Since $\hat H_0$ is diagonal, exponentiating it only requires exponentiating each of the diagonal elements, and the annihilation operators in the interaction picture are
\begin{equation}
\begin{split}
	U_\text{int}\hat{a}U_\text{int}^\dagger &=\sum_{\substack{n=0,1,.. \\ j=g,e,..}} \sqrt{j+1}e^{-(\epsilon_{n,j+1} - \epsilon_{n,j})t/\hbar}\ket{j,n}\bra{j+1,n}\\
	U_\text{int}\hat{b}U_\text{int}^\dagger &=\sum_{\substack{i=0,1,.. \\ j=g,e,..}} \sqrt{n+1}e^{-(\epsilon_{j,n+1} - \epsilon_{j,n})t/\hbar}\ket{j,n}\bra{j,n+1}\ .
\end{split}
\label{eq:ab_interaction_picture}
\end{equation}
Note that if the system were harmonic, these expressions would simplify to $e^{-i\omega_a t}\hat a$ and $e^{-i\omega_r t}\hat b$.
If we substitute Eqs.~(\ref{eq:ab_interaction_picture}) into Eq.~(\ref{eq:cooling_out_of_interaction_picture}), one of the terms we obtain is
\begin{equation}
\begin{split}
	-\hbar Je^{i\left(\omega_\text{p}-\left(2\omega_a-A_a/\hbar-\omega_r\right)\right)t}&\ket{g,1}\bra{f,0} +h.c.\ ,
\end{split}
\end{equation}
where we defined the interaction strength
\begin{equation}
	J=\sqrt{2}A_a^{\frac{3}{4}}A_r^{\frac{1}{4}}|\xi_\text{p}|/\hbar\ .
\end{equation}
By choosing the pump frequency $\omega_\text{p} = 2\omega_a - A_a/\hbar - \omega_r $, the term becomes time-independent, making it more relevant than the other terms of $\hat H_P$ as we will derive next.
More generally, we can engineer the cooling interactions
\begin{equation}
\begin{split}
	&-\hbar J\sqrt{n+1}\ket{f,n}\bra{g,n+1} +h.c.\ ,
\end{split}
\label{eq:cooling_interaction_n}
\end{equation}
by choosing the pump frequencies
\begin{equation}
	\omega_\text{p} = 2\omega_a-2n\chi/\hbar - A_a/\hbar - \omega_r\ .
\end{equation}

This is the interaction used in all expriments presented in the last three figures of the main text.
Cooling by driving the $|g,1\rangle\leftrightarrow|e,0\rangle$ transition may seem like a more natural choice, but it is a two pump-photon process (due to four-wave mixing selection rules), and hence requires higher pumping power. 
Additionally, due to its higher energy, the $|f,0\rangle$ state has a lower thermal occupation than $|e,0\rangle$.
As discussed below, high pump powers and thermal occupation of the transmon place strong limitations on the cooling efficiency.

Rather than lowered, the number of excitations in the resonator can also be raised using interactions of the form 
\begin{equation}
\begin{split}
	&-\hbar J\sqrt{n+1}\ket{f,n+1}\bra{g,n} +h.c.\ ,
	\label{eq:raising_interaction}
\end{split}
\end{equation}
which are realized by choosing the pump frequencies
\begin{equation}
	\omega_\text{p} = 2\omega_a-2(n+1)\chi/\hbar - A_a/\hbar + \omega_r\ .
\end{equation}
\subsubsection{Derivation of cooling rate}
\label{sec:cooling_rate}

In this section we focus on the cooling interaction of Eq.~(\ref{eq:cooling_interaction_n}), however the methodology described is generalizable to all interaction terms.
The objective of this section is to translate the interaction term derived previously into a cooling rate for the resonator.
We assume that this interaction is sufficiently weak to enable us to perform first-order perturbation theory, considering the transmon as a fluctuating quantum noise source $\hat{F}_H$ perturbing the resonator following App.~B.1 of Ref.~\cite{clerk2010introduction}.
An initial state of the resonator $|n\rangle$ will evolve following
\begin{equation}
\begin{split}
    |\psi(t)\rangle=|n\rangle
    +i\sqrt{n}J\left(\int_0^t d\tau e^{i\Delta \tau}\hat{F}_H (\tau) \ket{n-1}\bra{n}\right)|n\rangle\
\end{split}
\end{equation}
where $\hat{F}_H (\tau) = \left(\ket{f}\bra{g}\right) (\tau)$ is treated as an independent noise source acting on the Hilbert space of the transmon. 
We consider the transition is off-resonantly driven such that the time-dependence in the interaction picture is not completely eliminated and the interaction term rotates at 
\begin{equation}
	\Delta = \omega_\text{p} - \left(2\omega_a-2n\chi/\hbar - A_a/\hbar - \omega_r\right)\ ,
\end{equation}
The probability amplitude of finding the resonator  in $|n-1\rangle$ is
\begin{equation}
    \langle n-1|\psi(t)\rangle =i\sqrt{n}J\int_0^t d\tau e^{i\Delta \tau}\left(\ket{f}\bra{g}\right) (\tau) \ ,
\end{equation}
leading to a probability
\begin{equation}
\begin{split}
    &|\langle n-1|\psi(t)\rangle|^2 =\langle n-1|\psi(t)\rangle^\dagger \langle n-1|\psi(t)\rangle\\
    &=nJ^2\int_0^t\int_0^t d\tau_1 d\tau_2 e^{i\Delta (\tau_2-\tau_1)} \left(\ket{f}\bra{g}\right)^\dagger (\tau_1)\left(\ket{f}\bra{g}\right) (\tau_2)\ .
    \label{proba}
\end{split}
\end{equation}
%
%
Note that $|\langle n-1|\psi(t)\rangle|^2$ is still a quantum operator acting on the transmon Hilbert space.
To obtain a classical probability, we now calculate its expectation value $\langle . \rangle_H$, provided that the transmon evolves in steady-state under thermal effects and dissipation
\begin{equation}
\begin{split}
    &p_{n\rightarrow n-1}(t)=\langle|\langle n-1|\psi(t)\rangle|^2\rangle_H\\
    &=nJ^2\int_0^t\int_0^t d\tau_1 d\tau_2 e^{i\Delta (\tau_2-\tau_1)}\langle\left(\ket{f}\bra{g}\right)^\dagger (\tau_1)\left(\ket{f}\bra{g}\right) (\tau_2)\rangle_H\ .
\end{split}
\end{equation}
As in Appendix A.2 of \cite{clerk2010introduction}, we transform the double integral $S$ to
\begin{equation}
\begin{split}
    S &= \int_0^t d\tau_1 \int_0^t d\tau_2 e^{i\Delta (\tau_2-\tau_1)}\langle\left(\ket{g}\bra{f}\right) (\tau_1)\left(\ket{f}\bra{g}\right) (\tau_2)\rangle_H\\
    &=\int_0^t dT \int_{-B(T)}^{B(T)} d\tau e^{-i\Delta \tau}\langle\left(\ket{g}\bra{f}\right) (T + \tau/2)\\
    &\ \ \ \ \ \ \ \ \ \ \ \ \ \ \ \ \ \ \ \ \ \ \ \ \ \ \ \ \ \ \ \ \times\left(\ket{f}\bra{g}\right) (T - \tau/2)\rangle_H\ ,\\
	&\text{where }B(T) = 2T\text{ if }T< t/2\\
	        &\ \ \ \ \ \ \ \ \ \ \ \ \ \ \ \ = 2(t-T)\text{ if }T> t/2
\end{split}
\end{equation}
For time-scales larger than the decay rate of the transmon $\tau\gg 1/\gamma$, the two time-dependent transmon operators are not correlated and the integrand will vanish (see Appendix A.2 of \cite{clerk2010introduction}). 
We can therefore extend the range of the inner integral to $\pm \infty$ in estimating the probability at a time $t\gg1/\gamma$. 
\begin{equation}
\begin{split}
	S =\int_0^t dT \int_{-\infty}^{+\infty} d\tau e^{-i\Delta \tau}\langle&\left(\ket{g}\bra{f}\right) (T + \tau/2)\\
    \times&\left(\ket{f}\bra{g}\right) (T - \tau/2)\rangle_H\ .\\
\end{split}
\end{equation}
Using time-translation invariance, we can remove the dependence on $T$
\begin{equation}
\begin{split}
	S &=\int_0^t dT \int_{-\infty}^{+\infty} d\tau e^{-i\Delta \tau} \langle\left(\ket{g}\bra{f}\right) (\tau)\left(\ket{f}\bra{g}\right) (0)\rangle_H\\
	&=t\int_{-\infty}^{+\infty} d\tau e^{-i\Delta \tau}\langle\left(\ket{g}\bra{f}\right) (\tau)\left(\ket{f}\bra{g}\right) (0)\rangle_H\ ,
\end{split}
\end{equation}
such that the rate becomes time-independent
\begin{equation}
\begin{split}
	\Gamma_{n\rightarrow n-1} &= p_{n\rightarrow n-1}(t)/t\\
		&=ng^2\int_{-\infty}^{+\infty} d\tau e^{-i\Delta \tau} \langle\left(\ket{g}\bra{f}\right) (\tau)\left(\ket{f}\bra{g}\right) (0)\rangle_H\ .\\
\end{split}
\end{equation}
Using time-translation invariance, we find that for negative values of $\tau$,
\begin{equation}
\begin{split}
\langle\left(\ket{g}\bra{f}\right)& (-|\tau|)\left(\ket{f}\bra{g}\right) (0)\rangle_H\\
 & = \langle\left(\ket{g}\bra{f}\right) (0)\left(\ket{f}\bra{g}\right) (|\tau|)\rangle_H \\
 & = \langle\left(\ket{g}\bra{f}\right) (|\tau|)\left(\ket{f}\bra{g}\right) (0)\rangle^*_H\ ,
\end{split}
\end{equation}
leading to 
\begin{equation}
\begin{split}
	\Gamma_{n\rightarrow n-1} &=nJ^2\int_{0}^{\infty} d\tau e^{-i\Delta \tau}\langle\left(\ket{g}\bra{f}\right) (\tau)\left(\ket{f}\bra{g}\right) (0)\rangle_H\\
	&+nJ^2\int_{0}^{\infty} d\tau e^{-i\Delta \tau}\langle\left(\ket{g}\bra{f}\right) (\tau)\left(\ket{f}\bra{g}\right) (0)\rangle^*_H\\
	=2nJ^2 &\text{Re}\left(\int_{0}^{\infty} d\tau e^{-i\Delta \tau} \langle\left(\ket{g}\bra{f}\right) (\tau)\left(\ket{f}\bra{g}\right) (0)\rangle_H\right)\ .\\
\end{split}
\end{equation}
In the steady state of the system, the quantum regression theorem 
can be shown to reduce the expression to
\begin{equation}
\Gamma_{n\rightarrow n-1} = 2nJ^2\text{Re}\left(\int_{0}^{\infty} d\tau  e^{-i\Delta \tau}~\text{Tr}\left[\ket{g}\bra{f}e^{\mathcal{L}\tau}\ket{f}\bra{g}\hat\rho\right]\right)\ ,
\end{equation}
%
where $\hat\rho$ is the steady-state density matrix of the transmon and $e^{\mathcal{L}t}$ its propagator, a function which takes a density matrix as an input and evolves it up to a time $t$ following the Lindblad equation.
Reducing the transmon to a three-level system and considering dissipation and thermal effects, this trace can be calculated analytically using the QuantumUtils Mathematica library
\begin{equation}
	\text{Tr}\left[\ket{g}\bra{f}e^{\mathcal{L}t}\ket{f}\bra{g}\hat\rho\right]=P_ge^{-\gamma t(1+\frac{3}{2}n_\text{th}^{(a)})}\ .
\end{equation}
By only considering dissipation and thermalization, we made the assumption that an excitation could not be driven back from $\ket{f,n-1}$ to $\ket{g,n}$ under the effect of pumping, \textit{i.e. }we assume $2\gamma\gg \sqrt{n}J$, that we are far from the strong coupling regime.
After integration, we obtain
\begin{equation}
	\Gamma_{n\rightarrow n-1} = \frac{2nJ^2P_g}{\gamma(1+\frac{3}{2}n_\text{th}^{(a)})}\frac{1}{1+\left(\frac{\Delta}{\gamma}\right)^2}\ .
	\label{eq:rate_down}
\end{equation}
Following the same method, we also obtain for the hermitian conjugate of this interaction term
\begin{equation}
	\Gamma_{n-1\rightarrow n} = \frac{2nJ^2P_f}{\gamma(1+\frac{3}{2}n_\text{th}^{(a)})}\frac{1}{1+\left(\frac{\Delta}{\gamma}\right)^2}\ ,
	\label{eq:rate_up}
\end{equation}
if the $\ket{f}$ level is populated, we find that there is a probability for the pump to raise the number of excitations in the resonator rather than lower it.
We refer to the steady state population of the ground and second-excited state of the transmon as $P_g$ and $P_f$ respectively.
The same calculation can be performed for the raising interaction, which yields identical rates only with $P_g$ and $P_f$ interchanged.
A good figure of merit of the cooling efficiency is then to compare this rate with $\kappa$, yielding the cooperativity
\begin{equation}
	C = \frac{\Gamma_{1\rightarrow 0}}{\kappa} =\frac{2J^2}{\gamma\kappa(1+\frac{3}{2}n_\text{th}^{(a)})}\frac{1}{1+\left(\frac{\Delta}{\gamma}\right)^2}\ .
	\label{eq:full_cooperativity}
\end{equation}

\subsubsection{Semi-classical description of the cooling process}
With the cooling rate above, we can construct a semi-classical set of rate equations describing the competition between thermalization and cooling.
They would correspond to the diagonal part of a Lindblad equation, and equates the population leaving and arriving to a given state of the resonator.
We restrict ourselves to the driving of $\ket{f,0}\leftrightarrow\ket{g,1}$ as in the experiment of Fig.~\ref{fig:LFT_2}, where these equations can be written as
\begin{equation}
	\dot P_0 = P_1\left(\kappa CP_g + \kappa (n_\text{th}+1)\right) - P_0\left(\kappa CP_f + \kappa n_\text{th}\right),
\end{equation}
\begin{equation}
\begin{split}
	\dot P_1 &= -P_1\left(\kappa CP_g + \kappa (n_\text{th}+1)\right) + P_0\left(\kappa CP_f + \kappa n_\text{th}\right)\\
	&  -2P_1\kappa n_\text{th} + 2P_2\kappa (n_\text{th}+1)\ ,
\end{split}
\end{equation}
and, for $n\ge2$
\begin{equation}
\begin{split}
	\dot P_n &= -n \kappa (n_\text{th}+1) P_n + n\kappa n_\text{th}P_{n-1}\\
	& -(n+1)P_n\kappa n_\text{th} + (n+1)P_{n+1}\kappa (n_\text{th}+1)\ .
\end{split}
\end{equation}
In steady state ($\dot P=0$), the solution is a function of $P_0$
\begin{equation}
\begin{split}
	\frac{P_0}{P_1}&=\frac{CP_g + n_\text{th}+1}{CP_f + n_\text{th}} =A\ ,\\
	\frac{P_n}{P_{n+1}}&=\frac{n_\text{th}+1}{n_\text{th}} = B\ \text{for}\ n\ge1\ .
\end{split}
	\label{eq:cooling_formula}
\end{equation}
We reach a unique solution by imposing $\sum_nP_n=1$, which yields an expression for $P_0$ 
\begin{equation}
	P_0 = \frac{
A
\left(A-1\right)
\left(B-1\right)
}{
B(A^2-1)
+ A(1-A)
}\ .
\label{eq:P0_with_thermal}
\end{equation}
This expression is used in Fig.~\ref{fig:S_version_fig_2} to show the temperature limited evolution of $P_0$ as a function of cooperativity.
A more accurate description of the cooling process at high cooperativities comes from a numerical simulation taking the strong coupling limit and off-resonant driving of other four-wave mixing processes into account.
%

\subsection{Limiting factors to cooling}

\begin{figure}[]
\centering
\includegraphics[width=0.8\textwidth]{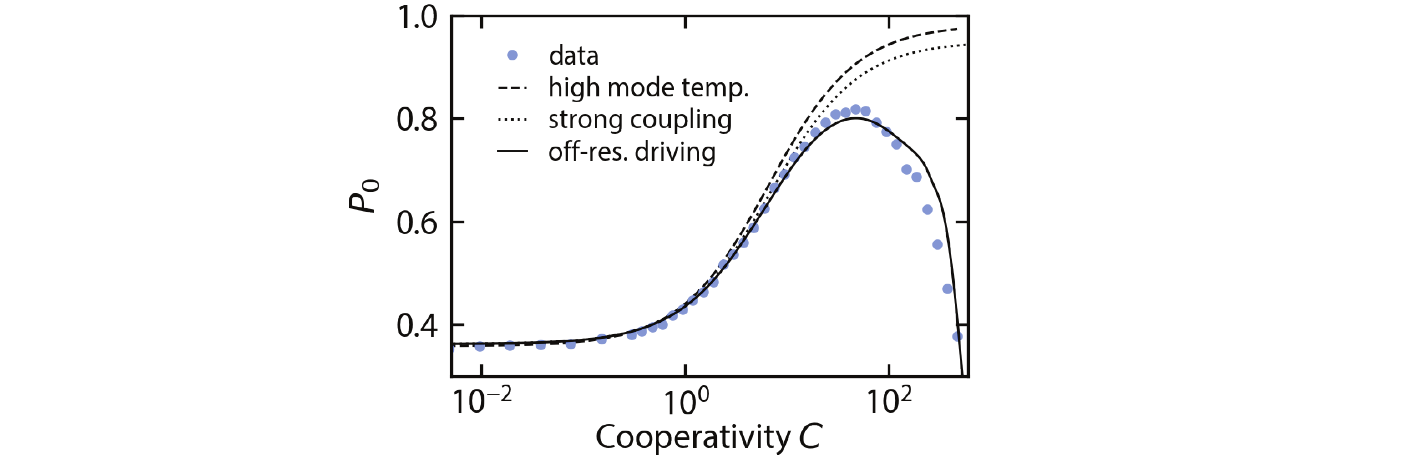}
\caption[Limiting factors to cooling.]{\textbf{Limiting factors to cooling. }
$P_0$ as a function of $C$, with dots showing data points identically to Fig.~\ref{fig:LFT_2}.
The decrease of $P_0$ at large $C$ is not captured by the cooling limitation due to thermal population of the $\ket{f}$ state (dashed line) or the limit imposed by strong coupling (dotted line), where the pump hybridizes the $\ket{g,1}$ and $\ket{f,0}$ states. 
The solid curve shows a prediction considering the off-resonant driving of other sideband transitions by the pump: as the cooling process starts to saturate due to the strong coupling limit, the driving rate of transitions that increase the photon number overpowers the cooling effect.
}
\label{fig:S_version_fig_2}
\end{figure}

Here, we discuss three limiting factors to the cooling experiment (Fig.~\ref{fig:LFT_2}), ending with some notes on how to improve the device cooling performance
The first limiting factor is the thermal occupation of the transmon. 
The pump tone drives the population from $|g,1\rangle$ to $|f,0\rangle$, but the reverse process also occurs since the $f$ level has a small thermal population $P_f\simeq0.006$ (see Sec.~\ref{sec:temperature}).
This leads to the limit $P_1/P_0 > P_f/P_g$ (dashed line in Fig.~\ref{fig:S_version_fig_2}) for which we have derived an exact analytical expression (Eq.~(\ref{eq:P0_with_thermal}))

The second limiting factor is that of strong coupling (similar to in optomechanical cooling~\cite{teufel2011circuit}), where the pump hybridizes the $|g,1\rangle$ and $|f,0\rangle$ states.
If $J$ exceeds the decay rate $2\gamma$, the population of state $|g,1\rangle$ will be driven to $|f,0\rangle$ and then transfered back to $|g,1\rangle$ without having the time to decay to $|e,0\rangle$.
To simulate this effect, we compute the steady state of the system by solving a Lindblad equation numerically (see ~\ref{sec:cooling_simulation}).
The result is shown as a dotted line in Fig.~\ref{fig:S_version_fig_2}, which additionally takes into account the population of the transmon.
As with the thermal effect, the strong coupling limit only imposes an upper bound on $P_0$, rather than predicting its decrease at high $C$.
%


%
When the cooling tone is detuned by $\Delta$ from its transition frequency, the cooperativity acquires a factor $1/\left(1+\Delta^2/\gamma^2\right)$ (Eq.~(\ref{eq:rate_up})).
A similar formula applies to all other four-wave mixing processes, including raising interactions (Eq.~(\ref{eq:raising_interaction})).
If the latter are far-detuned, their off-resonant driving will have little impact on the system.
However, as the cooling process starts to saturate due to the previously discussed limiting factors, the driving of other transitions is still far from saturation and can overpower the cooling effect.
What ensues is a competition between off-resonantly driven transitions that cool and raise the photon occupation.
We simulate this by following the bootstrap step of the adaptive rotating-wave approximation method of Ref.~\cite{baker2018adaptive}, which offers a way to include the most relevant off-resonantly driven transitions to the system Hamiltonian (see Sec.~\ref{sec:cooling_simulation}).
The result is shown as the solid curve of Fig.~\ref{fig:S_version_fig_2} which predicts the maximum $P_0$ and the strong cooperativity behavior.
We emphasize that, except for a small shift on the calibrated cooperativity-axis, the theoretical curves do not correspond to a fit to the data, but rather constitute a prediction based on the independently determined dissipation rates, thermal occupations and circuit parameters.
From this simulation we extract that, at maximum $P_0=0.82$, the average photon number in the cooled resonator is $\bar n =0.65$.
Note that 
\begin{equation}
\begin{split}
	\bar n &= 0\times(P_{g0}+P_{e0}+P_{f0}+...)\\
	&+1\times(P_{g1}+P_{e1}+P_{f1}+...)\\
	&+2\times(P_{g2}+P_{e2}+P_{f2}+...)\\
	&+...
\end{split}
\end{equation}
The first 10 most populated levels are: $P_{g0} = 0.736$, $P_{e0} = 0.067$, $P_{g1} = 0.036$, $P_{g2} = 0.028$, $P_{g3} = 0.028$, $P_{g4} = 0.022$, $P_{g5} = 0.017$, $P_{g6} = 0.014$, $P_{g7} = 0.011$, $P_{f0} = 0.009$, where  $P_{j,n}$ refers to the occupation of state $\ket{j,n}$.
Taking only the contribution of these states into account in the above formula already leads to $\bar n = 0.51$, and including the occupation of all 50 simulated levels leads to $\bar n =0.65$.


Determining the ideal system parameters to improve cooling (and Fock-state stabilization fidelity) is not straightforward.
One path to improvement could lie in determining values of $A_a$ and $\chi$ which minimize the effect of off-resonant driving by moving the most problematic transitions away from the cooling frequency.
Another is to reach a higher ground-state occupation before being limited by strong coupling, which can only be achieved by reducing the resonators dissipation $\kappa$.
Decreasing the transmon dissipation $\gamma$ is not necessarily beneficial: it diminishes off-resonant driving, but strong coupling would occur at smaller pump powers.
For our system, decreasing $\gamma$ in the simulation of Fig.~\ref{fig:LFT_2}(c) results in a lower ground-state occupation.

\section{Numerical procedures}

\subsection{Spectrum}

The eigenfrequencies of the system are determined by diagonalizing the system Hamiltonian.
%
%
Unless specified otherwise, we diagonalize the Hamiltonian of Eq.~\ref{eq:Hamiltonian_8th_order} with the junction non-linearity Taylor expanded to 8-th order.
We consider 10 excitations in the transmon and 20 in the resonator, and have verified that extending the Hilbert space further only leads to negligeable changes in the obtained spectrum.
This diagonalization also provides the dressed eigenstates $\ket{j,n}$, which are to be distinguished from the bare eigenstates $\widetilde{\ket{j,n}}_{\substack{n=0,1,2,.. \\ j=g,e,f,..}}$.

\subsection{Microwave reflection}
\label{sec:numerics_S11}

In order to compute the microwave reflection of the device, we solve a Lindblad equation using Qutip~\cite{johansson2012qutip}.
The Hamiltonian is written in the dressed basis defined above, it is hence diagonal with entries corresponding to the eigenfrequencies obtained in the diagonalization.
We consider 5 transmon excitations and 10 resonator excitations.
We add the drive term $i\hbar\epsilon_d(\hat a^\dagger -\hat a)$ defined in the dressed basis, and move to the frame rotating at the drive frequency $\omega_d$ by adding $-\hbar\omega_d\hat a ^\dagger \hat a$.
We add jump operators defined in the dressed basis by
\begin{equation}
\begin{split}
	(n_\text{th}^{(a)}\gamma)^{\frac{1}{2}}\hat a^\dagger\ ,\ ((n_\text{th}^{(a)}+1)\gamma)^{\frac{1}{2}}\hat a\ ,\\
	(n_\text{th}\kappa)^{\frac{1}{2}}\hat b^\dagger\ , \ ((n_\text{th}+1)\kappa)^{\frac{1}{2}}\hat b\ ,
	\label{eq:collapse_ops}
\end{split}
\end{equation}
to describe dissipation and thermal effects.
Finally, we compute the expectation value of $\hat S_{11}=1-\frac{\gamma_\text{ext}}{\epsilon_\text{d}}\hat{a}$ for different drive frequencies.
As shown in Fig.~\ref{fig:S_temperature}, this computation is in excellent agreement with the sum of Lorentzian formula of Eq.~(\ref{eq:sum_of_lorentzians}).

\subsection{Cooling simulation}
\label{sec:cooling_simulation}

We use a similar method for the adaptive rotating-wave approximation (aRWA) simulation of Fig.~\ref{fig:LFT_2}.
We start with the same diagonal Hamiltonian.
We denote by $\omega_{j,n}$ the eigenfrequency of the dressed eigenstates $\ket{j,n}$.
As a result of the collapse operators of Eq.~(\ref{eq:collapse_ops}), a dressed state of the system $\ket{j,n}$ will have a total decay rate to other states of the system 
\begin{equation}
\begin{split}
	\Gamma_{j,n} = (j+1)(n_\text{th}^{(a)}\gamma) +j((n_\text{th}^{(a)}+1)\gamma)\\ +(n+1)(n_\text{th}\kappa) +n((n_\text{th}+1)\kappa)\ .
\end{split}
\end{equation}
Following Ref.~\cite{baker2018adaptive}, we can then estimate the impact of a pump tone at a frequency $\omega_p$ and driving rate $\epsilon_p$ on the steady state of the system.
Two states $\ket{k}=\ket{j,n}$ and $\ket{k'}=\ket{j',n'}$ will be coupled by this pump.
And to first order in $\epsilon_p$, the only change in the steady state density matrix will be in its off-diagonal element 
\begin{equation}
	\rho_{kk'} = \frac{V_{kk'}(P_{k'}-P_k)}{(\omega_{k'}-\omega_k)-\omega_p+i(\Gamma_k+\Gamma_{k'})/2}\ ,
	\label{eq:ranking_parameter}
\end{equation}
where $P_{k}$ is the occupation of state $\ket{k}$ under the collapse operators of Eq.~(\ref{eq:collapse_ops}).
The dipole moment $V_{kk'} = \bra{k}\epsilon_p(\tilde a + \tilde a^\dagger)\ket{k'}$ is computed using annihilation $\tilde a$ and creation operators $\tilde a^\dagger$ defined in the bare basis. 
The transitions between all the states are then ranked with decreasing $|\rho_{kk'}|$ (\textit{i.e.} decreasing relevance).
The most relevant terms are added in the form $\hbar V_{kk'}\ket{k}\bra{k'}$ to the Hamiltonian which is moved to the rotating frame in which states $\ket{k}$ and $\ket{k'}$ are resonant.

In Fig.~\ref{fig:S_version_fig_2}, we perform this calculation for $\omega_p = \omega_{f,0}-\omega_{g,1}$.
We show both the result of including a maximum number of transitions (465) and a single transition.
It was only possible to include 465 transitions out of the 650 transitions which have a non zero dipole moment.
This is due to limitations in the construction of the rotating frame, for more details see Ref.~\cite{baker2018adaptive}.

\begin{figure}[]
\centering
\includegraphics[width=0.8\textwidth]{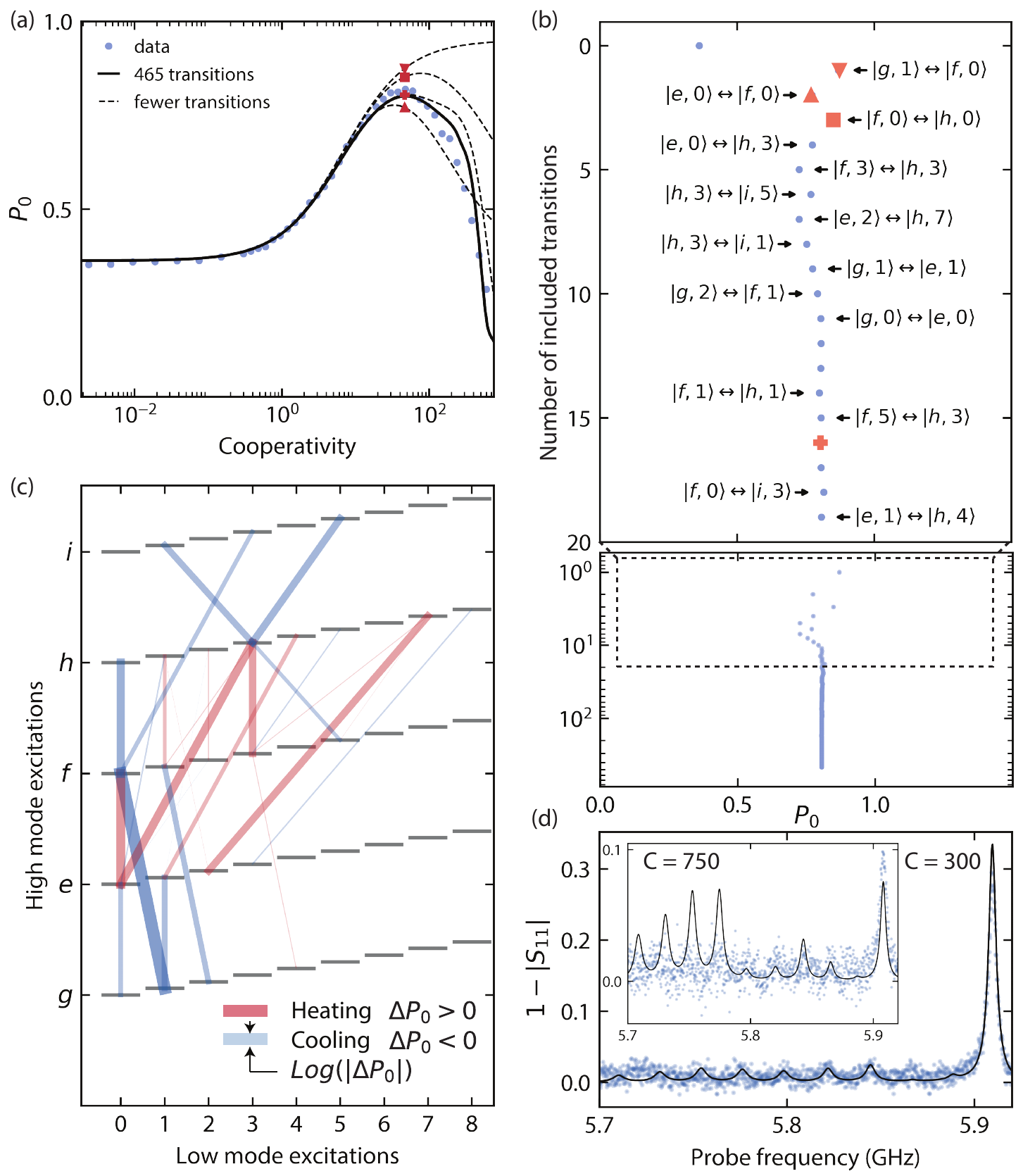}
\caption[Simulation of off-resonantly driving transitions in cooling experiment.]{\textbf{Simulation of off-resonantly driving transitions in cooling experiment. }
\textbf{(a) }Ground-state occupation $P_0$ as a function of cooperativity in the cooling experiment of Fig.~\ref{fig:LFT_2}.
We show data (blue dots) and a simulation including the driving of 1, 2, 3, 16 (dashed curves) and 465 transitions (solid curve) using the adaptive rotating-wave approximation~\cite{baker2018adaptive} (aRWA).
The red symbols in panels (a,b) correspond to identical simulation points.
\textbf{(b) }
Evolution of $P_0$ at a fixed cooperativity $C=46$.
For each point, we consider an additional transition.
\textbf{(c) }
The transitions leading to the largest change in $P_0$ are displayed in the system energy diagram.
The transitions are colored red (blue) if adding them causes an increase (decrease) of $P_0$.
This distinction should be interpreted with care since the change in $P_0$ may be the result of multiple transitions interacting.
The thickness of the lines is logarithmically related to the change in $P_0$ that comes from adding the transition.
\textbf{(d) }
Experimental spectrum (blue dots) and numerical predictions (sum of Lorentzian formula Eq.~(\ref{eq:sum_of_lorentzians})) at very high cooling powers.
We use the aRWA simulation to estimate the amplitudes of each Lorentzian peak. 
Up to a cooperativity of $C=300$, the data is consistent with aRWA predictions.
However, there is a clear deviation between data and simulation at the highest powers (inset).
}
\label{fig:S_tree}
\end{figure}

In Fig.~\ref{fig:S_tree}, we study how each transition affects the steady-state of the system.
Ranking using  $|\rho_{kk'}|$ does not take into account that multiple transitions may interact.
To rank the relevance of the transitions in a more realistic way, we further rank the transitions following their impact on $P_0$.
We add transitions one by one in the simulation, recording for each transition the change $\Delta P_0$ that ensues.
We then rank the transitions with decreasing $|\Delta P_0|$, and repeat: we add the transitions in the new order one by one, rank them and start again until reaching convergence.

This simulation is in good agreement with the data except at the very highest powers (see Fig.~\ref{fig:S_tree}(d)).
There are four possible limitations in our aRWA simulation that could be the cause of this discrepancy.
First, our implementation of aRWA does not take into account the AC-Stark shift of each level.
Present only at high powers, these AC-Stark shifts could bring certain transitions in or out of resonance with each-other, modifying the final steady-state of the system.
Secondly, we work with a Hilbert space of only 10 excitations in the resonator.
At the highest power, the simulation indicates an average resonator occupation of $~5$ and a larger Hilbert space may be needed to reach more accurate results.
Thirdly, only first order transitions were considered in the ranking of the transitions, so no higher order processes, such as those shown in Fig.~\ref{fig:S_all_transitions}C, are taken into account.
Fourthly, we rank transitions with Eq.~(\ref{eq:ranking_parameter}) using $P_{k}$ the occupation of states $\ket{k}$ under the collapse operators of Eq.~(\ref{eq:collapse_ops}).
However $P_{k}$ may change under the effect of the driving, modifying the relevance of a given transition.
This can be taken into account as described in Ref.~\cite{baker2018adaptive}, but is too computationally expensive with the Hilbert-space size used here.

\section{Background subtraction}

\subsection{Network analysis}

Most of our data analysis relies on fitting a sum of complex Lorentzians (see Eq.(~\ref{eq:sum_of_lorentzians})), to the measured microwave reflection $S_{11}$ in both phase and amplitude.
The signal we acquire is affected by the imperfections of the microwave equipment used to carry the signals to and from the device.

These can be modeled by a two port network with $s$ parameters $s_{11},s_{22}$, corresponding to the reflections at the VNA ports (reflected back to the VNA) and at the device (reflected back to the device) respectively, and $s_{21},s_{12}$, corresponding to the attenuation chain from the VNA to the device and the amplification chain from the device to the VNA respectively.
\begin{figure}[]
\centering
\includegraphics[width=0.8\textwidth]{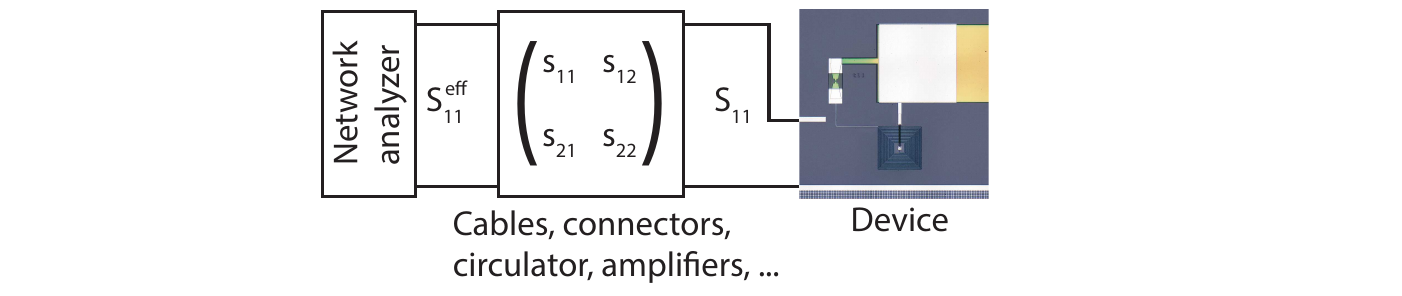}
\caption[Effective microwave network]{\textbf{Effective microwave network. }
We do not directly have access to the reflection at our device $S_{11}$.
We measure an effective reflection coefficient $S_{11}^{\text{eff}}$, affected by the imperfect microwave equipment between the network analyzer and device described by an $s-$matrix.
}
\label{fig:background_substraction}
\end{figure}
%

We hence measure with our VNA the effective microwave reflection
\begin{equation}
	S_{11}^{\text{eff}} = s_{11}+\frac{s_{12}s_{21}}{1-s_{22}S_{11}}S_{11}
\end{equation}
Note that these $s$ parameters are generally frequency dependent.
We make the approximation $s_{11},s_{22}\ll s_{12},s_{21}$, meaning we attribute most of the measured microwave background to the frequency dependent transmission of the attenuation and amplification chain.
The signal we want to measure is now proportional to a so-called ``microwave background''
\begin{equation}
	S_{11}^{\text{eff}} \simeq s_{12}s_{21}S_{11}\ ,
\end{equation}
which we have to experimentally measure.

\begin{figure}[]
\centering
\includegraphics[width=0.8\textwidth]{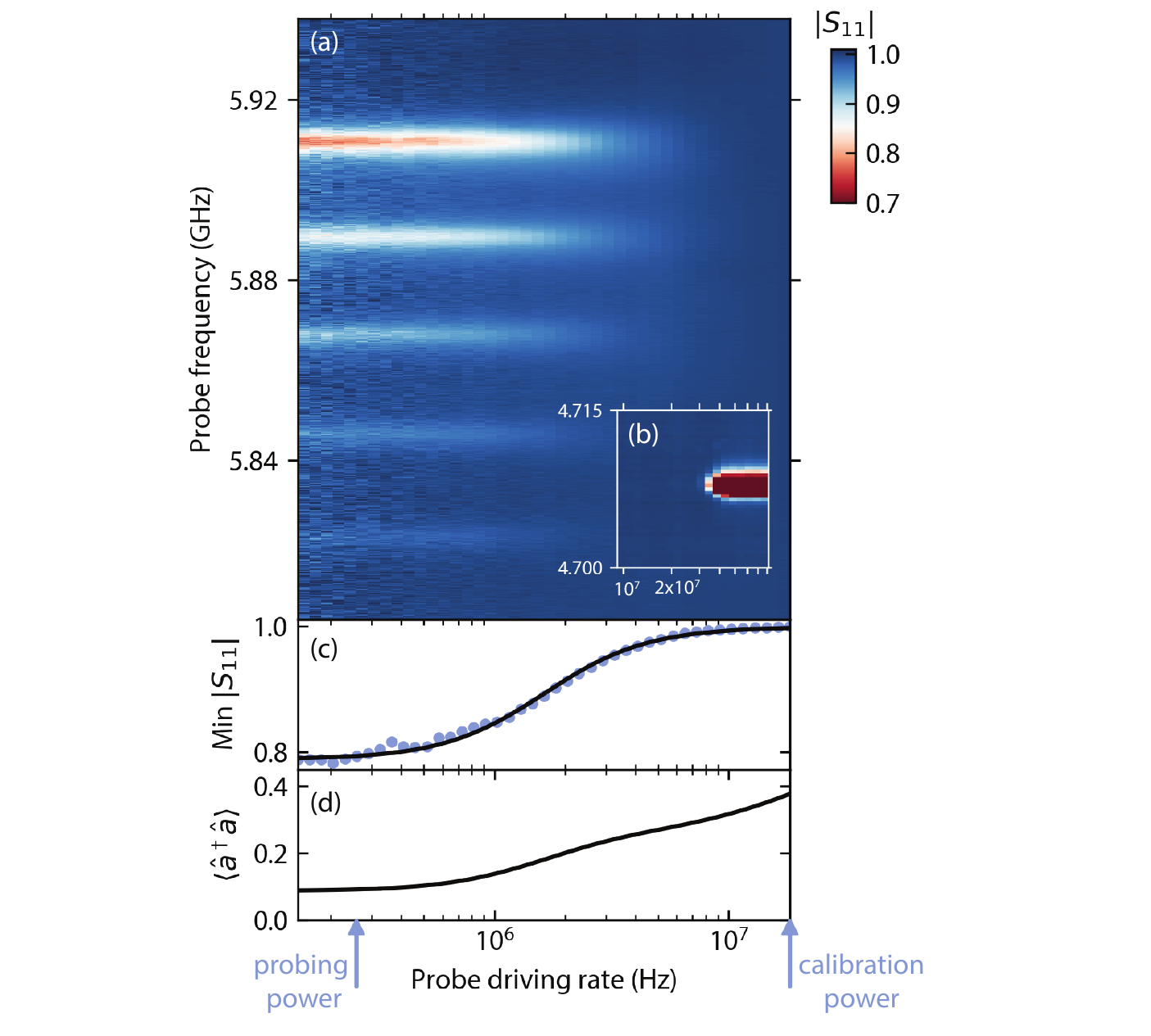}
\caption[Probe power dependence]{
	\textbf{High-probe-power behavior. (a) }
	$|S_{11}|$ as a function of probe frequency and probe power. 
	\textbf{(b) }
	At higher powers, the system starts to resonate at a different frequency, corresponding to the junction being replaced by an open circuit.
	\textbf{(c) }
	Depth of the $n=0$ peak extracted from data (blue dots) and numerical steady-state calculation (see Sec.~\ref{sec:numerics_S11}).
	As the probe driving rate exceeds $\gamma$, the peaks vanish.
	We use the disappearance of peaks at the high power indicated by an arrow to acquire a microwave background that is subtracted (divided) in phase (amplitude) from all datasets.
	\textbf{(d) }
	Population in the transmon as a function of probe power as extracted from simulation.
	We used this information to choose the probing power indicated by an arrow for all other experiments. 
	It is as high as possible to increase signal to noise ratio, but low enough to not populate the transmon.
}
\label{fig:S_power}
\end{figure}

\subsection{Measuring the microwave background}

As shown in Fig.~\ref{fig:S_power}, when probing the system at high power the device response is $S_{11}=1$, allowing us to extract the microwave background $s_{12}s_{21}$.
This phenomenon is a consequence of super-splitting as explained in \cite{Bishop2009a}, which we will briefly summarize here.

To understand super-splitting, we have to truncate the transmon to a two-level system constituted of its two first levels $\ket{g}$ and $\ket{e}$.
In the Bloch sphere, the probe tone will cause rotations around the y-axis and $1-S_{11}$ corresponds to the projection of the state vector on the x-axis.
For driving rates faster than $\gamma$, the state vector will rapidly rotate around the y-axis yielding a zero projection on the x-axis hence $S_{11}=1$ and no peak.
For driving rates slower than $\gamma$, random decays of the state vector will be very likely to occur before the state vector can rotate around the y-axis, yielding a non zero projection on the x-axis and a dip in the microwave reflection.
A signature of this effect is the splitting of the absorption peak in two for large probe powers.
Whilst our signal to noise does not allow the resolution of this feature, it is present in the fitted simulation, supporting this explanation.

At even higher power, the system starts to resonate at a different frequency, corresponding to the junction being replaced by an open circuit when the current traversing the junction exceeds the critical current.
%
This effect is shown in the inset, Fig.~\ref{fig:S_power}(b).

We use the disappearance of peaks at a high power indicated by the arrow ``calibrating power" to acquire a microwave background that is subtracted (divided) in phase (amplitude) to all datasets.
\begin{equation}
	\frac{S_{11}^{\text{eff}}}{s_{12}s_{21}} \simeq S_{11}\ .
\end{equation}

\section{Fitting}

Here, we summarize our fitting routine.
We start by extracting $\kappa$ from the time-domain data, which will be used in the formula for the linewidth $\gamma_n$ in all subsequent fits.
By fitting the microwave reflection $S_{11}$ to a sum of Lorentzians (see Eq.~(\ref{eq:sum_of_lorentzians})), we get access to the peaks linewidths and amplitudes which allows us to determine $\gamma$, $\gamma_\text{ext}$ and $n_\text{th}^{(a)}$.
By fitting $S_{11}$ to the eigenfrequencies obtained from a diagonalization of the Hamiltonian of Eq.~(\ref{eq:Hamiltonian_8th_order}), we determine the values of the circuit elements.
The occupation of the resonator $n_\text{th}$ is determined separately for each individual experiment.
Each step is detailed in the subsections below.

\subsection{Low-frequency mode dissipation}
\begin{figure}[]
\centering
\includegraphics[width=0.8\textwidth]{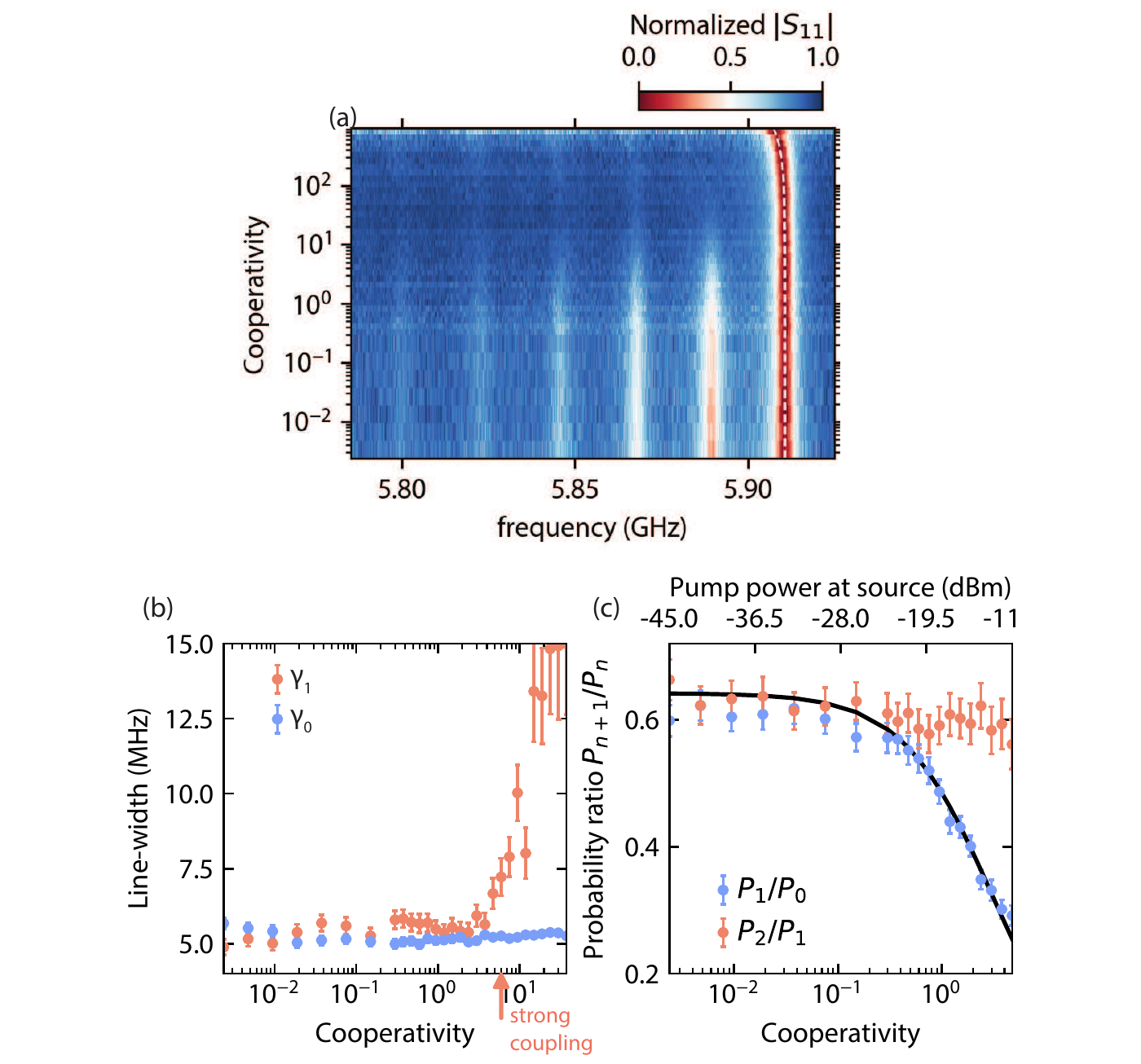}
\caption[AC-stark shift and cooperativity measurement]{\textbf{AC-stark shift and cooperativity measurement. (a) }
	Normalized $|S_{11}|$ as a function of probe frequency and cooperativity in a single-pump cooling experiment.
	The AC-Stark shift of the $n=0$ peak follows the fitted white dashed line.
	%
	%
	\textbf{(b) }
	Extracted line-widths of peaks $n=0,1$.
	Error bars correspond to the standard errors estimated from the least-squares fit.
	%
	%
	Fluctuations in $\gamma_n$ result from fluctuations in $n_\text{th}$.
	The onset of the strong-coupling regime, indicated by a red arrow, is seen through the increase in line-width of the $n=1$ peak.
	Below this value, we can accurately extract $P_1$ from the amplitude of the $n=1$.
	\textbf{(c) }
	Ratio of extracted probabilities $P_2/P_1$ and $P_1/P_0$.
	The former is constant $P_2/P_1 = n_\text{th}/(1+n_\text{th})$, whilst the latter is fitted to $P_1/P_0 = (n_\text{th}+CP_f)/(1+n_\text{th}+CP_g)$, allowing us to convert pump power (top x-axis) to cooperativity (bottom x-axis).
}
\label{fig:S_cooling}
\end{figure}

We start by fitting the thermalization from the ground-state measured in time-domain (Fig.~\ref{fig:S_time_domain}(a)) to determine $\kappa$.
Since the line-width of the $S_{11}(t)$ peaks is a function of $\kappa$ and $n_\text{th}$, we start by postulating these two values to extract a first estimate of the time evolution of $P_n$.
By fitting the evolution of $P_n$ to the rate equation of Eq.~(6), we extract a new value for $\kappa$ and $n_\text{th}$.
We then repeat this process many times, each time using the new values $\kappa$ and $n_\text{th}$ to fit $S_{11}(t)$, until we converge to $\kappa/2\pi=23.5\cdot 10^3s^{-1}$.
%
%

The low-frequency mode dissipation can also be measured without recourse to time-domain experiments.
The knowledge of the power dependent AC-stark shift and the cooperativity, measured in a single tone cooling experiment, is sufficient to extract $\kappa$.
We use this method to confirm our time-domain results, as well as verify the theory developed in Sec. ~\ref{sec:cooling_rate}.
First we measure the AC-stark shift of the $n=0$ peak, from which we extract the the proportionality factor $\xi^2/P$, between pumping rate $\xi$ and pump power $P$ (Fig.~\ref{fig:S_cooling}(a)).
Secondly we determine the power at which the strong coupling regime arises (Fig.~\ref{fig:S_cooling}(b)).
Above this power, the line-width of the $n=1$ peak will rise as the state $\ket{g,1}$ hybridizes with $\ket{f,0}$ under the effect of the cooling pump.
Below this power, the line-width of the $n=1$ peak is approximatively constant, and its height provides an accurate measure of $P_1$.
In this regime, we thirdly extract the ratio of probabilities $P_2/P_1$ and $P_1/P_0$
Following Eqs.~(\ref{eq:cooling_formula}), the former should remain constant $P_2/P_1 = n_\text{th}/(1+n_\text{th})$.
The latter, however, decreases with power, $P_1/P_0 = (n_\text{th}+CP_f)/(1+n_\text{th}+CP_g)$, and fitting this curve provides the conversion factor between cooperativity $C$ and power.
If we also know the anharmonicity $A_a$, cross-Kerr $\chi$, the transmon occupation $n_\text{th}^{(a)}$ and dissipation rate $\gamma$, we can estimate the resonator dissipation $\kappa = 2(\xi^2/P)/(C/P)A_a\chi/\gamma/(1+3n_\text{th}^{(a)}/2) \simeq 2\pi\cdot16\cdot 10^3s^{-1}$ close to the value obtained in time-domain.
The discrepancy is due to the inaccuracy of the relation $P_1/P_0 = (n_\text{th}+CP_f)/(1+n_\text{th}+CP_g)$, arising from the off-resonant driving of other four-wave mixing transitions.

\subsection{Transmon dissipation and device temperature}
\label{sec:temperature}
\begin{figure}[]
\centering
\includegraphics[width=0.8\textwidth]{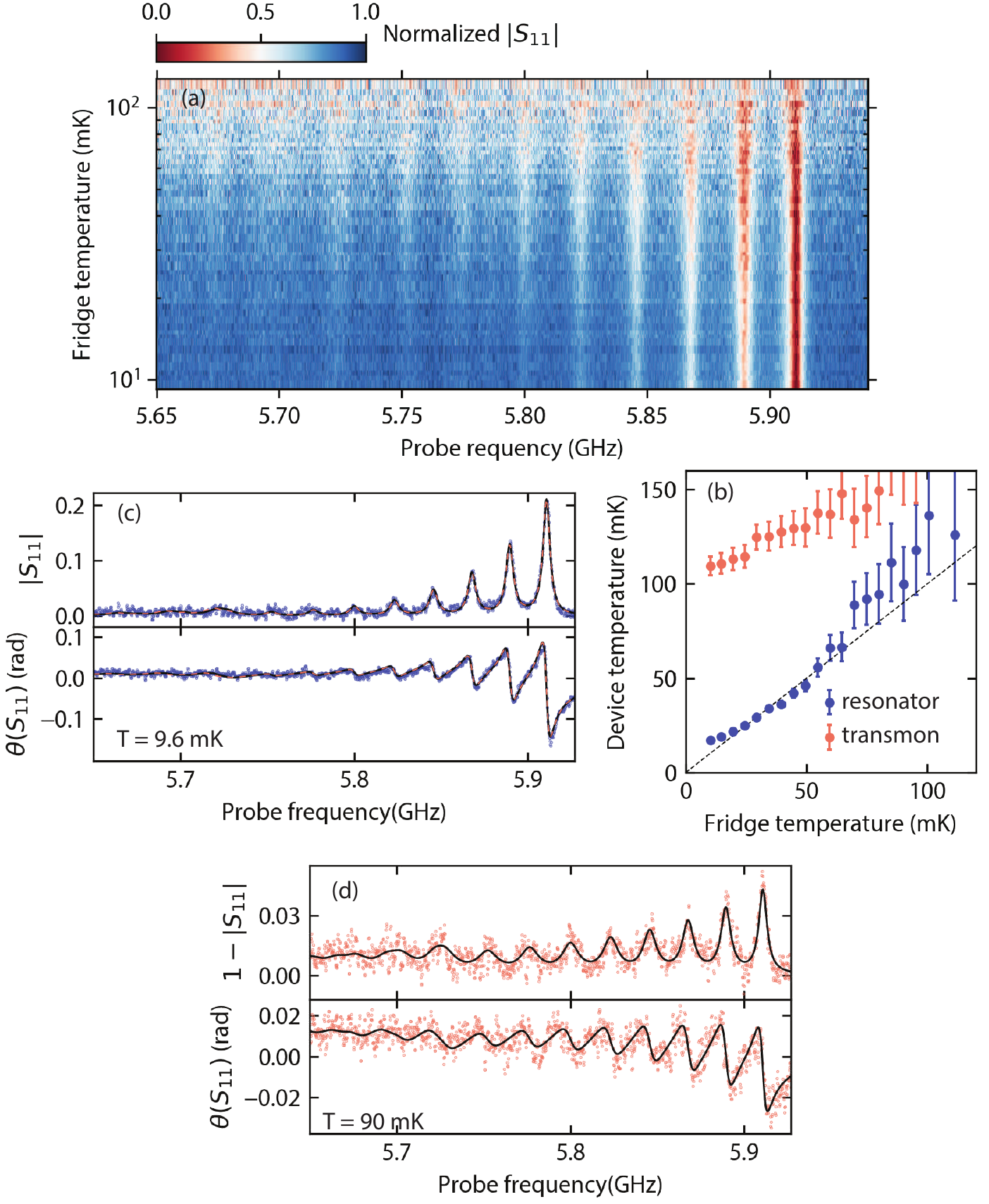}
\caption[Temperature dependence]{
	\textbf{Temperature dependence. (a) }
	Normalized $|S_{11}|$ as a function of probe frequency and fridge temperature. 
	\textbf{(b) }
	Temperature of both modes, fit using using Eq.~(\ref{eq:sum_of_lorentzians}), as a function of fridge temperature.
	\textbf{(c) }
	Lowest-temperature data (blue) and two fits, one using Eq.~(\ref{eq:sum_of_lorentzians}) (black), and another fitting a numerical model as described in Sec.~\ref{sec:numerics_S11} (dashed red line).
	Excellent agreement between both fits validates our method of fitting the spectrum with a sum of Lorentzian functions.
	\textbf{(d) }
	Higher-temperature data (red) and fit (black) using Eq.~(\ref{eq:sum_of_lorentzians}).
}
\label{fig:S_temperature}
\end{figure}

Using $\kappa/2\pi=23\cdot 10^3s^{-1}$, we fit the spectra shown in Fig.~\ref{fig:S_temperature} to fix $\gamma$, $\gamma_\text{ext}$ and $n_\text{th}^{(a)}$.
Here, the fridge temperature is varied, and from a fit of Eq.~(\ref{eq:sum_of_lorentzians}) we extract $\gamma,\gamma_\text{ext}$, $n_\text{th}$ and $n_\text{th}^{(a)}$ at each temperature.
We took care to let the system thermalize for $\sim 10$ minutes at each temperature before starting measurements.
The linear scaling of resonator temperature with fridge temperature, shown in Fig.~\ref{fig:S_temperature}(b), confirms that we can extract a realistic mode temperature from the Bose-Einstein distribution.
A large difference in temperature is measured between low and transmon, which could be explained by the difference in external coupling to the feedline of the two modes.
We fix the values of $\gamma$, $\gamma_\text{ext}$ and $n_\text{th}^{(a)}$ to the lowest fridge temperature fit (Fig.~\ref{fig:S_temperature}(c)).
We leave $n_\text{th}$ as free parameters in the other experiments as it was found to vary by 10 to 20 percent on a time-scale of hours.
In the main text we quote the value of $n_\text{th}$ of the lowest point in the temperature sweep ($n_\text{th}=1.62$), but in the Fock state stabilization measurement, we measured $n_\text{th}=1.40$, in the cooling experiment $n_\text{th}=1.81$ and in the time-domain $n_\text{th}=1.37$.
These fluctuations are much smaller than the uncertainty in fitting $n_\text{th}$.
In both the cooling and Fock state stabilization experiments, $n_\text{th}$ was extracted from an initial measurement of $S_{11}$ in absence of pump tones.

\subsection{Circuit parameters}
The frequency of the system transitions (and hence the circuit parameters) is determined by fitting a numerical steady-state calculation of $S_{11}$ to the lowest temperature data (Fig.~\ref{fig:S_cooling}(c)).
This simulation, described in \ref{sec:numerics_S11}, starts with a diagonalization of the Hamiltonian of Eq.~(\ref{eq:Hamiltonian_8th_order}).
In this fit we additionally impose that the transition frequency $\ket{g,0}\leftrightarrow\ket{g,1}$ match the value measured in two-tone spectroscopy (Fig.~\ref{fig:S_two_tone}(a)).
%
%

\begin{figure}[]
\centering
\includegraphics[width=0.8\textwidth]{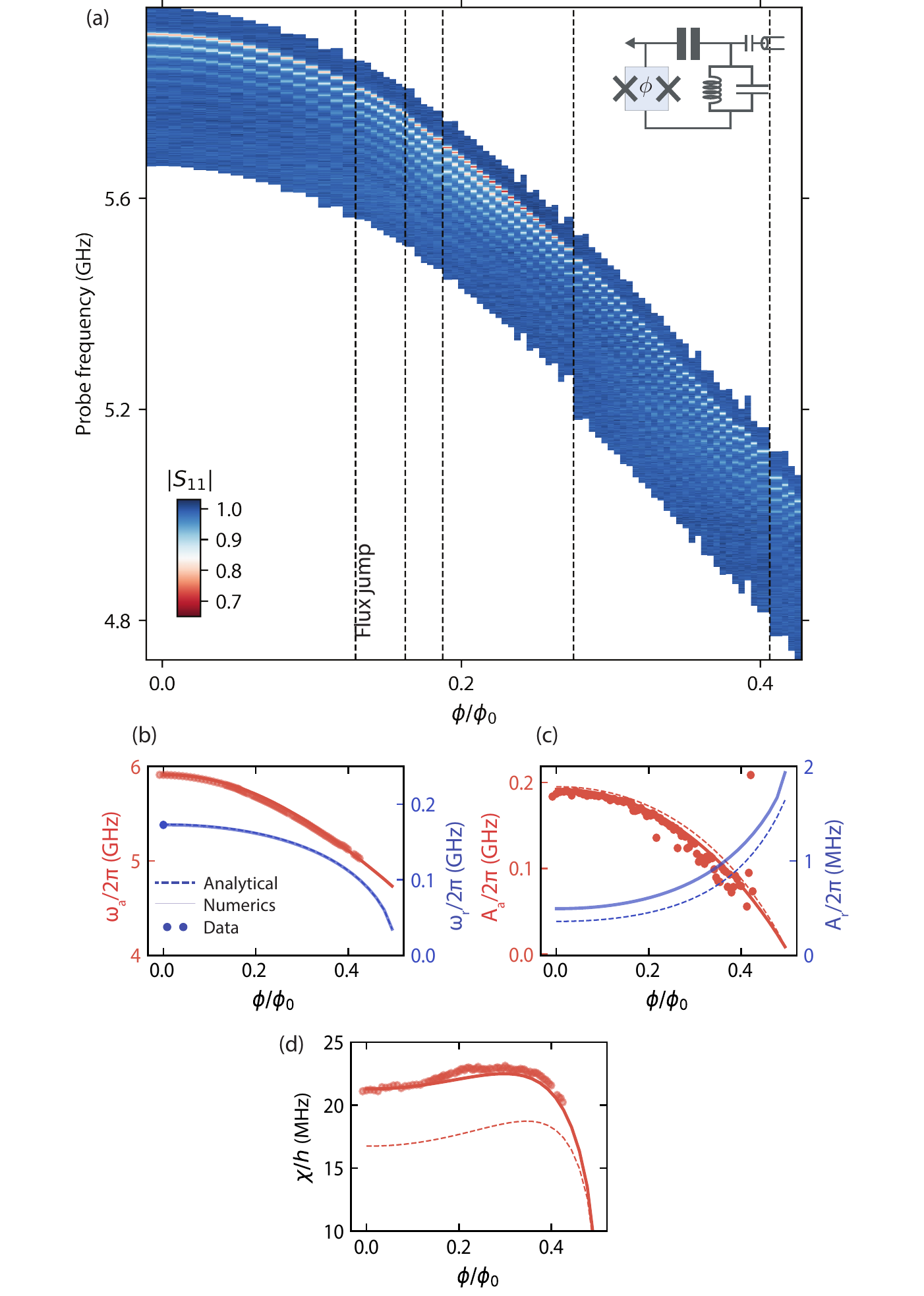}
\caption[Flux dependence of the system parameters]{
	\textbf{Flux dependence of the system parameters. (a) }
	$|S_{11}|$ as a function of probe frequency and flux through the SQUID. 
	Dashed black lines correspond to flux jumps.
	\textbf{(b,c,d) }
	Eigenfrequencies, anharmonicities and cross-Kerr coupling of the system as a function of flux.
	Dots are extracted through a sum-of-Lorentzians fit of the dataset in (a).
	Full curves correspond to a numerical diagonalization of the Hamiltonian of Eq.~(\ref{eq:Hamiltonian_8th_order}), Taylor expanded to the 8-th order.
	Dashed lines correspond to analytical formulas obtained from black-box quantization.
	The single data point corresponding to the resonator frequency is extracted from the sideband transition frequencies (Fig.~\ref{fig:S_two_tone}).
}
\label{fig:S_flux}
\end{figure}

We further verify the values of $C_H$, $C_L$, $L_\text{J}$ and $L$, as well as the black-box circuit analysis of Sec.~\ref{sec:black_box}, by extracting $A_a$, $\chi$ and $\omega_a$ for a varying $L_\text{J}$.
The junction or rather SQUID inductance is modified by sweeping the flux traversing it.
This is done by current-biasing a coil situated beneath our sample.
We show in Figs.~\ref{fig:S_flux}B,C,D the result of fitting a sum of Lorentzians to the flux-dependent spectrum (Fig.~\ref{fig:S_flux}(a)).
For each extracted parameter, we plot the theoretical evolution with flux obtained through a numerical diagonalization of the Hamiltonian of Eq.~(\ref{eq:Hamiltonian_8th_order}) (Taylor expanded to the 8-th order), as well as the analytical expressions obtained from black-box quantization (Eqs.~(\ref{eq:wl},\ref{eq:Al},\ref{eq:wh},\ref{eq:Ah})).
The only discrepancy is between the numerical and analytical estimation of $A_r$ and $\chi$.
It arises due to a term obtained from the quartic non-linearity of the junction proportional to: $(\hat a ^\dagger \hat a +1) (\hat a ^\dagger + \hat a )(\hat b ^\dagger + \hat b ) + h.c.$. 
This term resembles a beam-splitter interaction which typically makes an oscillator more anharmonic when coupled to an oscillator more non-linear than itself.

The asymmetry of the SQUID dictating the dependence of $L_\text{J}$ on flux was a fit parameter in the construction of this figure and was found to be $20\%$.
This experiment also suffered from a number of flux jumps, where the transition frequency of the circuit suddenly jumped to a different value.
The flux was then swept until we recovered the same frequency before continuing the scan.
This data-set is thus assembled from 6 different measurements.
Therefore, an entire flux periodicity was not successfully measured, making the conversion between the current fed into a coil under the sample and flux a free parameter.

{\renewcommand{\arraystretch}{1}
\begin{table}[]
\centering
\caption{Fitted system parameters}
\begin{tabular}{l l l l}
Quantity                                             & Symbol              & Value     & Equation \\ \hline
\\\textbf{Hamiltonian parameters}\\
Dressed transmon frequency \\($\ket{g,0}\rightarrow\ket{e,0}$) & $\omega_a$          & $2\pi\times$ 5.911~GHz & $\bar\omega_a+A_a/\hbar+\chi/2\hbar$ \\
Dressed resonator frequency \\($\ket{g,0}\rightarrow\ket{g,1}$)  & $\omega_r$          & $2\pi\times$ 173~MHz & $\bar\omega_r+A_r/\hbar+\chi/2\hbar$\\
Bare transmon frequency                                  & $\bar{\omega}_a$    & $2\pi\times$ 6.113~GHz          &  $\sqrt{\frac{L+L_J}{LL_JC_L}}$        \\
Bare resonator frequency                                   & $\bar{\omega}_r$    & $2\pi\times$ 182~MHz          &  $\frac{1}{\sqrt{(L+L_J)C_L}}$        \\ 
Transmon anharmonicity                               & $A_a$               &  $h\times$ 192~MHz  &  $\frac{e^2}{2C_H}\left(\frac{L}{L+L_J}\right)$,         \\
Resonator anharmonicity                                & $A_r$               &  $h\times$ 495 kHz  &  $\frac{e^2}{2C_L}\left(\frac{L_J}{L+L_J}\right)^3$,       \\
Cross-Kerr                                           & $\chi$              &  $h\times$ 21.29~MHz   &  $2\sqrt{A_rA_a}$        \\
\\\textbf{Dissipation rates}\\
Transmon dissipation rate                            & $\gamma$            &  $2\pi\times$ 3.70~MHz         &  \\
External coupling rate                               & $\gamma_\text{ext}$          &  $2\pi\times$ 1.63~MHz         &         \\
Resonator dissipation rate                             & $\kappa$            &  $2\pi\times$ 23.50 kHz         &         \\
Resonator external coupling rate                       & $\kappa_\text{ext}$          &  $2\pi\times$ 1.99 Hz         &        \\
\\
Transmon quality factor                              & $Q_a$               &  1599         &  $\omega_a/\gamma$        \\
Transmon external quality factor                     & $Q_a^{(\text{ext})}$         &  3617         &  $\omega_a/\gamma_\text{ext}$        \\
Resonator quality factor                               & $Q_r$               &  7348         &  $\omega_r/\kappa$        \\
Resonator external quality factor                      & $Q_r^{(\text{ext})}$         &  87 $\times 10^6$        &  $Z_0\sqrt{\frac{C_L}{L+L_J}}\left(\frac{C_\text{ext}}{C_L}\right)^2$        \\
\\\textbf{Thermal parameters}\\
Transmon temperature                                 & $T_a$               &  112~mK         &   \\
Resonator temperature                                  & $T_r$               &  17~mK         &    \\
\\
Transmon occupation number                           & $n_\text{th}^{(a)}$ &  0.09         &  $\frac{1}{e^{\frac{\hbar\omega_a}{k_BT_a}}-1}$        \\
Resonator occupation number                            & $n_\text{th}$       &  1.62        &  $\frac{1}{e^{\frac{\hbar\omega_r}{k_BT_r}}-1}$          
\\\textbf{Circuit parameters}\\
Josephson energy                                     & $E_J$               & $h\times$ 4.01~GHz &  $\frac{\hbar^2\bar{\omega}_a^2\bar{\omega}_r^2}{8\left(\bar{\omega}_a\sqrt{A_r}+\bar{\omega}_r\sqrt{A_a}\right)^2}$        \\
Josephson inductance                                 & $L_J$               & 41 nH          &  $\frac{\hbar^2}{4e^2E_J}$        \\
Resonator capacitance                                  & $C_L$               & 11.1 pF          &  $\frac{e^2 \sqrt{A_r}  \bar{\omega}_a^3}{2 \left(\bar{\omega}_r\sqrt{A_a}+\bar{\omega}_a\sqrt{A_r}\right)^3}$       \\
Transmon capacitance                                 & $C_H$               & 40.7~fF          &  $\frac{e^2 \bar{\omega}_r}{2 \left(\sqrt{A_aA_r} \bar{\omega}_a+A_a\bar{\omega}_r\right)}$        \\
Transmon inductance                                  & $L$                 & 28.2 nH          &  $\frac{2 \sqrt{A_a} \left(\bar{\omega}_r\sqrt{A_a}+\bar{\omega}_a\sqrt{A_r}\right)^2}{e^2\sqrt{A_r}  \bar{\omega}_a^3 \bar{\omega}_r}$        \\
Coupling capacitor                                   & $C_\text{ext}$               & 0.95~fF          &  $C_H\sqrt{\frac{\gamma_\text{ext}LL_J}{Z_0(L+L_J))}}   $    \\
Feedline impedance                                   & $Z_0$               & 50$\Omega$          &        \\
\end{tabular}
\label{tab:params}
\end{table}
}
\pagebreak

\section{Supplementary experimental data}

\subsection{Full time-dependent spectrum}
\begin{figure}
\centering
\includegraphics[width=0.8\textwidth]{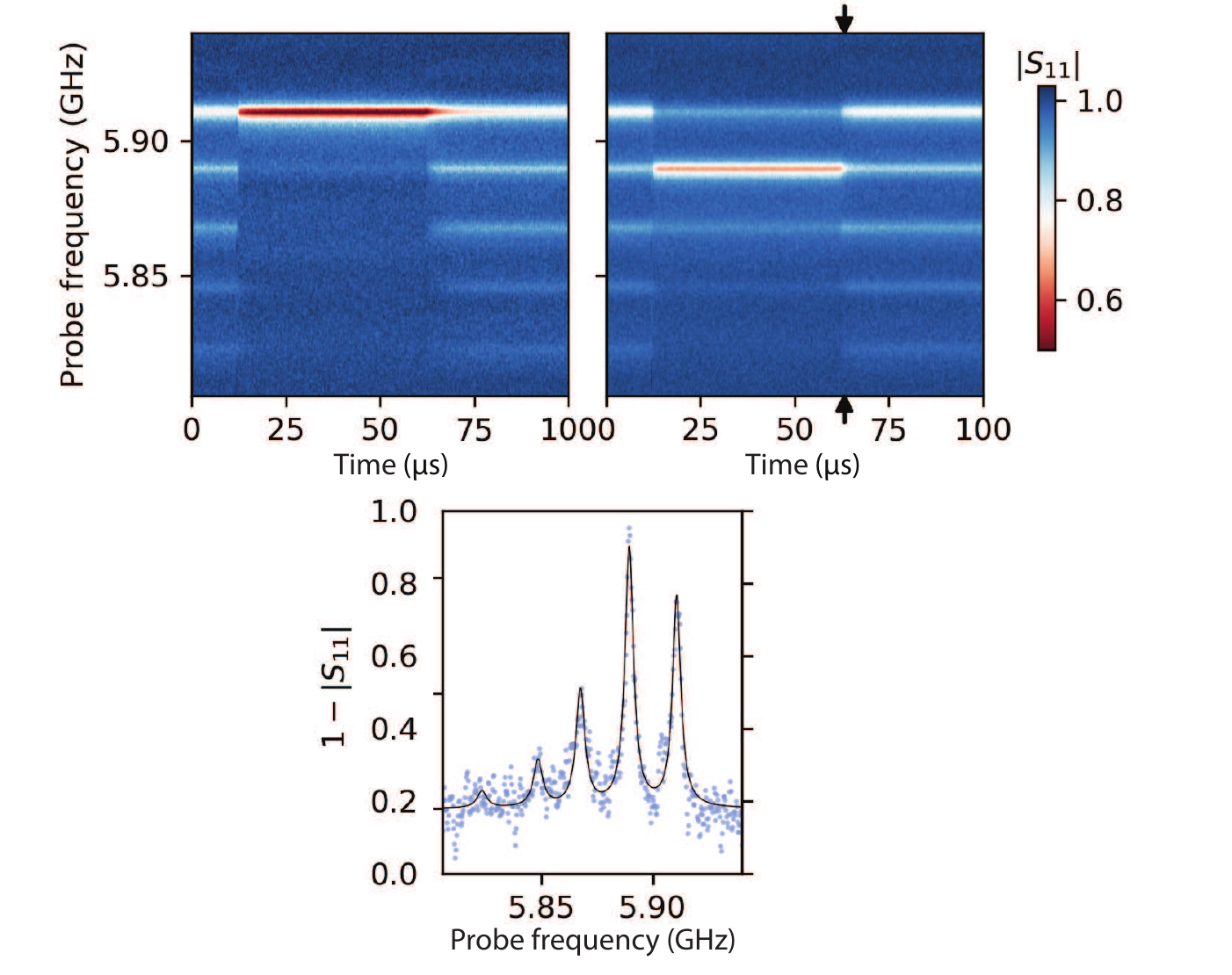}
\caption[Full time-dependent spectrum]{\textbf{Full time-dependent spectrum. }
	Time and probe frequency dependence of $|S_{11}|$ for both ground-state cooling (\textbf{(a)}) and the one-photon-state stabilization (\textbf{(b)}).
	By fitting these datasets using Eq.~(\ref{eq:sum_of_lorentzians}), in both frequency and time, we construct the plots shown in Fig.~\ref{fig:LFT_4}.
	\textbf{(c) }line cut of the data set (b) (indicated by arrows in (b)) is shown as blue dots, the black line corresponds to a fit.
	The relatively low signal-to-noise ratio is responsible for the large noise in probability of Fig.~\ref{fig:LFT_4}.
}
\label{fig:S_time_domain}
\end{figure}

See Fig.~\ref{fig:S_time_domain} for the full spectrum used to construct Fig.~\ref{fig:LFT_4}.

\subsection{Four-wave mixing spectrum}

\begin{figure}[]
\centering
\includegraphics[width=0.8\textwidth]{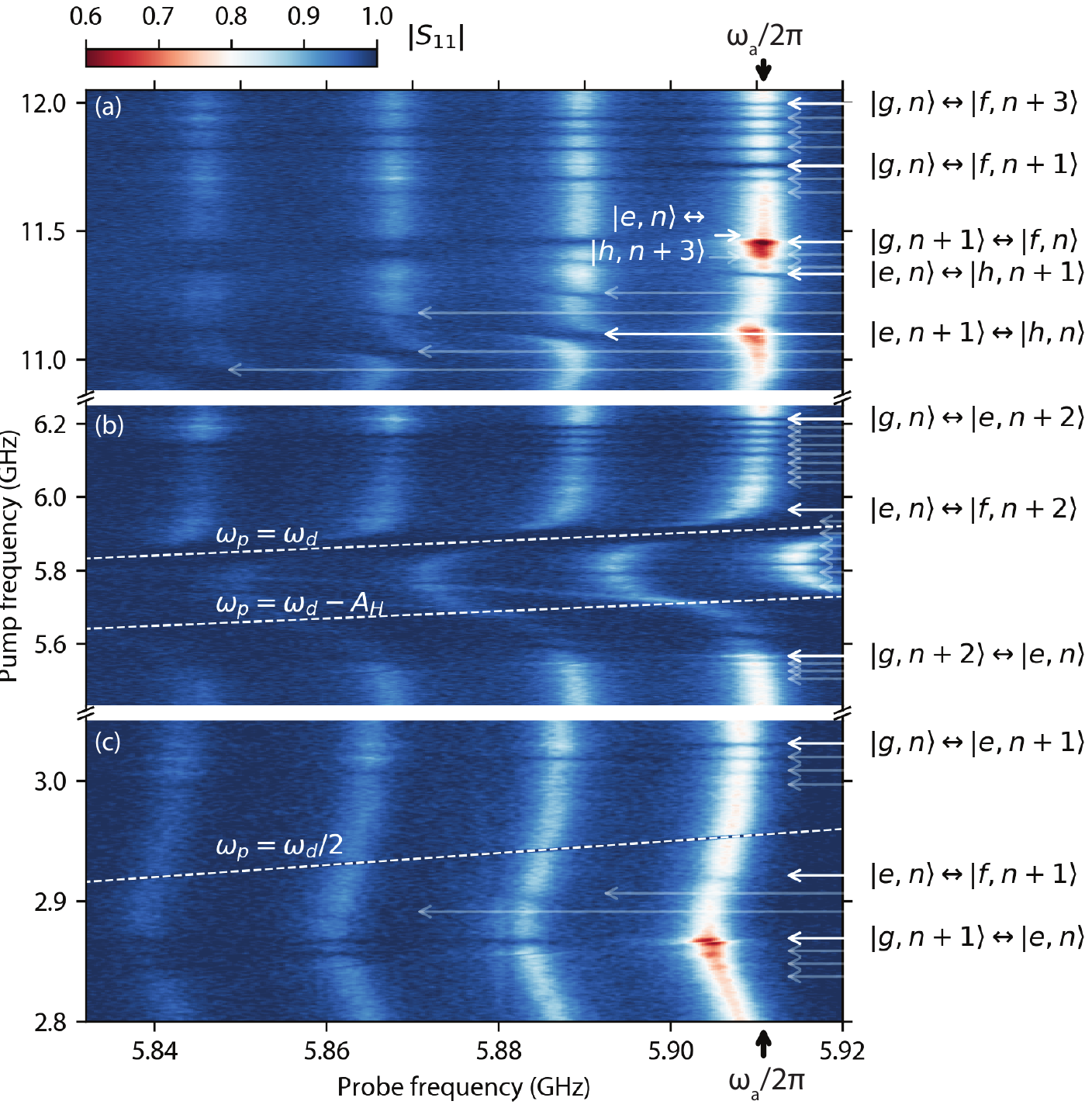}
\caption[Measurement of four-wave mixing processes]{
\textbf{Measurement of four-wave mixing processes.}
	$|S_{11}|$ as a function of probe frequency and of the frequency of a stronger pump tone.
	Features in the data are indicated by arrows with position generated from the eigenvalues of the Hamiltonian of Eq.~(\ref{eq:Hamiltonian_8th_order}), Taylor expanded to the 8-th order.
	%
	%
	The AC-Stark shift is not considered in the computation of transition energies, and a constant transmon frequency is taken as indicated by a black arrow in the x-axis, leading to slight mismatches between the transitions and placed arrows.
	In panel \textbf{(b)} dashed lines indicate the large avoided crossings observed when the pump tone is directly resonant with the transition frequencies of the transmon.
	In panel \textbf{(c)}, the dashed line indicates $\omega_p=\omega_d/2$, the features arising there being due to the first harmonic overtone of the pump (issuing from our microwave generator) driving the transmon.
}
\label{fig:S_all_transitions}
\end{figure}

By measuring the spectrum whilst sweeping the frequency of a pump tone, we show in Fig.~\ref{fig:S_all_transitions} the multitude of four-wave mixing processes possible in this system.
Panel (a) is particularly relevant to the cooling experiment, and is shown in Fig.~\ref{fig:LFT_3}(a), as one can see the relevant transitions lying next to the cooling transition $\ket{g,0}\leftrightarrow\ket{f,1}$.
We tested different combinations of raising and cooling four-wave mixing processes (panels (b) and (c)) for cooling and Fock-state stabilization, but these alternatives consistently produced lower state occupations than the results shown in the main text.

Two transitions in panel (a) are unexpected from a simple four-wave mixing approach to the system: $\ket{g,n}\leftrightarrow\ket{f,n+3}$ and $\ket{e,0}\leftrightarrow\ket{h,n+3}$.
These are six-wave mixing processes and one could expect them to have very weak effects.
However in this system the cross-Kerr $\chi$ is a considerable fraction of $\omega_r$.
The usually neglected term of the quartic non-linearity of the junction proportional to $\chi(2\hat a ^\dagger \hat a +1)(\hat b\hat b+\hat b^\dagger\hat b^\dagger)$ then leads to the dressed low Fock state $\ket{n}$ having a significant overlap with the bare states $\ket{n\pm 2k}$ where k is a positive integer.
The transition $\ket{g,0}\leftrightarrow\ket{f,3}$ is thus visible since $\ket{f,3}$ has a large overlap with $\ket{f,1}$ and $\ket{g,0}\leftrightarrow\ket{f,1}$ is an easily drivable four-wave mixing transition.

\subsection{Resonator spectrum}\label{sec:LFT_supp_low}

\begin{figure}[]
\centering
\includegraphics[width=0.93\textwidth]{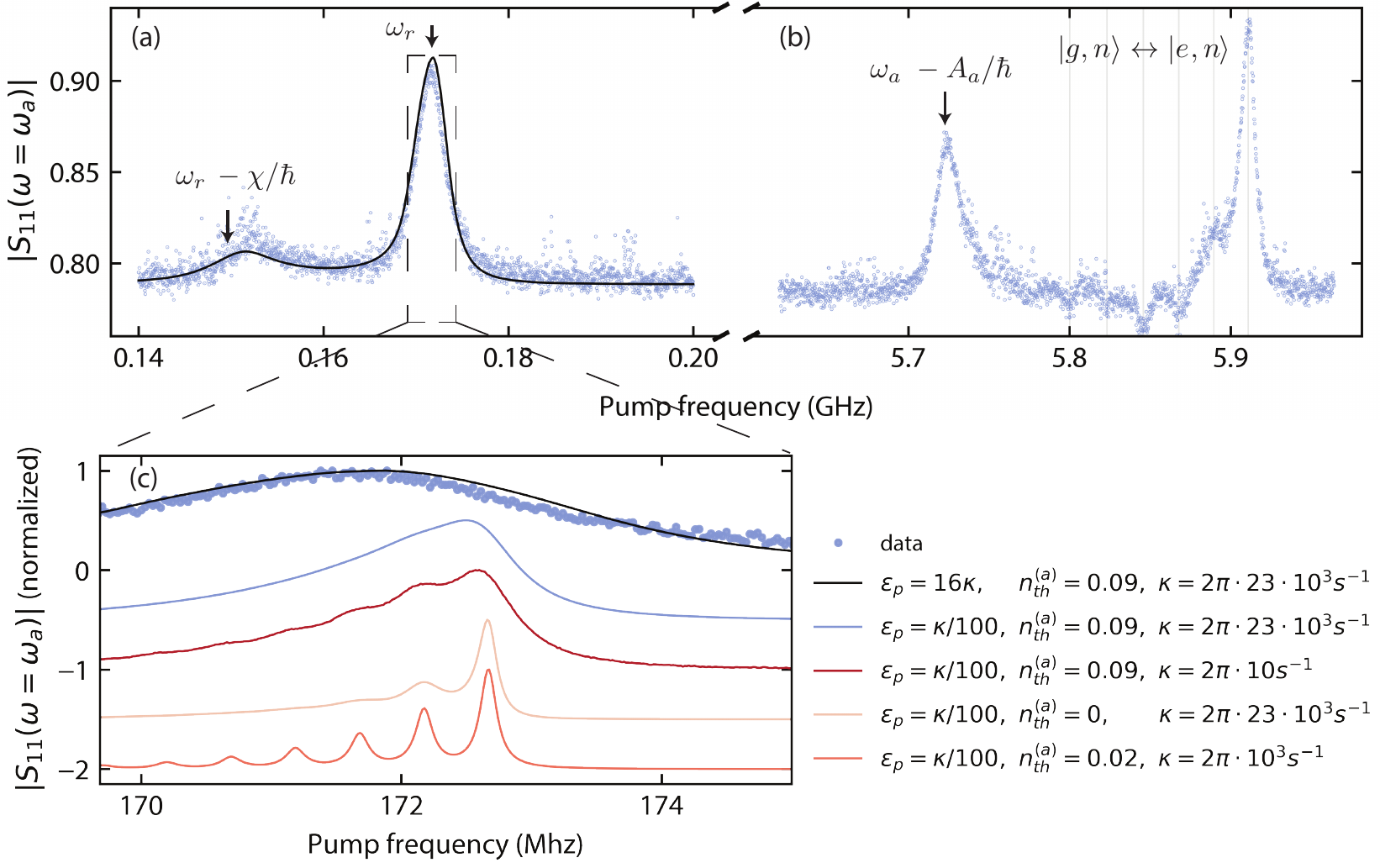}
\caption[Two-tone measurement of the anharmonicity and lower mode spectrum]{\textbf{Two-tone measurement of the anharmonicity and lower mode spectrum. }
\textbf{(a) }we sweep a pump tone around the resonator frequency (x-axis) whilst monitoring the depth of the $n=0$ peak $|S_{11}(\omega=\omega_a)|$ (y-axis).
We observe two peaks, separated by $\chi$ corresponding to the transitions $\ket{g,n}\leftrightarrow\ket{g,n+1}$ at $\omega_\text{p}=\omega_r$ and $\ket{e,n}\leftrightarrow\ket{e,n+1}$ at $\omega_\text{p}=\omega_r-\chi/\hbar$.
A steady-state numerical computation is shown as a black line and data as blue points.
\textbf{(b) }by performing the same measurement around the transmon frequency, we measure a peak corresponding to $\ket{e,0}\leftrightarrow\ket{f,0}$ at $\omega_\text{p}=\omega_a-A_a/\hbar$.
Compared to Fig.~\ref{fig:S_all_transitions}, these two datasets constitute a more direct measurement of $\omega_r$ and $A_a$.
\textbf{(c) }the line-width of the resonator was found to be significantly broader than $\kappa$, with no accessible photon-number resolution.
By varying simulation parameters as detailed in the legend, we explore the origin of this broad line-width.
}
\label{fig:S_two_tone}
\end{figure}

We monitor the height of the $\ket{g,0}\leftrightarrow\ket{e,0}$, whilst sweeping the frequency of a secondary pump tone.
As shown in Fig.~\ref{fig:S_two_tone}(a,b), this allows us to easily measure the anharmonicity of the transmon and the frequency of the resonator.
The line-width of the resonator peak is considerably larger than the previously determined resonator dissipation rate $\kappa/2\pi=23\cdot 10^3s^{-1}$.
If the line-width was equal to $\kappa$, we would expect to see photon number splitting, distinct peaks separated by the resonator anharmonicity $A_r$, corresponding to the transitions $\ket{g,n}\leftrightarrow\ket{g,n+1}$.
To understand why this is not the case, we fit a steady-state numerical computation of a pumped and probed Hamiltonian 
\begin{equation}
\begin{split}
	\hat H&=-A_a(\hat a^\dagger)^2\hat a^2 +\hbar(\omega_r-\omega_\text{p})\hat b^\dagger\hat b -A_r(\hat b^\dagger)^2\hat b^2\\ 
	&+ i\hbar\epsilon_\text{d}(\hat a^\dagger-\hat a)+ i\hbar\epsilon_\text{p}(\hat b^\dagger-\hat b)\ ,
\end{split}
\end{equation}
with the collapse operators of Eq.~(\ref{eq:probed_lindbald}).
The only free parameter is the pumping strength $\epsilon_\text{p}\sim16\times\kappa$, the probe strength was taken to be negligibly small with respect to all other rates in the model.
By varying simulation parameters, we can then explore the origin of this broad line-width. 
These results are summarized in Fig.~\ref{fig:S_two_tone}(c).
Reducing the pumping strength $\epsilon_\text{p}$ will suppress what is usually referred to as `power broadening', at the expense of the signal-to-noise ratio, but does not reveal photon-number splitting.
By reducing $\kappa$ to a negligibly small rate, photon number splitting can only be glimpsed behind a line-width broadening induced by the process $\ket{g,n}\rightarrow\ket{e,n}$ which occurs at a rate $\gamma n_\text{th}^{(a)}$.
This becomes clear if we instead keep $\kappa/2\pi=23\cdot 10^3s^{-1}$ and take the limit $n_\text{th}^{(a)}=0$, making the first two peaks apparent. 
As derived in Eq.~(\ref{eq:sum_of_lorentzians}), the line-width of a thermally populated anharmonic oscillator broadens significantly with its thermal occupation, which is responsible in this case for the disappearance (broadening) of peaks $n\ge2$.
By reducing both $\kappa$ and $n_\text{th}^{(a)}$, photon-number resolution would become visible.

\FloatBarrier\chapter{QuCAT: a Quantum Circuit Analyzer in Python}
\label{chapter-qucat}
\label{chapter_qucat}

\begin{abstract}
A large part of this thesis has focused on determining the correct circuit to implement a desired Hamiltonian.
Here we present QuCAT, or ``Quantum Circuit Analyzer Tool", an open-source framework to help in such a task.
This open-source Python library features an intuitive graphical or programmatical interface to create circuits, the ability to compute their Hamiltonian, and a set of complimentary functionalities such as calculating dissipation rates or visualizing current flow in the circuit.
\end{abstract}

\newpage

\noindent 
A modified version of this chapter was published as M. F. Gely and G. A. Steele, \textit{QuCAT: Quantum Circuit Analyzer Tool in Python}, \href{https://iopscience.iop.org/article/10.1088/1367-2630/ab60f6/meta}{New Journal of Physics \textbf{22}, 013025 (2020)}.
MFG developed QuCAT and the underlying theory and methods. 
MFG wrote the manuscript with input from GAS. 
GAS supervised the project.
The authors acknowledge Marios Kounalakis for useful discussions and for reading the manuscript.
This work was supported by the European Research Council under the European Union’s H2020 program [grant numbers 681476-QOM3D, 732894-HOT, 828826-Quromorphic].
The code used to generate the figures of this paper are available in Zenodo with the identifier 10.5281/zenodo.3298107.
Tutorials and examples, including those presented here are available on the QuCAT website at \url{https://qucat.org/}.
The latest version of the QuCAT source code, is available to download or to contribute to at \url{https://github.com/qucat}.

\section{Introduction}
Here we present QuCAT, which stands for ``Quantum Circuit Analyzer Tool", an open-source Python framework to help in designing, analyzing and understanding quantum circuits.
The toolbox provides an easy interface to create and visualize circuits, either programmatically or through a graphical user interface.
A Hamiltonian can then be generated for further analysis in QuTiP~\cite{johansson2012qutip,johansson2013qutip}.
The current version of QuCAT supports quantization in the basis of normal modes of the linear circuit~\cite{nigg_black-box_2012}, making it suited for the analysis of weakly anharmonic circuits with small losses.
The properties of these modes: their frequency, dissipation rates, anharmonicity and cross-Kerr couplings can be directly calculated.
The user can also visualize the current flows in the circuit associated with each normal mode.
The library covers lumped element circuits featuring an arbitrary number of Josephson junctions, inductors, capacitors and resistors.
Through equivalent lumped element circuits, certain distributed elements such as waveguide resonators can also be analyzed (see Sec.~\ref{sec:mmusc}).
The software relies on the symbolic manipulation of the circuits equations, making it reliable even for vastly different circuits and parameters.
It also results in efficient parameter sweeps, as analytical manipulations need not be repeated for different circuit parameters.
In a few seconds, circuits featuring 10 nodes (or degrees of freedom), corresponding to between 10 and 30 circuit elements can be simulated.

We first cover the functionalities of the software.
We start by showing how to create circuits, first using the graphical user interface, then programmatically.
We then demonstrate how to generate the corresponding Hamiltonian.
Lastly, we show how to extract the characteristics of the circuit modes: frequencies, dissipation, anharmonicity and cross-Kerr coupling and present a tool to visualize these modes. 
This first section will feature as an example the standard circuit of a transmon qubit coupled to a resonator~\cite{koch_charge-insensitive_2007}.
In the appendices, we will first use QuCAT to analyze some recent experiments: a tuneable coupler~\cite{kounalakis2018tuneable}, a multi-mode ultra-strong coupling circuit~\cite{bosman_multi-mode_2017}, a microwave optomechanics circuit~\cite{ockeloen2016low} and a Josephson-ring based qubit~\cite{roy2017implementation}.
We then provide an overview of the circuit quantization method used and the algorithmic methods which implement it.
The limitations of these methods regarding weak anharmonicity and circuit size will then be presented.
Finally we will explain how to install QuCAT and we provide a summary of all its functions.
More tutorials and examples are available on the QuCAT website \url{https://qucat.org/}.

\section{Circuit construction}
\begin{figure}[]
\centering
\includegraphics[width=0.8\textwidth]{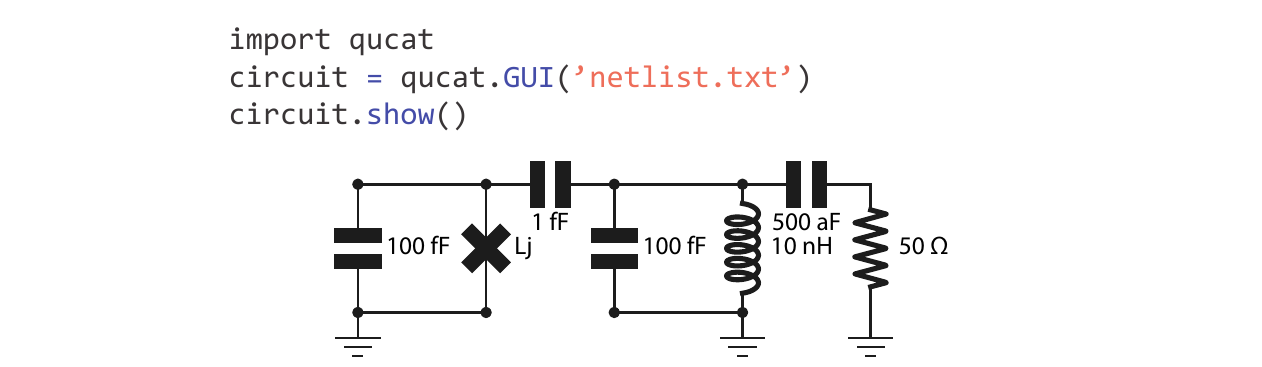}
\caption{
\textbf{Construction of a circuit: }
code and output.
The circuit used as an example in this section comprises of a transmon qubit on the left, coupled through a 1 fF capacitor to an LC-oscillator.
Dissipation arises from the capacitive coupling of the LC-oscillator to a 50 $\Omega$ resistor on the right.
After importing the \inline{qucat} package, the \inline{circuit} object is created manually through a graphical user interface (GUI) opened after calling \inline{qucat.GUI("netlist.txt")}.
All information necessary to construct the circuit is stored in the text file \inline{netlist.txt}.
After closing the GUI, this information is also stored in the variable \inline{circuit}.
The \inline{show} method finally displays the circuit.
}
\label{fig:QuCAT_1}
\end{figure}
Any use of QuCAT will start with importing the \inline{qucat} library
\begin{lstlisting}
import qucat
\end{lstlisting}
One should then create a circuit.
These are named \inline{Qcircuit}, short for ``quantum circuit" in QuCAT.
There are two ways of creating a \inline{Qcircuit}: using the graphical user interface (GUI), or programmatically.

\subsection{Creating a circuit with the GUI}

We first cover how to create a circuit with the GUI.
This is done through this command 
\begin{lstlisting}
circuit = qucat.GUI('netlist.txt')
\end{lstlisting}
which opens the GUI.
The GUI will appear as a separate window, which will block the execution of the rest of the Python script until the window is closed.
The user can drag-in and drop capacitors, inductors, resistors or Josephson junctions, or grounds.
These components can then be inter-connected with wires.
Each change made to the circuit will be automatically be saved in the \inline{'netlist.txt'} file.
After closing the GUI, the \inline{Qcircuit} object will be stored in the variable named \inline{circuit} which we will use for further analysis.

\subsection{Creating a circuit programmatically}
Alternatively, one can create a circuit with only Python code.
This is done by creating a list of circuit components with the functions \inline{J}, \inline{L}, \inline{C} and \inline{R} for junctions, inductors, capacitors and resistors respectively.
For the circuit of Fig.~\ref{fig:QuCAT_1}:
\begin{lstlisting}
circuit_components = [
qucat.C(0,1,100e-15), # transmon
qucat.J(0,1,'Lj'),
qucat.C(0,2,100e-15), # resonator
qucat.L(0,2,10e-9),
qucat.C(1,2,1e-15), # coupling capacitor
qucat.C(2,3,0.5e-15), # ext. coupl. cap.
qucat.R(3,0,50) # 50 Ohm load
]
\end{lstlisting}
All circuit components take as first two argument integers referring to the negative and positive node of the circuit components.
Here 0 corresponds to the ground node for example.
%
%
The third argument is either a float giving the component a value, or a string which labels the component parameter to be specified later.
Doing the latter avoids performing the computationally expensive initialization of the \inline{Qcircuit} object multiple times when sweeping a parameter.
By default, junctions are parametrized by their Josephson inductance $L_j = \phi_0^2/E_j$
where $\phi_0 = \hbar/2e$ is the reduced flux quantum, 
and $E_j$ (in Joules) is the Josephson energy.

Once the list of components is built, we can create a \inline{Qcircuit} object via the \inline{Network} function
\begin{lstlisting}
circuit = Network(circuit_components)
\end{lstlisting}
as with a construction via the GUI, the \inline{Qcircuit} object will be stored in the variable named \inline{circuit} which we will use for further analysis.

\section{Generating a Hamiltonian}

\begin{figure}[]
\centering
\includegraphics[width=0.8\textwidth]{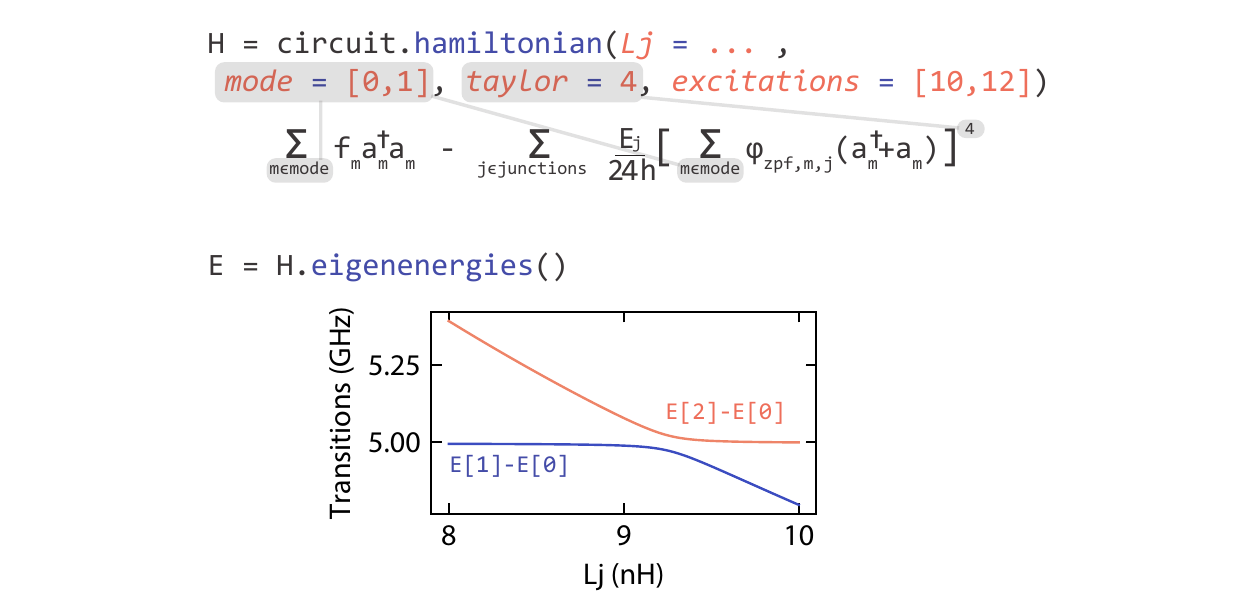}
\caption{
\textbf{Hamiltonian generation}
is done by applying the \inline{hamiltonian} method to the \inline{circuit} variable defined in Fig.~\ref{fig:QuCAT_1}.
The Hamiltonian is expressed in the basis of circuit normal modes $m$ with frequencies $f_m = \omega_m/2\pi$, annihilation operators $\hat{a}_m$, and zero-point phase fluctuations $\varphi_{\text{zpf},m,j}$ across junction $j$ with Josephson energy $E_j$.
The junction non-linearities are expressed through a Taylor expansion of the cosine potentials, where the user chooses the degree of Taylor expansion.
The other arguments are the list of modes to include, the number of excitations to consider for each of these modes, and any unspecified component value, here \inline{Lj}.
The returned Hamiltonian is a QuTiP object, giving the user access to an extensive set of tools for further analysis~\cite{johansson2012qutip,johansson2013qutip}.
As an example, we compute the eigenenergies of the Hamiltonian, and plot the two first transition frequencies, as a function of \inline{Lj}.
}
\label{fig:QuCAT_4}
\end{figure}
The Hamiltonian of a Josephson circuit, for small values of the phase fluctuations $\hat \varphi_j$ across its junctions, is given by 
\begin{equation}
\hat{H} = \sum_{m} \hbar\omega_m\hat{a}_m^\dagger\hat{a}_m + \sum_j\sum_{n\ge k\ge 2}E_j\frac{(-1)^{n+1}}{(2n)!}\hat \varphi_j^{2n}\ .
\label{eq:hamiltonian_taylor}
\end{equation}
It is written in the basis of its normal modes.
These have an angular frequency $\omega_m$ and we write the operator which creates (annihilates) photons in the mode $\hat a_m^\dagger$ ($\hat a_m$).
The cosine potential of each Josephson junction $j$ with Josephson energy $E_j$ has been Taylor expanded to order $2n$ for small values of its phase fluctuations $\hat \varphi_j$ across it.
The phase fluctuations are a function of the annihilation and creation operators of the modes $\hat\varphi_j=\sum_{m}\varphi_{\text{zpf},m,j}(\hat{a}_m^\dagger+\hat{a}_m)$.
For a detailed derivation of this Hamiltonian, and the method used to obtain its parameters, see Sec.~\ref{sec:circuit_quantization_overview}.
There are three different parameters that the user should fix
\begin{enumerate}
\item the set of modes to include
\item for each of these modes, the number of excitations to consider
\item the order of the Taylor expansion.
\end{enumerate}
The more modes and excitations are included, and the higher Taylor expansion order, the more faithful the Hamiltonian will be to physical reality.
The resulting increase in Hilbert space size will however make it more computationally expensive to perform further calculations.
Typically, larger degrees of anharmonicity require a larger Hilbert space, with a fundamental limitation on the maximum anharmonicity due to the choice of basis.
We expand on these topics in Sec.~\ref{sec:high_anharmonicity}.

Such a Hamiltonian is generated through the method \inline{hamiltonian}. 
More specifically, this function returns a QuTiP object~\cite{johansson2012qutip,johansson2013qutip}, enabling an easy treatment of the Hamiltonian.
All QuCAT functions use units of Hertz, so the function is actually returning $\hat H/h$.

As an example, we generate a Hamiltonian for the circuit of Fig.~\ref{fig:QuCAT_1} at different values of the Josephson inductance and use QuTiP to diagonalize it and obtain the eigen-frequencies of the system.
For a Josephson inductance of $8$ nH this is achieved through the commands
\begin{lstlisting}
H = circuit.hamiltonian(
        modes = [0,1],
        excitations = [10,12],
        taylor = 4,
        Lj = 8e-9)        
E = H.eigenenergies() # Eigenenergies (here in units of frequency) using the QuTiP function eigenenergies
\end{lstlisting}
With \inline{modes = [0,1]}, we are specifying that we wish to consider the first and second modes of the circuit.
Modes are numbered with increasing frequency, so here we are selecting the two lowest frequency modes of the circuit.
With \inline{excitations = [10,12]}, we specify that for mode \inline{0} (\inline{1}) we wish to consider \inline{10} (\inline{12}) excitations.
With \inline{taylor = 4}, we are specifying that we wish to expand the cosine potential to fourth order, this is the lowest order which will give an anharmonic behavior.
The unspecified Josephson inductance must now be fixed through a keyword argument \inline{Lj = 8e-9}. 
Doing so avoids initializing the \inline{Qcircuit} objects multiple times during parameter sweeps, as initialization is the most computationally expensive task.
We calculate these energies with different values of the Josephson inductance, and the first two transition frequencies are plotted in Fig.~\ref{fig:QuCAT_4}, showing the typical avoided crossing seen in a coupled qubit-resonator system.

\section{Mode frequencies, dissipation rates, anharmonicities and cross-Kerr couplings}
\begin{figure}[]
\centering
\includegraphics[width=0.8\textwidth]{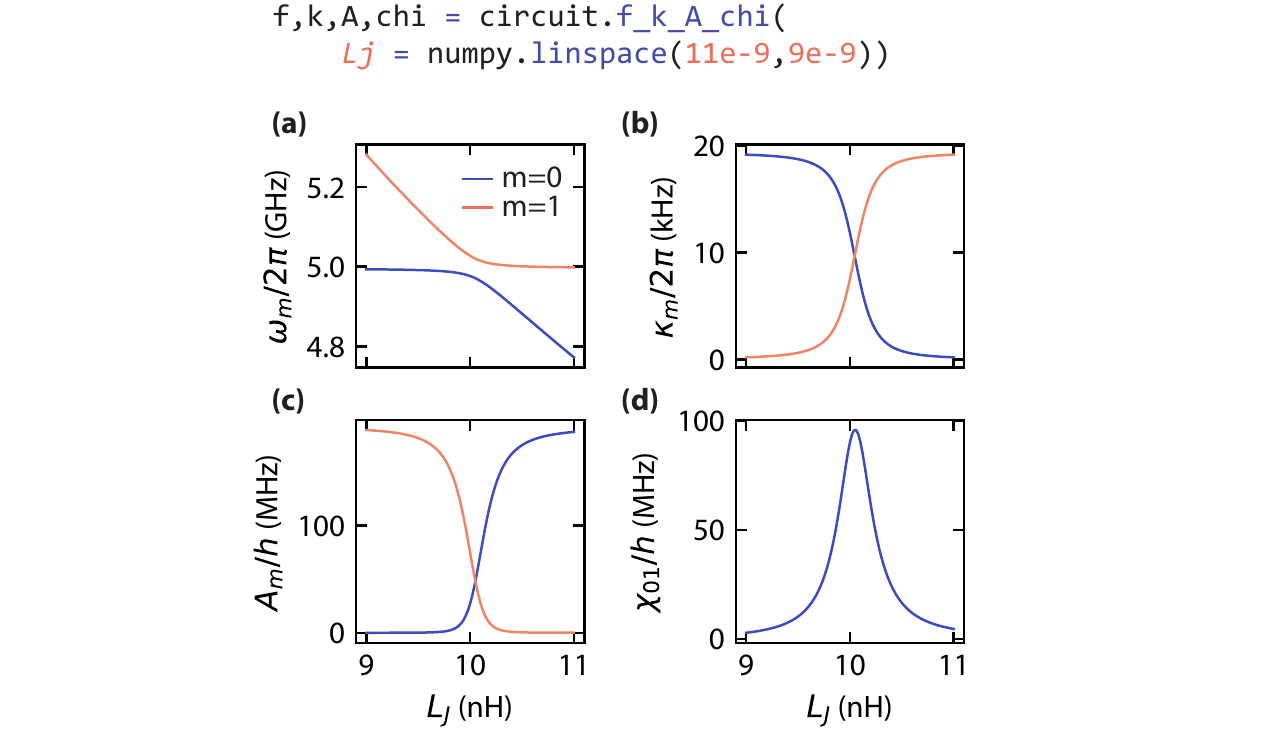}
\caption{
\textbf{Extracting eigenfrequencies, loss-rates, anharmonicities, and cross-Kerr couplings.}
We apply the \inline{f_k_A_chi} method to \inline{circuit} defined in Fig.~\ref{fig:QuCAT_1} to obtain a list of eigenfrequencies (\inline{f}), loss-rates (\inline{k}), anharmonicities (\inline{A}), and cross-Kerr couplings (\inline{chi}), for all the normal modes of the circuit.
There is one unspecified variable in the circuit, the Josephson inductance $L_j$, which is here specified with a list of values.
In \textbf{(a)}, we plot the eigenfrequencies of the two first modes \inline{f[0]} and \inline{f[1]}.
In \textbf{(b)}, we plot the loss-rates of the same modes \inline{k[0]} and \inline{k[1]}, and in \textbf{(c)} their anharmonicities \inline{A[0]} and \inline{A[1]}.
In \textbf{(d)}, we plot the cross-Kerr coupling between modes \inline{0} and \inline{1}: \inline{chi[0,1]}.
}
\label{fig:QuCAT_2}
\end{figure}
QuCAT can also return the parameters of the (already diagonal) Hamiltonian in first-order perturbation theory
\begin{equation}
\begin{split}
\hat{H} = \sum_m\sum_{n\ne m} (\hbar\omega_m-A_m-\frac{\chi_{mn}}{2})\hat{a}_m^\dagger\hat{a}_m \\
-\frac{A_m}{2}\hat{a}_m^\dagger\hat{a}_m^\dagger\hat{a}_m\hat{a}_m -\chi_{mn}\hat{a}_m^\dagger\hat{a}_m\hat{a}_n^\dagger\hat{a}_n
\label{eq:hamiltonian_first_order_maintext}
\end{split}
\end{equation}
valid for weak anharmonicity $\chi_{mn},A_m\ll \omega_m$.
The physics of this Hamiltonian can be understood by considering that an excitation of one of the circuit modes may lead to current traversing a Josephson junction. 
This will change the effective inductance of the junction, hence changing its own mode frequency, as well as the mode frequencies of all other modes.
This is quantified through the anharmonicity or self-Kerr $A_m$ and cross-Kerr $\chi_{mn}$ respectively.
When no mode is excited, vacuum-fluctuations in current through the junction give rise to shifted mode energies $\hbar\omega_m-A_m-\sum_n\chi_{mn}/2$.
In a circuit featuring resistors, these anharmonic modes will be dissipative.
A mode $m$ will lose energy at a rate $\kappa_m$.
If these rates are specified in angular frequencies, the relaxation time $T_{1,m}$ of mode $m$ is given by $T_{1,m} = 1/\kappa_m$.
A standard method to include the loss rates in a mathematical description of the circuit is through the Lindblad equation~\cite{johansson2012qutip}, where the losses would be included as collapse operators $\sqrt{\kappa_m}\hat a_m$.

The frequencies, dissipation rates, and Kerr parameters can all be obtained via methods of the \inline{Qcircuit} object.
These methods will return numerical values, and we should always specify the values of symbolically defined circuit parameters as keyword arguments.
Lists, or Numpy arrays, can be provided here making it easy to perform parameter sweeps.
Additionally, initializing the circuit is the most computationally expensive operation, so this will be by far the fastest method to perform parameter sweeps.
We will assume that we want to determine the parameters of the Hamiltonian~(\ref{eq:hamiltonian_first_order_maintext}) for the circuit of Fig.~\ref{fig:QuCAT_1} at different values of $L_j$.
The values for $L_j$ are stored as a Numpy array
\begin{lstlisting}
Lj_list = numpy.linspace(11e-9,9e-9,101)
\end{lstlisting}
We can assign the frequency, dissipation rates, self-Kerr, and cross-Kerr parameters to the variables \inline{f}, \inline{k}, \inline{A} and \inline{chi} respectively, by calling
\begin{lstlisting}
f = circuit.eigenfrequencies(Lj = Lj_list)
k = circuit.loss_rates(Lj = Lj_list)
A = circuit.anharmonicities(Lj = Lj_list)
chi = circuit.kerr(Lj = Lj_list)
\end{lstlisting}
or alternatively through a single function call:
\begin{lstlisting}
f,k,A,chi = circuit.f_k_A_chi(Lj = Lj_list)
\end{lstlisting}
All values returned by these methods are given in Hertz, not in angular frequency.
With respect to the conventional way of writing the Hamiltonian, which we have also adopted in~(\ref{eq:hamiltonian_first_order_maintext}), we thus return the frequencies as $\omega_m/2\pi$, the loss rates as $\kappa_m/2\pi$ and the Kerr parameters as $A_m/h$ and $\chi_{mn}/h$.
Note that \inline{f}, \inline{k}, \inline{A}, are arrays, where the index \inline{m} corresponds to mode $m$, and modes are ordered with increasing frequencies.
For example, \inline{f[0]} will be an array of length 101, which stores the frequencies of the lowest frequency mode as \inline{Lj} is swept from 11 to 9 nH.
The variable \inline{chi} has an extra dimension, such that \inline{chi[m,n]} corresponds to the cross-Kerr between modes \inline{m} and \inline{n}, and \inline{chi[m,m]} is the self-Kerr of mode \inline{m}, which has the same value as \inline{A[m]}.
These generated values are plotted in Fig.~\ref{fig:QuCAT_2}.
We can also print these parameters in a visually pleasing way to get an overview of the circuit characteristics for a given set of circuit parameters.
For a Josephson inductance of $9$ nH, this is done through the command
\begin{lstlisting}
circuit.f_k_A_chi(Lj = 10e-9, pretty_print = True)
\end{lstlisting}
which will print 
\begin{lstlisting}
mode |   freq.  |   diss.  |   anha.  |
   0 | 4.99 GHz | 9.56 kHz | 10.5 kHz |
   1 | 5.28 GHz |  94.3 Hz |  189 MHz |

Kerr coefficients 
diagonal = Kerr
off-diagonal = cross-Kerr
mode |     0    |    1    |
   0 | 10.5 kHz |         |
   1 | 2.82 MHz | 189 MHz |
\end{lstlisting}
We see that mode 1 is significantly more anharmonic than mode 0, whereas mode 0 has however a higher dissipation.
We would expect that mode 1 is thus the resonance which has current fluctuations mostly located in the junction, 
whilst mode 0 is located on the other side to the coupling capacitor, where it can couple more strongly to the resistor.

\begin{figure}[]
\centering
\includegraphics[width=0.8\textwidth]{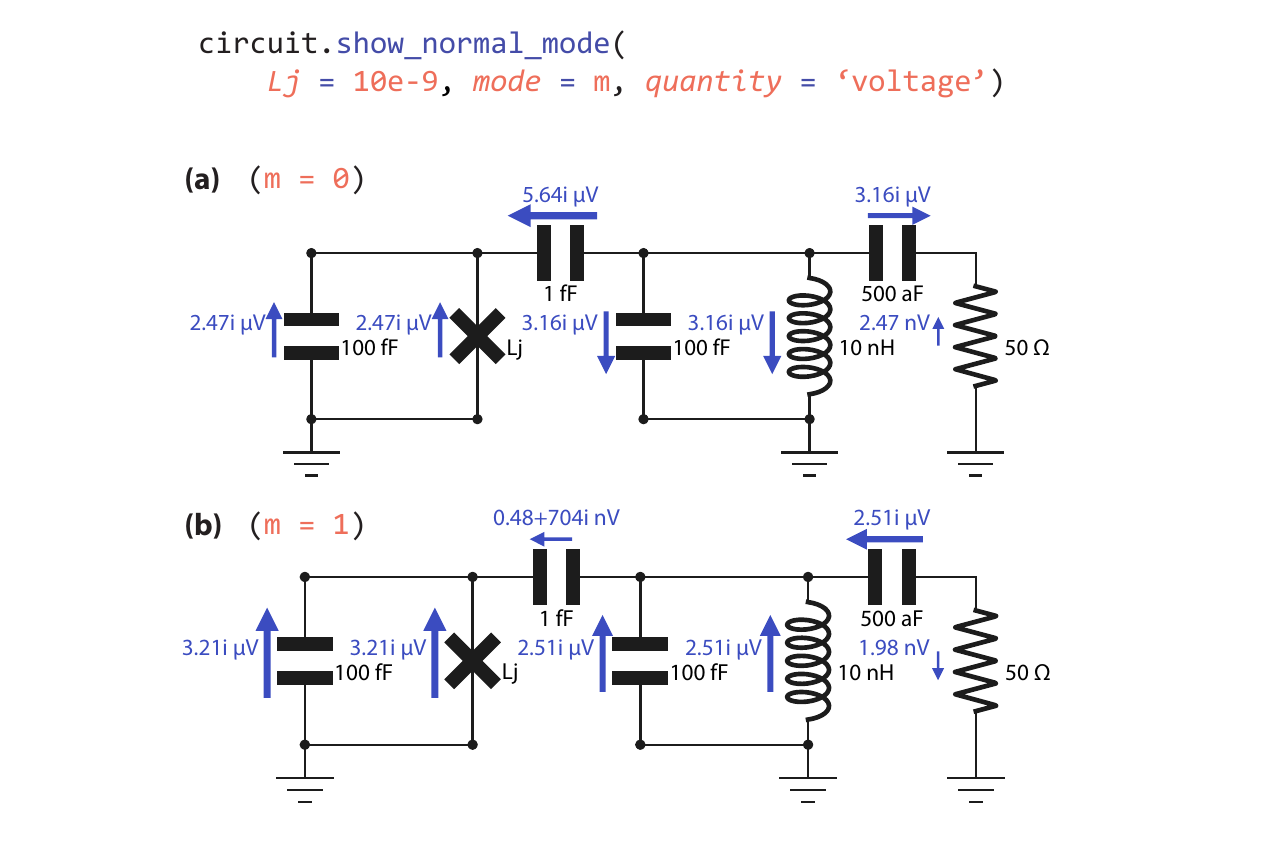}
\caption{
\textbf{Visualizing normal modes.}
The \inline{show_normal_mode} method overlays the circuit with arrows representing the voltage across components when the circuit is populated with a quarter-photon amplitude coherent state.
The arrows are annotated with the value of the complex voltage oscillating across a component, where the direction of the arrow indicates the direction of a phase 0 for that component.
The absolute value of this annotation corresponds to the zero-point fluctuations of the given quantity across the component.
The length and thickness of the arrows scale with the magnitude of the voltage.
\inline{show_normal_mode} takes as argument any unspecified circuit parameter, here we specify \inline{Lj=10e-9} where the two modes undergo an avoided crossing.
We plot each mode by specifying \inline{mode = 0} or \inline{mode = 1} and see that for mode \inline{0}, the anti-symmetric mode, the voltage has opposite signs on each side of the coupling capacitor, leading to a larger voltage across the coupler (and hence a larger effective capacitance and lower frequency) than the symmetric mode.
}
\label{fig:QuCAT_3}
\end{figure}

Such interpretations can be verified by plotting a visual representation of the normal modes on top of the circuit as explained below.
This can be done by plotting either the current, voltage, charge or flux distribution, overlaid on top of the circuit schematic.
As shown in Fig.~\ref{fig:QuCAT_3}, this is done by adding arrows, representing one of these quantities at each circuit component and annotating it with the value of that component.
The annotation corresponds to the complex amplitude, or phasor, of a quantity across the component, if the mode was populated with a quarter-photon amplitude coherent state.
The absolute value of this annotation corresponds to the contribution of a mode to the zero-point fluctuations of the given quantity across the component.
The direction of the arrows indicates what direction we take for 0 phase for that component.

We note that an independently developed Julia platform also allows the calculation of normal mode frequencies and dissipation rates for circuits~\cite{ScheerBlock2018}.

\section{Applications}

\subsection{Designing a microwave filter}
\begin{figure}[]
\centering
\includegraphics[width=0.8\textwidth]{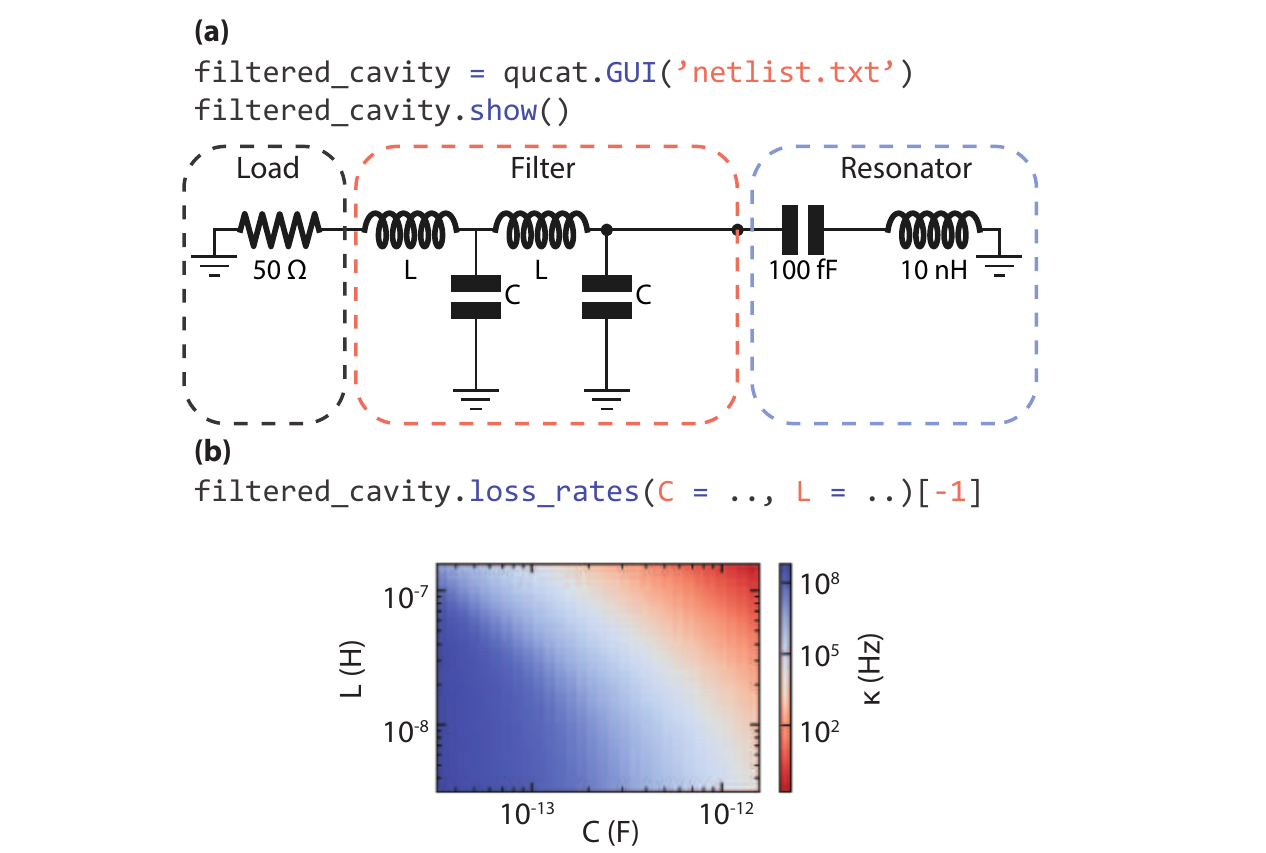}
\caption{
\textbf{Design of a microwave filter. (a) }
Using the QuCAT GUI, we build then plot a model of a filtered cavity.
A 50 $\Omega$ load, representing a cable, is connected to an LC resonator through two LC band pass filters.
\textbf{(b)} The dissipation rate of the resonator is plotted as a function of inductance and capacitance of the filter using the \inline{loss_rates} method.
}
\label{fig:QuCAT_filtered_cavity}
\end{figure}
In this application we show how QuCAT can be used to design classical microwave components. 
We study here a band pass filter made from two LC oscillators with the inductor inline and a capacitive shunt to ground.
Such a filter can be used to stop a DC bias line from inducing losses, whilst being galvanically connected to a resonator, see for example Ref.~\cite{Viennot2018}.
In this case we are interested in the loss rate $\kappa$ of a LC resonator connected through this filter to a 50 $\Omega$ load, which could emulate a typical microwave transmission line.
We want to study how $\kappa$ varies as a function of the inductance $L$ and capacitance $C$ of its components.

The QuCAT \inline{GUI} function can be used to open the GUI, the user will manually create the circuit, and upon closing the GUI a Qcircuit object is stored in the variable \inline{filtered_cavity}.
By calling the method \inline{show}, we display the circuit as shown in Fig.~\ref{fig:QuCAT_filtered_cavity}(a).
These steps are accomplished with the code
\begin{lstlisting}
# Open the GUI and manually build the circuit
filtered_cavity = qucat.GUI('netlist.txt')
# Display the circuit
filtered_cavity.show()
\end{lstlisting}
We can then access the loss rates of the different circuit modes through the method \inline{loss_rates}.
Since the values of $C$ and $L$ were not specified in the construction of the circuit, their values have to be passed as keyword arguments upon calling \inline{loss_rates}.
For example, the loss rate for a 1 pF capacitor and 100 nH inductor is obtained through
\begin{lstlisting}
# Loss rates of all modes
k_all = filtered_cavity.loss_rates(C = 1e-12, L = 100e-9)
# Resonator loss rate
k = k_all[-1]
\end{lstlisting}
Since the filter capacitance and inductance is large relative to the capacitance and inductance of the resonator, the modes associated with the filter will have a much lower frequency.
We can thus access the loss rate of the resonator by always selecting the last element of the array of loss rates with the command \inline{k_all[-1]}
The dissipation rates for different values of the capacitance and inductance are plotted in Fig.~\ref{fig:QuCAT_filtered_cavity}(b).
%

%

\subsection{Computing optomechanical coupling}
\begin{figure}[]
\centering
\includegraphics[width=0.8\textwidth]{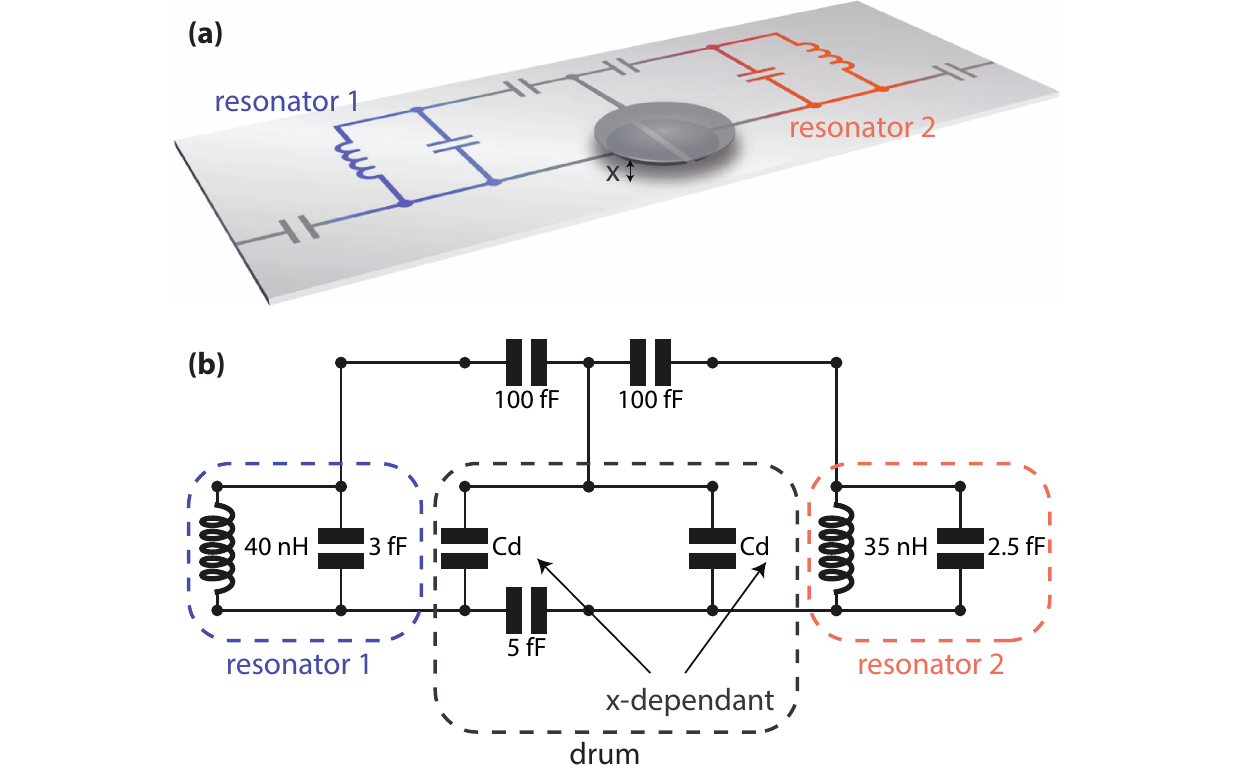}
\caption{
\textbf{Example of an optomechanical system}
\textbf{(a)} Schematic of the device, adapted from Ref.~\cite{ockeloen2016low} under a \href{https://creativecommons.org/licenses/by/3.0/}{CC BY 3.0} license. 
Two resonators are connected through a network of capacitances and a mechanically compliant capacitor (drum).
\textbf{(b)} QuCAT reconstruction of the circuit.
By specifying a label for the mechanically compliant capacitances, we have the possibility to compute the eigenfrequencies $\omega_m$ with the method \inline{eigenfrequencies} for small variation in $C_d(x)$.
This enables an easy computation of the optomechanical coupling $\propto d\omega_m/dx$.
}
\label{fig:QuCAT_optomechanics}
\end{figure}
In this application, we show how QuCAT can be used for analyzing another classical system, that of microwave optomechanics.
One common implementation of microwave optomechanics involves a mechanically compliant capacitor, or drum, embedded in one or many microwave resonators~\cite{Teufel2011}.
One quantity of interest is the single-photon optomechanical coupling.
This quantity is the change in mode frequency $\omega_m$ that occurs for a displacement $x_\text{zpf}$ of the drum (the zero-point fluctuations in displacement)
\begin{equation}
    g_0 = x_\text{zpf}\frac{\partial \omega_m}{\partial x}
\end{equation}
The change in mode frequency as the drum head moves $\partial \omega_m/\partial x$ is not straightforward to compute for complicated circuits.
One such example is that of Ref.~\cite{ockeloen2016low}, where two microwave resonators are coupled to a drum via a network of capacitances as shown in Fig.~\ref{fig:QuCAT_optomechanics}(a).
Here, we will use QuCAT to calculate the optomechanical coupling of the drums to both resonator modes of the circuit.
We start by reproducing the circuit of Fig.~\ref{fig:QuCAT_optomechanics}(a), excluding the capacitive connections on the far left and right.
This is done via the graphical user interface opened with the \inline{qucat.GUI} function.
Upon closing the graphical user interface, the resulting Qcircuit is stored in the variable \inline{OM}, and the \inline{show} method is used to display the schematic of Fig.~\ref{fig:QuCAT_optomechanics}(a).
These steps are accomplished with the code below
\begin{lstlisting}
# Open the GUI and manually build the circuit
OM = qucat.GUI('netlist.txt')
# Display the circuit
OM.show()
\end{lstlisting}
We use realistic values for the circuit components without trying to be faithful to Ref.~\cite{ockeloen2016low}, the aim of this section is to illustrate a method to obtain $g_0$.
Crucially, the mechanically compliant capacitors have been parametrized by the symbolic variable $C_d$.
We can now calculate the resonance frequencies of the circuit with the method \inline{eigenfrequencies} as a function of a keyword argument $C_d$.
The next step is to define an expression for $C_d$ as a function of the mechanical displacement $x$ of the drum head with respect to the immobile capacitive plate below it. 
\begin{lstlisting}
def Cd(x):
    # Radius of the drumhead
    radius = 10e-6
    # Formula for half a circular parallel plate capacitor
    return eps*pi*radius**2/x/2
\end{lstlisting}
where \inline{pi} and \inline{eps} have been set to the values of $\pi$ and the vacuum permittivity respectively.
We have divided the usual formula for parallel plate capacitance by 2 since, as shown in Fig.~\ref{fig:QuCAT_optomechanics}(a), the capacitive plate below the drum head is split in two electrodes.
We are now ready to compute $g_0$.
Following Ref.~\cite{Teufel2011}, we assume the rest position of the drum to be $D=50$ nm above the capacitive plate below.
And we assume the zero-point fluctuations in displacement to be $x_\text{zpf} = 4$ fm.
We start by differentiating the mode frequencies with respect to drum displacement using a finite differences formula
\begin{lstlisting}
# drum-capacitor gap
D = 50e-9
# difference quotient
h = 1e-18
# derivative of eigenfrequencies
G = (OM.eigenfrequencies(Cd = Cd(D+h))-OM.eigenfrequencies(Cd = Cd(D)))/h
\end{lstlisting}
\inline{G} is an array with values $2.3\times 10^{16}$ Hz.$\text{m}^{-1}$ and $3.6\times 10^{16}$ Hz.$\text{m}^{-1}$ corresponding to the lowest and higher frequency modes respectively.
Multiplying these values with the zero-point fluctuations 
\begin{lstlisting}
# zero-point fluctuations
x_zpf = 4e-15
g_0 = G*x_zpf
\end{lstlisting}
yields couplings of $96$ and $147$ Hz.
The lowest frequency mode thus has a $96$ Hz coupling to the drum.

If we want to know to which part of the circuit (resonator 1 or 2 in Fig.~\ref{fig:QuCAT_optomechanics}) this mode pertains, we can visualize it by calling
\begin{lstlisting}
OM.show_normal_mode(
    mode=0,
    quantity='current',
    Cd=Cd(D))
\end{lstlisting}
and we find that the current is majoritarily located in the inductor of resonator 1.
%
%
\subsection{Convergence in multi-mode cQED}\label{sec:mmusc}
\begin{figure*}[]
\centering
\includegraphics[width=0.8\textwidth]{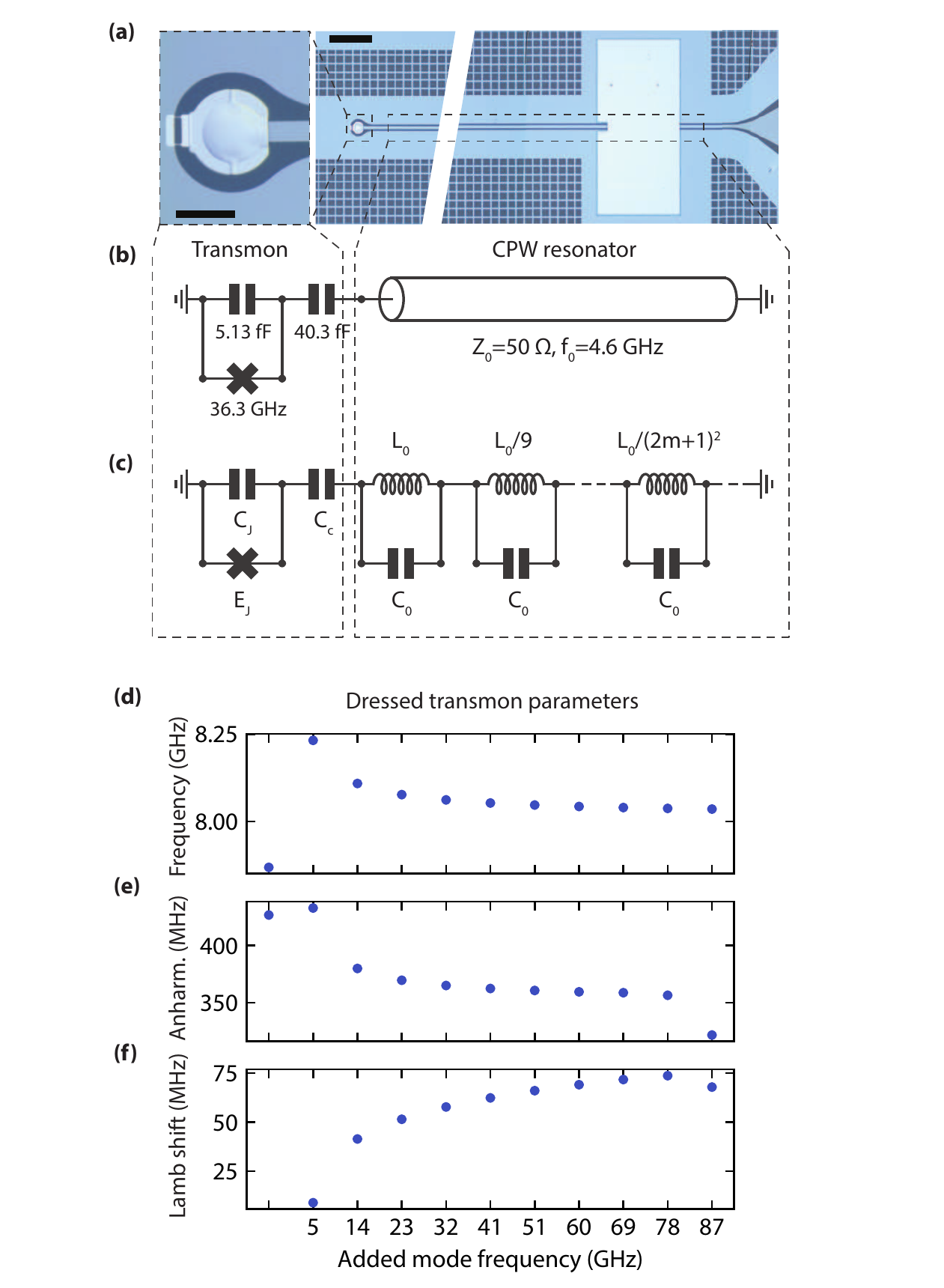}
\caption{
\textbf{Convergence in multi-mode cQED. (a) }
Optical micrograph of the device studied in this example.
Light-blue corresponds to superconductor, dark blue to an insulating substrate.
On the left we see a vacuum-gap transmon: a capacitor plate suspended over the end of a coplanar-waveguide (CPW) resonator shorted to ground through two Josephson junctions.
The scale bar corresponds to 30 $\mu$m.
On the right a CPW $\lambda/4$ resonator, capacitively coupled to the transmon on one side and shorted to ground through a large shunt capacitor on the other.
The scale bar corresponds to 100 $\mu$m.
\textbf{(b)} Circuit schematic of the device.
The CPW resonator hosts a number of modes, and is equivalent to a series assembly of LC oscillators shown in \textbf{(c)}.
This circuit is built programmatically in QuCAT, and the qubit parameters are extracted for different total numbers of modes.
In \textbf{(d)} and \textbf{(e)} we plot the transmon mode frequency $\omega_t/2\pi$ and anharmonicity $A_t/h$, where $t$ refers to the transmon-like mode, using the methods \inline{eigenfrequencies} and \inline{anharmonicities} respectively.
In \textbf{(f)} we plot the shift defined in Ref.~\cite{gely2017nature} as the Lamb shift: the shift in transmon frequency (following Eq.~\ref{eq:hamiltonian_first_order_maintext}) due solely to the vacuum-fluctuations in the other modes $\frac{1}{2}\sum_{m\ne t}\chi_{t,m}$, obtained with the \inline{kerr} method.
These calculations allow the user to gauge how many modes are relevant to the physics of the circuit.
}
\label{fig:QuCAT_mmusc}
\end{figure*}
In this section we use QuCAT to study the convergence of parameters in the first order Hamiltonian (Eq.~\ref{eq:hamiltonian_first_order_maintext}) of an ultra-strongly coupled multi-mode circuit QED system.
Using a length of coplanar waveguide terminated with engineered boundary conditions is a common way of building a high quality factor microwave resonator.
One implementation is a $\lambda/4$ resonator terminated on one end by a large shunt capacitor, acting as a near-perfect short circuit for microwaves such that only a small amount of radiation may enter or leave the resonator.
On the other end one places a small capacitance to ground: an open circuit.
The shunt capacitor creates a voltage node, and at the open end the voltage is free to oscillate.
This resonator hosts a number of normal modes, justifying its lumped element equivalent circuit: a series of LC oscillators with increasing resonance frequency~\cite{gely2017convergence}.
Here, we study such a resonator with a transmon circuit capacitively coupled to the open end.
In particular we consider this coupling to be strong enough for the circuit to be in the multi-mode ultra-strong coupling regime as studied experimentally in Ref.~\cite{bosman_multi-mode_2017} and theoretically in Ref.~\cite{gely2017convergence}.
The particularity of this regime is that the transmon has a considerable coupling to multiple modes of the resonator.
It then becomes unclear how many of these modes to consider for a realistic modeling of the system.
This regime is reached by maximizing the coupling capacitance of the transmon to the resonator and minimizing the capacitance of the transmon to ground.
The experimental device accomplishing this is shown in Fig.~\ref{fig:QuCAT_mmusc}(a), with its schematic equivalent in Fig.~\ref{fig:QuCAT_mmusc}(b), and the lumped-element model in Fig.~\ref{fig:QuCAT_mmusc}(c).

We will use QuCAT to track the evolution of different characteristics of the system as the number of considered modes $N$ increases.
For this application, programmatically building the circuit is more appropriate than using the GUI.
We start by defining some constants
\begin{lstlisting}
# fundamental mode frequency of the resonator
f0 = 4.603e9
w0 = f0*2.*numpy.pi
# characteristic impedance of the resonator
Z0 = 50
# Josephson energy (in Hertz)
Ej = 18.15e9
# Coupling capacitance
Cc = 40.3e-15
# Capacitance to ground
Cj = 5.13e-15

# Capacitance of all resonator modes
C0 = numpy.pi/4/w0/Z0
# Inductance of first resonator mode
L0 = 4*Z0/numpy.pi/w0
\end{lstlisting}
we can then generate a Qcircuit we name \inline{mmusc}, as an example here with $N=10$ modes.
\begin{lstlisting}
# Initialize list of components for Transmon and coupling capacitor
netlist = [
    qucat.J(12,1,Ej,use_E=True),
    qucat.C(12,1,Cj),
    qucat.C(1,2,Cc)]

# Add 10 oscillators
for m in range(10):
    # Nodes of m-th oscillator
    node_minus = 2+m
    node_plus = (2+m+1)
    # Inductance of m-th oscillator
    Lm = L0/(2*m+1)**2
    # Add oscillator to netlist
    netlist = netlist + [
        qucat.L(node_minus,node_plus,Lm),
        qucat.C(node_minus,node_plus,C0)]

# Create Qcircuit
mmusc = qucat.Network(netlist)
\end{lstlisting}
Note that \inline{12} is the index of the ground node.

We can now access some parameters of the system.
Only the first mode of the resonator has a lower frequency than the transmon. 
The transmon-like mode is thus indexed as mode \inline{1}.
Its frequency is given by
\begin{lstlisting}
mmusc.eigenfrequencies()[1]
\end{lstlisting}
and the anharmonicity of the transmon, computed from first order perturbation theory (see Eq.~\ref{eq:hamiltonian_first_order_maintext}) with
\begin{lstlisting}
mmusc.anharmonicities()[1]
\end{lstlisting}
Finally the Lamb shift, or shift in the transmon frequency resulting from the zero-point fluctuations of the resonator modes, is given following Eq.~(\ref{eq:hamiltonian_first_order_maintext}) by the sum of half the cross-Kerr couplings between the transmon mode and the others
\begin{lstlisting}
lamb_shift = 0
K = mmusc.kerr()
for m in range(10):
  if m!=1:
    lamb_shift = lamb_shift + K[1][m]/2
\end{lstlisting}
These parameters for different total number of modes are plotted in Figs~\ref{fig:QuCAT_mmusc}(d-f).
From this analysis, we find that as we reach 10, the plotted parameters are converging.
Surprisingly, adding even the highest modes significantly modifies the total Lamb shift of the Transmon despite large frequency detunings.

\subsection{Modeling a tuneable coupler}
\begin{figure*}[]
\centering
\includegraphics{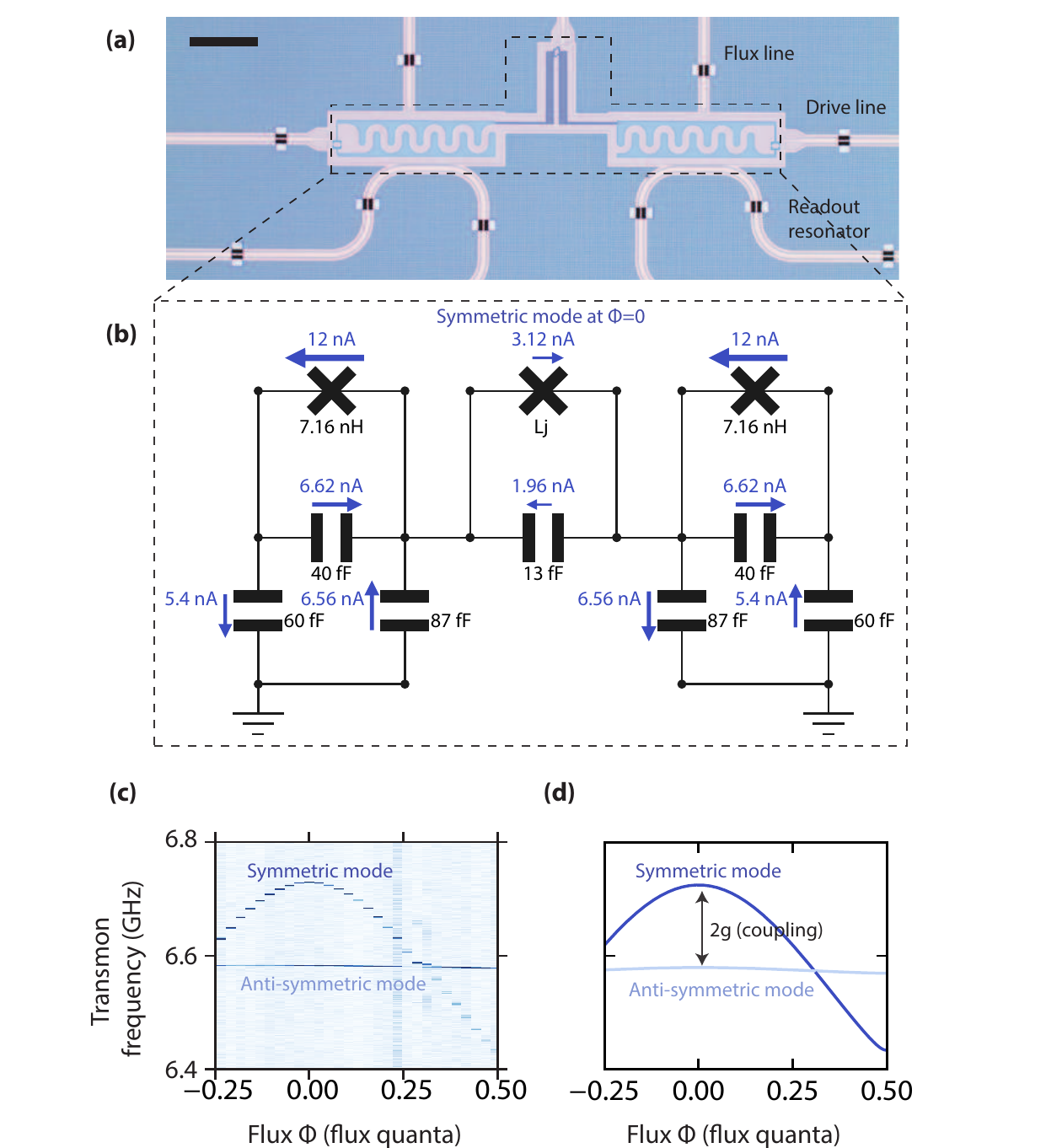}
\caption{
\textbf{Tuneable coupler circuit analysis. (a) }
Optical micrograph of the device studied in this example, adapted from Ref.~\cite{kounalakis2018tuneable} under a \href{https://creativecommons.org/licenses/by/4.0/}{CC BY 4.0} license.
We will omit the flux lines, drive lines and readout resonators for simplicity in this example, and concentrate on the part of the device in the dashed box.
The circuit consists of two near-identical transmon qubits coupled through a third ``coupler" transmon.
Scale bar corresponds to 200 $\mu$m.
\textbf{(b)} Equivalent lumped-element circuit constructed with the QuCAT GUI and displayed using the \inline{show_normal_mode} method.
This method has overlaid the circuit with the currents flowing through the components when the highest frequency mode is populated with a quarter-photon-amplitude coherent state.
Most of the current is located in the resonantly coupled transmons rather than the coupler, and the fact that the coupled transmons are identical leads to the symmetry on each side of the coupler.
This mode is called symmetric since the current in both coupled transmons flows in the same direction. 
The net current through the coupling junction makes the mode frequency sensitive to changes in the coupling junction inductance tuned with a superconducting quantum interference device or SQUID.
The change in symmetric mode frequency is shown in the experimental measure of the response frequencies in \textbf{(c)} (adapted from Ref.~\cite{kounalakis2018tuneable} under a \href{https://creativecommons.org/licenses/by/4.0/}{CC BY 4.0} license), and in the diagonalization of the Hamiltonian generated from QuCAT in \textbf{(d)}.
}
\label{fig:QuCAT_tuneable_coupler}
\end{figure*}
In this section, we study the circuit of Ref.~\cite{kounalakis2018tuneable} where two transmon qubits are coupled through a tuneable coupler.
This tuneable coupler is built from a capacitor and a Superconducting Quantum Interference Device, or SQUID.
By flux biasing the SQUID, we change the effective Josephson energy of the coupler, which modifies the coupling between the two transmons.
We will present how the normal mode visualization tool helps in understanding the physics of the device.
Secondly, we will show how a Hamiltonian generated with QuCAT accurately reproduces experimental measurements of the device.

We start by building the device shown in Fig.~\ref{fig:QuCAT_tuneable_coupler}(a).
More specifically, we are interested in the part of the device in the dashed box, consisting of the two transmons and the tuneable coupler.
The other circuitry, the flux line, drive line and readout resonator could be included to determine external losses, or the dispersive coupling of the transmons to their readout resonator.
We will omit these features for simplicity here.
After opening the GUI with the \inline{qucat.GUI} function, manually constructing the circuit, then closing the GUI, the resulting \inline{Qcircuit} is stored in a variable \inline{TC}.
\begin{lstlisting}
TC = qucat.GUI('netlist.txt')
\end{lstlisting}
The inductance $L_j$ of the junction which models the SQUID is given symbolically, and will have to be specified when calling \inline{Qcircuit} functions.
Since $L_j$ is controlled through flux $\phi$ in the experiment, we define a function which translates $\phi$ (in units of the flux quantum) to $L_j$
\begin{lstlisting}
def Lj(phi):
    # maximum Josephson energy
    Ejmax = 6.5e9
    # junction asymmetry
    d = 0.0769
    # flux to Josephson energy
    Ej = Ejmax*numpy.cos(pi*phi) *numpy.sqrt(1+d**2 *numpy.tan(pi*phi)**2)
    # Josephson energy to inductance
    return (hbar/2/e)**2/(Ej*h)
\end{lstlisting}
where \inline{pi},  \inline{h}, \inline{hbar}, \inline{e} were assigned the value of $\pi$, Plancks constant, Plancks reduced constant and the electron charge respectively.

By visualizing the normal modes of the circuit, we can understand the mechanism behind the tuneable coupler.
We plot the highest frequency mode at $\phi=0$, as shown in Fig.~\ref{fig:QuCAT_tuneable_coupler}(b)
\begin{lstlisting}
TC.show_normal_mode(mode = 2, 
    quantity = 'current',
    Lj=Lj(0))
\end{lstlisting}
This mode is called symmetric since the currents flow in the same direction on each side of the coupler.
This leads to a net current through the coupler junction, such that the value of $L_j$ influences the oscillation frequency of the mode.
Conversely, if we plot the anti-symmetric mode instead, where currents are flowing away from the coupler in each transmon, we find a current through the coupler junction and capacitor on the order of $10^{-21}$ A.
This mode frequency should not vary as a function of $L_j$.
When the bare frequency of the coupler matches the coupled transmon frequencies, the coupler acts as a band-stop filter, and lets no current traverse.
At this point, both symmetric and anti-symmetric modes should have identical frequencies.

In Fig.~\ref{fig:QuCAT_tuneable_coupler}(c) this effect is shown experimentally through a measure of the first transitions of the two non-linear modes.
One is tuned with flux (symmetric mode), the other barely changes (anti-symmetric mode).
We can reproduce this experiment by generating a Hamiltonian with QuCAT and diagonalizing it with QuTiP for different values of the flux.
For example, at 0 flux, the two first two transition frequencies \inline{f1} and \inline{f2} can be generated from
\begin{lstlisting}
# generate a Hamiltonian
H = TC.hamiltonian(Lj = Lj(phi = 0), 
    excitations = [7,7], 
    taylor = 4, 
    modes = [1,2])
# diagonalize the Hamiltonian
ee = H.eigenenergies()
f1 = ee[1]-ee[0]
f2 = ee[2]-ee[0]
\end{lstlisting}
\inline{f1} and \inline{f2} is plotted in Fig.~\ref{fig:QuCAT_tuneable_coupler}(d) for different vales of flux and closely matches the experimental data.
Note that we have constructed a Hamiltonian with modes 1 and 2, excluding mode 0, which corresponds to oscillations of current majoritarily located in the tuneable coupler.
One can verify this fact by plotting the distribution of currents for mode 0 using the \inline{show_normal_mode} method.

This experiment can be viewed as two ``bare" transmon qubits coupled by the interaction
\begin{equation}
    \hat H_\text{int} = g\sigma_x^L\sigma_x^R
\end{equation}
where left and right transmons are labeled $L$ and $R$ and $\sigma_x$ is the $x$ Pauli operator.
The coupling strength $g$ reflects the rate at which the two transmons can exchange quanta of energy.
If the transmons are resonant a spectroscopy experiment reveals a hybridization of the two qubits, which manifests as two spectroscopic absorption peaks separated in frequency by $2g$.
From this point of view, this experiment thus implements a coupling which is tuneable from an appreciable value to near 0 coupling.

\subsection{Studying a Josephson-ring-based qubit}
\begin{figure*}[]
\centering
\includegraphics{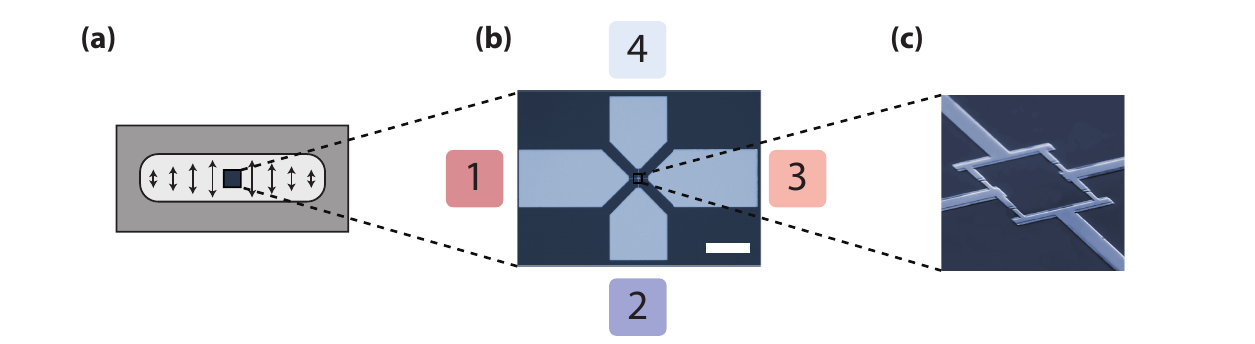}
\caption{
\textbf{Trimon device and Purcell-decay-protected mode visualization. (a) }
Schematic of the cross-cut of a 3D microwave cavity.
Dark gray shows metal whilst light gray show the hollowed out section forming the cavity.
Arrows represent the electric field of the $TE_{101}$, or ``vertical" cavity mode.
In the cavity is placed a chip hosting the trimon circuit shown in the optical micrograph \textbf{(b)}.
The circuit has 4 capacitive pads labeled from 1 to 4.
These pads are connected by the Josephson junction ring shown in the scanning electron microscope image \textbf{(c)}.
Scale bars correspond 200 and 2 $\mu$m for panels \textbf{(b)} and \textbf{(c)} respectively.
Reprinted figures with permission from \href{https://doi.org/10.1103/PhysRevApplied.7.054025}{T. Roy et al., Phys. Rev. Appl. \textbf{7} (5), 054025 (2017)}. Copyright 2017 by the American Physical Society.
}
\label{fig:QuCAT_trimon}
\end{figure*}
\begin{figure*}[]
\centering
\includegraphics{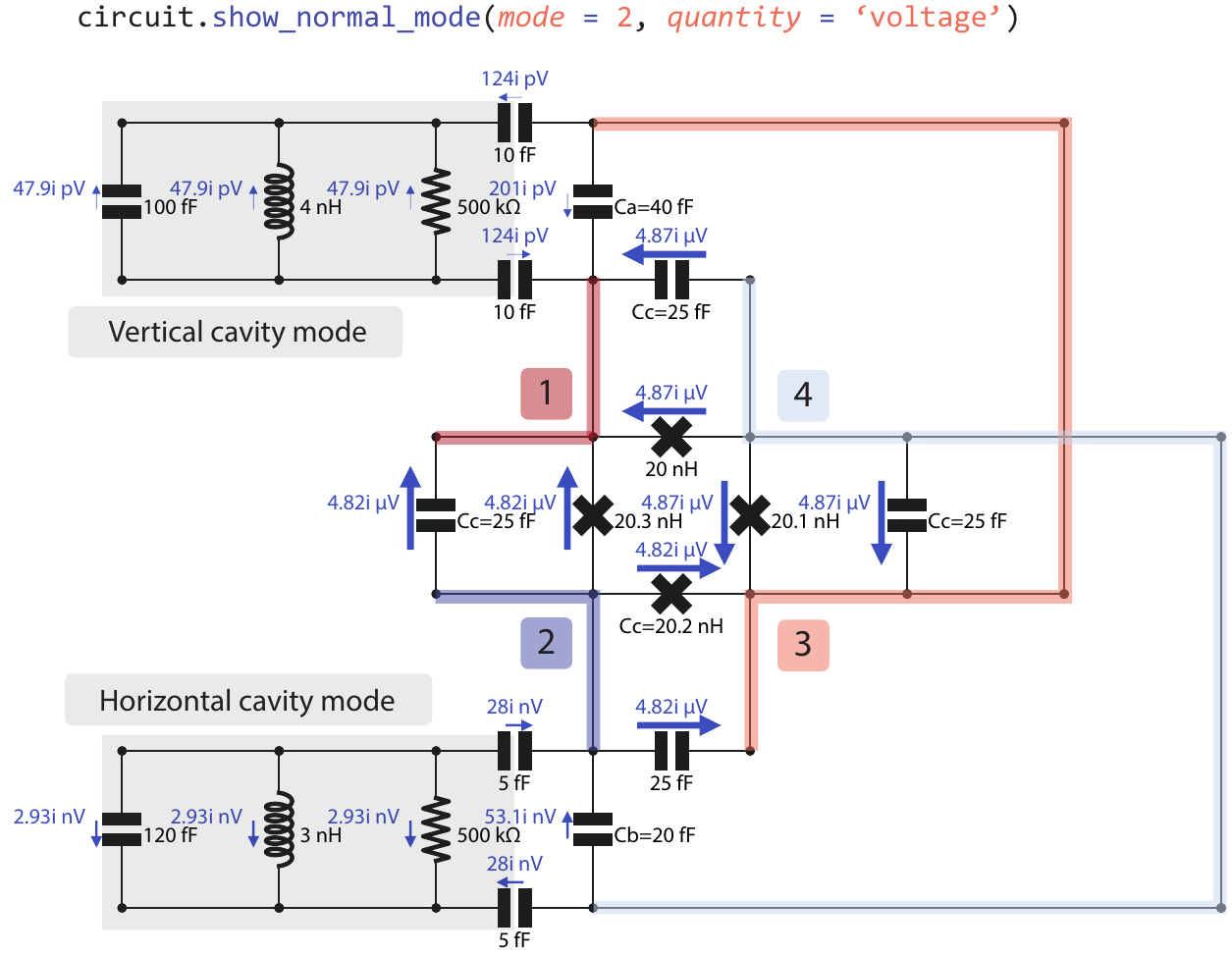}
\caption{
\textbf{} Lumped-element equivalent circuit of the device constructed using the QuCAT GUI and displayed with \inline{show_normal_mode}.
The four pads of the trimon are color-coded to match Fig.~\ref{fig:QuCAT_trimon}\textbf{(b)}.
The capacitor $C_a$ formed by pads 1 and 3 forms an electrical dipole which couples to a vertical cavity mode, and the capacitor $C_b$ formed by pads 2 and 4 forms an electrical dipole which couples to modes with horizontal electric fields.
The \inline{show_normal_mode} overlays the voltage across different components if a quarter-photon amplitude coherent state was populating mode 2.
This mode has a particularity that the voltage is concentrated across the junctions and their parallel capacitors without leading to a buildup of voltage across the capacitors $C_a$ or $C_b$.
This decouples mode 2 from the cavity mode decay (no Purcell effect) whilst the presence of voltage fluctuations across the junctions will ensure cross-Kerr coupling to the other modes of the system.
}
\label{fig:QuCAT_trimon_d}
\end{figure*}
\begin{figure}[]
\centering
\includegraphics{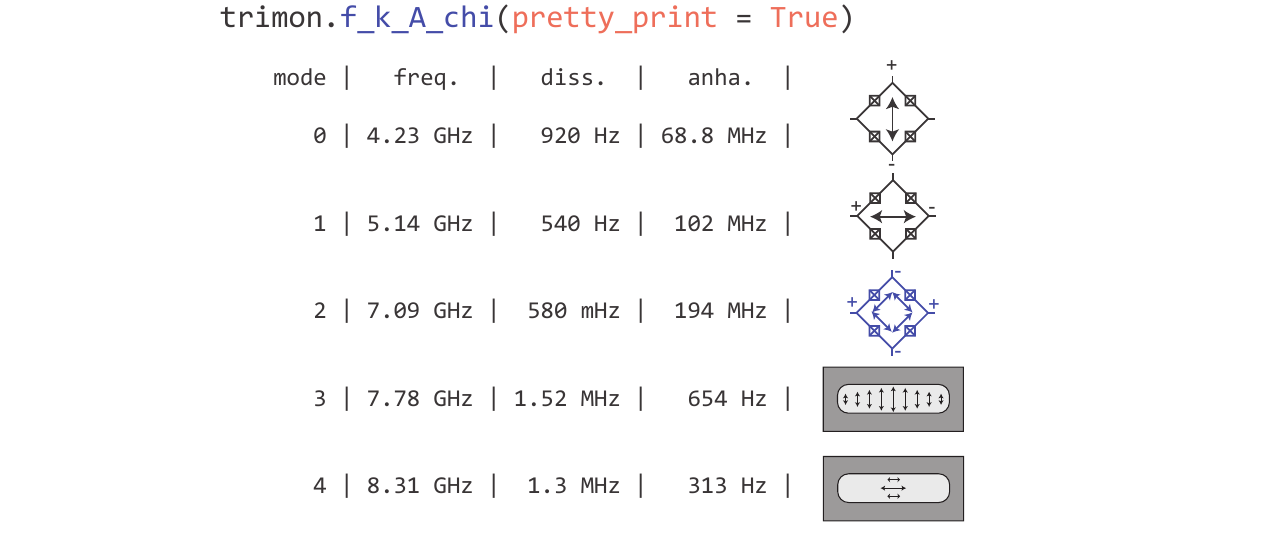}
\caption{
\textbf{Other modes of the Trimon. }
Using the \inline{f_k_A_chi} method together with the \inline{pretty_print} option gives the user an overview of the different modes frequencies, dissipations rates and levels of anharmonicity.
Here we have overlaid the output of the method with schematics of the corresponding trimon and cavity modes adapted from Ref.~\cite{roy2017implementation}.
One can identify a mode to the schematic by observing where the currents or voltages are mostly located in the circuit using the \inline{show_normal_mode} method as in Fig.~\ref{fig:QuCAT_trimon_d}.
Since the only resistors of the circuit are located in the cavity modes, all dissipation in transmon modes 0 through 2 are due to the Purcell effect. 
Mode 2 is better protected from this effect by 3 orders of magnitude with respect to the two other transmon modes.
Concerning schematics: reprinted figures with permission from \href{https://doi.org/10.1103/PhysRevApplied.7.054025}{T. Roy et al., Phys. Rev. Appl. \textbf{7} (5), 054025 (2017)}. Copyright 2017 by the American Physical Society.
}
\label{fig:QuCAT_trimon_2}
\end{figure}
In this section, we demonstrate the ability for QuCAT to analyze more complex circuits.
The experiment of Ref.~\cite{roy2017implementation} features a Josephson ring geometry, which is a Wheatstone-bridge-like circuit, typically difficult to analyze as it cannot be decomposed in series and parallel connections.
We consider the coupling of this ring to two lossy modes of a cavity, bringing the total number of modes in the circuit to 5.
We aim to understand of the key feature of this circuit: that one qubit-like mode acts as a quadrupole with little coupling to the resonator modes.
The studied device consists of a 3D cavity (Fig.~\ref{fig:QuCAT_trimon}(a)) hosting a number of microwave modes, in which is positioned a chip patterned with the trimon circuit.
The trimon circuit has four capacitive pads in a cross shape (Fig.~\ref{fig:QuCAT_trimon}(b)) which have an appreciable coupling between each other making up the capacitance of the trimon qubit modes.
The two vertically (horizontally) positioned pads will couple to modes of the 3D cavity featuring vertical (horizontal) electric fields.
We will consider both a vertical and a horizontal cavity mode in our model.
We number these pads from 1 to 4 as displayed in Fig.~\ref{fig:QuCAT_trimon}(b).
Each pad is connected to its two nearest neighbors by a Josephson junction (Fig.~\ref{fig:QuCAT_trimon}(c)), forming a Josephson ring.
Using the QuCAT GUI, we build a lumped element model of this device, generating a \inline{Qcircuit} object we store in the variable \inline{trimon}.
\begin{lstlisting}
trimon = qucat.GUI('netlist.txt')
\end{lstlisting}
The cavity modes are modeled as RLC oscillators with each plate of their capacitors capacitively coupled to a pad of the trimon circuit.
The junction inductances are assigned different values, first to reflect experimental reality, but also to avoid infinities arising in the QuCAT analysis.
Indeed, the voltage transfer function of this Josephson ring between nodes 1,3 and nodes 2,4 will be exactly 0, which will cause errors when initializing the \inline{Qcircuit} object.
Component parameters are chosen to only approximatively match the experimental results of Ref.~\cite{roy2017implementation}, the objective here is to demonstrate QuCAT features rather than accurately model the experiment.
The particularity of this circuit is that it hosts a quadrupole mode.
It corresponds here to the second highest frequency mode and can be visualized by calling
\begin{lstlisting}
trimon.show_normal_mode(
    mode = 2,
    quantity = 'voltage')
\end{lstlisting}
the result of which is displayed in Fig.~\ref{fig:QuCAT_trimon_d}.
The voltage oscillations are majoritarily located in the junctions, indicating this is not a cavity mode, but a mode of the trimon circuit.
Crucially, the polarity of voltages across the junctions is such that the total voltage between pads 1 and 3 and the total voltage across pads 2 and 4 is 0, warranting the name of ``quadrupole mode".
Due to the orientation of the chip in the cavity, the vertically and horizontally orientated cavity modes will only be sensitive to voltage oscillations across pads 1 and 3 or 2 and 4.
This ensures that the mode displayed here is decoupled from the cavity modes, and from any loss channels they may incur.
We can verify this fact by computing the losses of the different modes, and comparing the losses of mode 2 to the other qubit-like modes of the circuit.
We perform this calculation by calling
\begin{lstlisting}
trimon.f_k_A_chi(pretty_print=True)
\end{lstlisting}
which will calculate and return the loss rates of the modes, along with their eigenfrequencies, anharmonicities and Kerr parameters.
Setting the keyword argument \inline{pretty_print} to \inline{True} prints a table containing all this information, which is shown in Fig.~\ref{fig:QuCAT_trimon_2}.
To be succinct, we have not shown the table providing the cross-Kerr couplings.
By using the \inline{show_normal_mode} method to plot all the other modes of the circuit, and noting where currents or voltages are majoritarily located, we can identify each mode with the schematics provided in Fig.~\ref{fig:QuCAT_trimon_2}.
The three lowest frequency modes are located in the trimon chip, and we notice that as expected the quadrupole mode 2 has a loss rate (due to resistive losses in the cavity modes) which is three orders of magnitude below the other two.
Despite this apparent decoupling, the quadrupole mode will still be coupled to both cavity mode through the cross-Kerr coupling, given by twice the square-root of the product of the quadrupole and cavity mode anharmonicities.

\section{Circuit quantization overview}\label{sec:circuit_quantization_overview}
\begin{figure}[]
\centering
\includegraphics{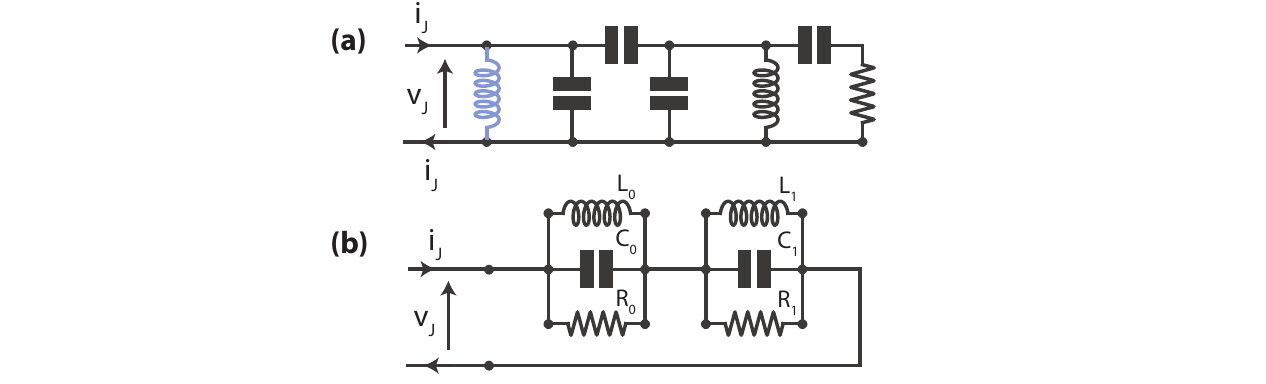}
\caption{
\textbf{Example of equivalent circuit construction to prepare for quantization.  }
We use the same example as used in Fig~\ref{fig:QuCAT_1}-\ref{fig:QuCAT_3}.
\textbf{(a)} The circuit is linearized by replacing the junction with an inductance $L_j$. The circuit is characterized at the nodes of the junction by its admittance $Y_j$.
\textbf{(b)} In the limit of small dissipation, this circuit is equivalent to a series combination of RLC resonators.
}
\label{fig:QuCAT_circuit_quantization}
\end{figure}
In this section we summarize the quantization method used in QuCAT, which is an expansion on the work of Ref.~\cite{nigg_black-box_2012}.
This approach is only valid in the weak anharmonic limit, where charge dispersion is negligible.
See~\cite{nigg_black-box_2012} or Sec.~\ref{sec:high_anharmonicity} for a detailed discussion of this condition.

The idea behind the quantization method is as follows.
We first consider the ``linearized" circuit.
This is a circuit where the junctions are replaced by their Josephson inductances $L_j = \phi_0^2/E_j$.
where $E_j$ is the Josephson energy and the reduced flux quantum is given by $\phi_0 = \hbar/2e$.
We determine the oscillation frequencies and dissipation rates of the different normal modes of this linearized circuit.
Then, we calculate the amplitude of phase oscillations across each junction when a given mode is excited.
This will determine how non-linear each mode is.
All this information will finally allow us to build a Hamiltonian for the circuit.

\subsection{Circuit simplification to series of RLC resonators (Foster circuit)}
The eigenfrequencies and non-linearity of each mode is obtained by transforming the linearized circuit to a geometry we can easily analyze.
We will first describe this process assuming there is only a single junction in the circuit, the case of multiple junctions will follow.
We consider the example circuit of Fig.~\ref{fig:QuCAT_1}.
After replacing the junction with its Josephson inductance, we determine the admittance $Y_j(\omega) = I_j(\omega)/V_j(\omega)$ evaluated at the nodes of the junction.
This admittance is the inverse of the impedance measured at the nodes of the junction.
It relates the amplitude $|V_j|$ and phase $\theta(V_j)$ of the voltage oscillating at frequency $\omega$ that would build up across the junction if one would feed a current oscillating at $\omega$ with amplitude $|I_j|$ and phase $\theta(I_j)$ to one of its nodes through a infinite impedance current source.
In Fig.~\ref{fig:QuCAT_circuit_quantization}(a) we show a schematic describing this quantity.
In the case where all normal modes of the circuit have small dissipation rates, this circuit has an approximate equivalent shown in Fig.~\ref{fig:QuCAT_circuit_quantization}(b), consisting of a series of RLC resonators~\cite{solgun2014blackbox}.
By equivalent, we mean that the admittance $Y_j$ of the circuit is approximatively equal to that of a series combination of RLC resonators
\begin{equation}
  Y_j(\omega) \simeq  \frac{1}{\sum_m1/Y_m(\omega)}\
  \label{eq:Y_as_sum}
\end{equation}
which each have an admittance
\begin{equation}
  Y_m(\omega) = \frac{1}{iL_m\omega}+iC_m\omega+\frac{1}{R_m}
\end{equation}
Each RLC resonator represents a normal mode of the circuit, with resonance frequency $\omega_m = 1/\sqrt{L_mC_m}$, and dissipation rate $\kappa_m = 1/R_mC_m$.
Since this equivalent circuit comes from an extension of Foster's reactance theorem~\cite{foster1924reactance} to lossy circuits, we call this the Foster circuit.

\subsection{Hamiltonian of the Foster circuit}
The advantage of this circuit form, is that it is easy to write its corresponding Hamiltonian following standard quantization methods (see Ref.~\cite{vool2017introduction}).
In the absence of junction non-linearity, it is given by the sum of the the Hamiltonians of the independent harmonic RLC oscillators:
\begin{equation}
  \sum_m \hbar \omega_m\hat a_m^\dagger \hat a_m\ .
\end{equation}
The annihilation operator $\hat a_m$ for photons in mode $m$ is related to the expression of the phase difference between the two nodes of the oscillator
\begin{equation}
\begin{split}
  \hat\varphi_{m,j} &= \varphi_{\text{zpf},m,j}(\hat a_m + \hat a^\dagger_m)\ ,\\
  \varphi_{\text{zpf},m,j} &= \frac{1}{\phi_0}\sqrt{\frac{\hbar}{2\omega_m C_m}}\ .
\label{eq:zpf}
\end{split}
\end{equation}
where $\varphi_{\text{zpf},m}$ are the zero-point fluctuations in phase of mode $m$.
The total phase difference across the Josephson junction $\hat\varphi_j$ is then the sum of these phase differences $\hat\varphi_j = \sum_m\hat\varphi_{m,j}$, and we can add the Junction non-linearity to the Hamiltonian
\begin{equation}
\hat{H} = \sum_m \hbar\omega_m\hat{a}_m^\dagger\hat{a}_m + E_j[1-\cos{\hat\varphi_j}-\frac{\hat\varphi_j^2}{2}]\ ,
\end{equation}
Since the linear part of the Hamiltonian corresponds to the circuit with junctions replaced by inductors, the linear part already contains the quadratic contribution of the junction potential $\propto\hat\varphi_j^2$, and it is subtracted from the cosine junction potential.

\subsection{Calculating Foster circuit parameters}
Both $\omega_m$ and $\kappa_m$ can be determined from $Y(\omega)$ since we have $Y(\omega_m+i\kappa_m/2)=0$ for low loss circuits.
This can be proven by noticing that the admittance $Y_m$ of mode $m$ has two zeros at
\begin{equation}
\begin{split}
  \zeta_m &= \frac{1}{\sqrt{L_mC_m}}\sqrt{1-\frac{1}{4Q^2}}+ i\frac{1}{2R_mC_m}\\
  &\simeq \omega_m + i\kappa_m/2\ .
\end{split}
\end{equation}
and $\zeta_m^*$.
The approximate equality holds in the limit of large quality factor $Q_m = R_m/\sqrt{L_m/C_m}\gg 1 $.
From Eq.~(\ref{eq:Y_as_sum}) we see that the zeros of $Y$ are exactly the zeros of the admittances $Y_k$.
The solutions of $Y(\omega)=0$, which come in conjugate pairs $\zeta_m$ and $\zeta_m^*$, thus provide us with both resonance frequencies $\omega_m = \text{Re}[\zeta_m]$ and dissipation rates $\kappa_m = 2\text{Im}[\zeta_m]$.
Additionally, we need to determine the effective capacitances $C_m$ in order to obtain the zero-point fluctuations in phase of each mode.
We focus on one mode $k$, and start by rewriting the admittance in Eq.~(\ref{eq:Y_as_sum}) as
\begin{equation}
  Y_j(\omega) =  Y_k(\omega)\frac{1}{1+\sum_{m\ne k}Y_k(\omega)/Y_m(\omega)}\ .
\end{equation}
Its derivative with respect to $\omega$ is
\begin{equation}
\begin{split}
  Y'_j(\omega) &=  Y'_k(\omega)\frac{1}{1+\sum_{m\ne k}Y_k(\omega)/Y_m(\omega)}\\
  &+Y_k(\omega)\frac{\partial}{\partial \omega}\left [\frac{1}{1+\sum_{m\ne k}Y_k(\omega)/Y_m(\omega)}\right ]\ .
\end{split}
\end{equation}
Evaluating the derivative at $\omega = \zeta_k$, where $Y_k(\zeta_k)=0$ yields
\begin{equation}
\begin{split}
  Y'_j(\zeta_k) &= Y'_k(\zeta_k) = iC_m\left(1+\frac{4Q^2}{\left(i+\sqrt{4Q^2-1}\right)^2} \right)\\
  &\simeq i2C_m\text{ for }Q_m\gg 1
\end{split}
\end{equation}
The capacitance is thus approximatively given by 
\begin{equation}
  C_m = \text{Im}\left[Y'_j(\zeta_k)\right]/2
\end{equation}

\subsection{Multiple junctions}

When more than a single junction is present, we start by choosing a single reference junction, labeled $r$.
All junctions will be again replaced by their inductances, and by using the admittance $Y_r$ across the reference junction, we can determine the Hamiltonian including the non-linearity of the reference junction through the procedure described above.
In this section, we will describe how to obtain the Hamiltonian including the non-linearity of all other junctions too
\begin{equation}
\hat H = \sum_m \hbar\omega_m\hat{a}_m^\dagger\hat{a}_m + \sum_j E_j[1-\cos{\hat\varphi_j}-\frac{\hat\varphi_j^2}{2}]\ ,
\label{eq:hamiltonian_no_taylor}
\end{equation}
where $\hat\varphi_j$ is the phase across the j-th junction.
This phase is determined by first calculating the zero-point fluctuations in phase $\varphi_{\text{zpf},m,r}$ through the reference junction $r$ for each mode $m$ given by Eq.~(\ref{eq:zpf}).
For each junction $j$, we then calculate the (complex) transfer function $T_{jr}(\omega)$ which converts phase in the reference junction to phase in junction $j$.
We can then calculate the total phase across a junction $j$ with respect to the reference phase of junction $r$, summing the contributions of all modes and both quadratures of the phase
\begin{equation}
\begin{split}
\hat\varphi_j = \sum_m\varphi_{\text{zpf},m,r}[\text{Re}\left (T_{jr}(\omega_m)\right )(\hat{a}_m+\hat{a}_m^\dagger)\\
+i\ \text{Im}\left (T_{jr}(\omega_m)\right )(\hat{a}_m-\hat{a}_m^\dagger)]
\end{split}
\label{eq:phi_j_from_phi_r}
\end{equation}

The definition of phase~\cite{vool2017introduction} $\varphi_j(t) = \phi_0^{-1}\int_{-\infty}^tv_j(\tau)d\tau$ where $v_j$ is the voltage across junction $j$ translates in the frequency domain to $\varphi_j(\omega) = i\omega\phi_0^{-1}V_j(\omega)$.
Finding the transfer function $T_{jr}$ for phase is thus equivalent to finding a transfer function for voltage $T_{jr}(\omega) = V_j(\omega)/V_r(\omega)$.
This is a standard task in microwave network analysis (see Sec.~\ref{sec:methods_network_transfer} for more details).

\subsection{Further treatment of the Hamiltonian}
The cosine potential in Eq.~(\ref{eq:hamiltonian_no_taylor}) can be expressed in the Fock basis by Taylor expanding it around small values of the phase.
This yields
\begin{equation}
\begin{split}
\hat{H} &= \sum_{m} \hbar\omega_m\hat{a}_m^\dagger\hat{a}_m\\
&+ \sum_j\sum_{n\ge 2}E_j\frac{(-1)^{n+1}}{(2n)!}\left[\sum_{m}\varphi_{\text{zpf},m,j}(\hat{a}_m^\dagger+\hat{a}_m)\right]^{2n}
\end{split}
\label{eq:hamiltonian_taylor_SI}
\end{equation}
which is the form returned by the QuCAT \inline{hamiltonian} method.
By keeping only the fourth power in the Taylor expansion and performing first order perturbation theory, we obtain 
\begin{equation}
\begin{split}
\hat{H} = \sum_m\sum_{n\ne m} (\hbar\omega_m-A_m-\frac{\chi_{mn}}{2})\hat{a}_m^\dagger\hat{a}_m \\
-\frac{A_m}{2}\hat{a}_m^\dagger\hat{a}_m^\dagger\hat{a}_m\hat{a}_m -\chi_{mn}\hat{a}_m^\dagger\hat{a}_m\hat{a}_n^\dagger\hat{a}_n
\label{eq:hamiltonian_first_order}
\end{split}
\end{equation}
Where the anharmonicity or self-Kerr of mode m is
\begin{equation}
A_m = \sum_j A_{m,j}
\end{equation}
as returned by the \inline{anharmonicites} method,
where
\begin{equation}
A_{m,j} = \frac{E_j}{2}\varphi_{\text{zpf},m,j}^4
\label{eq:Amj}
\end{equation}
is the contribution of junction j to the total anharmonicity of a mode m.
The cross-Kerr coupling between mode m and n is
\begin{equation}
\chi_{mn} = 2\sum_j \sqrt{A_{m,j}A_{n,j}}\ .
\end{equation}
Both self and cross-Kerr parameters are computed by the \inline{kerr} method.
Note in Eq.~\ref{eq:hamiltonian_first_order} that the harmonic frequency of the Hamiltonian is shifted by $A_m$ and $\sum_{n\ne m}\chi_{nm}/2$.
The former comes from the change in Josephson inductance induced by phase fluctuations of mode $m$. 
The latter is called the Lamb shift~\cite{gely2017nature} and is induced by phase fluctuations of the other modes of the circuit.

\section{Algorithmic methods}

There are three calculations to accomplish in order to obtain all the parameters necessary to write the circuit Hamiltonian. We need:
\begin{itemize}
\item the eigen-frequencies $\omega_m$ and loss rates $\kappa_m$ fulfilling $Y_r(\zeta_m=\omega_m+i\kappa_m/2) = 0$ where $Y_r$ is the admittance across a reference junction
\item the derivative of this admittance evaluated at $\zeta_m$
\item the transfer functions $T_{jr}$ between junctions $j$ and the reference junction $r$
\end{itemize}

In this section, we cover the algorithmic methods used to calculate these three quantities

\subsection{Resonance frequency and dissipation rate}\label{sec:methods_resonance_frequency}
\subsubsection{Theoretical background}

In order to obtain an expression for the admittance across the reference junction, we start by writing the set of equations governing the physics of the circuit.
We first determine a list of nodes, which are points at which circuit components connect.
Each node, labeled $n$, is assigned a voltage $v_n$.
We name $r_\pm$ the positive and negative nodes of the reference junction.

We are interested in the steady-state oscillatory behavior of the system.
We can thus move to the frequency domain, with complex node voltages $|V_n(\omega)| e^{i(\omega t+\theta(V_n(\omega_n)))}$, fully described by their phasors, the complex numbers $V_n = |V_n(\omega)| e^{i\theta(V_n(\omega_n))}$.
In this mathematical construct, the real-part of the complex voltages describes the voltage one would measure at the node in reality.
Current conservation dictates that the sum of all currents arriving at any node $n$, from the other nodes $k$ of the circuit should be equal to the oscillatory current injected at node $n$ by a hypothetical, infinite impedance current source.
This current is also characterized by a phasor $I_n$.
This can be compactly written as
\begin{equation}
    \sum_{k\ne n} Y_{nk}(V_n-V_k) = I_n
    \label{eq:kirchhoff_1}
\end{equation}
where $k$ label the other nodes of the circuit and $Y_{nk}$ is the admittance directly connecting nodes $k$ and $n$.
Note that in this notation, if a node $k_1$ can only reach node $n$ through another node $k_2$, then $Y_{nk_1} = 0$.
Inductors (with inductance $L$), capacitors (with capacitance $C$) and resistors (with resistance $R$) then have admittances $1/iL\omega$, $iC\omega$ and $1/R$ respectively.
Expanding Eq.~\ref{eq:kirchhoff_1} yields
\begin{equation}
    (\sum_{k\ne n} Y_{nk})V_n-\sum_{k\ne n} Y_{nk}V_k = I_n
\end{equation}
which can be written in matrix form as
\begin{equation}
    \begin{pmatrix}
    \Sigma_{k\ne 0}Y_{0k} &-Y_{01}  &\cdots  &-Y_{0N} \\ 
    -Y_{10} &\Sigma_{k\ne 1}Y_{1k} &\cdots  &-Y_{1N} \\ 
    \vdots &\vdots  &\ddots  & \vdots\\ 
    -Y_{N0} &-Y_{N1}  &\cdots  &\Sigma_{k\ne N}Y_{Nk} 
    \end{pmatrix}
    \begin{pmatrix}
    V_0\\ 
    V_1\\ 
    \vdots\\ 
    V_N
    \end{pmatrix}=
    \begin{pmatrix}
    I_0\\ 
    I_1\\ 
    \vdots\\ 
    I_N
    \end{pmatrix}
    \label{eq:admittance_matrix_no_ground}
\end{equation}
Since voltage is the electric potential of a node relative to another, we still have the freedom of choosing a ground node. 
Equivalently, conservation of currents imposes that current exciting that node is equal to the sum of currents entering the others, there is thus a redundant degree of freedom in Eq.(\ref{eq:admittance_matrix_no_ground}).
For simplicity, we will choose node 0 as ground.
Since we are only interested in the admittance across the reference junction, we set all currents to zero, except the currents entering the positive and negative reference junction nodes: $I_{r_+}$ and $I_{r_-} = -I_{r_+} $  respectively.
The admittance is defined by $Y_r = I_{r_+}/(V_{r_+}-V_{r_-})$.
The equations then reduce to
\begin{equation}
    \mathbf{Y}\begin{pmatrix}
    V_1\\ 
    V_2\\ 
    \vdots\\ 
    V_N
    \end{pmatrix}=Y_r
    \begin{pmatrix}
    \vdots\\ 
    0\\ 
    V_{r_+}-V_{r_-}\\ 
    0\\
    \vdots\\ 
    0\\ 
    V_{r_-}-V_{r_+}\\ 
    0\\
    \vdots
    \end{pmatrix}
    \label{eq:admittance_matrix_equation}
\end{equation}
Where $\mathbf{Y}$ is the admittance matrix
\begin{equation}
  \mathbf{Y} = 
    \begin{pmatrix}
    \Sigma_{k\ne 1}Y_{1k} &-Y_{12}  &\cdots  &-Y_{1N} \\ 
    -Y_{21} &\Sigma_{k\ne 1}Y_{2k} &\cdots  &-Y_{2N} \\ 
    \vdots &\vdots  &\ddots  & \vdots\\ 
    -Y_{N1} &-Y_{N2}  &\cdots  &\Sigma_{k\ne N}Y_{Nk} 
    \end{pmatrix}
    \label{eq:admittance_matrix}
\end{equation}

For $Y_r = 0$, Eq~\ref{eq:admittance_matrix_equation} has a solution for only specific values of $\omega = \zeta_m$.
These are the values which make the admittance matrix singular, i.e. which make its determinant zero
\begin{equation}
    \text{Det}\left[\mathbf{Y}(\zeta_m)\right] = 0
    \label{eq:det}
\end{equation}
The determinant is a polynomial in $\omega$, so the problem of finding $\zeta_m = \omega_m+i\kappa_m/2$ reduces to finding the roots of this polynomial.
Note that plugging $\zeta_m$ into the frequency domain expression for the node voltages yields $V_k(\zeta_m)e^{i\omega_mt}e^{-\kappa_mt/2}$, such that the energy $\propto v_k(t)^*v_k(t) \propto e^{-\kappa_mt}$ decays at a rate $\kappa_m$, which explains the division by two in the expression of $\zeta_m$.
Also note, that we would have obtained equation Eq.~(\ref{eq:det}) regardless of the choice of reference element.

\subsubsection{Algorithm}
\begin{figure}[]
\centering
\includegraphics{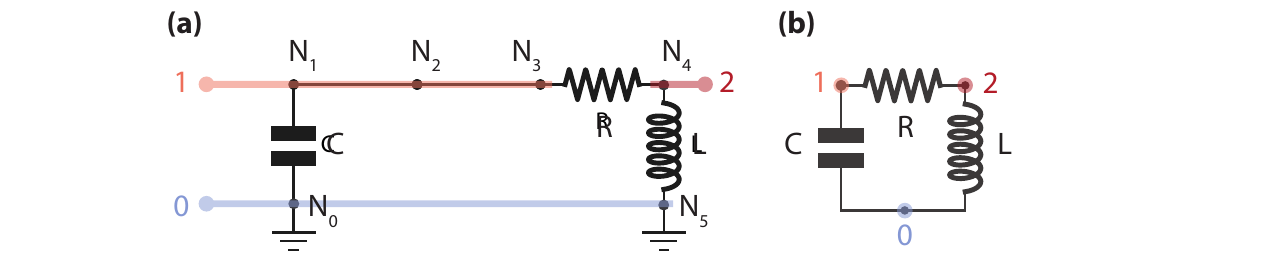}
\caption{
\textbf{Example circuit to illustrate the mode frequency finder algorithm. }
\textbf{(a)} Example of a circuit built through the GUI by a user.
\textbf{(b)} Application of the first step of the algorithm which removes the wires and grounds to obtain a minimal number of nodes without removing any components.
}
\label{fig:QuCAT_f_finder_example}
\end{figure}

We now describe the algorithm used to determine the solutions $\zeta_m = \omega_m+i\kappa_m/2$ of Eq.~\ref{eq:det}.
As an example, we consider the circuit of Fig.~\ref{fig:QuCAT_f_finder_example}(a) that a user would have built with the GUI.
The algorithm is as follows
\begin{enumerate}
    \item  Eliminate wires and grounds. In this case, nodes $N_0,N_5$ would be grouped under a single node labeled 0 and nodes $N_1,N_2,N_3$ would be grouped under node 1, we label node $N_4$ node 2, as shown in Fig.~\ref{fig:QuCAT_f_finder_example}(b).
    \item Compute the un-grounded admittance matrix. For each component present between the different couples of nodes, we append the admittance matrix with the components admittance. The matrix is then multiplied by $\omega$ such that all components are polynomials in $\omega$, ensuring that the determinant is also a polynomial. In this example, the matrix is
    \begin{equation}
        \begin{pmatrix}
        iC\omega^2+1/iL & -iC\omega^2  & -1/iL  \\
        -iC\omega^2 &iC\omega^2+\omega/R  &-\omega/R  \\
         -1/iL &-\omega/R  &1/iL +\omega/R
        \end{pmatrix}
    \end{equation}
    \item Choose a ground node. The node which has a corresponding column with the most components is chosen as the ground node (to reduce computation time). These rows and columns are erased from the matrix, yielding the final form of the admittance matrix
    \begin{equation}
    \mathbf{Y}=
        \begin{pmatrix}
        iC\omega^2+\omega/R  &-\omega/R  \\
        -\omega/R  &1/iL +\omega/R
        \end{pmatrix}
    \end{equation}
    \item Compute the determinant. Even if the capacitance, inductance and resistance were specified numerically, the admittance matrix would still be a function of the symbolic variable $\omega$. We thus rely on a symbolic Berkowitz determinant calculation algorithm~\cite{berkowitz1984computing,kerber2009division} implemented in the Sympy library through the \inline{berkowitz_det} function. In this example, one would obtain
    \begin{equation}
        \text{Det}[\mathbf{Y}] = LC\omega^2-iRC\omega -1\ .
    \end{equation}
    \item Find the roots of the polynomial. Whilst the above steps have to be performed only once for a given circuit, this one should be performed each time the user edits the value of a component. The root-finding is divided in the following steps as prescribed by Ref.~\cite{press2007numerical}.
    \subitem Diagonalize the polynomials companion matrix~\cite{horn1985cr} to obtain an exhaustive list of all roots of the polynomial. This is implemented in the NumPy library through the \inline{roots} function.
    \subitem Refine the precision of the roots using multiple iterations of Halley's gradient based root finder~\cite{press2007numerical} until iterations do not improve the root value beyond a predefined tolerance given by the \inline{Qcircuit} argument \inline{root_relative_tolerance}. The maximum number of iterations that may be carried out is determined by the \inline{Qcircuit} argument \inline{root_max_iterations}. If the imaginary part relative to the real part of the root is lower than the relative tolerance, the imaginary part will be set to zero. The relative tolerance thus sets the highest quality factor that QuCAT can detect.
    \subitem Remove identical roots (equal up to the relative tolerance), roots with negative imaginary or real parts, 0-frequency roots, roots for which $Y_l'(\omega_m)<0$ for all $l$, where $Y_l$ is the admittance evaluated at the nodes of an inductive element $l$, and roots for which $Q_m<$\inline{Qcircuit.Q_min}. The user is warned of a root being discarded when one of these cases is unexpected.
\end{enumerate}

The roots $\zeta_m$ obtained through this algorithm are accessed through the method \inline{eigenfrequencies} which returns the oscillatory frequency in Hertz of all the modes ${\text{Re}[\zeta_m]/2\pi}$ or \inline{loss_rates} which returns ${2\text{Im}[\zeta_m]/2\pi}$.

\subsection{Derivative of the admittance}\label{sec:methods_network_dY}
The zero-point fluctuations in phase $\varphi_{\text{zpf},m,r}$ for each mode $m$ across a reference junction $r$ is the starting point to computing a Hamiltonian for the non-linear potential of the Junctions.
As expressed in Eq.~(\ref{eq:zpf}), this quantity depends on the derivative $Y_r'$ of the admittance $Y_r$ calculated at the nodes of the reference element.
In this section we first cover the algorithm used to obtain the admittance at the nodes of an arbitrary component.
From this admittance we then describe the method to obtain the derivative of the admittance on which $\varphi_{\text{zpf},m,r}$ depends
Finally we describe how to choose a (mode-dependent) reference element.

\subsubsection{Computing the admittance}
\begin{figure}[]
\centering
\includegraphics[width=0.8\textwidth]{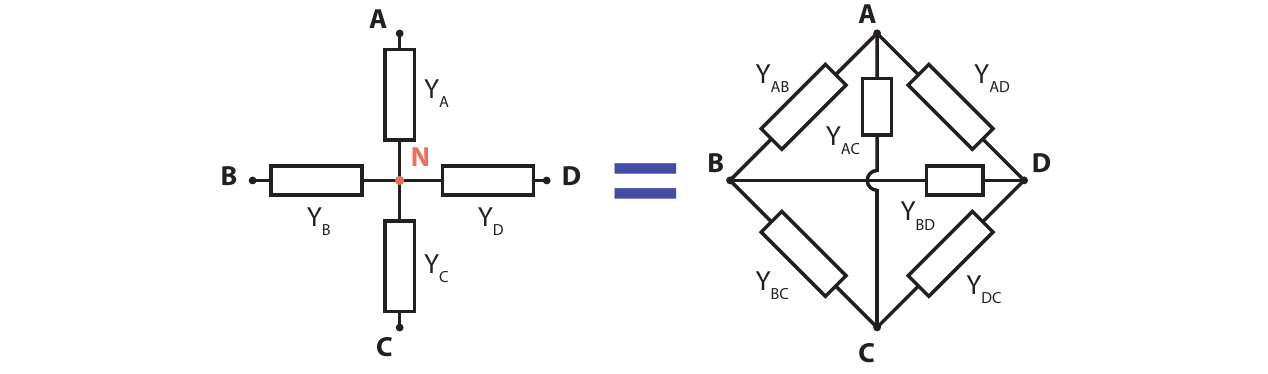}
\caption{
\textbf{Star-mesh transform.}
A node $N$ connected to nodes $A,B,C,..$ through admittances $Y_A,Y_B,...$ can be eliminated if we interconnect nodes $A,B,C,..$ with impedances $Y_{AB},Y_{AC},Y_{BC},...$ given by $Y_{XY} = Y_XY_Y/\sum_MY_M$.
We show the 5 node case, the initial network on the left is called the ``star", which is then transformed to the ``mesh" on the right, reducing the total number of nodes by 1.
}
\label{fig:QuCAT_star_mesh}
\end{figure}

\begin{figure}[]
\centering
\includegraphics{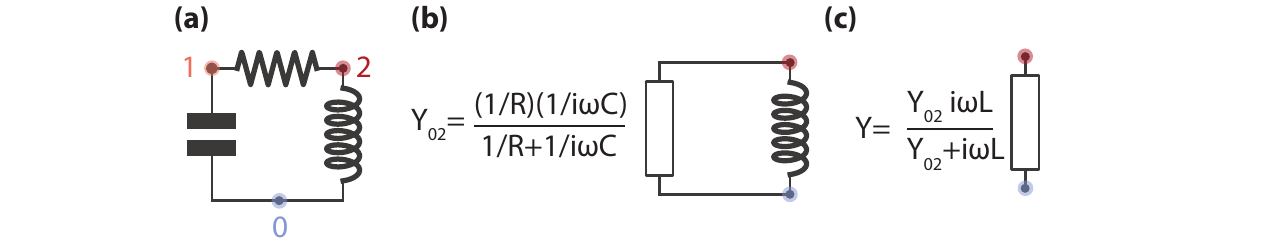}
\caption{
\textbf{Example to illustrate the admittance calculation algorithm. }
\textbf{(a)} Example of a circuit built through the GUI by a user, after removal of all wires and grounds.
\textbf{(b)} Application of the star-mesh transformation to remove node 1.
\textbf{(c)} After each application of the star-mesh transformation, parallel connections are grouped together. Only the two nodes across which we want to compute the admittance remain, the admittance is that of the remaining ``admittance component".
}
\label{fig:QuCAT_Y_computation_example}
\end{figure}

Here we describe a method to compute the admittance of a network between two arbitrary nodes.
We will continue using the example circuit of Fig.~\ref{fig:QuCAT_f_finder_example}, assuming we want to compute the admittance at the nodes of the inductor.
\begin{enumerate}
    \item Eliminate wires and grounds as in the resonance finding algorithm, nodes $N_0,N_5$ would be grouped under a single node labeled 0 and nodes $N_1,N_2,N_3$ would be grouped under node 1, we label node $N_4$ node 2. We thus obtain Fig.~\ref{fig:QuCAT_Y_computation_example}(a)
    \item Group parallel connections. Group all components connected in parallel as a single ``admittance component" equal to the sum of admittances of its parts. In this way two nodes are either disconnected, connected by a single inductor, capacitor, junction or resistor, or connected by a single ``admittance component".
    \item Reduce the network through star-mesh transformations. Excluding the nodes across which we want to evaluate the admittance, we utilize the star-mesh transformation described in Fig.~\ref{fig:QuCAT_star_mesh} to reduce the number of nodes in the network to two. If following a star-mesh transformation, two components are found in parallel, they are grouped under a single ``admittance component" as described previously. For a node connected to more than 3 other nodes the star-mesh transformation will increase the total number of components in the circuit. So we start with the least-connected nodes to maintain the total number of components in the network to a minimum. In this example, we want to keep nodes 0 and 2, but remove node 1, a start-mesh transform leads to the circuit of Fig.~\ref{fig:QuCAT_Y_computation_example}(b) then grouping parallel components leads to (c).
    \item The admittance is that of the remaining ``admittance component" once the network has been completely reduced to two nodes.
\end{enumerate}
The symbolic variables at this stage (Sympy \inline{Symbols}) are $\omega$, and the variables corresponding to any component with un-specified values.


\subsubsection{Differentiating the admittance}
The expression for the admittance obtained from the above algorithm will necessarily be in the form of multiple multiplication, divisions or additions of the admittance of capacitors, inductors or resistors.
It is thus possible to transform $Y$ to a rational function of $\omega$
\begin{equation}
    Y(\omega) = \frac{P(\omega)}{Q(\omega)} = \frac{p_0+p_1\omega+p_2\omega^2+...}{q_0+q_1\omega+q_2\omega^2+...}
\end{equation}
with the sympy function \inline{together}.
It is then easy to symbolically determine the derivative of Y, ready to be evaluated at $\zeta_m$ once the coefficients $p_i$ and $q_i$ have been extracted
\begin{equation}
\begin{split}
    Y'(\zeta_m) &= \left(P'(\zeta_m)Q(\zeta_m) -P(\zeta_m)Q'(\zeta_m)\right)/Q(\zeta_m)^2\\
      &=(p_1+2p_2\zeta_m+...)/(q_0+q_1\zeta_m+...)\ ,
\end{split}
\end{equation}
taking advantage of the property $P(\zeta_m)\propto Y(\zeta_m)=0$.

\subsubsection{Choice of reference element}
\begin{figure}[]
\centering
\includegraphics{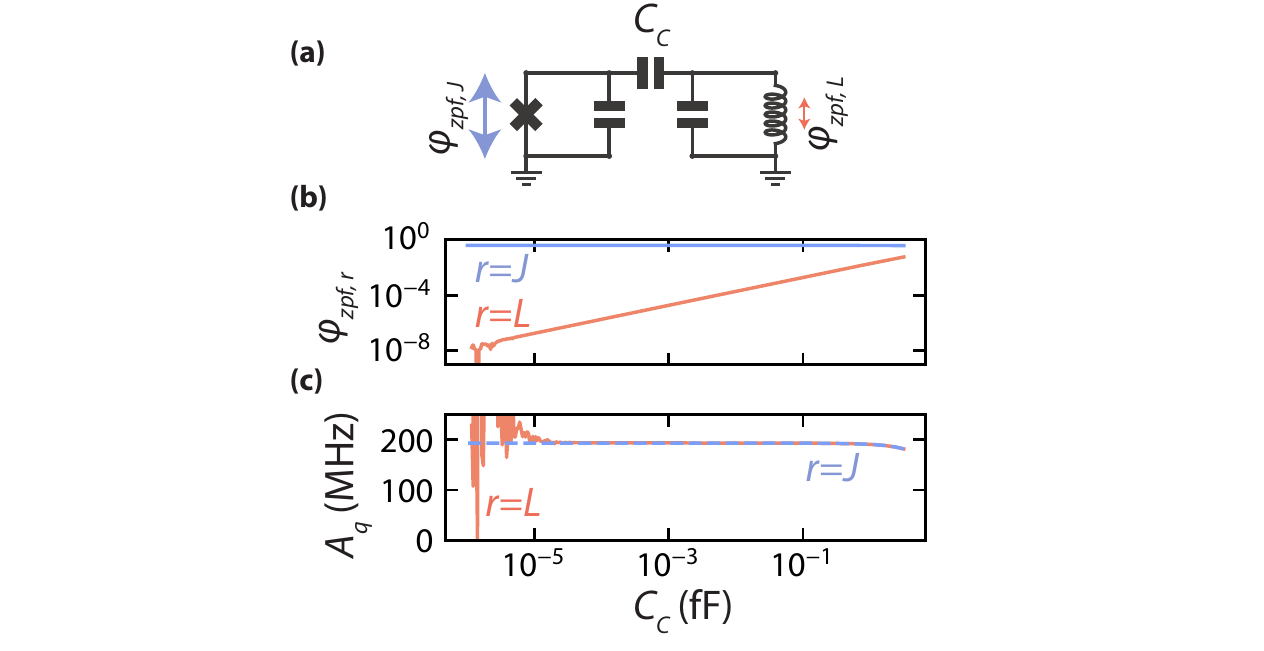}
\caption{
\textbf{Impact of the choice of reference element. }
\textbf{(a)} Schematic of the circuit used in this example.
We have used a 100 fF capacitances, a 10 nH inductor and a 8 nH Josephson inductance, we will vary the coupling capacitance.
The zero-point fluctuations in phase $\varphi_{\text{zpf},r}$ across the inductor ($r=L$) and junction ($r=J$) for most anharmonic mode are drawn on the schematic and plotted in \textbf{(b)} for different values of the coupling capacitor $C_C$.
The phase oscillations associated with this mode are mostly located in the junction, so as the coupling capacitor is lowered, the amplitude of phase oscillations diminishes in the inductor.
Below $C_C\sim 10^{-5}$ fF, numerical accuracies lead to unreliable values of the phase fluctuations in the inductor.
This results in the anharmonicity of the qubit-like mode $A_q$, plotted in \textbf{(c)}, to be incorrectly estimated if the inductor is chosen as a reference element and the anharmonicity is computed using Eq.~(\ref{eq:phi_j_from_phi_r}).
}
\label{fig:QuCAT_best_ref_elt}
\end{figure}

For each mode $m$, we use as reference element $r$ the inductor or junction which maximizes $\varphi_{\text{zpf},m,r}$ as specified by Eq.~\ref{eq:zpf}.
This corresponds to the element where the phase fluctuations are majoritarily located.
We find that doing so considerably increases the success of evaluating $Y'(\omega_m)$.
As an example, we plot in Fig.~\ref{fig:QuCAT_best_ref_elt} the zero-point fluctuations in phase $\varphi_{\text{zpf},m,r}$ of the transmon-like mode, calculated for the circuit of Fig.~\ref{fig:QuCAT_1}, with the junction or inductor as reference element.
What we find is that if the coupling capacitor becomes too small, resulting in modes which are nearly totally localized in either inductor or junction, choosing the wrong reference element combined with numerical inaccuracies leads to unreliable values of $\varphi_{\text{zpf},m,r}$.

\subsection{Transfer functions}\label{sec:methods_network_transfer}

\begin{figure}[]
\centering
\includegraphics{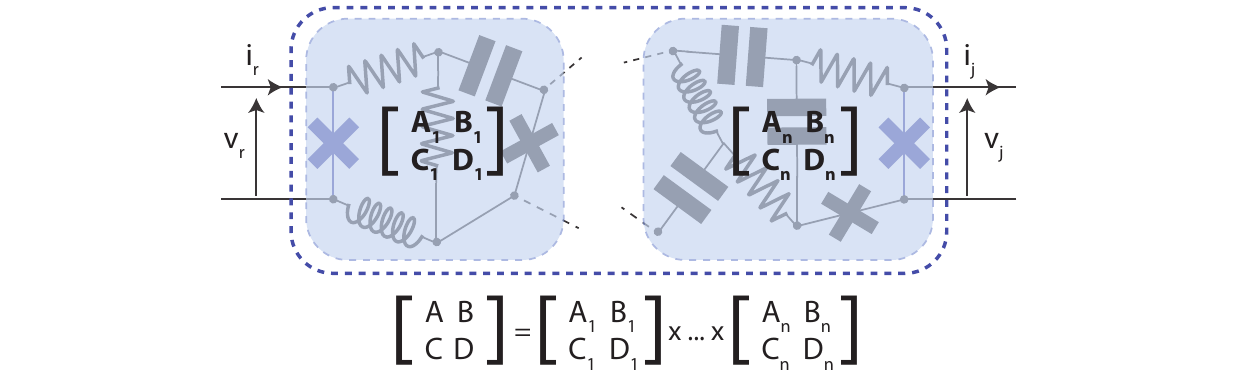}
\caption{
\textbf{Visual summary of the notations and properties of the ABCD matrix applied to the calculation of $T_{rj}$. }
The transfer function $T_{rj} = 1/A$ is the inverse of the first coefficient of the ABCD matrix which relates the voltages and currents on either end of a network.
These currents and voltages are defined as shown above, with the reference junction on the left and junction $j$ on the right.
Currents are defined as entering and exciting on the left and right respectively.
If the circuit is constituted of a cascade of two port sub-networks, the product of the sub-network ABCD matrices are equal to the ABCD matrix of the total network.
}
\label{fig:QuCAT_two_port_network}
\end{figure}

\begin{figure}[]
\centering
\includegraphics[width=0.8\textwidth]{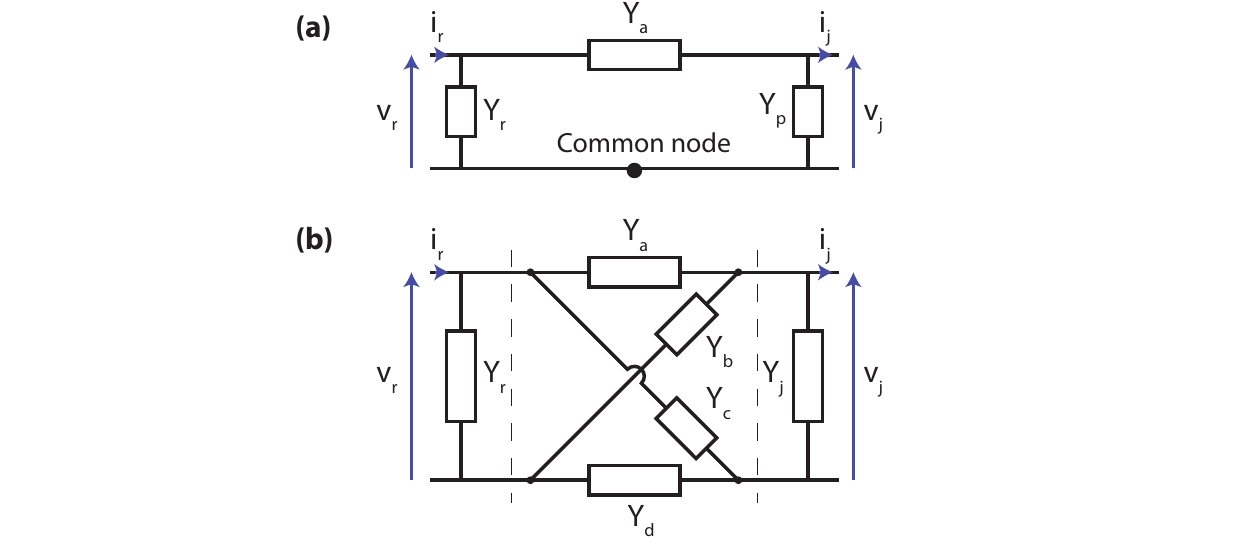}
\caption{
\textbf{Networks after star-mesh reduction.}
The two non-trivial situations reached after applying star-mesh transformations to a network to obtain $T_{rj}$.
}
\label{fig:QuCAT_network}
\end{figure}

In this section, we describe the method used to determine the transfer function $T_{jr}$ between a junction $j$ and the reference junction $r$.
This quantity can be computed from the ABCD matrix~\cite{pozar2009microwave}.
The ABCD matrix relates the voltages and cu{}rrents in a two port network
\begin{equation}
    \begin{pmatrix}
    V_r \\
    I_r
    \end{pmatrix}=
    \begin{pmatrix}
    A&B \\
    C&D
    \end{pmatrix}
    \begin{pmatrix}
    V_j\\
    I_j\\
    \end{pmatrix}
\end{equation}
where the convention for current direction is described in Fig.~\ref{fig:QuCAT_two_port_network}.
By constructing the network as in Fig.~\ref{fig:QuCAT_two_port_network}, with the reference junction on the left and junction $j$ on the right, the transfer function is given by
\begin{equation}
 T_{jr}(\omega) = \frac{V_j(\omega)}{V_r(\omega)} = \frac{1}{A}
\end{equation}
To determine $A$, we first reduce the circuit using star-mesh transformations (see Fig.\ref{fig:QuCAT_star_mesh}), and group parallel connections as described in the previous section, until only the nodes of junctions $r$ and $j$ are left.
If the junctions initially shared a node, the resulting circuit will be equivalent to the network shown in Fig.\ref{fig:QuCAT_network} (a).
In this case, 
\begin{equation}
    A = \left(1+\frac{Y_p}{Y_a}\right)\ .
\end{equation}
%
%
If the junctions do not share nodes, the resulting circuit will be equivalent to the network shown in Fig.\ref{fig:QuCAT_network}(b), where some admittances may be equal to $0$ to represent open circuits.
To compute the ABCD matrix of this resulting circuit, we make use of the property illustrated in Fig.~\ref{fig:QuCAT_two_port_network}: the ABCD matrix of a cascade connection of two-port networks is equal to the product of the ABCD matrices of the individual networks.
We first determine the $ABCD$ matrix of three parts of the network (separated by dashed line in Fig.\ref{fig:QuCAT_network}) such that the $ABCD$ matrix of the total network reads
\begin{equation}
\begin{bmatrix}
    A & B   \\
    C & D  \\
\end{bmatrix} = 
\begin{bmatrix}
    1 & 0   \\
    Y_r & 1  \\
\end{bmatrix}
\begin{bmatrix}
    \tilde A  & \tilde B   \\
    \tilde C & \tilde D  \\
\end{bmatrix}
\begin{bmatrix}
    1 & 0   \\
    Y_j & 1  \\
\end{bmatrix}\ .
\end{equation}
\begin{equation}
  A = \tilde A+\tilde B Y_j\ ,
\end{equation}
where the $A$ and $B$ coefficients of the middle part of the network are
\begin{equation}
\begin{split}
    \tilde A&=(Y_a + Y_b)(Y_c + Y_d)/(Y_a Y_d-Y_b Y_c)\\
    \tilde B&=(Y_a + Y_b + Y_c + Y_d)/(Y_a Y_d-Y_b Y_c)\ .
\end{split}
\end{equation}
The ABCD matrix for the middle part of the circuit is derived in Sec. 10.11 of Ref.~\cite{arshad2010network}, and the ABCD matrices for the circuits on either sides are provided in Ref.~\cite{pozar2009microwave}.

This method is also applied to calculate the transfer function to capacitors, inductors and resistors, notably to visualize the normal mode with the \inline{show_normal_mode} function.

\subsection{Alternative algorithmic methods}

Since symbolic calculations are the most computationally expensive steps in a typical use of QuCAT, we cover in this section some alternatives to the methods previously described, and the reasons why they were not chosen.
\subsubsection{Eigen-frequencies from the zeros of admittance}
One could solve $Y_r(\omega)=0$ where $Y_r$ is the admittance computed as explained in Sec.~\ref{sec:methods_network_dY}.
Providing good initial guesses for all values of the zeros $\zeta_m$ can be provided, a number of root-finding algorithms can then be used to obtain final values of $\zeta_m$.
A set of initial guesses could be obtained by noticing that $Y_r$ is a rational function of $\omega$.
Roots of its numerator are potentially zeros of Y, and a complete set of them is easy to obtain through a diagonalization of the companion matrix as discussed before.
Note that if these roots are roots of the denominator with equal or higher multiplicity, then they are not zeros of Y.
They can, however, make good initial guesses of a root-finding algorithm run on $Y_r$.
This requires a simplification of $Y_r$, as computed through star-mesh transforms, to its rational function form.
We find this last step to be as computationally expensive as obtaining a determinant.

A different approach, which does not require using a root-finding algorithm on $Y_r$, is to simplify the rational-function form of Y such that the numerator and denominator share no roots.
This can be done by using the extended Euclidian algorithm to find the greatest common polynomial divisor (GCD) of the numerator and denominator.
However, the numerical inaccuracies in the numerator and denominator coefficients may make this method unreliable.
The success of both of these approaches is dependent on determining a good reference component $r$, which may be mode-dependent (see Fig.~\ref{fig:QuCAT_best_ref_elt}).
This reference component is difficult to pick at this stage, when the mode frequencies are unknown.

\subsubsection{Finite difference estimation of the admittance derivative}
Rather than symbolically differentiating the admittance, one could use a numerical finite difference approximation, for example
\begin{equation}
     Y_r'(\omega)\simeq \frac{Y_r(\omega+\delta\omega/2)-Y_r(\omega-\delta\omega/2)}{\delta\omega}\ .
     \label{eq:finite_difference_dY}
 \end{equation} 
$Y_r$ can be obtained through star-mesh reductions, or from a resolution of Eq.~\ref{eq:admittance_matrix_equation}.

\begin{figure}[]
\centering
\includegraphics[width=0.8\textwidth]{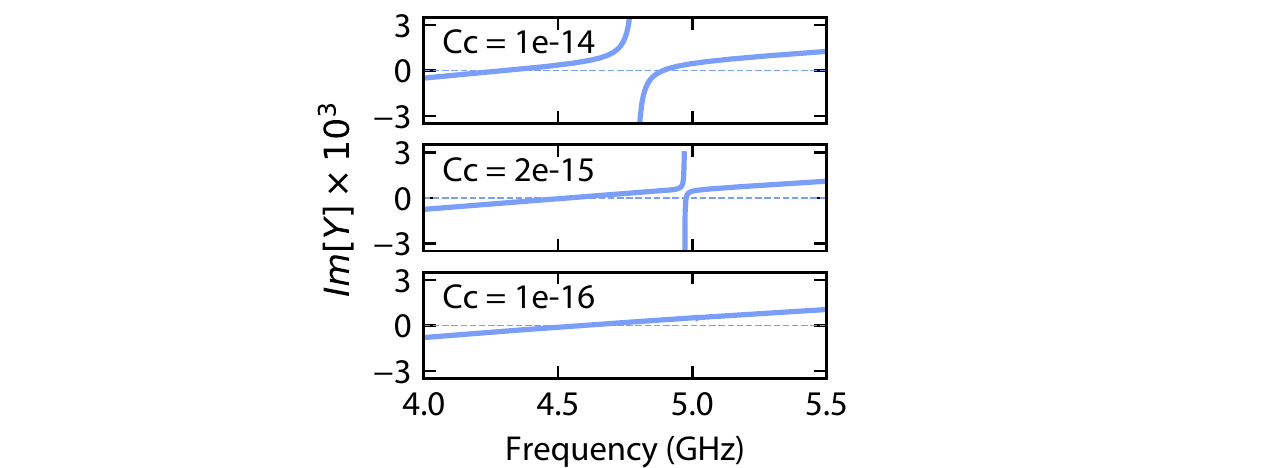}
\caption{
\textbf{Determining zero-point fluctuations from differentiating the admittance in decoupled circuits.}
Imaginary part of the admittance $\text{Im}[Y]$ across the junction of the circuit of Fig.~\ref{fig:QuCAT_1} for different values of the coupling capacitor $Cc$ and for $L_j=12$ nH.
Resonances correspond{} to the frequencies $\omega$ at which the admittance crosses 0, and the calculation of zero-point fluctuations depends on the derivative $\text{Im}[Y']$ at that point.
As the resonator and transmon parts of the circuit decouple, $\text{Im}[Y']$ becomes larger, requiring a lower $\delta\omega$ if the admittance is to be determined through Eq.~(\ref{eq:finite_difference_dY}).
In extreme cases (see lower panel), when the derivative is very large, the smaller variation in the background slope may be mistaken for the slope at a resonance.
}
\label{fig:QuCAT_dY_from_differentiation}
\end{figure}

But finding a good value of $\delta\omega$ is no easy task.
As an example, we consider the circuit of Fig.~\ref{fig:QuCAT_1}, where we have taken as reference element the junction.
As shown in Fig.~\ref{fig:QuCAT_dY_from_differentiation}, when the resonator and transmon decouple through a reduction of the coupling capacitor, a smaller and smaller $\delta\omega$ is required to obtain $Y_r'$ evaluated at the $\zeta_1$ of the resonator-like mode.
We have tried making use of Ridders method of polynomial extrapolation to try and reliably approach the limit $\delta x\rightarrow 0$~\cite{press2007numerical}.
However at small coupling capacitance, it always converges to the slower varying background slope of $Y_r$, without any way of detecting the error.

\subsubsection{Transfer functions from the admittance matrix}
Calculating the transfer function $T_{ij}$ could alternatively be carried out through the resolution of the system of equations (\ref{eq:admittance_matrix}).
The difference in voltage of a reference elements nodes would first have to be fixed to the zero-point fluctuations computed with the method of Sec.~(\ref{sec:methods_network_dY}).
These equations would have to be resolved at each change of system parameters and for each mode, with $\omega$ replaced in the admittance matrix by its corresponding value for a given mode.
This is to be balanced against a single symbolic derivation of $T_{ij}$ through star-mesh transformations, and fast evaluations of the symbolic expression for different parameters.

\section{Performance and limitations}
\subsection{Number of nodes}
In this section we ask the question: how big a circuit can QuCAT analyze?
To address this, we first consider the circuit of Fig.~\ref{fig:QuCAT_mmusc}(c), and secondly the same circuit with resistors added in parallel to each capacitor.
As the number of (R)LC oscillators representing the modes of a CPW resonator is increased, we measure the time necessary for the initialization of the Qcircuit object.
This is typically the most computationally expensive part of a QuCAT usage, limited by the speed of symbolic manipulations in Sympy.

These symbolic manipulations include:
\begin{itemize}
\item Calculating the determinant of the admittance matrix
\item Converting that determinant to a polynomial
\item Reducing networks through star-mesh transformations both for admittance and transfer function calculations
\item Rational function manipulations to prepare the admittance for differentiation
\end{itemize}
Once these operations have been performed, the most computationally expensive step in a Qcircuit method is finding the root of a polynomial (the determinant of the admittance matrix) which typically takes a few milliseconds.

\begin{figure}[]
\centering
\includegraphics[width=0.8\textwidth]{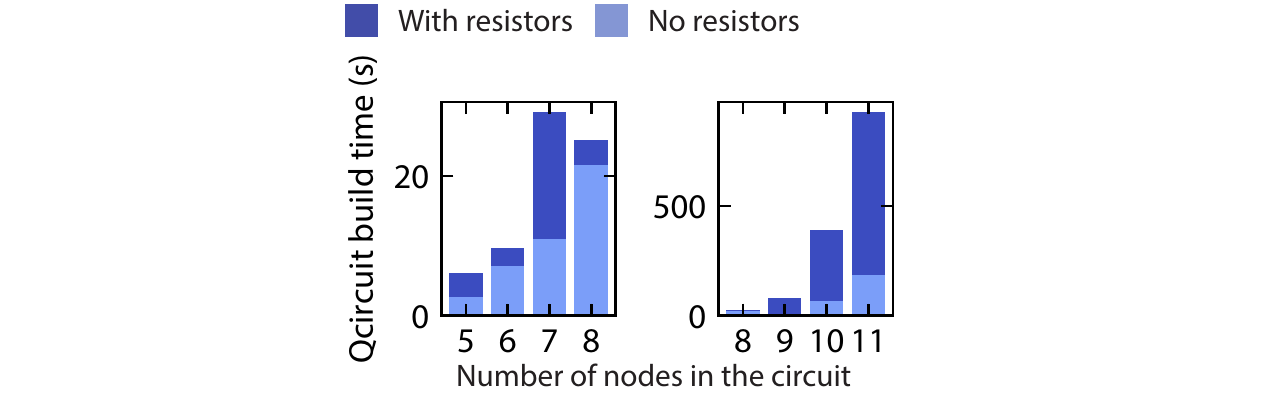}
\caption{
\textbf{Computation time with increasing circuit size. }
On the vertical axis, we show the time necessary to initialize the \inline{Qcircuit} object, which is the computationally expensive part of a typical QuCAT user case.
This is plotted as a function of the number of nodes in the circuit.
The test circuit used here is the multi-mode circuit of Fig.~\ref{fig:QuCAT_mmusc}(c), optionally with a resistor in parallel of each capacitor.
The number of nodes are increased by adding modes to the circuit.
Most of the computational time is spent in the symbolic manipulations performed with the sympy library.
}
\label{fig:QuCAT_performance}
\end{figure}

The results of this test are reported in Fig.~\ref{fig:QuCAT_performance}. 
We find that relatively long computation times above 10 seconds are required as one goes beyond 10 circuit nodes.
Due to an increased complexity of symbolic expressions, the computation time increases when resistors are included.
For example, the admittance matrix of a non-resistive circuit will have no coefficients proportional to $\omega$, only $\omega^2$ and only real parts, translating to a polynomial in $\Omega = \omega^2$ which will have half the number of terms as a resistive circuit.
However, we find that this initialization time is also greatly dependent on the circuit connectivity, and this test should be taken as only a rough guideline.
Making QuCAT compatible with the analysis of larger circuits will inevitably require the development of more efficient open-source symbolic manipulation tools.
The development of the open-source C++ library SymEngine \url{https://github.com/symengine/symengine}, together with its Python wrappers, the symengine.py project \url{https://github.com/symengine/symengine.py}, could lead to rapid progress in this direction.
An enticing prospect would then be able to analyze the large scale cQED systems underlying modern transmon-qubit-based quantum processors~\cite{arute2019quantum}.
One should keep in mind that an increase in circuit size translates to an increase in the number of degrees of freedom of the circuit and hence of the Hilbert space size needed for further analysis once a Hamiltonian has been extracted from QuCAT.

\subsection{Degree of anharmonicity}\label{sec:high_anharmonicity}
\begin{figure}[]
\centering
\includegraphics[width=0.8\textwidth]{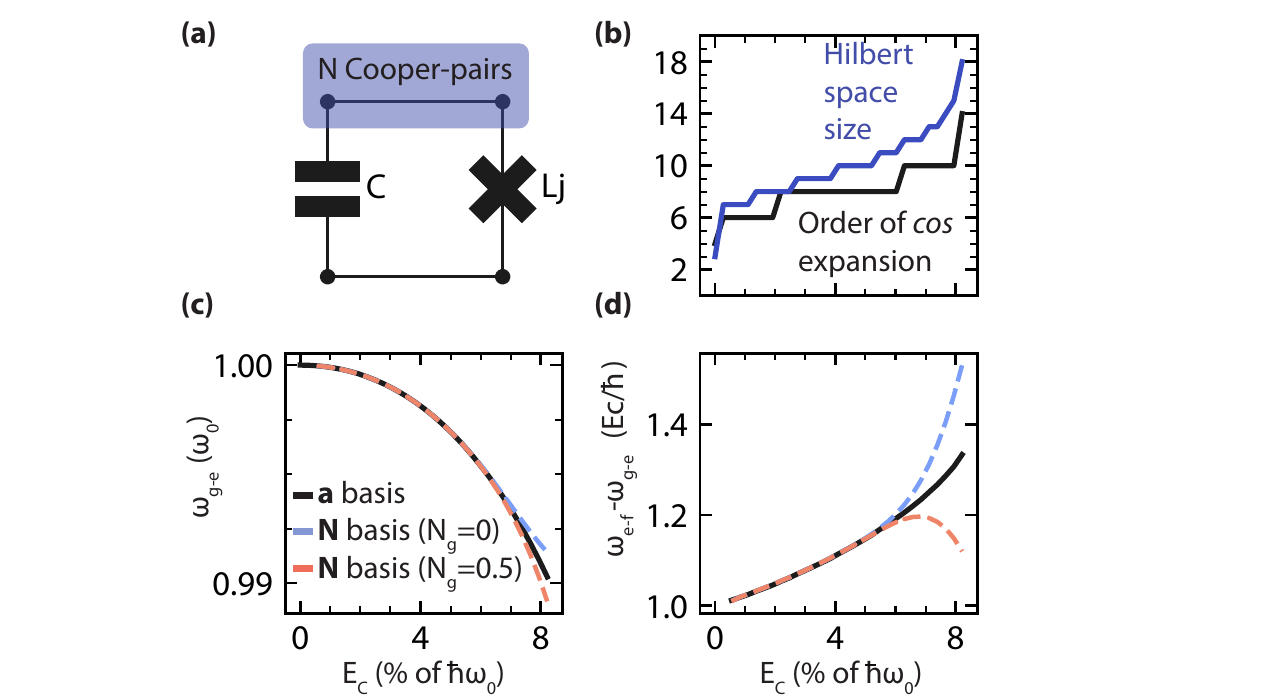}
\caption{
\textbf{Applicability of the harmonic Fock basis.}
\textbf{(a)} Transmon or Cooper-pair-box circuit.
%
%
\textbf{(b)} On the x-axis, we vary the approximate anharmonicity $E_C= e^2/2C$ with respect to the frequency $\omega_0=1/\sqrt{L_jC}$.
For each value, we plot a Hilbert space size, and order of Taylor-expansion of the junction cosine-potential.
Incrementing these values produces less than a 0.1 percent change in the first two transition frequencies obtained by diagonalizing the Hamiltonian.
Beyond a relative anharmonicity of 8, convergence is no longer reached, even for Hilbert space sizes and Taylor expansions up to 100.
\textbf{(c)}
Frequency of the first transition  $\omega_{g-e}$ obtained from Hamiltonian diagonalization, relative to the value expected from first order perturbation theory: $\omega_0-E_C/\hbar$.
\textbf{(d)}
Anharmonicity $\omega_{e-f}-\omega_{g-e}$ obtained from Hamiltonian diagonalization relative to the value expected from first order perturbation theory $E_C/\hbar$.
Black lines correspond to a diagonalization in the harmonic Fock basis (Eq.~(\ref{eq:hamiltonian_taylor_SI})), blue and orange dashed lines correspond to a diagonalization of the Cooper-pair box Hamiltonian with gate charges of $0$ and $1/2$ respectively.
The harmonic Fock basis provides reliable results up to approximatively 6 percent anharmonicity.
%
}
\label{fig:QuCAT_high_anharmonicity}
\end{figure}
In this section we study the limits of the current quantization method used in QuCAT.
More specifically, we study the applicability of the basis used to express the Hamiltonian, that of Fock-states of harmonic normal modes of the linearized circuit.
To do so, we use the simplest circuit possible (Fig.~\ref{fig:QuCAT_high_anharmonicity}(a)), the parallel connection of a Josephson junction and a capacitor.
As the anharmonicity of this circuit becomes a greater fraction of its linearized circuit resonance, the physics of the circuit goes from that of a Transmon to that of a Cooper-pair box~\cite{koch_charge-insensitive_2007}, and the Fock-state basis becomes inadequate.
This test should be viewed as a guideline for the maximum acceptable amount for  anharmonicity.
We find that when the anharmonicity exceeds 6 percent of the eigenfrequency, a QuCAT generated Hamiltonian will not reliably describe the system.

In this test, we vary the ratio of Josephson inductance $L_j$ to capacitance $C$, increasing the anharmonicity expected from first-order perturbation theory (see Eq.~\ref{eq:hamiltonian_first_order}), called charging energy $E_C = e^2/2C $.
The resonance frequency of the linearized circuit $\omega_0 = 1/\sqrt{L_jC}$ is maintained constant.
For each different charging energy, we use the \inline{hamiltonian} method to generate a Hamiltonian of the system. 
We are interested in the order of the Taylor expansion of the cosine potential, and the size of the Hilbert space, necessary to obtain realistic first and second transition frequencies of the circuit, named $\omega_{g-e}$ and $\omega_{e-f}$ respectively.
To do so, we increase the order of Taylor expansion, and for each order we sweep through increasing Hilbert space sizes.
In Fig.~\ref{fig:QuCAT_high_anharmonicity}(b), we show the values of these parameters at which incrementing them would not change $\omega_{g-e}$ and $\omega_{e-f}$ by more than 0.1 percent.
Beyond a relative anharmonicity $E_C/\hbar\omega_0$ of 8 percent, such convergence is no longer reached, even for cosine expansion orders and Hilbert space sizes up to 100.

Up to the point of no convergence, we compare the results obtained from the diagonalization in the harmonic Fock basis (Hamiltonian generated by QuCAT), with a diagonalization of the Cooper-pair box Hamiltonian.
In regimes of higher anharmonicity, the system becomes sensitive to the preferred charge offset between the two plates of the capacitor $N_g$ (expressed in units of Cooper-pair charge $2e$) imposed by the electric environment of the system.
The Cooper-pair box Hamiltonian takes this into account
\begin{equation}
\begin{split}
    \hat H_\text{CPB} = 4E_C(\sum_N\ket{N}\bra{N}-N_g)^2\\-\sum_NE_j(\ket{N+1}\bra{N}+\ket{N}\bra{N+1})
\end{split}
\end{equation}
where $\ket{N}$ is the quantum state of the system where N Cooper-pairs have tunneled across the junction to the node indicated in Fig.~\ref{fig:QuCAT_high_anharmonicity}(a).
For more details on Cooper-pair box physics and the derivation of this Hamiltonian, refer to Ref.~\cite{schuster2007circuit}.
We diagonalize this Hamiltonian in a basis of 41 $\ket{N}$ states.
We find that beyond 6 percent anharmonicity, the Cooper-pair box Hamiltonian becomes appreciably sensitive to $N_g$ and diverges from the results obtained in the Fock basis.
This corresponds to $E_j/E_C \simeq 35$ at which the charge dispersion (the difference in frequency between $0$ and $0.5$ charge offset) is $4\times10^{-5}$ and $1\times10^{-3}$ for the first two transitions respectively.

Beyond 8 percent anharmonicity, one cannot reach convergence with the Fock basis and just before results diverge considerably from that of the Cooper-pair box Hamiltonian.
This corresponds to $E_j/E_C \simeq 20$ at which the charge dispersion is $1.5\times10^{-3}$ and $3\times10^{-2}$ for the first two transitions respectively.
A possible extension of the QuCAT Hamiltonian could thus include handling static offsets in charge and different quantization methods, for example quantization in the charge basis to extend QuCAT beyond the scope of weakly-anharmonic circuits.

\section{Installing QuCAT and dependencies}
The recommended way of installing QuCAT is through the standard Python package installer by running \inline{pip install qucat} in a terminal.
Alternatively, all versions of QuCAT, including the version currently under-development is available on github at \url{https://github.com/qucat}.
After downloading or cloning the repository, one can navigate to the \inline{src} folder and run \inline{pip install .} in a terminal.

QuCAT and its GUI is cross-platform, and should function on Linux, MAC OS and Windows.
QuCAT requires a version of Python 3, using the latest version is advised.
QuCAT relies on several open-source Python libraries: Numpy, Scipy, Matplotlib, Sympy and QuTiP~\cite{johansson2012qutip,johansson2013qutip}, installation of Python and these libraries through Anaconda is recommended.
The performance of Sympy calculations can be improved by installing Gmpy2.
\section{List of QuCAT objects and methods}
\setlength\parindent{0pt}    

\textbf{QuCAT objects }

\inline{Network} -- Creates a \inline{Qcircuit} from a list of components

\inline{GUI}  -- Opens a graphical user interface for the construction of a \inline{Qcircuit}

\inline{J}  -- Creates a Josephson junction object

\inline{L} -- Creates a inductor object

\inline{C} -- Creates a capacitor object

\inline{R} -- Creates a resistor object

\vspace{5pt}
\textbf{Qcircuit methods}

\inline{eigenfrequencies}  -- Returns the normal mode frequencies 

\inline{loss_rates}  -- Returns the normal mode loss rates

\inline{anharmonicities}  -- Returns the anharmonicities or self-Kerr of each normal mode 

\inline{kerr} -- Returns the self-Kerr and cross-Kerr for and between each normal mode  

\inline{f_k_A_chi} -- Returns the eigenfrequency, loss-rates,  anharmonicity, and Kerr parameters of the circuit  

\inline{hamiltonian} -- Returns the Hamiltonian of Ref.~\ref{eq:hamiltonian_taylor_SI}  

\vspace{5pt}
\textbf{Qcircuit methods}  \textit{(only if built with GUI)}  

\inline{show}  -- Plots the circuit 

\inline{show_normal_mode}  -- Plots the circuit overlaid with the currents,  voltages, charge or fluxes through each component when a normal mode is populated with a quarter-photon coherent state  

\vspace{5pt}
\textbf{J,L,R,C methods}

\inline{zpf}  -- Returns contribution of a mode to the zero-point fluctuations  in current, voltages, charge or fluxes 

\vspace{5pt}
\textbf{J methods}

\inline{anharmonicity}  -- Returns the contribution of this junction to the anharmonicity of a given normal mode (Eq.~(\ref{eq:Amj}))


\FloatBarrier\chapter{Gravitational effects in quantum superpositions of mechanical oscillators.}
\label{chapter_gravity}
\label{sec:quantum_gravity}

\begin{abstract}
Attempting to reconcile general relativity with quantum mechanics is one of the great undertakings of contemporary physics.
In this chapter, we will first present how the incompatibility between the two theories arises in the simple thought experiment of preparing a heavy object in a quantum superposition.
Through this thought experiment, and an analysis of the different theoretical approaches to the problem, we then quantify the mass and coherence times required to perform experiments where both theories interplay in an unknown way.
Finally we will look at concrete implementations of such an experiment using a micro-mechanical oscillators.
In particular, we focus on oscillators which can be coupled to superconducting circuits. 
\end{abstract}

\newpage

\noindent 

\section{The problem of combining time-dilation and quantum mechanics}
\begin{figure}[h!]
\centering
\includegraphics[width=0.8\textwidth]{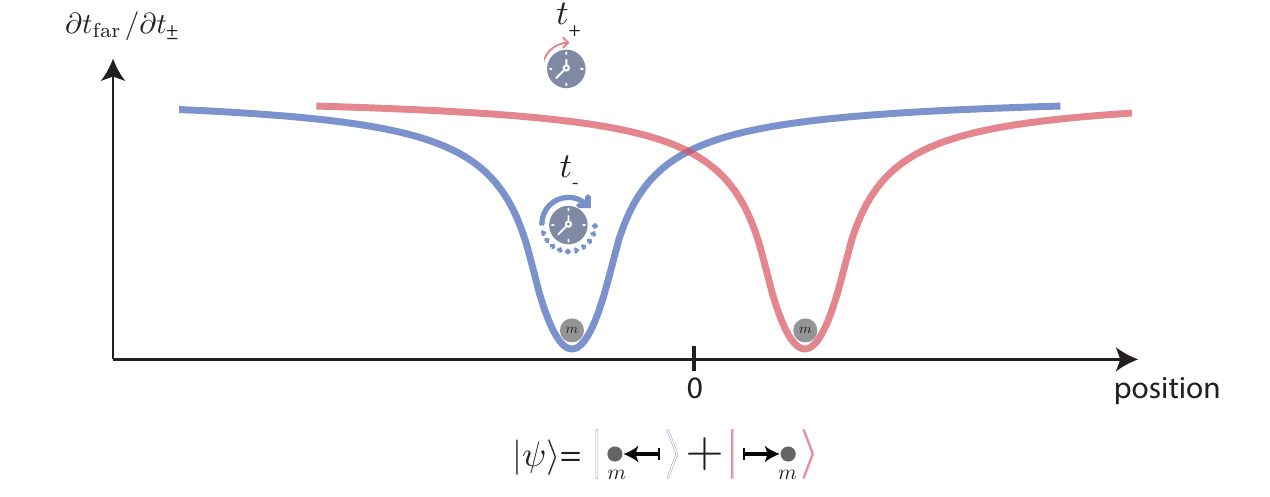}
\caption{\textbf{Quantum superposition of a massive object and its impact on the definition of time.}
We consider a ball of mass $m$ in a quantum superposition $\ket{\psi}$ of being displaced in two opposite directions.
The two parts of the superposition are given a different color, blue and red.
We plot the rate at which the proper time $t_\pm$ corresponding the left (right) displaced ball evolves with respect to a the proper time $t_\text{far}$ of a reference location far away from the ball.
In the vicinity of one of the parts of the superposition, there is thus an ambiguity in the local time one should consider for the quantum evolution of the state quantified by the difference between $t_-$ and $t_+$.
}
\label{fig:time_dilation}
\end{figure}
One problem in writing a consistent theory of nature from which would stem general relativity (GR) and quantum mechanics (QM), is that of time.
In GR, the rate at which time flows can be distorted by the presence of a massive object, a phenomenon called time-dilation.
Consider an incompressible ball with mass $m$ and radius $R$.
Solving Einsteins field equations~\cite{schwarzschild1999gravitational} reveals that the proper time $t$ measured by a stationary clock at a distance $r$ from the center of the ball is related to the time $t_\text{far}$ measured by a clock infinitely far from the ball through
\begin{equation}
	t = t_\text{far}\sqrt{1-\frac{2mG}{rc^2}}\simeq t_\text{far}\left(1-\frac{mG}{rc^2}\right)\ ,
\end{equation}
where $G$ is the gravitational constant, $c$ is the speed of light and $r>R$.
The approximation holds for the small masses we will consider here $m\ll Rc^2/G$.
%
%
The incompatibility with QM comes when one considers such an object in a superposition, where it is simultaneously displaced left and right by at least its radius as schematically shown in Fig.~\ref{fig:time_dilation}.
There is then no longer a unique definition of time -- which is required to describe this system with Schrodingers equation.
For example, there is an ambiguity in the time that a clock sitting on the left-displaced part of the superposition would record: does it experience the space time defined by the left- or right- displaced ball?
If it experiences the left-displaced time, then $t_- \simeq t_\text{far}(1-mG/Rc^2)$, if it experiences the left-displaced  $t_+ \simeq t_\text{far}$ (for a displacement which greatly exceeds $R$). 
The fact that there is an issue in describing this situation with our current physical theories is undeniable. 
What is currently debatable is for what system parameters this effect starts to become noticeable, and we report in the next section on the tentative guesses that have been made.
\section{Masses and coherence times necessary for detection of such effects}
We now try and estimate what parameters $m,R$ which are necessary for time-dilation to have a noticeable effect in this superposition.
We assume that the ball displaced to the left and right $\psi_\pm$ are eigenstates of a hypothetical general-relativistic-Hamiltonian describing this experiment, with eigenenergies $E_\pm = E$.
According to Schrodingers equation, the state of the system would then evolve following 
\begin{equation}
	\psi(\ \stackrel{?}{t}\ ) = \psi_+e^{i\stackrel{?}{t}E/\hbar}+\psi_-e^{i\stackrel{?}{t}E/\hbar}\ .
\end{equation}
Since time has no clear definition, we denoted it by $\stackrel{?}{t}$. 
We can however approximate the uncertainty in time 
\begin{equation}
	\Delta t = t_+-t_- \simeq t_\text{far}mG/Rc^2
\end{equation}
which would result in an uncertainty in the relative phase of the two parts of this superposition $\Delta\theta = \Delta tE/\hbar$.
Most of our ignorance in determining a gravitational time-scale $t_G$ lies in choosing of a value for $E$.
This absolute energy of the state would usually be a neglected global phase which plays no role in the physics of the system.
Here it sets the time at which general-relativistic effects come into play.
One guess is that that this energy is the relativistic rest-energy of the mass $E = mc^2$.
In this assumption, we arrive at a time-scale for general-relativistic effects
\begin{equation}
	t_G = h\frac{R}{m^2G}\ ,
	\label{eq:t_G_heuristic}
\end{equation}
defined by $\Delta\theta(t_\text{far}=t_G) = 2\pi$.
Observing such effects would require $t_G\ll t_\text{coh}$, with $t_\text{coh}$ the coherence time of the system as determined by its interaction with the environment following conventional decoherence mechanisms~\cite{zurek2003decoherence}.
Note that the time $t_G$ is crucially dependent on the guess $E = mc^2$ which is a very large number.
What will occur at $t_G$ is still up for debate, the fact that an uncertainty in phase seems to emerge from time-dilation points towards decoherence. 
Whilst only experiments can validate or invalidate this guess, it is worth noting that other approaches to this problem have led to a similar estimates of $t_G$.
Whilst the heuristic arguments presented here are inspired by Ref.~\cite{oosterkamp2013clock}, such ideas were first put forth by Karolyhazy~\cite{karolyhazy1966gravitation} and later by Di{\'o}si~\cite{DIOSI1987377} and then Penrose~\cite{penrose1996gravity}.
Karolyhazy~\cite{karolyhazy1966gravitation} first studied the impact of the Heisenberg's uncertainty in the position of a macroscopic object.
Since a large mass distribution will distort the structure (metric) of space-time, he argues that the uncertainty in its position should translate into an uncertainty in this metric.
More specifically, he estimates that the resulting fluctuations in the metric should in turn result in the decoherence of parts of the quantum states which extend out of a critical volume of space.
For a macroscopic ball of radius $R$ and mass $m$, this results in the decoherence of the phase of the center of mass wave function on time scales~\cite{bassi2013models}
\begin{equation}
	t_K = \frac{1}{m}\left(\frac{\hbar R^4}{G^2}\right)^\frac{1}{3}\ .
\end{equation}
Di{\'o}si approached the problem from a different standpoint.
%
%
Di{\'o}si and Luk{\'a}cs~\cite{diosi1987favor} estimate the quantum fluctuations of a gravitational field in a given volume from the quantum-induced measurement imprecision of an accelerating ball in the field.
They then consider that these fluctuations in the gravitational field act as a noise source acting on massive objects which will cause decoherence~\cite{DIOSI1987377}.
The characteristic time of this effect is~\cite{bassi2013models}
\begin{equation}
\begin{split}
	t_D^{-1} = &\frac{G}{2\hbar}\left(2E_{1,2}-E_1-E_2\right) \\
	E_{i,j}=&\int\int d\vec{r}_1 d\vec{r}_2 \frac{\rho_i(\vec{r}_1)\rho_j(\vec{r}_2)}{|\vec{r}_1-\vec{r}_2|}
\end{split}
\end{equation}
where $\rho_{1,2}$ correspond to the mass distribution of the different parts of a superposition.
Pensrose~\cite{penrose1996gravity} quantifies the difference of free-falls through the space-times of two distinct parts of a superposition to reach a near identical formula
\begin{equation}
\begin{split}
	t_P^{-1} = &\frac{G}{\hbar}\left(2E_{1,2}-E_1-E_2\right)
	\label{eq:t_G_penrose_full}
\end{split}
\end{equation}
By considering the case of a ball of mass $m$, radius $R$, and uniform density, with a center of a mass superposed in position by a distance $\Delta x\gg R$, 
we reach 
\begin{equation}
	t_P = \hbar\frac{5R}{12Gm^2}\ 
\end{equation}
using the Penrose formula.
Note that the result is nearly identical to that obtained heuristically in Eq.~(\ref{eq:t_G_heuristic}).
The intensity with which these ideas are being pursued experimentally, as well as the criticism they suffer reflect the importance of this topic.
If it turns out that gravity induces decoherence, it would give a fundamental, mass-related, origin to the quantum-to-classical transition, and possibly address foundational questions in the interpretation of quantum mechanics.
More importantly, experimental data in this regime could shed light on the details of a theory unifying general relativity and quantum mechanics.

\section{Implementation with micro-mechanical oscillators}

\begin{figure}[h!]
\centering
\includegraphics[width=0.8\textwidth]{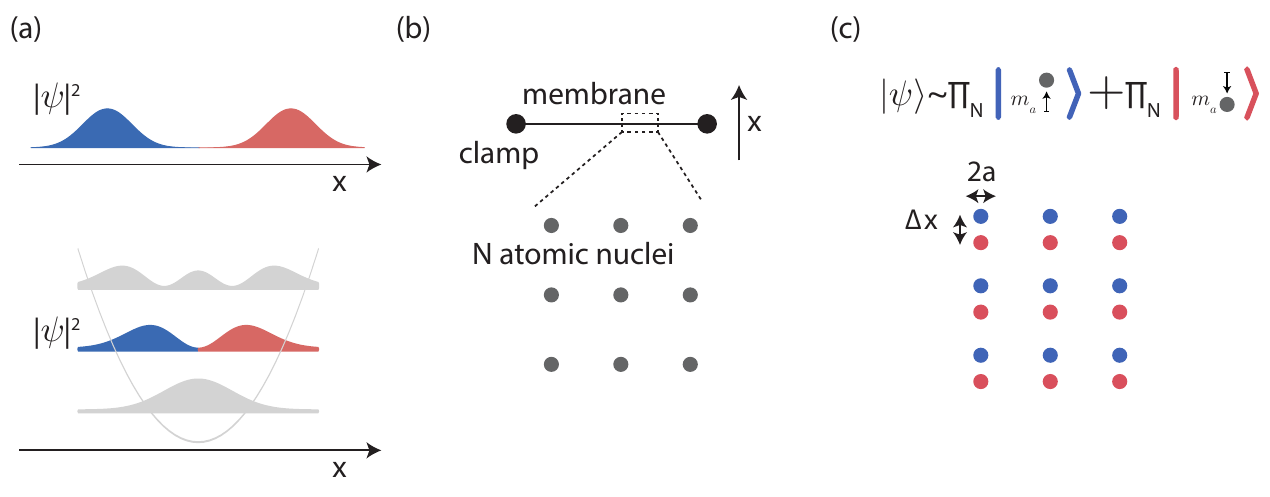}
\caption{\textbf{Implementation of a massive superposition with a membrane.}
The probability distribution $|\psi|^2$  of a cat state (top) and of the first three eigenstates of a harmonic oscillator (bottom) are plotted in in the $x-$ basis \textbf{(a)}.
The colored states have two maxima, and correspond approximatively to the gedanken experiment of Fig.~\ref{fig:time_dilation}.
The spread in position of such a state can be on the order of the atomic nucleus.
Considering the nuclei of such a membrane, schematically depicted in \textbf{(b)}, the first eigenstate of the membrane then resembles a quantum superposition where the nuclei are simultaneously displaced up and down \textbf{(c)}.
}
\label{fig:superposition}
\end{figure}
Although we previously gave the example of a statically displaced ball, we focus on realizing this experiment here with a mechanical oscillator.
This mechanical oscillator is given a total mass $m$, and an angular frequency $\omega_m$.
With the ball, we assumed that these displacements realized eigenstates of a hypothetical general-relativistic Hamiltonian, which allowed us to estimate a gravitational time-scale.
In reality, we have no idea what such a Hamiltonian or its eigenstates are, all we can say is that we are looking to prepare a pure state with a large spread in the position basis.
We will consider two such states: cat states and Fock states.
Cat states are the closest analogue to the ball displaced in two directions.
A cat state is a superposition of coherent states oscillating out of phase such that when one part of the superposition is moving in one direction, the other part is moving in the other.
At maximum displacement, the center of mass is localized in two locations separated by $\Delta x = 2\sqrt{n}x_\text{zpf}$
where $n$ is the size of the cat given in number of phonons, and the zero-point fluctuations in position are given by 
\begin{equation}
	x_\text{zpf} = \sqrt{\frac{\hbar}{2\omega_m m}}\ .
\end{equation}
At $n=0$, the quantum fluctuations in position of the ground state wave packet $x_\text{zpf}$, give a lower bound to $\Delta x$.
We can thus compactly write $\Delta x = \sqrt{4n+1}x_\text{zpf}$.
If we assume that decoherence is limited by interaction with a thermal bath at temperature $T$, the time-scale on which a cat state will decohere to a mixture of two Gaussian states is given by~\cite{asjad2014reservoir} 
\begin{equation}
	t_\text{coh} = \frac{1}{2(2n_\text{th}+1)n\gamma_m}
\end{equation}
where $\gamma_m$ is the intrinsic damping rate of the mechanical oscillator
and $n_\text{th} = 1/[\exp (\hbar\omega_m/k_BT)-1]$ the average number of phonons in it after thermalization with the environment.
The underlying assumption here is that there are no major sources of dephasing beyond the interaction with the thermal bath characterized by $\gamma_m$ and $n_\text{th}$.
In the limit $\hbar\omega_m\ll k_BT$, typical for the mechanical oscillators we will be considering, we have
\begin{equation}
	t_\text{coh} = \frac{Q_m}{4n+1}\frac{\hbar}{k_BT}\simeq \frac{Q_m}{4n+1} 8\times 10^{-10}\text{s}
\end{equation}
where the approximate equality holds for $T=10$ mK
and where $Q_m = \omega_m/\gamma_m$ is the quality factor of the mechanical oscillator.
For a Fock state, similar results can be derived.
The center of mass is not localized in two locations, but rather spread out with an uncertainty $\Delta x = x_\text{zpf}\sqrt{2n+1}$.
The rate at which a Fock state decoheres is
\begin{equation}
	t_\text{coh} = \frac{1}{(n_\text{th}+1)n\gamma_m+n_\text{th}(n+1)\gamma_m}\ .
	\label{eq:tcoh_fock_state}
\end{equation}
where the two parts of the denominator correspond to the loss of a phonon (with rate $(n_\text{th}+1)n\gamma_m$) and the gain of a phonon (with rate $n_\text{th}(n+1)\gamma_m$).
In the limit $\hbar\omega_m\ll k_BT$ we have
\begin{equation}
	t_\text{coh} = \frac{Q_m}{(2n+1)}\frac{\hbar}{k_BT}\simeq \frac{Q_m}{(2n+1)} 8\times 10^{-10}\text{s}
\end{equation}
where the approximate equality holds for $T=10$ mK.
These decoherence rates are to be compared with the gravitational timescale $t_G$\footnote{The Karolyhazy time $T_K$ gives time-scales 10 to 20 orders of magnitude larger than the Penrose time-scale used here.}.
When calculating this time for mechanical oscillators, it is unclear what form the mass distribution should take.
The unambiguous situation would be to have the displacement $\Delta x$ be larger than the size of the mechanical oscillator in the direction of the  displacement. 
However these oscillators are typically at least tens of nanometers thick, so this unambiguous situation will be very far from reality.
As an alternative, it is commonly argued~\cite{kleckner2008creating} that since the mass is mostly located in the nuclei of the atoms, it is sufficient for all the nuclei to be displaced by at least twice the nucleus radius
\begin{equation}
	a = A^{\frac{1}{3}}R_0
\end{equation}
where A is the mass number of the atom (total number of protons and neutrons in the atom), and $R_0 = 0.9$ fm~\cite{mohr2000codata}.
Having $\Delta x\gg2a$ translates to $\sqrt{2n+1}x_\text{zpf}\gg2a$ and $\sqrt{4n+1}x_\text{zpf}\gg2a$ for Fock and cat-states respectively.
By writing the zero-point fluctuations explicitely, these two conditions write
\begin{equation}
	\frac{(2,4)n+1}{m}\gg f_m\frac{8a^2}{h}\simeq f_m\ 2\times 10^5 \text{ kg}^{-1}\text{ Hz}^{-1}\ ,
\end{equation}
here (2,4) are multiplicative factors for Fock and cat states respectively.
The approximate equality is calculated for the mass number $A$ of common isotopes of silicon ($A=28$) and aluminum ($A=27$), typical materials for the fabrication of micro-mechanical oscillators.
If one nucleus is in such a superposition, the gravitational time scale associated with that quantum state following Eq.~\ref{eq:t_G_penrose_full} is~\cite{kleckner2008creating}
\begin{equation}
	t_G = \frac{\hbar}{\Delta E},\ \Delta E = 2Gm_a^2\left(\frac{6}{5a}-\frac{1}{\Delta x}\right)
\end{equation}
valid for $\Delta x \ge 2a$.
Here $m_a = A m_u$ refers to the mass of the nucleus with $m_u\simeq1.6\times 10^{-27}$.
Multiplying this energy difference by the number of nuclei in the mechanical oscillator and assuming $\Delta x \gg a$ 
we obtain
\begin{equation}
	t_G = \frac{1}{m}\frac{\hbar R_0}{2GA^{\frac{2}{3}}m_u}\simeq \frac{1}{m}\ 7\times 10^{-14}\text{kg.s}
	\label{eq:gravitational_timescale}
\end{equation}
for aluminum or silicon.
Note that if $\Delta x \gtrsim 2a$, the difference in the final formula will be of order unity.
Grouping the condition of $\Delta x$ being much larger than the nucleus with $t_G\ll t_\text{coh}$ we obtain
\begin{equation}
	f_m\frac{8a^2}{h} \ll \frac{(2,4)n+1}{m}\ll {Q_m}\frac{2A^{\frac{2}{3}}Gm_u}{k_BT R_0}\ .
	\label{eq:ravity_condition_ana}
\end{equation}
where (2,4) are multiplicative factors for Fock and cat states respectively. For aluminum or silicon oscillators at $10$ mK this relation translates to
\begin{equation}
	 f_m\ 2\times 10^5 \text{s.kg}^{-1}\ll \frac{(2,4)n+1}{m}\ll  Q_m 10^{4}\text{ kg}^{-1}\ .
	 \label{eq:gravity_condition_numerical}
\end{equation}

\section{Candidate micro-mechanical systems to control with superconducting circuits}
\begin{figure}[]
\centering
\includegraphics[width=0.8\textwidth]{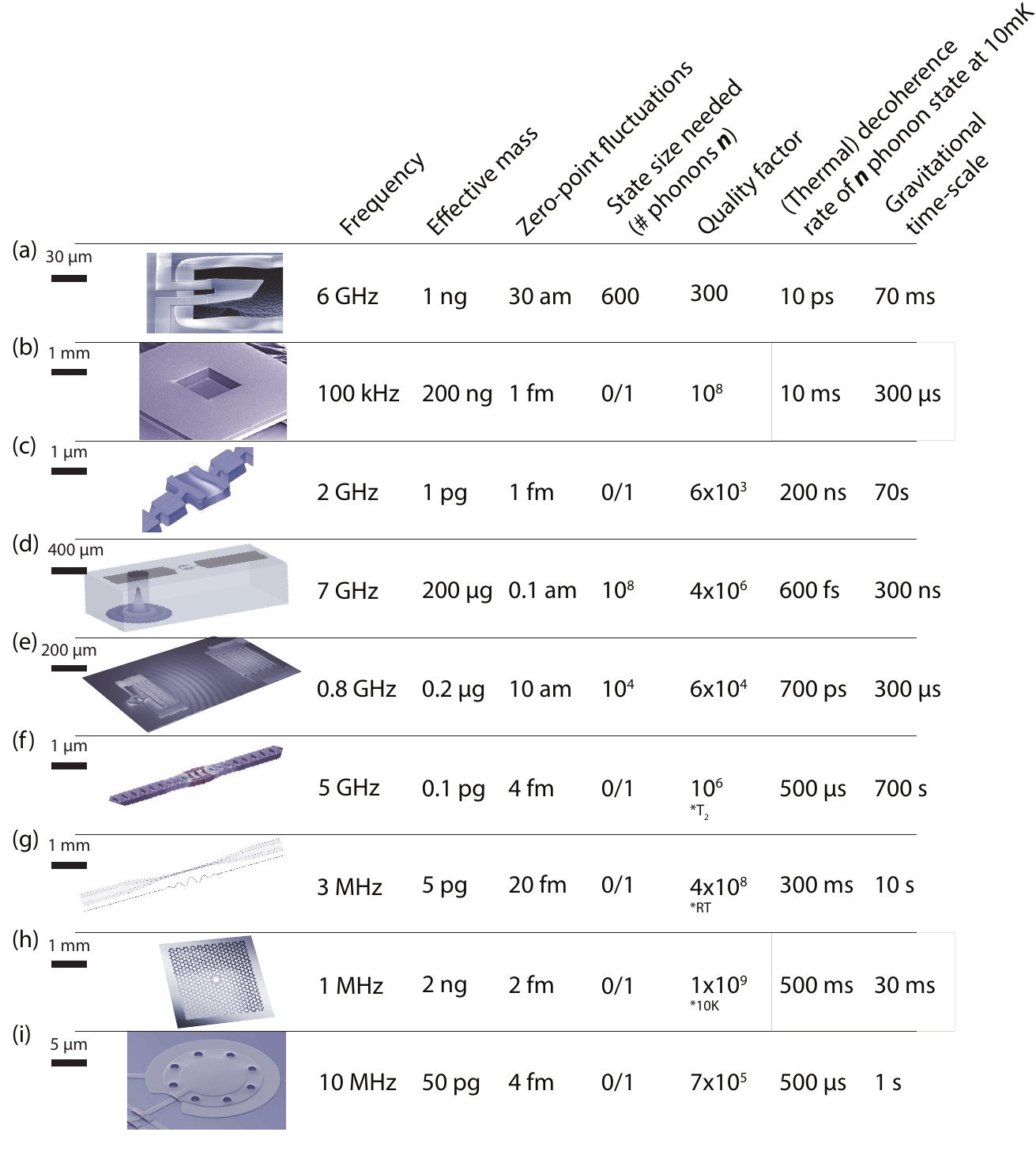}
\caption{
\textbf{Candidate mechanical oscillators for quantum gravity experiments with superconducting qubits.}
We provide an overview of the different types of mechanical oscillators which have successfully been coupled to superconducting qubits. 
The aim is to evaluate their relevance in the measurement of the quantum gravitational effects described previously.
From left to right, we show an illustration of the mechanical oscillator, its resonance frequency $\omega_m/2\pi$ and its effective mass $m_\text{eff}$.
From these two quantities we compute the zero-point fluctuations of motion $x_\text{zpf} = \sqrt{\hbar/2\omega_m m_\text{eff}}$ and the size of the Fock state, in number of phonons $n$, to implement the gedanken experiment schematically shown in Fig.~\ref{fig:superposition}. 
The state size $n$ should be large enough to lead to a quantum spread in position larger than the size of the atomic nucleus.
We then quote the highest measured quality factor $Q$, and the resulting decoherence time of the $n$-th Fock state given by Eq.~\ref{eq:tcoh_fock_state} (very close to the decoherence rate of a size $n$ cat-state).
Note that the higher $n$ is required, the shorter the decoherence time.
Finally, we display the gravitational decoherence time, following Eq.~\ref{eq:gravitational_timescale}.
We find that the two most relevant systems are membranes in the 0.1-1 MHz regime.
Special considerations:
(a,c) Mechanically compliant volume and effective mass extracted from the thickness of the deposited materials, and an estimate of the cantilever/defect area of motion. 
(b) Numbers quoted from Ref.~\cite{yuan2015silicon}.
(d) Mass obtained from the volume under the piezoelectric disk. Quality factor quoted from  Ref.~\cite{chu2018creation}.
(e) Effective mass extracted from the zero-point fluctuations and frequency quoted in Ref.~\cite{noguchi2017qubit} . Quality factor quoted from Ref.~\cite{shao2019phononic}.
(f) Numbers quoted from Ref.~\cite{maccabe2019phononic}. For this system, the decoherence rate was measured, in addition to the decay rate. The former is the relevant number quoted here.
(g) System measured at room temperature.
(h) Numbers quoted from Ref.~\cite{chen2020entanglement}, measured at 10 Kelvin.
(a-i) Illustrations and quoted numbers (unless specified otherwise above) credited to (top to bottom) Refs.~\cite{o2008microwave},\cite{yuan2015large},\cite{arrangoiz2019resolving},\cite{chu2017quantum}, Philip Krantz - Krantz NanoArt~\cite{delsing20192019},\cite{riedinger2016non},\cite{ghadimi2018elastic},\cite{schliesser_press_release},\cite{clark2017sideband}.
}
\label{fig:candidate_oscillators}
\end{figure}
In this section we consider the suitability of different micro-mechanical oscillators to accomplish a sufficiently large and long-lived superposition of space-time.
We will solely focus on mechanical oscillators which have successfully been coupled to superconducting circuits: acoustic oscillators, beams and membranes or drums.
Acoustic resonators, including phononic crystal defect~\cite{arrangoiz2019resolving}, surface~\cite{satzinger2018quantum} and bulk~\cite{o2010quantum,chu2018creation} acoustic waves or optomechanical crystals~\cite{forsch2019microwave}, are typically coupled to superconducting circuits through piezoelectricity.
The above-mentioned mechanical oscillators have frequencies in the GHz range, such that there coherence times are no longer limited by thermal effects when cooled down $10$ mK in a dilution refrigerator.
Their coherence time is then given by
\begin{equation}
	t_\text{coh} = \frac{1}{(1,2)n\gamma_m}\ ,
\end{equation}
where the factor (1,2) correspond to Fock and cat states of size $n$ phonons.
This modifies the previously derived conditions of Eq.~\ref{eq:gravity_condition_numerical} following
\begin{equation}
	 f_m\ 2\times 10^5 \text{Hz}^{-1}\text{kg}^{-1}\ll \frac{(2,4)n+1}{m}\ll  \frac{Q_m}{f_m} 2\times 10^{12}\text{ Hz}\text{ kg}^{-1}\ .
\end{equation}
By plugging in $f_m = 1$ GHz, we find that the quality factor should be $Q_m\gg 10^{11}$ which is currently far from reality.

Another class of mechanical oscillators are beams, which can be coupled capacitively~\cite{regal2008measuring,barzanjeh2019stationary} or through superconducting quantum interference~\cite{rodrigues2019coupling} to superconducting circuits.
The state of the art are beams made from non-uniform phononic cystals~\cite{ghadimi2018elastic}, which are representative of typical beams in terms of size, but feature extremely high quality factors.
These are characterized by $m = 5$ pg, $f_m = 3$ MHz, $Q = 4\times 10^8$ (at room temperature).
Plugging these numbers into Eq.~\ref{eq:gravity_condition_numerical} reveals that the ratio of mass to frequency is sufficient for $\Delta x \gg 2a$ even in the ground-state.
However, the mass is too low to have a sufficiently short gravitational time-scale, and $t_G\ll t_\text{coh}$ would require $Q_m \gg 2\times 10^{10}$.
The last type of oscillator that can be coupled to superconducting circuits are membranes or drums.
High-stress SiNx membranes have cryogenic quality factors up to $10^8$~\cite{yuan2015silicon}.
The fundamental mode of such membranes resonates at $f_m \sim 100$ kHz, and they have a total mass $m=200$~ng~\cite{yuan2015large}
The gravitational time-scale is $t_G\sim 0.3$ ms, much shorter than the thermal decoherence time $t_\text{coh}\sim 10$ ms of small amplitude Fock or cat states.
And the zero-point fluctuations $x_\text{zpf}\sim $ fm are on the order of the size of a Si nucleus $3.7$ fm.
Manipulated into small Fock or cat states, these membranes make good candidates for measuring quantum gravitational effects.
Higher frequencies typically translate to lower quality factors, but through the use of phononic shields~\cite{yu2014phononic}, soft clamping and dissipation dilution~\cite{tsaturyan2017ultracoherent}, this can be mitigated.
An order of magnitude higher frequency resonators,  $f_m = 1$ MHz, with an effective mass $m=2$~ng were demonstrated to have $10^9$ quality factors at 10 Kelvin~\cite{chen2020entanglement}. 
Their gravitational time-scale is $t_G\sim 30$ ms, much shorter than the thermal decoherence time $t_\text{coh}\sim 0.5$ s of low amplitude Fock or cat states.
And the zero-point fluctuations $x_\text{zpf}\sim 2$ fm are on the order of the size of a Si nucleus $3.7$ fm.
Again, manipulated into small Fock or cat-states, these membranes make good candidates for measuring quantum gravitational effects.
Higher frequency membranes, that we will call drums, can be made from by suspending thin films of superconductor, usually Aluminum~\cite{Teufel2011}.
These typically resonate at $f_m\sim 10$ MHz, with a mass $m\sim 50 pg$~\cite{Teufel2011} and state of the art drums have quality factors $Q \sim 7\times 10^5$~\cite{clark2017sideband}.
Whilst their gravitational time-scale $t_G\sim 1$ s, remains inaccessible considering the typical their thermal decoherence time $t_\text{coh}\sim 0.5$ ms, these systems may form a good starting point to test coupling methods to Josephson junction based circuit.
Indeed, they have already been successfully integrated in Josephson junction circuits~\cite{lecocq2015resolving}, they have a frequency closer to the typical operating frequency of Josephson circuits and feature less thermal phonons than lower frequency membranes.
In the later discussions, we will consider even smaller drums with frequencies $f_m\sim 100$ MHz and a factor 100 reduction in weight $m\sim 0.5 pg$, in order to explore how circuits can couple to membranes spanning four orders of magnitude in frequency.

Coupling these types of membranes to superconducting circuits is typically done via optomechanical coupling~\cite{teufel2011circuit,yuan2015large}, which relies on large microwave pump signals injecting millions of photons into the circuit.
Closely connecting this optomechanical system to Josephson junction based quantum circuits, which operate with at amplitudes of a few photons is challenging.
One way to address this is to use optomechanics as a means to convert quantum states emitted from an isolated source (made out of a Josephson qubit) to the mechanical system.
This is a very promising approach to creating quantum states in mechanics which has been realized in Ref.~\cite{reed2017faithful} and definitely deserves to be pursued further.
Another option, which will be explored in the following chapter, is to apply a voltage bias to the moving membrane, such that its oscillations create currents in a coupled circuit, enabling the circuit to measure and manipulate the quantum state of the membrane.

\FloatBarrier
\chapter{Prospects for phonon number resolution of voltage-biased membranes using weakly-anharmonic superconducting circuits}
\label{chapter_phonon_res}

\begin{abstract}
In the previous chapter, we have established the fundamental importance of studying quantum superpositions of mechanical oscillators resonating in the MHz frequency range.
In this chapter, we explore the requirements for realizing this experimentally.
We focus on achieving phonon-resolution in the spectroscopy of voltage-biased membranes coupled to weakly-anharmonic superconducting circuits.
We start by presenting the coupling concept, and deriving the system Hamiltonian.
This is done through the use of an equivalent electrical circuit emulating the voltage biased membrane.
We then consider coupling membranes to GHz circuits, and expose the harsh conditions that are required to achieve phonon-resolution in this case.
We then consider membranes coupled to MHz circuits.
These are either single-mode circuits or two-mode circuits featuring a MHz and a GHz mode (a RFcQED system).
Regardless of the scheme, we demonstrate that strong coupling to the MHz mode is required, which currently constitutes the main roadblock to bringing this idea to fruition.
\end{abstract}

\newpage

\noindent

In this chapter we study voltage biased drums, or membranes, coupled to weakly anharmonic superconducting circuits.
As explored in the previous chapter, there is fundamental interest in creating quantum states of such macroscopic objects, to study the interplay between quantum mechanics and general relativity.
Here, we will explore the conditions to spectroscopically resolve phonon-states of the drum.
With phonon-number resolution comes the ability to measure and control the quantum state of the mechanical motion, following techniques presented in chapter~\ref{chapter_RFcQED}.
With control over the quantum state of the drum, we will be able to prepare the superpositions of space time covered in the previous chapter.

Such electro-mechanical systems have already been implemented in multiple experiments, however the coupling was always too small to gain full control over the quantum state of mechanical oscillators resonating at less than a 100 MHz~\cite{Viennot2018,lahaye2009nanomechanical,pirkkalainen_hybrid_2013}.
This limitation stems in the large frequency span between the mechanical element and the gigahertz resonance frequency of the electrical mode.
The details of these limitations are explored in the first section of this chapter.
In a second part, we explore resonant coupling of mechanics and transmon-type circuits.

\section{Coupling mechanism and coupling rate}
\subsection{Intuitive picture of the coupling mechanism}
The coupling mechanism between the mechanical motion and the circuit is illustrated in Fig.~\ref{fig:drum_and_transmon}(a).
The drum plays a role of a capacitor, where one of its plates is a metallic membrane free to oscillate.
By voltage biasing the drum with a voltage $V_0$, a charge $q$ will accumulate on the plates following 
\begin{equation}
	q=C_d(x)V_0,
\end{equation}
where $C_d$ is the drum capacitance, a function of the displacement of the drumhead $x$.
If the drum-head moves $\dot x \ne 0$, a current will start flowing in the leads supplying the voltage
\begin{equation}
	i = \dot q = \dot x \frac{\partial C_d}{\partial x} V_0\ .
\end{equation}
If a Josephson junction is placed in series with the drum and voltage supply, then the movement of the drum will be dragging oscillating charges through the junction, effectively changing the junction inductance.
As with a resonator coupled to a transmon, the junction inductance has then distinct values for different Fock states in the drum.
Probing the frequency of the transmon composed of the junction and the drum capacitance could reveal different peaks corresponding to different Fock states of the drum, similarly to the measurements shown in Fig~\ref{fig:LFT_1}, which we call phonon number resolution.
\subsection{Coupling rate and system Hamiltonian}

\begin{figure}[]
\centering
\includegraphics[width=0.8\textwidth]{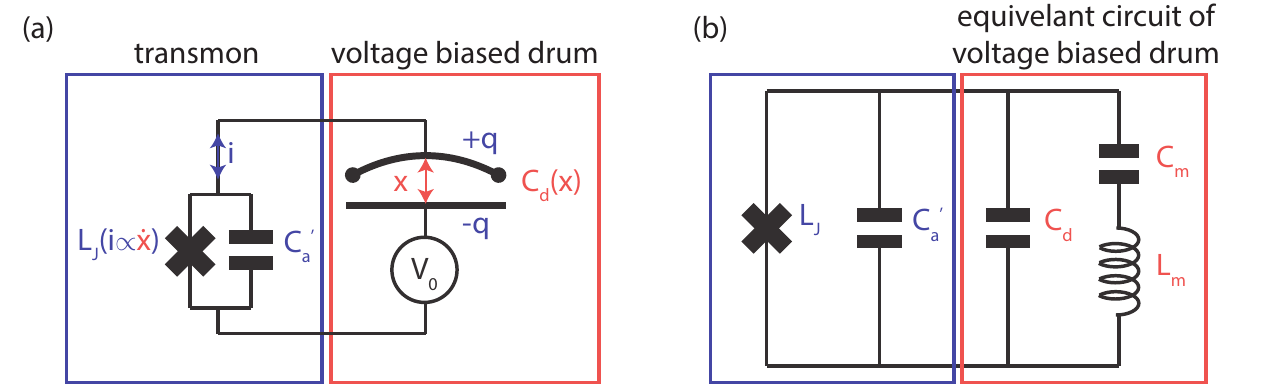}
\caption{
\textbf{Equivalent circuit of a voltage bias drum coupled to a transmon.}
\textbf{(a)} A transmon with Josephson inductance $L_J$ and capacitance $C_a'$ is coupled to a drum with capacitance $C_d$ biased with a voltage $V_0$.
Movement of the drum will induce a current through the junction, which in turn translates in a shift of the transmon frequency, realizing a cross-Kerr coupling between the two systems.
\textbf{(b)} In a circuit equivalent of this system, the voltage biased drum is replaced with a capacitor $C_d$ in parallel with a series $L_m-C_m$ resonator representing the mechanical degree of freedom.
The electrical equivalent allows us to readily apply circuit quantization to describe the system quantum mechanically.
}
\label{fig:drum_and_transmon}
\end{figure}

In this section, we derive the coupling rate $g$ between a voltage biased drum and a transmon, and the Hamiltonian of the combined electro-mechanical system.
As derived in Sec.~\ref{sec:equiv_circuit_drum}, the voltage biased drum has a circuit equivalent shown in Fig.~\ref{fig:drum_and_transmon}(b).
Here $C_d$ corresponds to the capacitance formed by the drum (including the static displacement of the drumhead induced by the voltage).
The series composition of $L_m$ and $C_m$ represent the mechanical degree of freedom, and this part of the circuit resonates at the mechanical frequency $1/ \sqrt{L_mC_m} = \omega^{V_0}_m = \sqrt{k_\text{eff}(V_0)/m}$, where $m$ is the mass of the mechanical oscillator.
The effective spring constant of the drum $k_\text{eff}$ varies with the applied bias voltage $V_0$ following
\begin{equation}
	k_\text{eff}(V_0) = k-\frac{V_0^2C_d}{D(V_0)}\ ,
	\label{eq:k_eff}
\end{equation}
where $k$ is the spring constant of the unbiased drum and $D(V_0)$ corresponds to the distance separating the two capacitive plates of the drum incorporating the static displacement imposed by the bias.
This effect is called electrostatic spring softening.
The condition $k_\text{eff}(V_0)>0$ dictates the maximum applicable voltage.
Increasing the bias voltage will reduce the impedance of the equivalent circuit 
\begin{equation}
	{Z_m = \sqrt{\frac{L_m}{C_m}} = \frac{D^2}{V_0^2C_d^2}\sqrt{k_\text{eff}m}}
\end{equation}
which is infinite in the absence of an applied bias.
We consider the voltage biased drum coupled to a transmon composed of a junction $L_J$ and a capacitance $C_a'$, such that the total capacitance in parallel of the junction is $C_a = C_a'+C_d$ (see Fig.~\ref{fig:drum_and_transmon}(b)).
Since this circuit is identical to that of Fig.~\ref{fig:chapter-2_circuit_notations}(b) with a single, infinite impedance mode, the Hamiltonian is readily given by
\begin{equation}
\begin{split}
  \hat{H} &= \hbar\bar\omega_a \hat{a}^\dagger\hat{a}-\frac{E_C}{12} \left(\hat{a}+\hat{a}^\dagger\right)^4\\
  &+\hbar\tilde\omega_m \hat{c}^\dagger\hat{c}-\hbar g \left(\hat{a}-\hat{a}^\dagger\right)\left(\hat{c}-\hat{c}^\dagger\right)\ .
\end{split}
\end{equation}
Here $\hat{c}$ is the annihilation operator for phonons in the drum, $\omega_m$ its frequency and $g$ the coupling strength.
The parameters of the Hamiltonian are given by

\begin{equation}
	\begin{split}
		\bar\omega_a &= 1/\sqrt{L_JC_a}\\
		E_C &= e^2/2C_a\\
		\tilde\omega_m &= \frac{1}{\sqrt{L_m\tilde C_m}} \\ 
		\tilde C_m &= C_mC_a/(C_m+C_a)\\
		g &= \frac{\tilde\omega_m\bar\omega_a}{2}\sqrt{\frac{\tilde C_m }{C_a}}\\
		\label{eq:drum_modes_hamiltonian parameters}
	\end{split}
\end{equation}
%
%
Note that $\omega_a = \bar\omega_a-E_C/\hbar$ is the frequency of the $\ket{g}\leftrightarrow\ket{e}$ transition of the transmon.
Using Eq.~(\ref{eq:k_eff}), we can rewrite the limiting factor in the coupling as
\begin{equation}
	\frac{\tilde C_m }{C_a} 
	= \frac{1-\left(\omega_m^{V_0}/\omega_m^{V_0=0}\right)^2}{1+\frac{C_a'}{C_d}\left(\omega_m^{V_0}/\omega_m^{V_0=0}\right)^2}
\end{equation}

\section{Experimental schemes and their requirements for phonon-resolution of a MHz drum}
\begin{figure}[]
\centering
\includegraphics[width=0.8\textwidth]{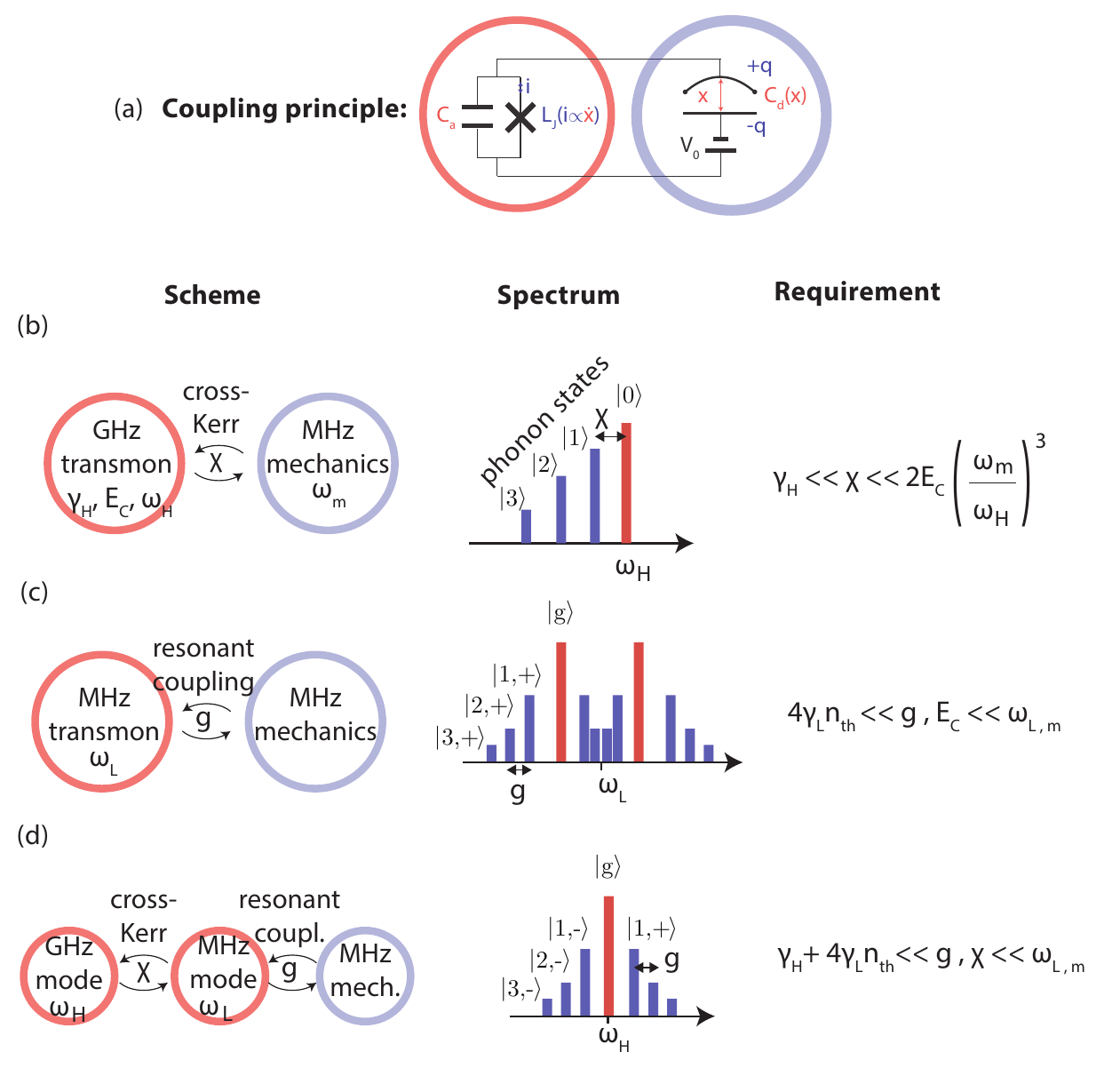}
\caption{
\textbf{Overview of coupling schemes and their requirements.}
\textbf{(a)}
A voltage biased membrane is connected to a transmon such that motion translates to an oscillating current through the junction or an oscillating voltage on the capacitor.
We distinguish three cases.
\textbf{(b)} MHz mechanics coupled to a GHz transmon.
In this case, phonon-resolution is achieved when the cross-Kerr interaction $\chi$ is larger than the transmon line-width $\gamma_H$. 
As shown in the main text, $\chi$ will bound by $2E_C(\omega_m/\omega_H)^3$ where the transmon has a frequency $\omega_H$ and an anharmonicity $E_C$ and $\omega_m$ the mechanical frequency.
\textbf{(c)} 
Mechanics resonantly coupled to a transmon.
The separation between phonon-dependent spectral peaks is given by the smallest energy scale: the coupling rate $g$ or the transmon anharmonicity $E_C$.
Here we schematically show the case where $g$ is the smallest, leading to a Jaynes-Cummings spectrum.
Phonon-resolution then relies on these energy scales being larger than the spectral line-width, given by $4\gamma_L n_\text{th}$ if the transmon has the dominating decay rate.
Here $\gamma_L$ refers to the low-frequency mode dissipation rate and $n_\text{th}$ to its thermal occupation
\textbf{(d)} 
A higher, GHz mode is added to the previous setup, in the spirit of the RFcQED device.
The difference with RFcQED, is that hybridized electro-mechanical mode rather than a single low-frequency electrical mode is readout through the high-frequency spectrum.
The spectrum one could obtain depends on the hierarchy between the coupling rate $g$, the cross-Kerr interaction $\chi$ and the low-frequency mode anharmonicity. 
As a general condition, phonon-number resolution comes when the high-frequency line-width, broadened by the low-frequency thermal occupation $\gamma_H + 4\gamma_L n_\text{th}$, is lower than the coupling rate $g$ and cross-Kerr coupling.
The spectrum shown here corresponds to the case where the coupling is smaller than the cross-Kerr coupling.
}
\label{fig:requirements}
\end{figure}
\subsection{GHz transmon -- MHz drum}
We first explore the conditions to obtain phonon-number resolution in the spectrum of a GHz transmon coupled to a MHz drum as schematically summarized in Fig.~\ref{fig:requirements}(b).

\subsubsection{Requirement: high-frequency mechanical oscillator or extremely high electrical Q}

If one wishes to use a GHz transmon and a MHz drum, then the condition $\bar\omega_a-\tilde\omega_m\sim\bar\omega_a+\tilde\omega_m$ excludes making the RWA, and the formula for the cross-Kerr interaction of Eq~(\ref{eq:shifts_beyond}) is applicable
\begin{equation}
	\chi_m \simeq 8 E_C g^2\frac{ \tilde\omega_m^2}{\bar\omega_a^4}\ .
\end{equation}
The anharmonicity $E_C$ has an upper limit $\hbar\bar\omega_a/20$ in order to remain in the transmon regime.
So the quantity of interest is the cross-Kerr relative to the transmon frequency at this highest possible anharmonicity
\begin{equation}
	\chi_m/\hbar\bar\omega_a \simeq \frac{8}{20} g^2\frac{ \tilde\omega_m^2}{\bar\omega_a^4}\ 
\end{equation}
which should exceed the quality factor of the transmon.
Using Eq.~(\ref{eq:k_eff}) to express $\tilde\omega_m$ and $g$ as a function of the un-biased and effective spring constants, we get
\begin{equation}
	\chi_m/\hbar\bar\omega_a \simeq \frac{1}{10}\left(\frac{\omega_m^{V_0=0}}{\omega_a}\right)^3\sqrt{\frac{C_d^2(k-k_\text{eff})^2(C_d k+C_a'k_\text{eff})}{(C_a'+C_d)^3k^3}}\ .
\end{equation}
This expression has a maximum (determined using Mathematica)
\begin{equation}
\begin{split}
	\chi_m/\hbar\bar\omega_a &= \frac{1}{10}\left(\frac{\omega_m^{V_0=0}}{\omega_a}\right)^3\left(\frac{C_d}{C_a'+C_d}\right)^{\frac{3}{2}}<\frac{1}{10}\left(\frac{\omega_m^{V_0=0}}{\omega_a}\right)^3\\
	&\text{ at }k_\text{eff}=0 \text{ if }C_a'<2C_d
\end{split}
\end{equation}
and 
\begin{equation}
\begin{split}
	\chi_m/\hbar\bar\omega_a &= \frac{1}{10}\left(\frac{\omega_m^{V_0=0}}{\omega_a}\right)^3\frac{C_d}{15\sqrt{3}C_a'}<\frac{1}{10}\left(\frac{\omega_m^{V_0=0}}{\omega_a}\right)^3\\
	&\text{ at }k_\text{eff}=k\frac{C_a'-2C_d}{3C_a'} \text{ if }C_a'>2C_d
\end{split}
\end{equation}
So the condition for phonon-number resolution $Q_a\gg\hbar\bar\omega_a/\chi$ can be written
\begin{equation}
	10\left(\frac{\omega_a}{\omega_m^{V_0=0}}\right)^3<\frac{\hbar\bar\omega_a}{\chi}\ll Q_a
	\label{eq:GHz_transmon_requirement}
\end{equation}
where $Q_a$ is the quality factor of the transmon.

For a typical transmon dephasing rate $T_2=30\ \mu$s~\cite{Devoret2013}, the quality factor of a transmon is given by $Q_a=2\pi T_2 \times 6$ GHz $=10^6$.
For such a quality factor, the line-width of the transmon would be at least an order of magnitude smaller than the cross-Kerr on the condition $\omega_m^{V_0=0}>\omega_a/20$ following Eq.~(\ref{eq:GHz_transmon_requirement}).
Even by pushing the transmon frequency to low values $\sim 1$ GHz one would require at least $50$ MHz for the mechanical oscillator frequency.
This is however a theoretical optimum, when the bias voltage is such that $k_\text{eff}=0$, resulting in a 0 frequency mechanical oscillator.
If the bias voltage only changes the mechanical frequency by a small amount $\omega_m^{V_0}=0.9\omega_m^{V_0=0}$ ($k_\text{eff}=0.9^2\ k$), the optimal cross-Kerr is obtained for $C_a'=0$ and is given by 
\begin{equation}
	\frac{\chi_m}{\hbar\bar\omega_a}\simeq \frac{1}{50}\left(\frac{\omega_a}{\omega_m^{V_0=0}}\right)^3
\end{equation}
such that for $Q_a=10^6$ and a 1 GHz transmon, the mechanical frequency should be at least $250$ MHz and constitute the sole capacitance of the transmon.
%
%

\subsubsection{Possible solutions}

There are a few options from this point on, which have been taken by different research groups.
First, one could move away from the transmon, to a more anharmonic system, such that the expression for $\chi$ is favorably modified by the irrelevance of higher levels of the electrical system.
Recent successes have been demonstrated using a Cooper-pair box~\cite{Viennot2018}.
Secondly, one could increase the frequency of the mechanical resonator.
This is by far the most explored option, by making use of acoustic resonators interfaced through piezoelectricity~\cite{o2010quantum,chu2018creation,satzinger2018quantum,arrangoiz2019resolving}.
However this route moves away from the motives described in chapter~\ref{sec:quantum_gravity}, as an increase in frequency is equivalent to a decrease in mass and/or zero-point fluctuations of displacement.

The third option is to reduce the frequency of the transmon and bring it closer to resonance with the drum.
This is the option we will now explore theoretically.

\subsection{MHz transmon -- MHz drum}

\subsubsection{Requirement: strong coupling}

In this section we summarize the required system parameters for phonon-number resolution using resonantly coupled drum and transmon, with two different schemes summarized in Fig.~\ref{fig:requirements}(c,d).
Detailed calculations are provided in Sec.~
\ref{sec:requirements_details_2body} and Sec.~
\ref{sec:requirements_details_3body}, and only highlights are presented here.
We define phonon-number resolution as being able to spectroscopically resolve the $\ket{0},\ket{1}$ manifold of mechanical Fock states.
We will assume that the limiting dissipation rate is that of the electrical degree of freedom.
First, we study phonon-number resolution of a voltage-biased drum with mode frequency $\omega_m$ coupled through a coupling rate $g$ to a low frequency (LF) electrical mode with frequency $\omega_L$ close to that of the mechanical mode and anharmonicity $A_L$
\begin{equation}
\begin{split}
\hat H =&\hbar\omega_L\hat a^\dagger\hat a-\frac{A_L}{2} \hat a^\dagger\hat a^\dagger\hat a\hat a\\
  &+\hbar\omega_m \hat{c}^\dagger\hat{c}-\hbar g \left(\hat{a}-\hat{a}^\dagger\right)\left(\hat{c}-\hat{c}^\dagger\right)\ 
\end{split}
\end{equation}
where $\hat a$ and $\hat c$ correspond to the electrical and mechanical degrees of freedom.
In Sec.~\ref{sec:requirements_details_2body}, we first established that the dispersive regime will always give more stringent conditions than the resonant regime.
We then studied two different regimes of resonant coupling.
First when the anharmonicity dominates $A_L/\hbar\gg g$ secondly when the coupling dominates $g \gg A_L/\hbar$, reaching the condition
\begin{equation}
	(4,8)\gamma_L n_\text{th} \ll g, A_L/\hbar
		\label{eq:main_ineq_2body}
\end{equation}
where the factors (4,8) correspond to the cases $A_L/\hbar\gg g$ and $g \gg A_L/\hbar$ respectively, $\gamma_L$ is the LF mode dissipation, and $n_\text{th}$ its thermal photon number.
In this regime, the hybridization of drum and LF mode gives rise to two anharmonic, electromechanical degrees of freedom.
Quantum control of the mechanics could then be enabled by directly driving specific transitions.
Readout could occur by probing the absorption of the system, or through an optomechanical setup.

Secondly, we studied the coupling of a drum to an RFcQED device.
\begin{equation}
\begin{split}
	\hat H &= \hbar \omega_H \hat a ^\dagger \hat a - A_H\hat a ^\dagger\hat a ^\dagger \hat a \hat a \\
	&+ \hbar \omega_L \hat b ^\dagger \hat b - A_L\hat b ^\dagger\hat b ^\dagger \hat b \hat b + \hbar \omega_m\hat c ^\dagger \hat c\\
	&+ \hbar g (\hat b\hat c ^\dagger +\hat b^\dagger\hat c )- \chi \hat a ^\dagger\hat a \hat b ^\dagger \hat b\\
\end{split}
\end{equation}
where $\hat a$, $\hat b$ and $\hat c$ correspond to the high frequency (HF) electrical mode, the LF mode, and the drum respectively.
We study different relations between $g$, $A_L$, and the cross-Kerr coupling between the two electrical modes $\chi$, to reach the conditions
\begin{equation}
		(15,2,3,1)\times\left(\gamma_H+4\gamma_L n_{\text{th}}\right)\ll\chi/\hbar,g\ll\omega_m,\omega_L\ 
		\label{eq:main_ineq_3body}
\end{equation}
where $\gamma_H$ is the dissipation rate of the HF mode.
The multiplicative factors $(15,2,3,1)$ correspond to the cases ($A_L\gg \hbar g\gg \chi$,$A_L,\chi\gg \hbar g$, $\hbar g\gg\chi\gg A_L$ , $\chi\gg \hbar g\gg A_L$) respectively.

\subsubsection{The challenge of achieving strong coupling.}

Currently the biggest common limitation of the two above schemes is to achieve strong coupling $4\gamma_L n_{\text{th}}\ll g$.
What we will derive here is that the achievable coupling with a transmon circuit is limited by its capacitance, fixed by the condition of low anharmonicity.
Considering a transmon with maximum anharmonicity $E_C/\hbar\omega_a = 1/20$ on near resonance with the mechanical oscillator, we calculate in the table of Fig.~\ref{fig:resonant_couplings} the achievable coupling of the transmon to different mechanical oscillators.
We have assumed that the applied voltage brings the oscillator on the cusp of instability $\omega_m^{V_0}/\omega_m^{V_0=0}=0.9$.
The studied mechanical oscillators are 100 kHz and 1 MHz membranes, which have and area of $A = (1 \text{mm})^2$ and $A = (250 \mu\text{m})^2$ respectively and can be suspended 300 nm above another capacitive plate~\cite{noguchi2016ground}.
Additionally, we consider 10 MHz, $15\mu$m diameter and 100 nm thick drums suspended 50 nm above an electrode~\cite{Teufel2011}, and hypothetical smaller drums with a $100$ MHz frequency achieved by reducing the mass a hundredfold through reducing both the area and thickness by a factor 10.
What we find is that for resonant coupling, we always have have $C_a'\gg C_d$: the capacitance of the transmon (large to stay in the transmon regime) greatly exceeds that of the drum or membrane.
In this regime, the coupling is approximated by 
\begin{equation}
	g = \frac{\sqrt{\tilde\omega_m\bar\omega_a}}{2} \sqrt{\frac{C_d}{C_a'}\left(\left(\frac{\omega_m^{V_0=0}}{\omega_m^{V_0}}\right)^2-1\right)}\ ,
\end{equation}
limited by the small quantity $C_d/C_a'$
Increasing this ratio is equivalent to increasing the impedance of the transmon.
With the circuit of Fig.~\ref{fig:drum_and_transmon} however, where the anharmonicity is only dependent on the value of $C_a'+C_d\simeq C_a'$, this cannot be achieved whilst staying in the transmon regime.
We discuss two routes to strong coupling, one being increasing the quality factor of the electrical part of the circuit, the other finding a different circuit for increased impedance and hence coupling.

\subsubsection{Going to strong coupling via increasing the quality factor}
\begin{figure}[]
\centering
\includegraphics[width=0.8\textwidth]{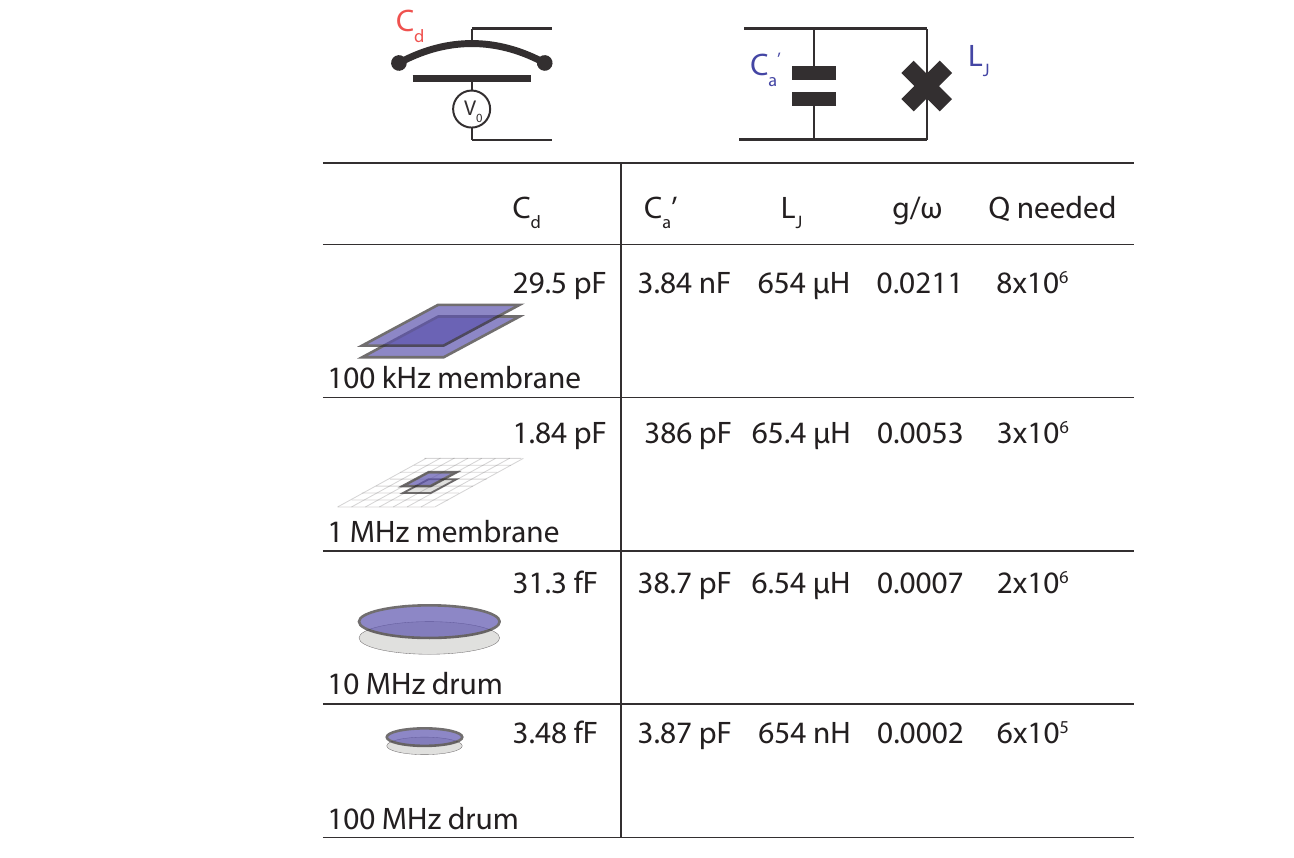}
\caption{
\textbf{Coupling rates in the resonant coupling of a drum and transmon.}
We consider the resonant coupling of a voltage biased drum and a transmon.
The rows of the table corresponds to one of four mechanical oscillator, each separated by at least an order of magnitude in frequency from the other.
In the first column we provide a schematic of the mechanical oscillator and the capacitance it implements considering the smallest experimentally achieved gap between the membrane and its corresponding electrode.
In the second column, we provide the capacitance and inductance which ensures resonance with the drum and maximum coupling, whilst remaining in the transmon regime.
We then provide the coupling relative to the frequency, assuming that the bias voltage applied to the drum only reduces its frequency through electrostatic spring softening by 10 percent.
Finally we provide the quality factor of the transmon required for the line-width to be ten times smaller than the coupling (strong coupling).
}
\label{fig:resonant_couplings}
\end{figure}

If one maintains the transmon circuit, the only way to achieve strong coupling is to increase the quality factor of the transmon.
Note that this concerns the quality factor of a low-frequency transmon on near-resonance with the drum.
The required quality factors to obtain a line-width ten times smaller than the coupling rate is displayed in the table of Fig.~\ref{fig:resonant_couplings}.
These are still far from our measured quality factor of $3\times 10^3$, likely limited by the dielectric losses in the parallel plate capacitor.
Whereas these can be mitigated in GHz transmon circuits by constructing capacitors on a single plane, with electric fields traversing extremely clean crystalline substrates, this approach may be harder for the large capacitances needed here (see Fig.~\ref{fig:resonant_couplings}).
Obtaining both large capacitors in conjunction with low losses is likely to be the next challenge to bring this idea to fruition.


\subsubsection{Increasing the coupling by increasing the transmon impedance.}

The only way to increase the coupling by changing the transmon part of the circuit seems to be to reduce $C_a'$.
The only way to do this whilst maintaining the same frequency is to increase the inductance. 
Increasing the coupling tenfold requires again decreasing the capacitance a hundredfold and increasing the inductance of the circuit a hundredfold.
This corresponds to inductances of 100 $\mu$H to 100mH for frequencies from 100 MHz to 100 kHz. 
If this is done by simply increasing $L_J$, then the anharmonicity will increase such that the transmon regime is exited. 
One could thus add some linear inductance $L$ to the circuit.
There are two ways to do so.
The first is to add inductance in series with the junction.
The anharmonicity is then given by
\begin{equation}
	A = \frac{e^2}{2C_a}\left(\frac{L_J}{L+L_J}\right)^3
\end{equation}
such that the increase in anharmonicity resulting from the decrease of $C_a$ can be compensated by choosing the ratio of inductances $L \simeq 4L_J$.
This circuit is however limited by the inevitable presence of parasitic capacitance $C_\text{para}$ in parallel of the added inductor, which will introduce an additional mode in the circuit with an anharmonicity ( relative to its frequency) given by 
\begin{equation}
	\alpha_\text{para} = \frac{e^2}{2h}\sqrt{\frac{L_J}{C_\text{para}}}\frac{1}{(1+\frac{L_J}{L})^\frac{3}{2}}\simeq \sqrt{\frac{L_J}{C_\text{para}}}\times10^{-5}
\end{equation}
%
%
Shunting this parasitic capacitance to make this mode weakly-anharmonic ($\alpha_\text{para}<1/20$) requires capacitances larger than the $C_a'$ (specifically 100 pF to 100 nF for 100 MHz to 100 kHz resonance frequencies), which cancels any gain one would have otherwise achieved in the electro-mechanical coupling.
A smaller capacitance would result in a highly anharmonic mode who's impact on the physics of the system is out of the scope of this work, but could, if tuned to GHz frequencies, be useful.
Note that this circuit is exactly the one used in RFcQED, however in the RFcQED case, the mode which we now want to couple to the mechanics was designed to be low-impedance (i.e. near-harmonic), which translated to less inductance in the circuit, and made it possible to have a GHz mode with a larger capacitance, in the transmon regime.
Conversely, one could add inductance in parallel to the junction.
In this case, the anharmonicity is given by
\begin{equation}
	A = \frac{e^2}{2C_a}\frac{L}{L+L_J}
\end{equation}
and the frequency by
\begin{equation}
\omega_a=\sqrt{\frac{L+L_J}{C_aLL_J}}\ .
\end{equation}
Maintaining weak anharmonicity then requires a ratio $L/(L+L_J)\simeq L/L_J = 1/100$ which translates to Josephson inductances of 10 mH to 10 H for frequencies from 100 MHz to 100 kHz. 
Constructing such large inductances seem at first glance experimentally challenging.

\subsubsection{Additional challenges}
Here we present a few additional challenges we foresee.
\textbf{Large inductances}
The requirement of strong coupling, together with the low frequencies, translates to very large inductances.
From the table of Fig.~\ref{fig:resonant_couplings}, we read Josephson inductances approaching miliHenries, which is out of the circuit QED comfort zone, and may be challenging to fabricate or operate.
\textbf{Thermally excited current through the junction}
Especially if one maintains the transmon circuit for this scheme, increasing the coupling relies on increasing the impedance and hence the anharmonicity of the transmon.
The higher the anharmonicity, the more current is traversing the junction for each thermal photon the environment is able to excite.
It is unclear to us what occurs if on average this current exceeds the critical current of the junction, but it is likely to make the operation of the device challenging or even impossible.

\FloatBarrier
\section{Supporting calculations}
\subsection{Equivalent circuit of a voltage-biased drum}
\label{sec:equiv_circuit_drum}

We derive the equivalent circuit of a voltage biased drum in the limit of small mechanical motion. 

\begin{figure}[]
\centering
\includegraphics[width=0.8\textwidth]{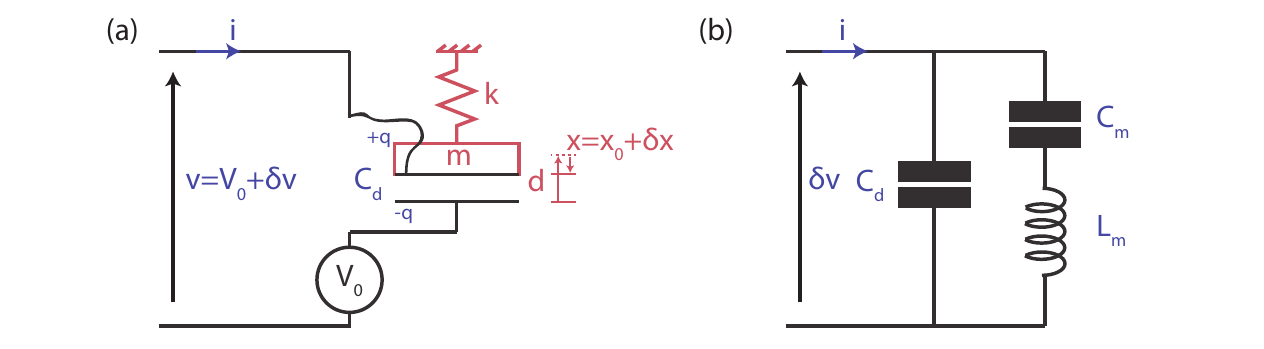}
\caption{
{Equivalent circuit (b) of a voltage biased drum (a).}
}
\label{fig:equivalent_circuit}
\end{figure}

\subsubsection{Mechanical equation of motion}
The Lagrangian of the system shown in Fig.~\ref{fig:equivalent_circuit}a is 
\begin{equation}
	L = +\frac{1}{2}C_d(x)v^2+\frac{1}{2}m\dot x^2 - \frac{1}{2}kx^2\ ,
	\label{eq:lagrangian}
\end{equation}
where the position dependent capacitance is
\begin{equation}
	C_d(x)=\frac{\epsilon_0A}{d-x}
\end{equation}
Equations of motion are given by 
\begin{equation}
	\frac{d}{dt}\left(\frac{\partial L}{\partial\dot q_i}\right) = \frac{\partial L}{\partial q_i}\ 
	\label{eq:lagrange_eqs}
\end{equation}
with ${q_i} = {x,v}$.
Plugging (\ref{eq:lagrangian}) into (\ref{eq:lagrange_eqs}) for $q_i = x$ yields the mechanical equation of motion
\begin{equation}
	\frac{d}{dt}\left[m\dot x\right] = m\ddot x = -kx+\frac{v^2}{2}\frac{dC_d(x)}{dx} = -kx+\frac{v^2}{2}\frac{\epsilon_0A}{(d-x)^2}
\end{equation}
equivalent to
\begin{equation}
	m\ddot x+kx=\frac{v^2}{2}\frac{\epsilon_0A}{(d-x)^2}
	\label{eq:mech_eq_motion}
\end{equation}

\subsubsection{Electrical equation of motion}
From the definition of capacitance, we have
\begin{equation}
	q = C_d(x)v\ .
	\label{eq:qcv}
\end{equation}
By deriving this relation with respect to time, we get 
\begin{equation}
	\dot q = i = v\frac{dC_d(x)}{dt}+C_d(x)\dot v
\end{equation}
where
\begin{equation}
	\frac{dC_d(x(t))}{dt} = \epsilon_0A\frac{d}{dt}\left[\frac{1}{d-x(t)}\right] = \epsilon A \left[\frac{\dot x(t)}{(d-x(t))^2}\right]\ .
\end{equation}
Finally we get the electrical equation of motion
\begin{equation}
	i = \dot v \frac{\epsilon_0A}{d-x}+v\dot x \frac{\epsilon_0A}{(d-x)^2}
	\label{eq:elec_eq_motion}
\end{equation}

\subsubsection{Perturbative analysis}
We now write (\ref{eq:mech_eq_motion}) in a static limit $\frac{d}{dt}=0$ given an initial DC voltage $v = V_0$ leading to a constant displacement $x=x_0$
\begin{equation}
	kx_0=\frac{V_0^2}{2}\frac{\epsilon_0A}{(d-x_0)^2}\ .
	\label{eq:static}
\end{equation}
A DC voltage (positive or negative), will thus increase $x_0$, pulling the two plates of the capacitor closer together.
We rewrite (\ref{eq:mech_eq_motion}) considering small variations of voltage and position with respect to their static values $x\rightarrow x_0+\delta x$, $v\rightarrow V_0+\delta v$
\begin{equation}
	m\ddot {\delta x}+kx_0+k\delta x=\frac{(V_0+\delta v)^2}{2}\frac{\epsilon_0A}{(d-x_0-\delta x)^2}
\end{equation}
Taylor expanding to first order in $\delta x, \delta v$
\begin{equation}
\begin{split}
	&m\ddot {\delta x}+{kx_0}+k\delta x={\frac{V_0^2}{2}\frac{\epsilon_0A}{(d-x_0)^2}}\\
	&+V_0^2\frac{\epsilon_0A}{(d-x_0)^2}\frac{\delta x}{d-x_0}+\delta vV_0\frac{\epsilon_0A}{(d-x_0)^2}
\end{split}
\end{equation}
where the second term on the left hand side cancels with the right hand side term due to (\ref{eq:static}).
Equivalent to
\begin{equation}
	m\ddot {\delta x}+k_\text{eff}(V_0)\delta x=\delta vV_0\frac{\epsilon_0A}{(d-x_0)^2}
	\label{eq:mech_eq_motion_pert}
\end{equation}
where the effective spring constant $k_\text{eff}(V_0)$ is
\begin{equation}
	k_\text{eff}(V_0) = k-\frac{V_0^2\epsilon_0A}{(d-x_0)^3}\ .
\end{equation}
For large enough DC voltage, the spring constant can become negative, making the mechanical oscillator unstable.
We thus necessarily have
\begin{equation}
		V_0 < \sqrt{k\frac{(d-x_0)^{3}}{\epsilon_0A}}
	\label{eq:max_V}
\end{equation}

We now Taylor expand the electrical equation of motion to first order in $\delta x, \delta v$
\begin{equation}
	i = \dot {\delta v} \frac{\epsilon_0A}{d-x_0}+\dot {\delta x} V_0\frac{\epsilon_0A}{(d-x)^2}
	\label{eq:elec_eq_motion_pert}
\end{equation}

\subsubsection{Oscillatory steady-state}

We introduce the phasors $\delta v\rightarrow  Ve^{j\omega t}$, $i\rightarrow  Ie^{j\omega t}$ and $\delta x\rightarrow   Xe^{j\omega t}$ where $ V,  X, I$ are time-independent complex number.
Additionally, we introduce the (voltage dependent) rest-position of the drum $D = d-x_0$ and the corresponding capacitance $C_d = \frac{\epsilon_0A}{d-x_0}$.
Making the above substitutions in Eqs~(\ref{eq:mech_eq_motion_pert}),(\ref{eq:elec_eq_motion_pert}), we get
\begin{equation}
	-\omega^2mX+k_\text{eff}X = V\frac{V_0C_d}{D}
	\label{eq:mech_eq_motion_sinu}
\end{equation}
\begin{equation}
	I = j\omega C_d V + j\omega C_d V_0\frac{X}{D}
	\label{eq:elec_eq_motion_sinu}
\end{equation}
Eq.~(\ref{eq:mech_eq_motion_sinu}) gives the conversion between mechanical motion amplitude and the amplitude of voltage oscillations
\begin{equation}
	{X = V\frac{V_0C_d/D}{k_\text{eff}-\omega^2m}}
\end{equation}
Which plugged into Eq.~(\ref{eq:elec_eq_motion_sinu}) provides the equivalent admittance of the voltage biased drum
\begin{equation}
	Y(\omega) = \frac{I}{V} = j\omega C_d + j\omega\frac{V_0^2C_d^2/D^2}{k_\text{eff}-\omega^2m}
\end{equation}

\subsubsection{Equivalent circuit}
The admittance above can be rewritten 
\begin{equation}
	Y(\omega) = \frac{I}{V} = j\omega C_d + \frac{1}{\frac{1}{j\omega C_m}+j\omega L_m}\ ,
\end{equation}
the admittance of the circuit shown in Fig.~\ref{fig:equivalent_circuit}b,
where the capacitance and inductance representing the mechanical mode are
\begin{equation}
	{C_m = \frac{V_0^2C_d^2}{D^2}\frac{1}{k_\text{eff}}}
\end{equation}

\begin{equation}
	{L_m = \frac{D^2}{V_0^2C_d^2}m}
\end{equation}

The voltaged biased drum resonates at 
\begin{equation}
	{\omega_m^{V_0} = \frac{1}{\sqrt{L_mC_m}} = \sqrt{\frac{k_\text{eff}}{m}}}
\end{equation}
and has a characteristic impedance
\begin{equation}
	{Z_m = \sqrt{\frac{L_m}{C_m}} = \frac{D^2}{V_0^2C_d^2}\sqrt{k_\text{eff}m}}
\end{equation}

\subsection{Derivation of requirements for drum to MHz transmon coupling}
\label{sec:requirements_details_2body}

In this section, we study the requirements for phonon resolution for a voltage biased drum coupled to a transmon, or low frequency (LF) mode.
The Hamiltonian of interest is 
\begin{equation}
\begin{split}
\hat H =&\hbar\omega_L\hat a^\dagger\hat a-\frac{A_L}{2} \hat a^\dagger\hat a^\dagger\hat a\hat a\\
  &+\hbar\omega_m \hat{c}^\dagger\hat{c}-\hbar g \left(\hat{a}-\hat{a}^\dagger\right)\left(\hat{c}-\hat{c}^\dagger\right)\ ,
\end{split}
\end{equation}
where $\hat a$ and $\hat c$ correspond to the electrical and mechanical degrees of freedom, with frequency $\omega_L$ and $\omega_m$ respectively.
The low mode has an anharmonicity $A_L$.
The dissipation rate of the mechanics will be denoted by $\gamma_m$, and the dissipation rate of the electrical mode is denoted by $\gamma_L$. 
We will assume that the electrical dissipation dominates $\gamma_L\gg\gamma_m$.

\subsubsection{Requirements for dispersive coupling}
We first study the dispersive limit $g\ll |\Delta|$, $g\ll |\Delta - A_L|$.
We also assume that the rotating wave approximation applies $|\Delta| = |\omega_m-\omega_L| \ll \Sigma = \omega + \omega_L$, such that the cross-Kerr interaction is given by~\cite{koch_charge-insensitive_2007}
\begin{equation}
	\chi_{m,L} = 2A_L\frac{g^2}{\Delta(\Delta-A_L)}
\end{equation}
and the Kerr induced in the mechanics is
\begin{equation}
	A_m =  \chi^2/4A_L = A_L\frac{g^4}{\Delta^2(\Delta-A_L)^2}
\end{equation}
Here we derive the conditions to have $A_m\gg\hbar \gamma_m$ or $\chi\gg \hbar \gamma_L$
\textbf{Requirements to attain A$_m\gg\gamma_m$.}
Here we explore the conditions required for the mechanical degree of freedom to acquire enough anharmonicity $A_m$ to behave as an artificial atom $A_m\gg \hbar\gamma_m$.
The drum dissipation rate will be broadened through its interaction with the mode, acquiring an effective dissipation rate
\begin{equation}
	\gamma_m^\text{eff} = \gamma_m + \gamma_L\frac{g^2}{\Delta(\Delta-A_L)}\ ,
\end{equation}
which can be derived from Fermi's golden rule~\cite{koch_charge-insensitive_2007}.
The difference in powers of $(g/\Delta)$ between this dissipation rate and $A_m$ arise since the dissipation rate is proportional to the current traversing a resistor squared whilst the anharmonicity is to first order proportional to the fourth power of the current.
If the drum is to behave like an atom, the line-width will further be broadened with the average number of thermal phonons $n_\text{th}$.
We want the anharmonicity to be at least larger than the line-width of the second transition ($\ket{1}\leftrightarrow\ket{2}$), given by (see Eq.~(\ref{eq:sum_of_lorentzians}))
\begin{equation}
	\left(3+8n_\text{th}\right)\left(\gamma_m + \gamma_L\frac{g^2}{\Delta(\Delta-A_L)}\right)\ ,
\end{equation}
Assuming $n_\text{th}\gg1$, we then have two requirements
\begin{equation}
\begin{split}
&8n_\text{th}\gamma_m\ll A_L/\hbar\left(g^4/\Delta^2(\Delta-A_L)^2\right)\\
&8n_\text{th}\gamma_L\ll A_L/\hbar\left(g^2/\Delta(\Delta-A_L)\right)
\end{split}
\end{equation}

\textbf{Requirements to attain $\chi_{m,L}/\hbar\gg\gamma_L$.}
Alternatively, we could try and achieve a large cross-Kerr coupling such that the state of the drum can be read out and controlled through interactions with the LF mode.
One then needs a cross-Kerr shift $\chi_{m,L}/\hbar$ which is larger than the effective line-width of the LF mode $\gamma_L^\text{eff}$, broadened notably by its thermal population $n_\text{th}$ following $\gamma_L^\text{eff} = (1+4n_\text{th})\gamma_L$ (derived in Eq.~(\ref{eq:sum_of_lorentzians})).
The condition $\chi_m/\hbar \gg \gamma_L^\text{eff}$, assuming $n_\text{th}\gg 1$ roughly writes
\begin{equation}
2n_\text{th}\gamma_L\ll A_L/\hbar(g^2/\Delta(\Delta-A_L))
\end{equation}

\textbf{Conclusion}
Both approaches lead to a similar requirement
\begin{equation}
(2,8)n_\text{th}\gamma_L\ll A_L/\hbar\left(g^2/\Delta(\Delta-A_L)\right)
\end{equation}
with only varying constants (2,8).
However the dispersive regime conditions $g\ll |\Delta|$, $g\ll |\Delta - A_L|$ impose an upper bound on the dispersive shift, such that\footnote{To prove this, first impose $g < \epsilon |\Delta|$, $g < \epsilon|\Delta - A_L|$, where $\epsilon$ is a small quantity, which establishes a domain $D$. Then distinguish two cases: $2g/\epsilon <A_L$ and $2g/\epsilon >A_L$. In the former (latter) case $|\chi_{m,L}|$ has 4 (2) local maxima on the domain $D$ which are easy to find graphically. For each maximum, it is easy to prove that $|\chi|<4g\epsilon$} $\chi_{m,L}\ll g$.
We will thus summarize the requirements as follows
\begin{equation}
(2,8)n_\text{th}\gamma_L\lll g\ , 
\end{equation}
meaning that the requirements are even harsher than strong coupling, which is what is required for example with resonant coupling.

\subsubsection{Resonant interaction}
\textbf{Case $A_L\gg \hbar g$.}
\begin{figure}[]
\centering
\includegraphics[width=0.8\textwidth]{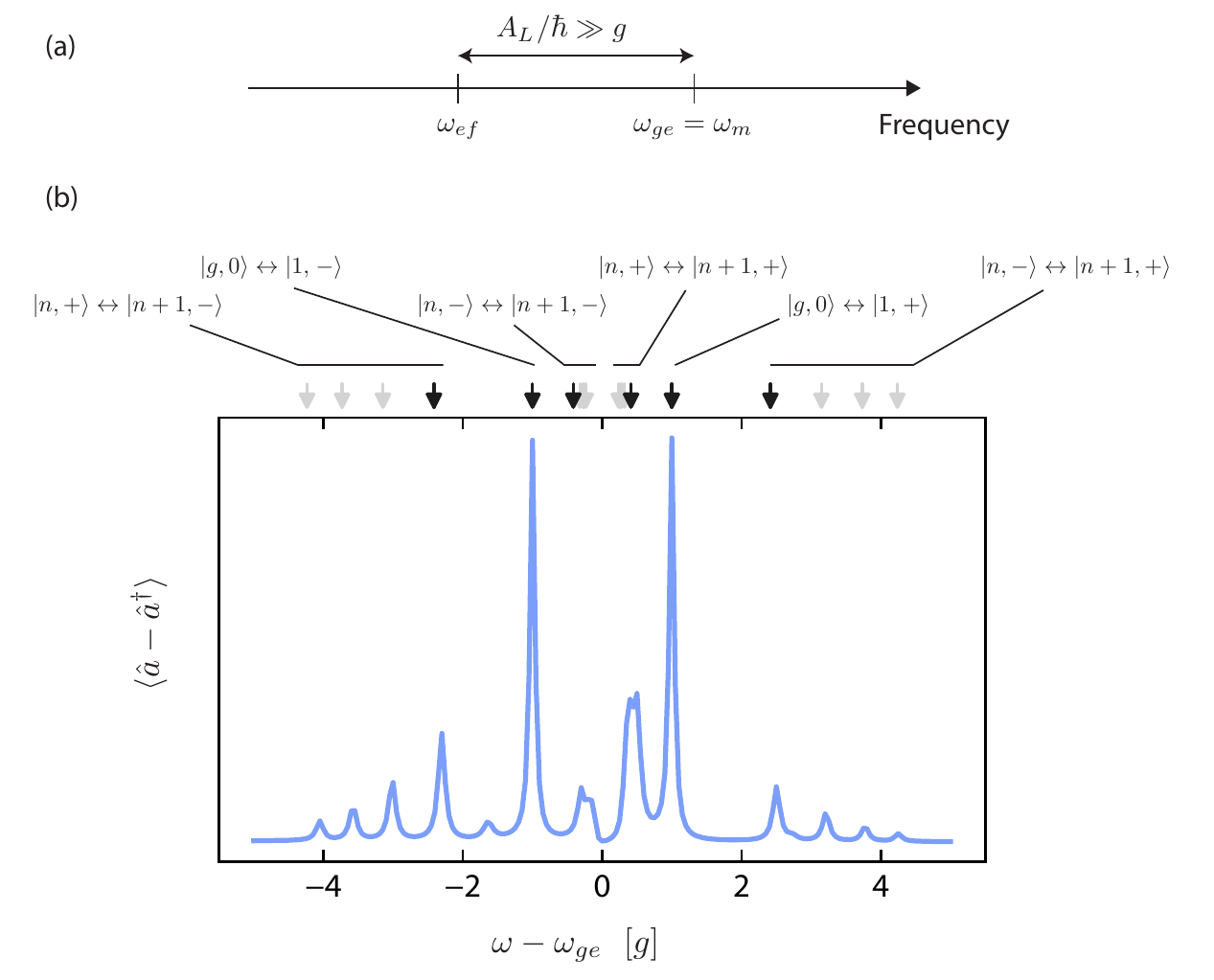}
\caption{
\textbf{Spectrum of an transmon coupled to a drum.} (case $A_L\gg \hbar g$)
\textbf{(a) Frequency landscape.}
We focus on the regime where the LF mode anharmonicity $A_L$ dominates over the drum-to-LF-mode coupling $g$ such that the $\ket{e}\leftrightarrow\ket{f}$ transition of the LF mode is effectively decoupled from the drum.
\textbf{(b) Spectrum.} 
We plot the expectation value $\langle\hat a - \hat a^\dagger\rangle$ whilst weakly driving the system at a frequency $\omega$.
This corresponds to one quadrature in a homodyne measurement of the LF mode.
With the condition $2n_{\text{th}}\gamma_L\ll g/2$, we can resolve the Jaynes-Cummings spectrum of the coupled drum and LF mode.
The exact simulation parameters and numerical methods are provided in Sec.~\ref{sec:chapter-gravity_numerical_methods}
}
\label{fig:g_A}
\end{figure}
In this case, the eigenstates are that of the Jaynes-Cummings Hamiltonian (see Eq.~\ref{eq:chapter-2_jaynes_cummings}).
These are given by
\begin{equation}
	\ket{n,\pm} = \frac{\ket{g,n}\pm\ket{e,n-1}}{\sqrt{2}}
\end{equation}
with eigenenergies\footnote{The groundstate $\ket{g,0}$ has 0 energy.}
\begin{equation}
	\hbar\omega_{n,\pm} =  n\hbar\omega_{m,L}\pm\sqrt{n}\hbar g\ .
\end{equation}
Addressing the $\ket{0,g}\leftrightarrow \ket{1,\pm}$ transition independently of the $\ket{1,\pm}\leftrightarrow \ket{2,\pm}$ is a requirement for phonon-resolution.
These transitions are separated in frequency by $g(2-\sqrt{2})\simeq g/2$.
The line-width of these transitions is the average of the line-width of the drum and LF mode~\cite{Bishop2009a}, given by $(\gamma_m + \gamma_L(1+4 n_\text{th}))/2$ where the LF mode line-width follows from Eqs.~\ref{eq:sum_of_lorentzians}.
Assuming the thermally broadened LF mode line-width dominates over that of the drum and $n_\text{th}\gg1$, the condition for phonon resolution, such as in the spectrum of Fig.~\ref{fig:g_A}, is
\begin{equation}
	4\gamma_L n_\text{th} \ll g \ll A_L/\hbar\ .
\end{equation}

\textbf{Case $\hbar g\gg A_L$.}
\begin{figure}[]
\centering
\includegraphics[width=0.8\textwidth]{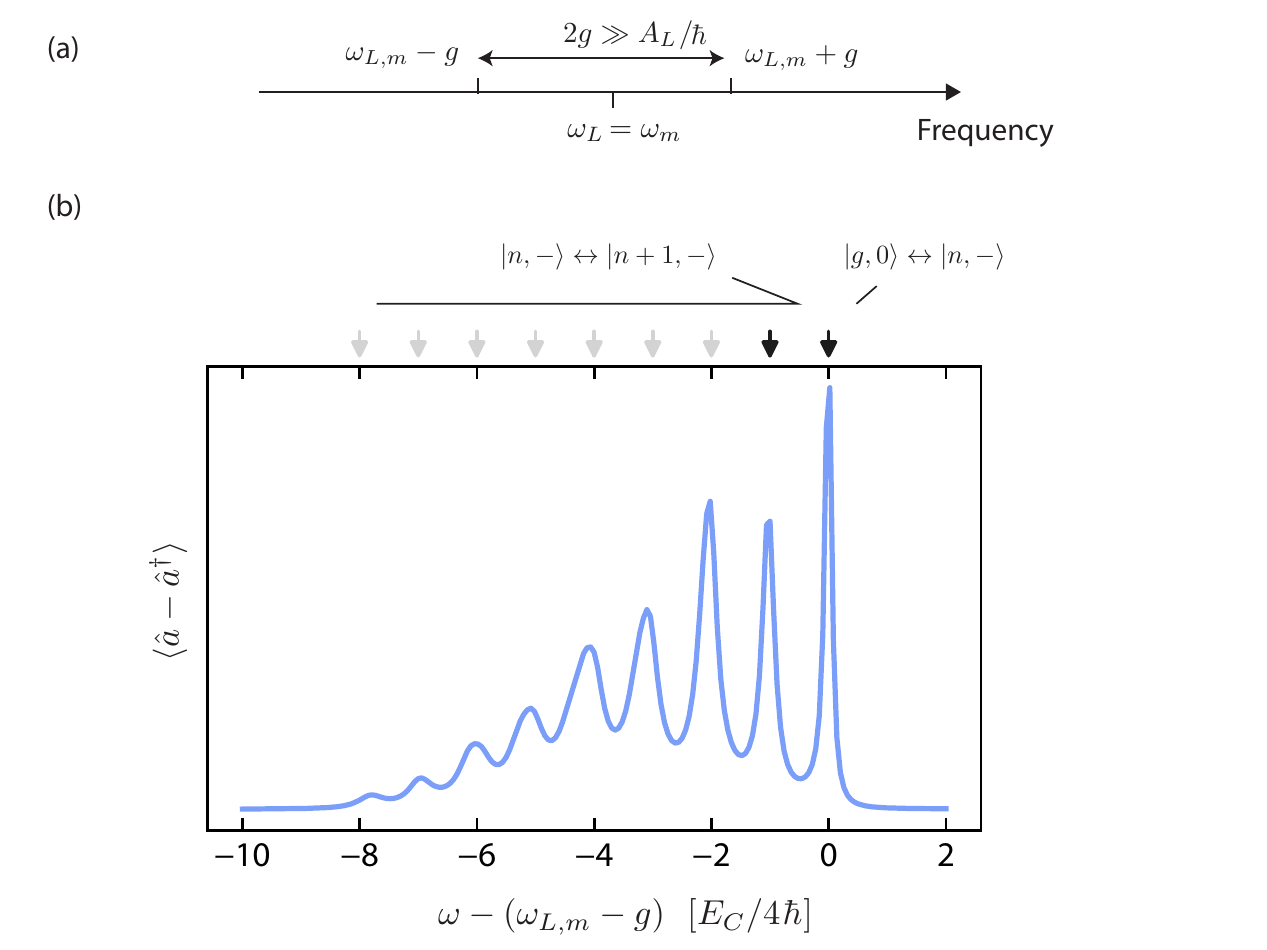}
\caption{
\textbf{Spectrum of an transmon coupled to a drum.}
\textbf{(a) Frequency landscape.} (case $A_L\ll \hbar g$)
We focus on the regime where the drum-to-LF-mode coupling $g$ dominates over the LF mode anharmonicity $A_L$ such that the two systems hybridize into two electromechanical modes separated in frequency by $2g$.
\textbf{(b) Spectrum.} 
We plot the expectation value $\langle\hat a - \hat a^\dagger\rangle$ whilst weakly driving the system at a frequency $\omega$.
This corresponds to one quadrature in a homodyne measurement of the LF mode.
With the condition $4n_{\text{th}}\gamma_L/2\ll A_L/4\hbar$, we can resolve the anharmonic spectrum of one of the electro-mechanical modes.
The exact simulation parameters and numerical methods are provided in Sec.~\ref{sec:chapter-gravity_numerical_methods}
}
\label{fig:A_g}
\end{figure}
As explored in Sec.~\ref{sec:chapter-2_resonant_large_coupling}, in this case the LF mode and drum will hybridize into two electro-mechanical modes with anharmonicities $A_L/4$.
Following Fermi's golden rule, these will have dissipation rates $(\gamma_L+\gamma_m)/2$~\cite{koch_charge-insensitive_2007}.
To resolve at least the transition between the ground and first-excited state of the electro-mechanical modes, the line-width of the first-to-second excited state, dressed by thermal effects following Eq.~(\ref{eq:sum_of_lorentzians}) should be smaller than the mode anharmonicity.
Assuming the electrical mode has a dominating dissipation rate, and that the modes are thermally populated $n_\text{th}\gg1$, this condition writes (including the large coupling condition)
\begin{equation}
	8n_\text{th}\gamma_L\ll A_L/\hbar \ll g\ .
\end{equation}
An example of the obtainable spectrum is given in Fig.~\ref{fig:A_g}.

\subsection{Derivation of requirements for coupling a drum to an RFcQED system}
\label{sec:requirements_details_3body}

We now study a system consisting of a voltage biased drum near-resonantly coupled with a coupling rate $g$ to a low frequency (LF) electrical mode with dissipation rate $\gamma_L$, and anharmonicity $A_L$ coupled through cross-Kerr interaction to a HF frequency mode with dissipation rate $\gamma_H$, and anharmonicity $A_H$.
This is described by the Hamiltonian
\begin{equation}
\begin{split}
	\hat H &= \hbar \omega_H \hat a ^\dagger \hat a - A_H\hat a ^\dagger\hat a ^\dagger \hat a \hat a \\
	&+ \hbar \omega_L \hat b ^\dagger \hat b - A_L\hat b ^\dagger\hat b ^\dagger \hat b \hat b + \hbar \omega_m\hat c ^\dagger \hat c\\
	&+ \hbar g (\hat b\hat c ^\dagger +\hat b^\dagger\hat c )- \chi \hat a ^\dagger\hat a \hat b ^\dagger \hat b\\
\end{split}
\end{equation}
where $\hat a,\hat b,\hat c$ refer to the annihilation operators of the HF, LF and mechanical modes, with angular frequencies $\omega_H, \omega_L, \omega_m $ respectively.
We have neglected many of the terms which arise from the quartic non-linearity of the junction notably under the assumptions $A_H\ll\hbar \omega_H$ and $A_L,\chi\ll\hbar \omega_{L,m}$ as well as the counter-rotating terms of the mechanics-low mode coupling under the assumption $g\ll\omega_{L,m}$.
Note that we necessarily have $\chi = 2\sqrt{A_HA_L}$.

\subsubsection{Case $A_L\gg\hbar  g\gg \chi$}
\begin{figure}[]
\centering
\includegraphics[width=0.8\textwidth]{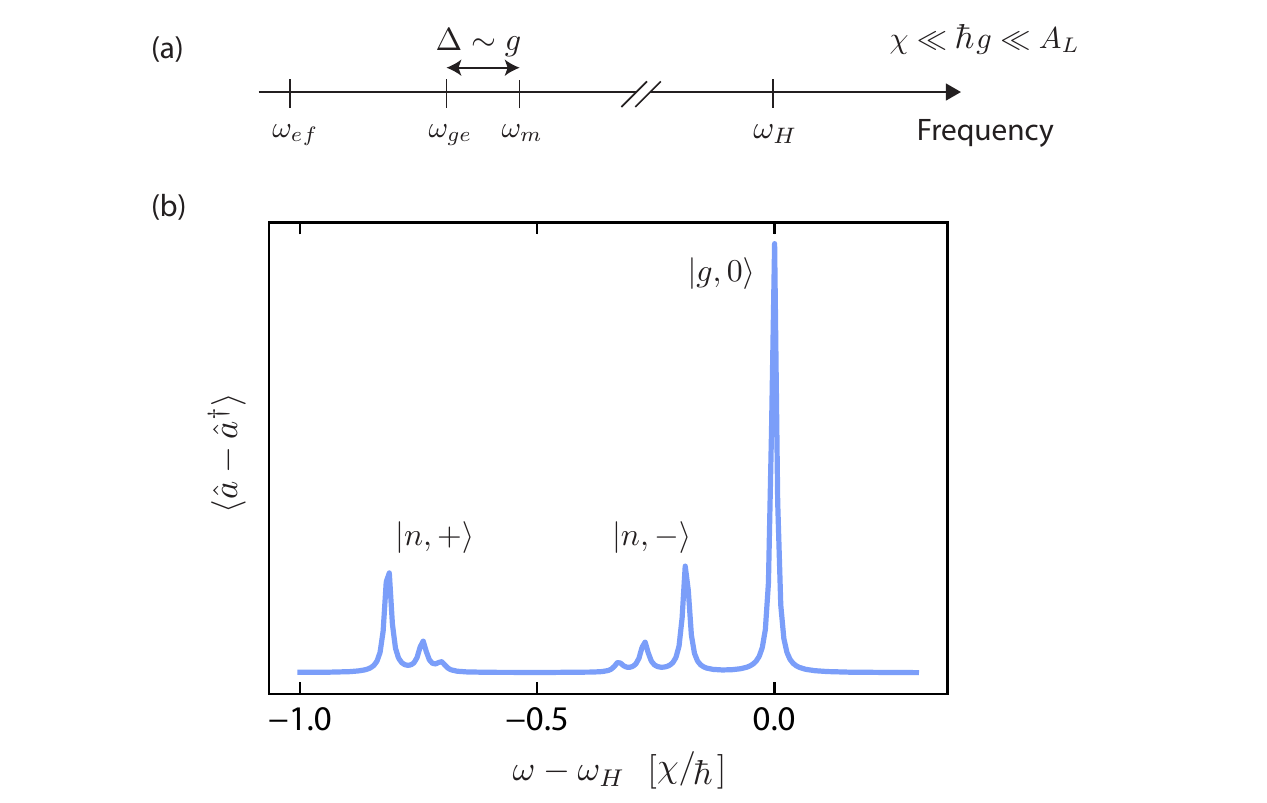}
\caption{
\textbf{High-frequency spectrum of an RFcQED coupled to a drum. }(case $A_L\gg \hbar g\gg \chi$)
\textbf{(a) Frequency landscape.}
In the regime where $A_L$ dominates over the coupling rate, the $\ket{e}\leftrightarrow\ket{f}$ transition of the LF mode is effectively decoupled of the drum.
The near resonant drum and $\ket{g}\leftrightarrow\ket{e}$ transition of the LF mode then hybridize following Jaynes-Cummings physics.
\textbf{(b) Spectrum.} 
We plot the expectation value $\langle\hat a - \hat a^\dagger\rangle$ whilst weakly driving the system at a frequency $\omega$.
This corresponds to one quadrature in a homodyne measurement of the HF mode.
With the condition $\gamma_H + 4n_{\text{th}}\gamma_L\ll \chi/\hbar/15$, we can resolve phonon-number dependent transitions.
The exact simulation parameters and numerical methods are provided in Sec.~\ref{sec:chapter-gravity_numerical_methods}
}
\label{fig:chi_g_A}
\end{figure}
As expanded upon in Sec.~\ref{sec:chapter-2_resonant_small_coupling}, since only the first transition of the LF mode will be resonant with the mechanical oscillator, the only relevant coupling terms in $g$ are those which involve both ground $\ket{g}$ and excited $\ket{e}$ states of the LF mode.
We may thus rewrite the Hamiltonian as
\begin{equation}
\begin{split}
	\hat H &= \hbar \omega_H \hat a ^\dagger \hat a - A_H\hat a ^\dagger\hat a ^\dagger \hat a \hat a \\
	&+\sum_{j\ge 2}(E_j-j\chi \hat a ^\dagger\hat a) \ket{j}\bra{j}\\
	&+ \hbar \omega_L \ket{e}\bra{e} + \hbar (\omega_L+\Delta)\hat c ^\dagger \hat c \\
	&+ \hbar g (\ket{g}\bra{e}\hat c ^\dagger +\ket{e}\bra{g}\hat c )- \sum_n\chi \hat a ^\dagger \hat a \ket{e,n}\bra{e,n}
\end{split}
\end{equation}
Note that multiplied the last term by identity $\sum_n\ket{n}\bra{n}$.
We may move to the eigen-basis of the Jaynes-Cummings Hamiltonian~\cite{jaynes_comparison_1963,bishop2010circuit}
\begin{equation}
\begin{split}
\ket{n,+}&= \cos \theta_n\ket{e,n-1} + \sin \theta_n\ket{g,n}\ ,\\
\ket{n,-}&= -\sin \theta_n\ket{e,n-1} + \cos \theta_n\ket{g,n}\ ,\\
\tan 2\theta_n &= -2g\sqrt{n}/\Delta\ ,\\
\end{split}
\end{equation}
where the states $\ket{n,\pm}$ have energies $\hbar\omega_{n,\pm}$ with\footnote{Constant energy terms are subtracted such that the ground state $\ket{0,g}$ has 0 energy.}
\begin{equation}
\omega_{n,\pm} =  (\omega_L+\Delta) n-\frac{\Delta}{2} \pm  \frac{1}{2}\sqrt{4g^2 n+\Delta^2}\\
\end{equation}  
The resulting Hamiltonian is
\begin{equation}
\begin{split}
	\hat H &= \hbar \omega_H \hat a ^\dagger \hat a - A_H\hat a ^\dagger\hat a ^\dagger \hat a \hat a  \\
	&+\sum_{j\ge 2}(E_j-j\chi \hat a ^\dagger\hat a) \ket{j}\bra{j}\\
	&+ \sum_{n\ge1,s=\pm}\hbar \omega_{n,s}\ket{n,s}\bra{n,s} \\
	&-\chi\sum_{n\ge1} \cos(\theta_n)^2\hat a ^\dagger \hat a\ket{n,+}\bra{n,+}\\
	&-\chi\sum_{n\ge1} \sin(\theta_n)^2\hat a ^\dagger \hat a\ket{n,-}\bra{n,-}\\
\end{split}
\end{equation}
where we neglected the term
\begin{equation}
\chi\cos(\theta_n)\sin(\theta_n)\sum_ n \left(\ket{n,+}\bra{n,-}+\ket{n,-}\bra{n,+}\right)
\end{equation}
valid in the limit which couple terms separated in frequency by $\sqrt{4g^2 n+\Delta^2}$ which is much smaller than $\chi$ under the initial assumption $\chi\ll\hbar g$.
The HF spectrum reveals the transitions
\begin{equation}
	\ket{g,n,\pm}\leftrightarrow \ket{e,n,\pm}
\end{equation}
at frequencies
\begin{equation}
	\omega_H-\chi_{n,\pm}/\hbar
\end{equation}
where $\chi_{n,+}  = \chi\cos(\theta_n)^2$ and $\chi_{n,-}  = \chi\sin(\theta_n)^2$.
Different values of the ratio $g/\Delta$ lead to different frequencies, note that we always have $0<\chi_{n,\pm}<\chi$.
If $g/\Delta\ll 1$, $\chi_{n,-}\sim0$ and $\chi_{n,+}\sim \chi$, and the HF mode is insensitive to $n$ and only sensitive to the state $g,e$ of the LF mode.
If $g/\Delta\gg 1$, $\chi_{n,\pm}\sim\chi/2$, and again sensitivity to $n$ is lost.
%
Numerically, we find that both $|\chi_{1,+}-\chi_{2,+}|$ and  $|\chi_{1,-}-\chi_{2,-}|$ have a maximum at $\sim \chi/15$ for $g/\Delta\simeq 0.6$ yielding
\begin{equation}
	\begin{split}
\chi_{1,+}&= 0.82\chi\\
\chi_{1,-}&= 0.18\chi\\
\chi_{2,+}&= 0.75\chi\\
\chi_{2,-}&= 0.25\chi\ .\\
	\end{split}
\end{equation}
In order to resolve the $\ket{g,1,\pm}\leftrightarrow \ket{e,1,\pm}$ transitions, the detuning to the transition $\ket{g,2,\pm}\leftrightarrow \ket{e,2,\pm}$, should exceed the line-width of the latter transition that we will call $\gamma_{H,\text{eff}}$
\begin{equation}
	\gamma_{H,\text{eff}}\ll\chi/15\hbar\ .
\end{equation}
An example of the obtainable spectrum is given in Fig.\ref{fig:chi_g_A}

\subsubsection{Case $A_L,\chi\gg \hbar g $}
\begin{figure}[]
\centering
\includegraphics[width=0.8\textwidth]{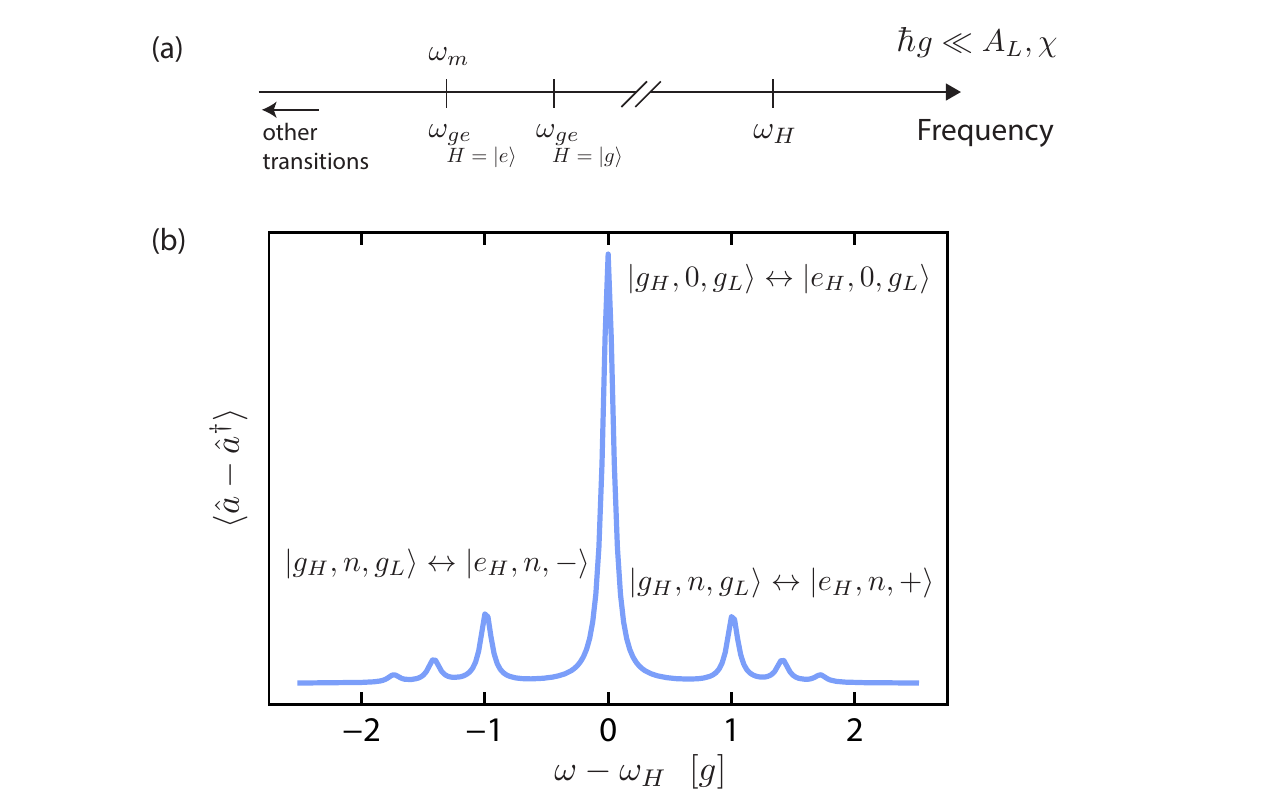}
\caption{
\textbf{High-frequency spectrum of an RFcQED coupled to a drum. }
\textbf{(a) Frequency landscape.} (case $A_L,\chi\gg \hbar g $)
In the regime where $\chi$ dominates the drum-to-low-mode coupling $g$, the LF mode can only be coupled to the drum for a given state of the HF mode.
As when the HF mode is in a different state, the frequency of the LF mode is shifted by $\chi\gg \hbar g$.
We focus on the case where the drum is resonant with the LF mode if the HF mode is in state $\ket{e}$.
Additionally, if $A_L$ dominates over the coupling rate, the $\ket{e}\leftrightarrow\ket{f}$ transition of the LF mode is effectively decoupled of the drum.
The near resonant drum and $\ket{g}\leftrightarrow\ket{e}$ transition of the LF mode then hybridize when the HF mode is in its excited state following Jaynes-Cummings physics.
\textbf{(b) Spectrum.} 
We plot the expectation value $\langle\hat a - \hat a^\dagger\rangle$ whilst weakly driving the system at a frequency $\omega$.
This corresponds to one quadrature in a homodyne measurement of the HF mode.
With the condition $\gamma_H + 4n_{\text{th}}\gamma_L\ll g/2$, we can resolve phonon-number dependent transitions at frequencies $\omega_H \pm g\sqrt{n}$, where $n$ is the number of phonons in the drum.
The exact simulation parameters and numerical methods are provided in Sec.~\ref{sec:chapter-gravity_numerical_methods}
}
\label{fig:g_A_chi}
\end{figure}
We write the Hamiltonian in the basis of eigenstates of the HF and LF modes.
\begin{equation}
\begin{split}
	\hat H =& \sum_{j_H}\ket{j_H}\bra{j_H}\otimes\left[E_{j_H}\right.\\
	+& \sum_{j_L}(E_{j_L}-j_H\chi) \ket{j_L}\bra{j_L} +  (\hbar\omega_L-\chi) \hat c ^\dagger \hat c\\
	-& g(\ket{j_L+1}\bra{j_L} \hat c+\ket{j_L}\bra{j_L+1}  \hat c^\dagger)\left.\right]\\
\end{split}
\end{equation}
Here the frequency of the LF mode depends on the state of the HF mode.
Since $\chi\gg \hbar g$, the LF mode will only be resonant with the drum when the HF mode is in a specific state.
We study the case where the LF mode and mechanical mode are resonant if the HF frequency mode is in its excited state.
Additionally, since $A_L\gg \hbar g$, only a single transition of the LF mode will be on resonance with the drum, the ground to excited state transition in this case.
We can thus rewrite the Hamiltonian as
\begin{equation}
\begin{split}
	\hat H = \ket{g_H}\bra{g_H}\otimes&\bigg[\sum_{j_L} E_{j_L} \ket{j_L}\bra{j_L} +  (\hbar\omega_L-\chi) \hat c ^\dagger \hat c \bigg]\\
	+\ket{e_H}\bra{e_H}\otimes&\bigg[\hbar\omega_H + \sum_{j_L\ge2} (E_{j_L}-\chi) \ket{j_L}\bra{j_L} \\
	&+(\hbar\omega_L-\chi) \ket{e_L}\bra{e_L}+  (\hbar\omega_L-\chi) \hat c ^\dagger \hat c \\
	&- g(\ket{e_L}\bra{g_L} \hat c+\ket{g_L}\bra{e_L}  \hat c^\dagger)\bigg]+...\\
\end{split}
\end{equation}
We now apply the unitary transformation $\Sigma_{j_H}\hat U_{j_H}$, where $U_{j_H}$ is identity except for $j_H=1$, when it brings the LF mode and drum to the Jaynes-Cummings basis
\begin{equation}
\ket{n,\pm}= \left(\ket{g_L,n}\pm \ket{e_L,n-1}\right)/\sqrt{2}\ ,\\
\end{equation}
where the states $\ket{n,\pm}$ have energies $\hbar\omega_{n,\pm}$\footnote{Constant energy terms are subtracted such that the ground state $\ket{0,g}$ has 0 energy.}
\begin{equation}
\omega_{n,\pm} =  (\omega_L-\chi/\hbar) n \pm g\sqrt{n}\\
\end{equation}  
The Hamiltonian writes
\begin{equation}
\begin{split}
	\hat H = \ket{g_H}\bra{g_H}\otimes&\bigg[\sum_{j_L} E_{j_L} \ket{j_L}\bra{j_L} + \hbar (\omega_L+\Delta) \hat c ^\dagger \hat c \bigg]\\
	+\ket{e_H}\bra{e_H}\otimes&\bigg[\hbar\omega_H + \sum_{j_L\ge2} E_{j_L} \ket{j_L}\bra{j_L} \\
	&+\sum_{n,s=\pm}\hbar\omega_L\ket{n,s}\bra{n,s} \bigg] + ...\\
\end{split}
\end{equation}
Probing the HF spectrum will reveal the following transitions
\begin{equation}
\begin{split}
	\ket{g_H,n,g_L}\leftrightarrow \ket{e_H,n,\pm}\\
	\ket{g_H,n,e_L}\leftrightarrow \ket{e_H,n+1,\pm}\\
\end{split}
\end{equation}
with frequencies
\begin{equation}
\begin{split}
&(\omega_H + \omega_{n,\pm})-(n(\omega_L-\chi/\hbar))\\
&(\omega_H + \omega_{n+1,\pm})-(n(\omega_L-\chi/\hbar)+\omega_L)\\
\end{split}
\end{equation}
\begin{equation}
\begin{split}
&\omega_H \pm g\sqrt{n}\\
&\omega_H -\chi/\hbar \pm g\sqrt{n}\\
\end{split}
\end{equation}
In order to resolve the $\ket{g,1,\pm}\leftrightarrow \ket{e,1,\pm}$ transitions, the detuning to the transition $\ket{g,2,\pm}\leftrightarrow \ket{e,2,\pm}$, given by $g\sqrt{2}-g\sim g/2$, should exceed the line-width of the latter transition $\gamma_{H,\text{eff}}$
\begin{equation}
	\gamma_{H,\text{eff}}\ll g/2\ .
\end{equation}
An example of the obtainable spectrum is given in Fig.\ref{fig:g_A_chi}.

\subsubsection{Case $\hbar g\gg \chi\gg A_L$}

In this regime, we write the LF mode as harmonic
\begin{equation}
\begin{split}
	\hat H &= \hbar \omega_H \hat a ^\dagger \hat a - A_H\hat a ^\dagger\hat a ^\dagger \hat a \hat a \\
	&+ \hbar \omega_L \hat b ^\dagger \hat b+ \hbar (\omega_L+\Delta) \hat c ^\dagger \hat c\\
	&+ \hbar g (\hat b\hat c ^\dagger +\hat b^\dagger\hat c )- \chi \hat a ^\dagger\hat a \hat b ^\dagger \hat b\\
\end{split}
\end{equation}
which comes under the condition that its anharmonicity represents a perturbation to the Hamiltonian smaller than the other interaction rates.
\begin{figure}[]
\centering
\includegraphics[width=0.8\textwidth]{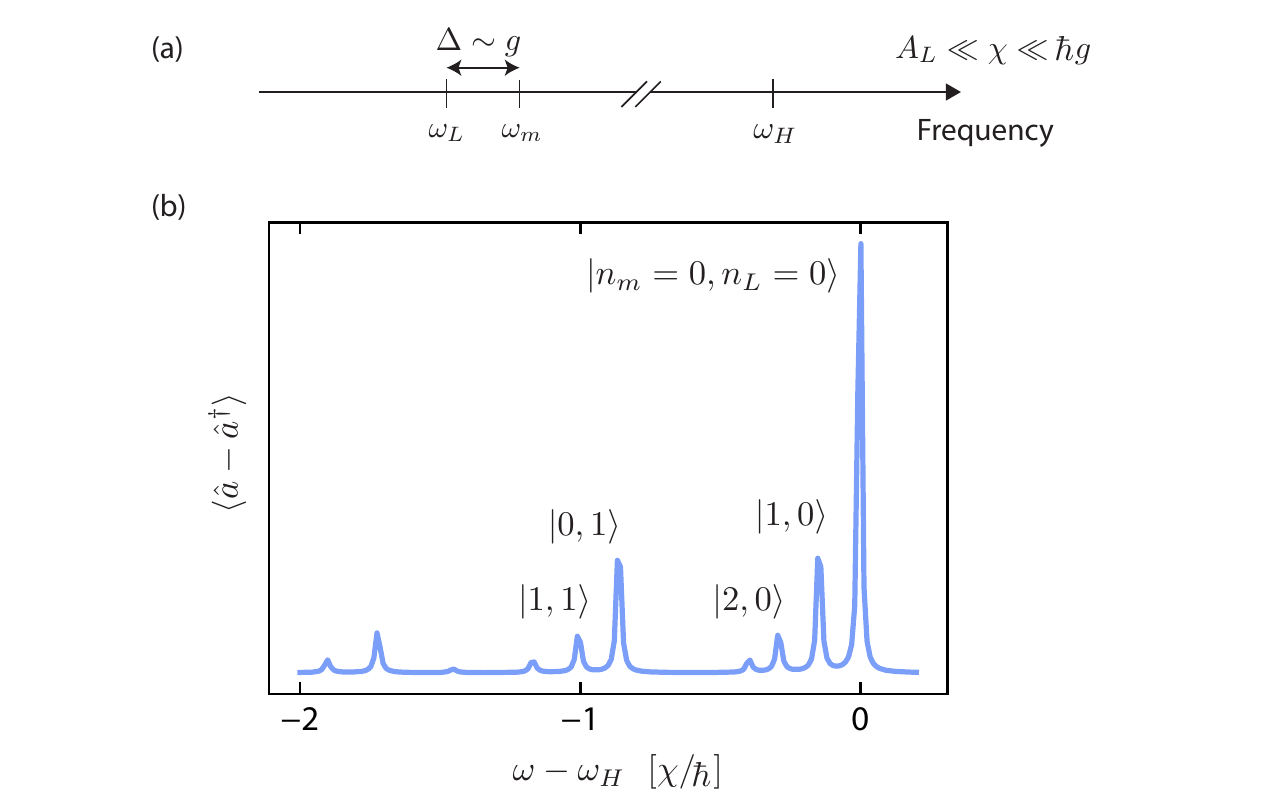}
\caption{
\textbf{High-frequency spectrum of an RFcQED coupled to a drum. }(case $\hbar g\gg \chi\gg A_L$)
\textbf{(a) Frequency landscape.}
In the regime where the drum-to-low-mode coupling $\hbar g$ dominates over the LF mode anharmonicity $A_L$, the LF mode and drum hybridize into two electromechanical modes.
\textbf{(b) Spectrum.} 
We plot the expectation value $\langle\hat a - \hat a^\dagger\rangle$ whilst weakly driving the system at a frequency $\omega$.
This corresponds to one quadrature in a homodyne measurement of the HF mode.
With the condition $\gamma_H + 4n_{\text{th}}\gamma_L\ll \chi_m/\hbar$, we can resolve phonon-number dependent transitions at frequencies $\omega_H - n_L\chi_L/\hbar- n_m\chi_m/\hbar$, where $n_m$ is the number of phonons in the most mechanical electromechanical mode, and $n_L$ the number of photons in the most electrical electro-mechanical mode, and $\chi_m+\chi_L=\chi$.
The exact simulation parameters and numerical methods are provided in Sec.~\ref{sec:chapter-gravity_numerical_methods}
}
\label{fig:A_chi_g}
\end{figure}
We first perform a first basis change to the normal-modes resulting from the $g$ coupling, leading to electromechanical modes indexed by $1$, with annihilation operators defined by
\begin{equation}
\begin{split}
\hat b &= f(g/\Delta)\hat b_1 + h(g/\Delta) \hat c_1\\
\hat c &= f(g/\Delta)\hat c_1 - h(g/\Delta) \hat b_1\\
\label{eq:chapter-gravity_mode_1_annihilation}
\end{split}
\end{equation}
where
\begin{equation}
\begin{split}
	f(x) &=  \frac{1+\sqrt{1+4x^2}}{\sqrt{8x^2+2\left(1+\sqrt{1+4x^2}\right)}} \\
	h(x) &= \frac{2x}{\sqrt{8x^2+2\left(1+\sqrt{1+4x^2}\right)}}
\end{split}
\end{equation}
and mode frequencies
\begin{equation}
\begin{split}
\omega_1+\Delta_1 &= \omega_L+\frac{\Delta}{2}\left(1+\sqrt{1+4\frac{g^2}{\Delta^2}}\right)\\
\omega_1 &= \omega_L+\frac{\Delta}{2}\left(1-\sqrt{1+4\frac{g^2}{\Delta^2}}\right)\\
\label{eq:chapter-gravity_mode_1_freqs}
\end{split}
\end{equation}
as calculated with the method described in Sec.~\ref{sec:normalmodebasisLS2}.
The resulting Hamiltonian is
\begin{equation}
\begin{split}
	\hat H &= \hbar \omega_H \hat a ^\dagger \hat a - A_H\hat a ^\dagger\hat a ^\dagger \hat a \hat a \\
	&+ \hbar \omega_1 \hat b_1^\dagger \hat b_1+ \hbar (\omega_1+\Delta_1) \hat c_1 ^\dagger \hat c_1\\
	&- \chi f(g/\Delta)^2 \hat a ^\dagger\hat a \hat b_1 ^\dagger \hat b_1 - \chi h(g/\Delta)^2 \hat a ^\dagger\hat a \hat c_1 ^\dagger \hat c_1\\
	&\\
\end{split}
\label{eq:chapter-gravity_mode_1_Hamiltonian_full}
\end{equation}
where we neglected the term
\begin{equation}
	- \chi f(g/\Delta)h(g/\Delta)\hat a ^\dagger\hat a(\hat b_1 ^\dagger \hat c_1+\hat b_1  \hat c_1^\dagger)
\end{equation}
valid in the limit $\hbar g\gg\chi$ where the interaction $\hat b_1 ^\dagger \hat c_1$ couples states separated in energy by $2g$ (near resonance), assumed to be much larger than the strength of this interaction, $\chi f(g/\Delta)h(g/\Delta)<\chi$.
We now look for a reasonable choice for the parameter $g/\Delta$.
In the case $\Delta=0$, the cross-Kerr interaction between the electromechanical modes $\chi f(g/\Delta)^2$ and $\chi h(g/\Delta)^2$ will be identical and equal to $\chi/2$.
The HF spectrum will then feature the transition frequencies
\begin{equation}
	\hbar\omega_H-\chi(n_++n_-)/2
\end{equation}
where $n_\pm$ refers to the occupation of each electromechanical mode.
Each measured peak would then correspond to multiple states, which restricts the level of control attainable over the quantum states of each electromechanical mode.
Alternatively, with $g/\Delta\ll1$, the LF mode would weakly hybridize, with a very small cross-Kerr coupling to the more mechanical degree of freedom.
An in-between is thus desireable, with $g/\Delta \sim 1$ giving rise to two electromechanical modes, one dominantly electrical, with a cross-Kerr coupling $\chi_L=\chi f(g/\Delta)^2$ and another more mechanical, with a cross-Kerr coupling $\chi_m = \chi h(g/\Delta)^2$.
This gives rise to a HF spectrum
\begin{equation}
	\hbar\omega_H-\chi_Ln_L-\chi_mn_m
\end{equation}
For example, with $g/\Delta = 1/4, 1/2, 1$, we obtain $\chi_m/\chi_L \simeq 0.06,0.17,0.38$.
Different values allow the resolution of more mechanical peaks between two electrical peaks, and the optimum will depend on the specifics of an experiment.
To conclude this section, resolution of the first mechanical Fock states is possible for $g/\Delta\sim 1$ such that $\chi_m = \chi /3$ and $\chi_L = 2\chi /3$ if
\begin{equation}
	\gamma_{H,\text{eff}} \ll \chi /3\hbar
\end{equation}
where $\gamma_{H,\text{eff}}$ is the effective line-width of the HF mode.
An example of the obtainable spectrum is given in Fig.\ref{fig:A_chi_g}.

\subsubsection{Case $\chi\gg \hbar g\gg A_L$}

In this limit, the cross-Kerr should impose the relelevant basis, and the coupling be treated as a perturbation only.
We write the Hamiltonian as
\begin{equation}
\begin{split}
	\hat H =& \sum_j\ket{j}\bra{j}\otimes\left[E_j\right.\\
	+&  (\hbar\omega_L-j\chi) \hat b^\dagger \hat b +  (\hbar\omega_L-\chi) \hat c ^\dagger \hat c\\
	-& \hbar g(\hat b ^\dagger \hat c+\hat b  \hat c^\dagger)\left.\right]\\
\end{split}
\end{equation}
where $j$ denotes the state of the HF mode.
What is emphasized here, is that for $\chi\gg \hbar g$, the two LF modes will only couple for certain values of $j$.
We explore the case where the mechanical mode is resonant\footnote{Contrary to the previous case, resonant interaction is possible, as the symmetry between electrical and mechanical mode is lifted by the fact that when the HF mode is in the ground state, it only couples to the electrical LF mode} with the LF electrical mode when the HF mode is in its first excited state\footnote{The case where the two low frequency modes are resonant for $j=0$ yields similar results, the advantage here is that the HF spectrum reflects the occupation of the un-coupled mechanical and LF mode, rather than a hybridized one.}.
We now apply the unitary transformation $\sum_j\ket{j}\bra{j}\hat U_j$, where the unitary transformation $\hat U_j$ acts upon the hilbert space of the two low frequency electromechanical modes.
The transformation $\hat U_j$ should be the ones which move the two coupled frequency electromechanical modes to a new normal-mode basis (one for each state of the HF mode $j$), with annihilation operators and frequencies defined as in Eqs.~(\ref{eq:chapter-gravity_mode_1_annihilation},\ref{eq:chapter-gravity_mode_1_freqs}).
For $j=0$ and $j\le 2$ the two modes are off-resonant by at least $ \chi \gg \hbar g$, such that we can apply the approximation $g\ll\Delta$ in the Eqs.~(\ref{eq:chapter-gravity_mode_1_annihilation},\ref{eq:chapter-gravity_mode_1_freqs}) leading to un-altered modes $\hat b_j = \hat b$ and $\hat c_j = \hat c$.
For $j=1$, the two modes are near resonance, leading to two normal modes $\hat \beta_\pm$ defined by
\begin{equation}
\hat \beta_\pm = (\hat b \pm \hat c)/\sqrt{2}\\
\end{equation}
and mode frequencies $\omega_L-\chi/\hbar\pm g$.
The Hamiltonian becomes
\begin{equation}
\begin{split}
	\hat H = \ket{g}\bra{g}\otimes&\left[\hbar \omega_L \hat b^\dagger \hat b +  (\hbar\omega_L-\chi+\hbar\Delta) \hat c ^\dagger \hat c\right]\\
	+ \ket{e}\bra{e}\otimes&\left[\hbar\omega_H +  (\hbar\omega_L-\chi - \hbar g) \hat \beta_-^\dagger \hat \beta_-+  (\hbar\omega_L-\chi + \hbar g) \hat \beta_+ ^\dagger \hat \beta_+\right]+...\\
\end{split}
\end{equation}
Probing the spectrum around $\omega_H$ will reveal peaks at the following frequencies
\begin{align}
\begin{split}
&(\omega_H + n_L(\omega_L-\chi/\hbar-g)+n_m(\omega_L-\chi/\hbar+g))-( n_L\omega_L+n_m(\omega_L-\chi/\hbar))\ ,\\
&(\omega_H + n_m(\omega_L-\chi/\hbar-g)+n_L(\omega_L-\chi/\hbar+g))-(n_L\omega_L+n_m(\omega_L-\chi/\hbar))\ ,
\label{eq:chapte-5_RFcQED_spectrum_small_g}
\end{split}
\end{align}
which couple the only states with some overlap
\begin{align}
\begin{split}
\ket{g,n_L,n_m}&\leftrightarrow\ket{e,n_{-}=n_L,n_{+}=n_m}\ ,\\
\ket{g,n_L,n_m}&\leftrightarrow\ket{e,n_{-}=n_m,n_{+}=n_L}
\end{split}
\end{align}
respectively.
Here the eigenstates $\ket{g,n_L,n_m}$ correspond the the HF mode in the ground state and the LF and mechanical modes populated with $n_L$,$n_m$ photons or phonons respectively.
The eigenstates $\ket{e,n_+,n_-}$ correspond the the HF mode in the excited state and the \textit{hybridized} low frequency electro-mechanical modes populated with $n_+$,$n_-$ excitations respectively.
Eq.~(\ref{eq:chapte-5_RFcQED_spectrum_small_g}) can be re-written as
\begin{align}
\begin{split}
&\omega_H - n_L(\chi/\hbar-g)-n_mg\ ,\\
&\omega_H - n_L(\chi/\hbar+g)+n_mg\ ,
\end{split}
\end{align}

\begin{figure}[]
\centering
\includegraphics[width=0.8\textwidth]{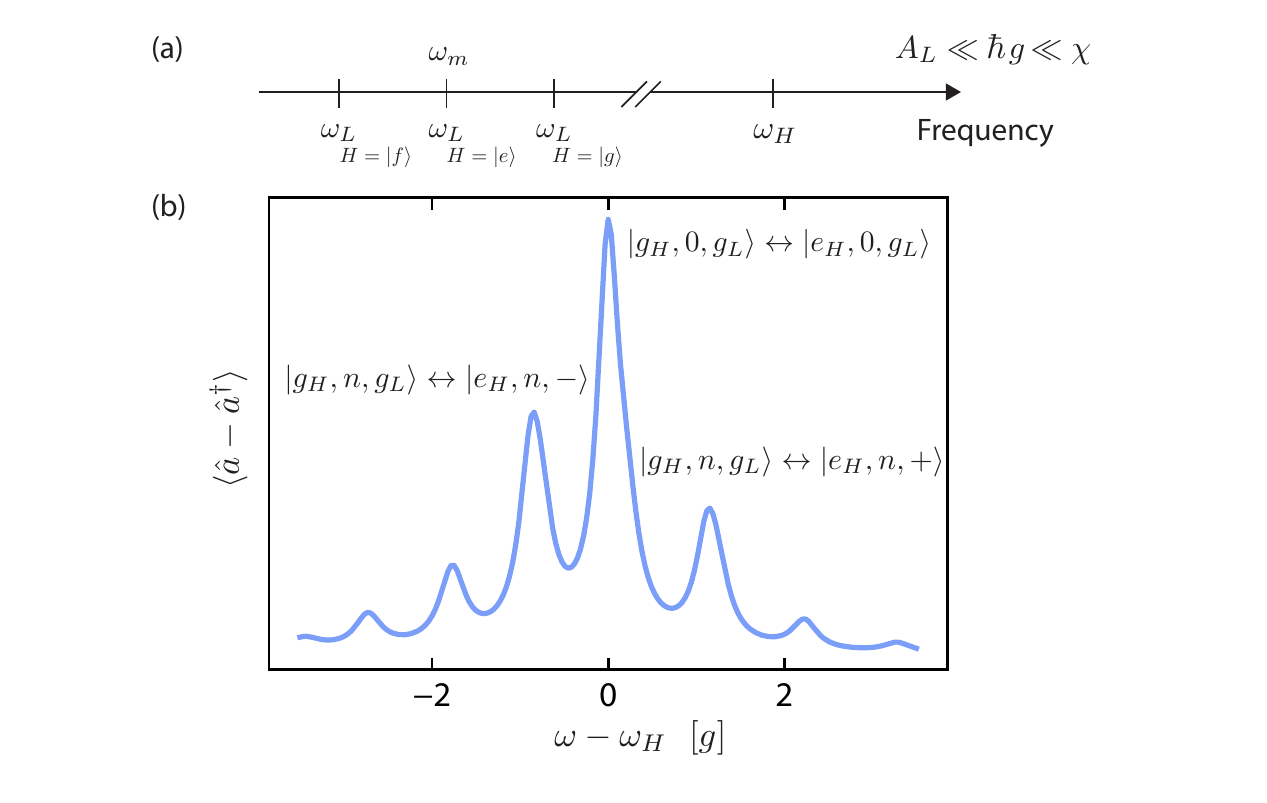}
\caption{
\textbf{High-frequency spectrum of an RFcQED coupled to a drum. }
\textbf{(a) Frequency landscape.}
The regime where $\chi$ dominates over the drum-to-low-mode coupling $g$ and the low frequency (LF) mode anharmonicity $A_L$ such that $A_L\ll \hbar g\ll \chi$ gives the most lax requirements on dissipation rates.
With dominating $\chi$, the LF mode frequency is a function of the state of the high frequency (HF) mode.
We focus on the case where the drum is resonant with the LF mode if the HF mode is in state $\ket{e}$.
\textbf{(b) Spectrum.} 
We plot the expectation value $\langle\hat a - \hat a^\dagger\rangle$ whilst weakly driving the system at a frequency $\omega$.
This corresponds to one quadrature in a homodyne measurement of the HF mode.
With the condition $\gamma_H + 4n_{\text{th}}\gamma_L\ll g$, with $\gamma_{H}$,$\gamma_{L}$ the dissipation rates of the HF and LF mode respectively, we can resolve phonon-number dependent transitions at frequencies $\omega_a \pm ng$, where $n$ is the number of phonons in the drum.
The exact simulation parameters and numerical methods are provided in Sec.~\ref{sec:chapter-gravity_numerical_methods}
}
\label{fig:A_g_chi}
\end{figure}

In this regime, the HF mode is mostly sensitive to the LF mode, and detection of mechanical Fock states necessitates
\begin{equation}
	\gamma_{H,\text{eff}}\ll g \ll \chi/\hbar\ ,
\end{equation}
where $\gamma_{H,\text{eff}}$ is the effective line-width of the HF mode.
An example of the obtainable spectrum is given in Fig.\ref{fig:A_g_chi}.

\textbf{Effective HF mode line-width $\gamma_{H,\text{eff}}$}.
In order to determine an expression for $\gamma_{H,\text{eff}}$, we make the assumption that the contribution coming from the coupling to the lower frequency electro-mechanical modes is dominated by the line-width of the electrical losses.
By extending Eq.~(\ref{eq:sum_of_lorentzians}), we then have
\begin{equation}
	\gamma_{H,\text{eff}} = \gamma_H + 2\gamma_L(\langle n_L\rangle+(1+2\langle n_L\rangle)n_{\text{th}}) 
\end{equation}
Where $\langle n_L\rangle$ measures the average occupation of the states involved in a transition of interest, not the average number of photons populating the low frequency mode.
In order to derive a compact set of requirements valid for all above cases, we make the assumption that a representative measure of $\gamma_{H,\text{eff}}$ corresponds to transitions between states with $\langle n_L\rangle \sim 1/2$ \textit{i.e.} mechanical Fock states which are half hybridized with the LF mode.
We also assume that the thermal occupation of the low frequency modes is significant $n_{\text{th}}\gg1$, yielding 
\begin{equation}
	\gamma_{H,\text{eff}} \simeq \gamma_H + 4n_{\text{th}}\gamma_L
\end{equation}

\subsection{Numerical methods}
\label{sec:chapter-gravity_numerical_methods}

Here we provide the code used to generate Figs.~\ref{fig:g_A},\ref{fig:A_g},\ref{fig:g_A_chi},\ref{fig:A_g_chi},\ref{fig:chi_g_A},\ref{fig:A_chi_g}.
We have used QuTiP~\cite{johansson2012qutip,johansson2013qutip} to construct the Hamiltonians defined in the previous section.
We then aim to emulate the spectrum that one would measure in spectroscopy experimentally.
This would be done by probing either the LF mode in the case where the drum is only coupled to a single transmon, or the HF mode in the case where the drum is coupled to an RFcQED system.
This read-out mode is denoted by "a", and the Hamiltonian is appended by a drive term $\left(\kappa_a/10^3\right)\left(\hat a+\hat a^\dagger \right)$ expressed in the rotating frame of the drive with frequency $\omega_d$ by subtracting $\omega_d \hat a^\dagger\hat a$.
In the case of a single transmon coupled a drum, we also move the mechanical degree of freedom to this rotating frame by subtracting $\omega_d \hat c^\dagger\hat c$, such that the coupling term $g\left(\hat a\hat c^\dagger + \hat a^\dagger\hat c \right)$ remains time-independent.
For each frequency, the spectrum is calculated through the expectation value of $(\hat a-\hat a^\dagger)$ after having reached the steady state.
The drive amplitude $\kappa_a/10^3$ is sufficiently low to not significantly populate the "a" mode.
We use relatively small thermal populations as increasing this number further requires larger Hilbert space sizes which rapidly make the simulation impossible to treat numerically.

\subsubsection{Drum and single transmon}

The code used to generate the figure in~\ref{fig:g_A}, which enforces the condition $\hbar g\ll A_L$, is
\begin{lstlisting}
from qutip import *
import numpy as np

# All frequencies and energies are in units
# of the transmon freqency

# Transmon parameters
Na = 4 # Hilbert space size
wa = 1 # Frequency
Ec = 0.05 # Anharmonicity
ka = 1e-4 # dissipation rate
ntha = 1.2 # Thermal occupation

# Mechanical parameters
Nc = 6
wc = wa
kc = 1e-7
nthc = ntha

# Coupling
g = Ec/10

# Hamiltonian
a = tensor(destroy(Na),qeye(Nc))
c = tensor(qeye(Na),destroy(Nc))
H = 0*a
H += (wa)*a.dag()*a 
H -= Ec/2*a.dag()*a.dag()*a*a 
H += (wc)*c.dag()*c
H += g*(a.dag()*c+a*c.dag()) 

# Collapse operator
c_ops = [
    np.sqrt(ka * (1 + ntha)) * a, 
     np.sqrt(ka * ntha) * a.dag(), 
     np.sqrt(kc * (1 + nthc)) * c, 
     np.sqrt(kc * nthc) * c.dag(), 
]

wdlist = np.linspace(wa-5*g,wa+5*g, 201)
spectrum = []
for wd in wdlist:
    rho = steadystate(H-(wd)*a.dag()*a-(wd)*c.dag()*c+ka/1e3*(a+a.dag()),c_ops, solver = 'scipy')
    spectrum += [expect(rho,1j*(a-a.dag()))]
\end{lstlisting}

The same code was used for \ref{fig:A_g}, with the following differences in parameters, which enforce the opposite condition $\hbar g\ll A_L$
\begin{lstlisting}
Na = 6 
Ec = 0.005 
ka = 4e-5 
g = Ec*15

wdlist = np.linspace(wa-g-2.5*Ec,wa-g+0.5*Ec, 201)
\end{lstlisting}

\subsubsection{Drum and RFcQED}

In the simulations describing the coupling of a drum to an RFcQED device, we denoted the HF mode, the LF mode and the drum by "a", "b" and "c" respectively.
The code used to generate the spectrum in Fig.~\ref{fig:A_chi_g}, where we enforced the condition $A_L\ll \chi \ll \hbar g$ is given below

\begin{lstlisting}
from qutip import *
import numpy as np

# All frequencies and energies are expressed 
# as a function of the low mode frequency

# Coupling parameters
chi = 0.01
g = 0.1

# Low frequency mode parameters
Nb = 4 # Hilbert space size
wb = 1 # Frequency
Al = 0.001 # Anharmonicity
kb =  0.001*chi # Dissipation rate
nthb = 0.5 # Thermal population

# High frequency mode parameters
Na = 3
wa = 50*wb
Ah = 0.05*wa
ka = 0.01*chi
ntha = 0

# Drum parameters
Nc = Nb
delta = 2*g
wc = wb+delta
kc =  0.001*chi
nthc = 0.5

a = tensor(destroy(Na),qeye(Nb),qeye(Nc))
b = tensor(qeye(Na),destroy(Nb),qeye(Nc))
c = tensor(qeye(Na),qeye(Nb),destroy(Nc))
H = 0*a
H += (wa)*a.dag()*a 
H -= Ah/2*a.dag()*a.dag()*a*a 
H -= chi*a.dag()*a *b.dag()*b 
H += (wb)*b.dag()*b 
H -= Al/2*b.dag()*b.dag()*b*b 
H += (wc)*c.dag()*c 
H += g*(c.dag()*b+c*b.dag()) 
H += g*(c.dag()*b.dag()+c*b)

c_ops = [
    np.sqrt(ka * (1 + ntha)) * a, 
     np.sqrt(ka * ntha) * a.dag(), 
     np.sqrt(kb * (1 + nthb)) * b, 
     np.sqrt(kb * nthb) * b.dag(), 
     np.sqrt(kc * (1 + nthc)) * c, 
     np.sqrt(kc * nthc) * c.dag()
]

wdlist = np.linspace(wa-2*chi,wa+0.2*chi, 201)
spectrum = []
for wd in wdlist:
    rho = steadystate(H-(wd)*a.dag()*a+ka/1e3*(a+a.dag()),c_ops, solver = 'scipy')
    spectrum += [expect(rho,1j*(a-a.dag()))]
\end{lstlisting}

To generate the spectrum in Fig.~\ref{fig:g_A_chi}, light modifications were made to the input parameters, to enforce the condition $\hbar g\ll A_L, \chi$.
These are provided below:

\begin{lstlisting}
# Coupling parameters
chi = 0.1
g = 0.001

# Low frequency mode parameters
Nb = 4 # Hilbert space size
wb = 1 # Frequency
Al = 0.05 # Anharmonicity
kb =  0.001*g # Dissipation rate
nthb = 0.5 # Thermal population

# High frequency mode parameters
Na = 3
wa = 50*wb
Ah = 0.05*wa
ka = 0.1*g
ntha = 0

# Drum parameters
Nc = Nb
delta = 0
wc = wb+delta
kc =  0.001*g
nthc = 0.5

wdlist = np.linspace(wa-2.5*g,wa+2.5*g, 201)
\end{lstlisting}

The spectrum in Fig.~\ref{fig:chi_g_A} follows the same method, but with $\chi\ll \hbar g\ll A_L$ realized with the parameters below

\begin{lstlisting}
# Coupling parameters
chi = 0.0005
g = 0.005

# Low frequency mode parameters
Nb = 4 # Hilbert space size
wb = 1 # Frequency
Al = 0.05 # Anharmonicity
kb =  0.001*chi # Dissipation rate
nthb = 0.5 # Thermal population

# High frequency mode parameters
Na = 3
wa = 50*wb
Ah = 0.05*wa
ka = 0.01*chi
ntha = 0

# Drum parameters
Nc = Nb
delta = g/0.6
wc = wb+delta-chi
kc =  0.001*chi
nthc = 0.5

wdlist = np.linspace(wa-chi,wa+0.3*chi, 201)
\end{lstlisting}

Finally, in Fig.~\ref{fig:A_g_chi}, the condition $A_L\ll \hbar g\ll \chi$ was realized with the following parameters

\begin{lstlisting}
from scipy.constants import k,h
T = 20e-3
f = 100e6
nth = 1/(np.exp(h*f/k/T)-1)

# Coupling parameters
chi = 15e6
g = 2e6

# High frequency mode parameters
Na = 3 # Hilbert space size
wa = 5e9 # Frequency
Ah = 0.05*wa # Anharmonicity
ka = 500e3 # Dissipation rate
ntha = 0 # Thermal population

# Low frequency mode parameters
Nb = 4
wb = 100e6
Al = chi**2/4/Ah
kb  =  wb/1e4
nthb = nth

# Drum parameters
Nc = Nb
wc = wb-chi
kc =  wc/1e5
nthc = nthb

wdlist = np.linspace(wa-3.5*g,wa+3.5*g, 201)
\end{lstlisting}

\FloatBarrier\chapter{Outlook}
\label{conclusion}

\begin{abstract}
\end{abstract}

\newpage

\noindent 
Although this thesis has focused on the different steps which led to radio-frequency circuit QED, and its applications to controlling micro-mechanical oscillators, this PhD work~\cite{yanai2017mechanical,gely2017convergence,bosman2017approaching,bosman_multi-mode_2017,gely2017nature,ockeloen2019sideband,gely2019observation,gely2020qucat,schmidt2020current} spans a broader set of applications and topics which are reflected in the multiple themes of this outlook.

\section{Extended discussion of the cQED Hamiltonian}
This section constitutes an extended discussion of chapter~\ref{chapter_2}.
The topics treated here constitute the main focus points of our published theoretical works: Ref~\cite{gely2017convergence} and Ref~\cite{gely2017nature}.

\subsection{Renormalization in the multi-mode cQED Hamiltonian}\label{sec:outlook_renormalization}

One consequence of the interaction between light and atoms is the Lamb shift of the atomic transition frequencies~\cite{lamb_fine_1947,heinzen1988}. %
Early attempts at calculating this shift led to the first shortcomings of QED theory, mainly that the transition energies of the atom diverge as the infinite number of electromagnetic modes are considered. 
Efforts to address these issues gave birth to renormalization theory~\cite{bethe1947electromagnetic}. 
In earlier days of circuit QED, ``ad hoc'' multi-mode extensions of the Rabi model~\cite{rabi1936process} suffer from divergences when considering the limit of infinite modes in a waveguide resonator~\cite{houck2008controlling,filipp2011multimode}.
The derivation from first-principles of the circuit QED Hamiltonian presented in chapter~\ref{sec:chapter-2-Hamiltonian_derivation} and published in Ref.~\cite{gely2017convergence} addresses this issue.

The key result is the renormalization of parameters such as the charging energy $E_C^{(M)}$, atomic frequency $\omega_a^{(M)}$ and the coupling rate $g^{(M)}$ with the number of modes included in the model $M$.
This renormalization arises because of the definition of the bare atom as current oscillations flowing only through the junction. 
In the extreme case of $C_a=0$, as the number of considered modes $M$ tends to infinity, the impedance path through only the series capacitors of the resonator equivalent circuit diverges. 
Charge from currents through the junction can no longer oscillate on $C_c$ and $\omega_a^{(M)}$ diverges.
The experimentally accessible parameters of the system show however no such divergence, as the shift arising from the presence of higher resonator modes exactly counteracts the changes in atomic frequency and anharmonicity once the Hamiltonian is diagonalized.
The details of this are expanded upon in Ref.~\cite{gely2017convergence}, and are an important result of the calculations presented in section~\ref{sec:chapter-2-Hamiltonian_derivation}.
It provides a useful framework for an intuitive understanding and modeling of experiments in the multi-mode ultra-strong coupling regime of section~\ref{sec:MMUSC}.
This formulation of the multi-mode quantum Rabi model in the context of circuits also hints at an intuitive picture on how this renormalization can arise physically, and it suggests the study of how this proposed physical picture could be applied to other problems in quantum field theory.

\subsection{Nature of the cQED Lamb shift}
\begin{figure*}[t!]
\centering
\includegraphics[width=0.85\textwidth]{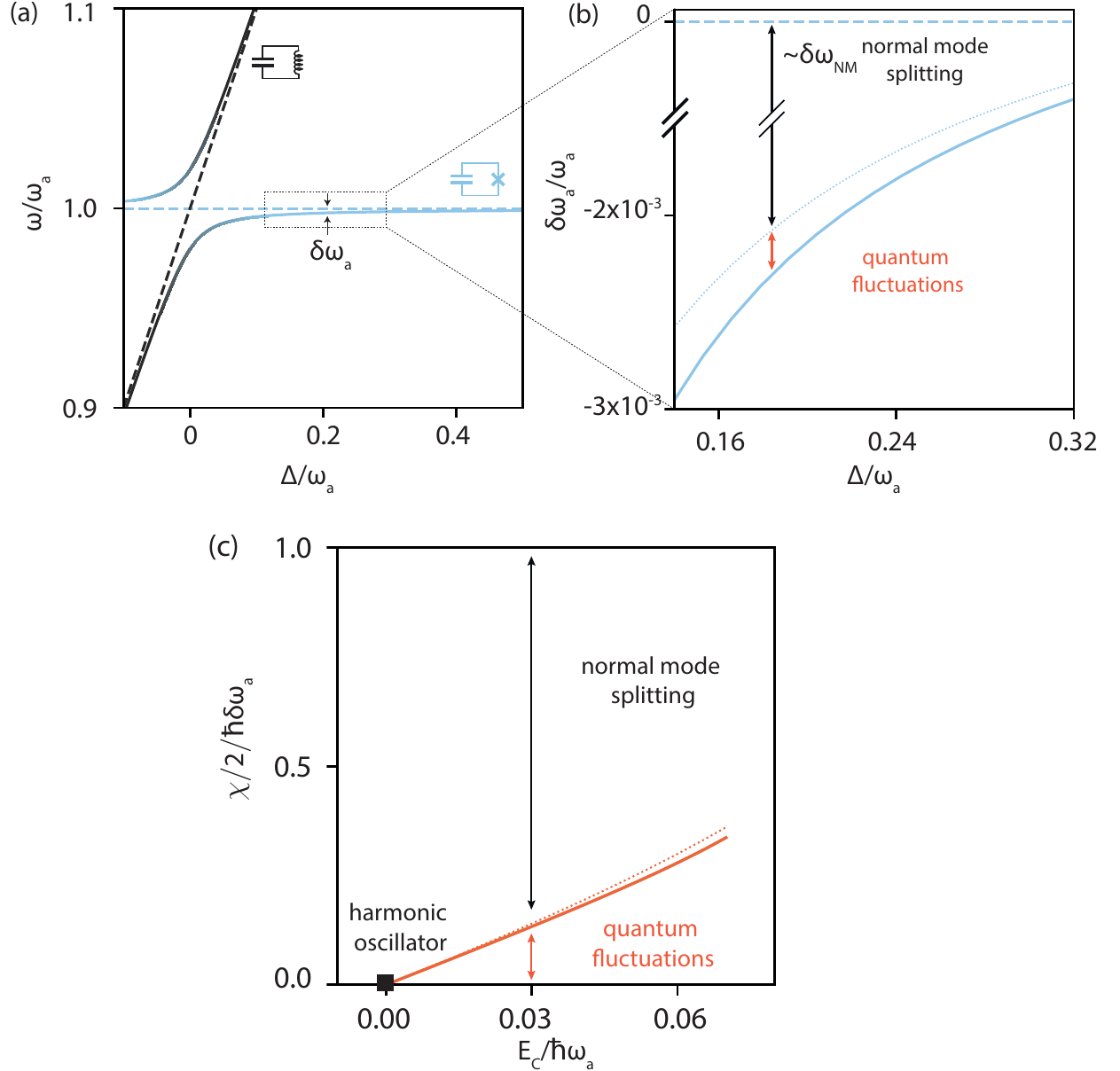}
\caption{\textbf{Fraction of the atomic energy shift due to quantum vacuum fluctuations.}
(a) Dressed frequency of the ground-to-first-excited state transitions of the harmonic oscillator (black) and atom (blue) as a function of detuning $\Delta = \omega_r-\omega_a$.
Bare frequencies ($g=0$) are shown as dashed lines,
We fixed $E_C/\hbar\omega_a=0.01$ and $g/\omega_a=0.02$.
(b) Total frequency shift $\delta\omega_a$ of the atom, decomposed into its two main components: normal-mode splitting $\delta_\text{NM}$ and a shift resulting from vacuum fluctuations $\chi/2$.
Coupling also changes the anharmonicity $A_a$, this results in a small shift absorbed here in $\delta_\text{NM}$.
(c) The vacuum-fluctuations-induced shift $\chi/2$ as a fraction of the total frequency shift of the atom  $\delta \omega_a$ for increasing anharmonicity $E_C$ and fixed detuning $\Delta = \omega_a/4$.
For a TLS, all of the energy shift arises from quantum fluctuations, $\chi/2/\hbar\delta\omega_a = 1$.
In all panels, the dotted lines are computed from Eqs.~(\ref{eq:shifts_beyond}), full lines correspond to a numerical diagonalization of Eq.~(\ref{eq:Hamiltonian_a_b_int_anh}). In (c), $\chi$ is computed from numerics as half the shift resulting from adding a photon in the oscillator.}
\label{fig:fig2}
\end{figure*}

Quantum theory predicts that vacuum is never at rest. 
On average, the electromagnetic field of vacuum has no amplitude, but quantum vacuum fluctuations impose a fundamental uncertainty in its value. 
This is notably captured in the ground-state energy of a harmonic oscillator (HO) $\hbar\omega_r/2$. 
When an atom couples off-resonantly to an electromagnetic mode, equivalent to a HO, the quantum vacuum fluctuations of the mode shift the transition frequencies between states of the atom~\cite{lamb_fine_1947}. 
This effect is called the Lamb shift. 
If the atom can be modeled as a two level system (TLS), this interaction is described in the rotating wave approximation (RWA) by the Jaynes-Cummings Hamiltonian~\cite{jaynes_comparison_1963}. 
The so-called Lamb shift is then given by $-g^2/\Delta$ in the dispersive regime $g\ll |\Delta|$ where $g$ is the coupling strength and $\Delta = \omega_r-\omega_a$ is the frequency detuning between the mode ($\omega_r$) and atom ($\omega_a$). 

If one replaces the TLS with a HO, a similar effect occurs from normal-mode splitting, where in the dispersive regime, each oscillator acquires a frequency shift due to the presence of the other oscillator. 
This similarity is not only qualitative: in the RWA parameter regime, a classical calculation of the normal mode splitting of two HOs also predicts this shift to be $-g^2/\Delta$.
A quantum calculation for two HOs will also give the same result: this shift for HOs is not influenced by the presence of quantum fluctuations.
Extending this further, one can replace the TLS atom with a weakly-anharmonic oscillator, such as a transmon qubit, and couple it to an LC resonator.
In this system, a shift $-g^2/\Delta$ of the transmon frequency was also observed, has been attributed to being induced by vacuum fluctuations, and is commonly referred to as the Lamb shift~\cite{fragner_resolving_2008}. 
However, normal mode splitting of two HOs, which includes no effect of quantum fluctuations, also leads to a shift of the same size.
This then raises the following question: how much of the dispersive shift in weakly-anharmonic atoms arises from quantum fluctuations? 
Or equivalently, how much of this shift persists if quantum fluctuations are neglected?
Following the derivation of Eqs.~(\ref{eq:shifts_rwa}) from chapter~\ref{chapter_2}, published in Ref.~\cite{gely2017nature}, we are in a position to answer these questions.
Compared to an isolated LC oscillator, the energy levels of a transmon coupled to a resonator are shifted by: \textit{(1)} normal-mode splitting $\delta_\text{NM}$, \textit{(2)} its own anharmonicity $A_a$ which arises from the quantum fluctuations of its eigen-states, and \textit{(3)} the shift proportional to $\chi$ arising from the quantum fluctuations of the resonator it is coupled to.
In Fig.~\ref{fig:fig2}(a,b), we show how these shifts manifest in a typical experimental setting where the detuning between the atom and resonator is varied, without explicitly showing contribution \textit{(2)}.
Off resonance, both modes are slightly shifted with respect to their un-coupled frequencies, and the total shift acquired when the resonator is in its ground-state $\delta\omega_a=\lambda-\delta_\text{NM}-A_a-\chi/2$, is indeed equal to $-g^2/\Delta$ following Eqs.~(\ref{eq:shifts_rwa}). 
There is a contribution from normal-mode splitting and from vacuum fluctuations which can both be quantified, and the former is much larger than the latter for a weakly-anharmonic system.
This also explains why earlier work~\cite{fragner_resolving_2008} found the Stark shift per photon to be smaller than the Lamb shift: vacuum fluctuations was not the only measured effect, normal-mode splitting also greatly contributed to the measured shift.
The proportion to which the total shift is due to vacuum fluctuations, as a function of anharmonicity, is shown in Fig.~\ref{fig:fig2}(c).

\section{What circuits could teach us about atoms}
One interesting avenue for future research would be to try and apply the concepts learned from circuits to QED with natural atoms in cavities or vacuum.

\subsection{Nature of the atomic Lamb shift}
A first topic of interest could be to determine if there exists an equivalent to the normal-mode basis for more highly anharmonic atoms.
One could then determine to what extent the atomic Lamb shift is due to quantum fluctuations.
We showed that for weakly anharmonic atoms, this small fraction of the total atomic shift is equal to half the Stark shift per photon.
QED experiments with atoms show that the Lamb shift of natural atoms can be 40\% larger than half the Stark shift per photon~\cite{brune1994lamb}. 
Since the original picture of the Lamb shift is of a phenomenon driven by quantum fluctuations, our results also raise questions about the terminology, and interpretation of, experiments in cavity and circuit QED.
In particular, should one reserve the terminology "Lamb shift" for only the part of the dispersive shift that arises from quantum fluctuations?

\subsection{Renormalization and cutoffs}\label{sec:outlook_atom_renormalization}
One could also try and apply to QED the concepts learned from quantizing circuits from first principles, as performed in sections~\ref{sec:chapter-2-Hamiltonian_derivation} and \ref{sec:circuit_quantization_overview}.
We distinguish the following questions.
\textit{Based on the first result of~\cite{gely2017convergence} -- } can we build the QED Hamiltonian from first principles mode by mode, observing the consequence of adding a mode on the parameters of the atom? Does renormalization of the system parameters then emerge in a non-ad-hoc way?
\textit{Based on the second result of~\cite{gely2017convergence} -- } is there a natural cutoff in the relevant modes of the vacuum which arises from the equivalent of the internal capacitance of the atom, \textit{i.e.} from the fact that the Coulomb interaction of the atom occurs through electric fields permeating the electromagnetic vacuum or cavity field?
\textit{Based on the results of~\cite{nigg_black-box_2012}, expanded upon in section \ref{sec:circuit_quantization_overview} -- } is there a version of the QED Hamiltonian which does not distinguish modes of the vacuum and the positional degree of freedom of the atom? If not, can one describe the vacuum as hybridized with the atom? Does such a model solve the divergence issues of the Lamb shift without requiring a cutoff?

\section{Utra-strong and deep-strong coupling with a transmon}

Circuit QED has opened up a wealth of new experimental possibilities by enabling much stronger light-matter coupling than in analogue experiments with natural atoms. 
One example is the interaction between an (artificial) atom and a resonator where the coupling rate $g$ becomes a considerable fraction to the atomic or resonator frequency.
This ultra-strong coupling (USC) regime, described by the quantum Rabi model, shows the breakdown of excitation number as conserved quantity, resulting in a significant theoretical challenge~\cite{casanova2010deep,braak2011integrability}.
In the regime of $g/\omega_{a,r} \simeq 1$, known as deep-strong coupling (DSC), a symmetry breaking of the vacuum is predicted~\cite{garziano2014vacuum} (\textit{i.e.} qualitative change of the ground state), similar to the Higgs mechanism or Jahn-Teller instability. 
To date, U/DSC with superconducting circuits has only been realized with flux qubits~\cite{forn2010observation,niemczyk2010circuit,yoshihara2017superconducting,forn2017ultrastrong,yoshihara2018inversion} or in the context of quantum simulations~\cite{langford_experimentally_2016,braumuller2017analog}. 
Whilst DSC is inaccessible to the transmon since $g<\sqrt{\omega_a\omega_r}/2$, using transmon qubits for USC could be interesting due to their potentially lower decoherence rates.
Also, transmon qubits are currently a standard in quantum computing efforts, implementing USC in a transmon architecture could also have technological applications by decreasing gate times~{\cite{romero2012ultrafast}} as well as the performance of quantum memories~{\cite{stassi2018long}}.
One limitation of the transmon however is that higher excitation levels become increasingly relevant with higher coupling rates, and in the USC the transmon cannot be considered as a two-level system. 
A transmon coupled to a resonator is therefore not a strict implementation of the Rabi model, yet is still expected to bear many of the typical USC features and a proposal has been made to measure them~\cite{andersen2017ultrastrong}.
The two devices presented in chapter~\ref{chapter_3} approach and exceed the $g/\omega_{a,r} = 0.1$ mark.
In the most strongly coupled, vacuum-gap transmon device, this leads to a resolvable Bloch-Siegert shift.
To understand the meaning of this shift, we divide the atom-resonator interaction $\hbar g(\hat a-\hat a^\dagger)(\hat b-\hat b^\dagger)$ into two parts.
The first terms conserve excitations $-\hbar g(\hat a\hat b^\dagger + \hat b\hat a^\dagger)$, and remain after applying the rotating-wave approximation (RWA). 
The second terms, called counter-rotating terms, add and extract excitations from the transmon and resonator in a pair-wise fashion $\hbar g(\hat a^\dagger\hat b^\dagger + \hat b\hat a)$.  
For sufficiently small couplings $g/\omega_{a,m}\ll 0.1$ the non-RWA terms can be neglected reducing the Rabi model to the Jaynes-Cummings model~\cite{jaynes_comparison_1963}.
For higher couplings $g/\omega_{a,m}\gtrsim0.1$ the RWA is no longer applicable and the excitation number conservation of the JC model is replaced by a conservation of excitation number parity~\cite{braak2011integrability,niemczyk2010circuit}. 
In this regime, making the RWA would lead to a deviation in the energy spectrum of the transmon known as Bloch-Siegert shift $\chi_\text{BS}$, marking the entry into the USC regime~\cite{forn2010observation}.
In the device of section~\ref{sec:MMUSC}, this shift is $62$ MHz at its largest.
One should however consider the implications of this result with care.
\textbf{Choice of basis -- }
the physics predicted in the USC or DSC regime depends on the basis in which the system is viewed.
For example, by tracing out the qubit in the ground state of the Rabi model, one finds that the resonator is in a cat state when in the DSC regime.
But this state is only accessible if we can measure the system in a basis which distinguishes the uncoupled resonator and atomic eigenstates.
This for example requires quenching the interaction on a time scale faster than $1/g$.
In a similar fashion, the Bloch-Siegert shift arises as a result of writing the Hamiltonian in a certain basis, where the coupling is expressed as $\hbar g(\hat a-\hat a^\dagger)(\hat b-\hat b^\dagger)$.
If we write the Hamiltonian in the normal-mode basis, the coupling is a cross-Kerr interaction between two anharmonic modes of the system, and the Bloch-Siegert shift is difficult to identify.
\textbf{How quantum are U/DSC features? -- }Another important question to ask is if quantum mechanics are involved in the appearance of U/DSC features at all.
This is of particular relevance when considering U/DSC with weakly-anharmonic atoms such as the transmon.
As an example, consider the Bloch-Siegert shift, as defined in the previous paragraph.
If we calculate the spectrum of the system for a given coupling, with and without the RWA, for diminishing values of the transmon anharmonicity, we find that the Bloch-Siegert shift only marginally changes value.
As with the Lamb shift, the quantum nature of this shift is then questionable.

\section{Radio-frequency circuit QED}
Our radio-frequency circuit QED architecture presented in this thesis enables the readout and manipulation of a radio-frequency resonator at the quantum level.
Here we summarize some of the applications of this platform.
Utilizing more elaborate readout methods available to cQED, such as single-shot readout or the tracking of quantum trajectories, we could enable even finer resolution of thermodynamic effects at the quantum scale.
This would extend the results of Fock-state-resolved tracking of quantum states thermalization presented in this work.
Another prospect would be to couple many RFcQED devices together.
This could for example enable the exploration of many-body effects in Bose-Hubbard systems with dynamically tunable temperatures~\cite{rigol2008thermalization,sorg2014relaxation}.
RFcQED could also be used to interface circuit QED with different physical systems in the MHz frequency range, such as spin systems~\cite{ares2016sensitive} or macroscopic mechanical oscillators (see Sec.~\ref{sec:conclusion_gravity}).
Another potential application is in sensing, where this circuit could be coupled to incoming radio frequency radiation.
This could enable quantum resolved sensing~\cite{murch2018single,chen2011microwave,inomata2016single,besse2018single,kono2018quantum} in a critical frequency range for a number of applications, such as nuclear magnetic resonance imaging or radio astronomy.

\section{Automation of circuit quantization}
\subsection{Scope of QuCAT}
The scope of QuCAT is threefold.
First, it enables quickly building a design for a superconducting circuit.
This is the case where the circuit architecture is know, for example one wants to build an LC oscillator coupled to a transmon, weakly coupled to a feedline.
But the unknowns are the values of the circuit components (inductances, capacitances and resistors), to reach required specifications.
Secondly is makes the design of novel circuits a lot easier.
Without the availability of such a library, and given a circuit idea, one would have to calculate, by hand, the corresponding Hamiltonian.
Depending on the circuit topology, this can be tedious or even challenging.
QuCAT allows one to quickly test different ideas, and potentially innovate, inventing circuits such as the RFcQED system presented in this thesis.
Lastly, it provides tools to gain a deeper understanding of the physical mechanisms taking place in the circuit, notably by plotting the distribution of currents corresponding to the normal modes of the circuit.
Multiple examples of this were provided in the application section of chapter~\ref{chapter-qucat}.
\subsection{Possible extensions}
We distinguish the extensions which could provide additional functionality to QuCAT from those which lead to better performance.
\subsubsection{Additional features}
Possible extensions of the QuCAT features could include 
\begin{itemize}
\item black-box impedance components to model distributed components~\cite{nigg_black-box_2012}, 
\item more precisely modeling lossy circuits~\cite{solgun2014blackbox,solgun2015multiport}
\item additional elements such as coupled inductors or superconducting quantum interference devices (SQUIDS)
\item different quantization methods, enabling for example quantization in the charge or flux basis 
\item handling static offsets in flux or charge through DC sources
\end{itemize}
The latter two items would extend QuCAT beyond the scope of weakly-anharmonic circuits.
\subsubsection{Increased performance}
In terms of performance, QuCAT would benefit from delegating analytical calculations to a more efficient, compiled language, with the exciting prospect of simulating large scale circuits~\cite{arute2019quantum}. 
This inevitably requires the development of more efficient open-source symbolic manipulation tools.
The development of the open-source C++ library SymEngine \url{https://github.com/symengine/symengine}, together with its Python wrappers, the symengine.py project \url{https://github.com/symengine/symengine.py}, could lead to rapid progress in this direction.
One should keep in mind that an increase in circuit size translates to an increase in the number of degrees of freedom of the circuit and hence of the Hilbert space size needed for further analysis once a Hamiltonian has been extracted from QuCAT.

\section{Towards quantum superpositions of space-time}
\label{sec:conclusion_gravity}

What we have exposed in chapter~\ref{chapter_gravity}, is that there is an undeniable gap in our knowledge of the physical laws of nature.
Indeed, we are not able to describe the physics of a massive object prepared in a quantum superposition of "being in two locations at the same time".
The uncertainty such a situation introduces in the definition of time, in virtue of relativistic time-dilation, means that we can not write Schrodinger's equation to study the evolution of the system.
The experimental requirements to observe a deviations from non-general-relativistic quantum mechanics, in terms of mass and spread in space of the quantum superposition, are however unclear.
All we have are (educated) guesses that now need to be verified through experiment.

One such experiment would be to prepare quantum states of oscillating micro-mechanical systems.
In such states, each nucleus of the oscillator is ideally in a quantum superposition spread over an area of space much larger than the nucleus size.
The challenge is to observe the coherence of such a state over a time-scale much larger than the time-scale of time-dilation-induced effects.
This requires oscillators with both high quality factors and a large mass.
Note that extending the size of achievable superposition states (i.e. zero-point fluctuations) beyond the size of the nucleus, whilst increasing the mass of the oscillator translates to lowering the frequency of the oscillator.
High quality factor membranes, constructed from high-stress silicon nitride, and oscillating between 100 kHz and a MHz, seem to be the best candidates for this experiment as they comfortably satisfy all the above requirements.
The challenge now lies in preparing these membranes in quantum states.
In chapter~\ref{chapter_phonon_res}, we have explored how to achieve this using weakly anharmonic superconducting circuits.
We find that megahertz circuits are a better option than the usual gigahertz circuits, but strong electro-mechanical coupling is inevitable.
We discuss that considerable experimental and theoretical work is still needed to reach strong coupling.
These developments could take the shape of a novel circuit architecture or an increase of the quality factor of megahertz circuits.
Other options are also worth exploring.
One option is to rely on optomechanics to both prepare the membrane in its ground-state~\cite{teufel2011circuit,yuan2015large}, and to swap remotely prepared quantum states into the mechanical oscillator~\cite{reed2017faithful}.
A second option is to move away from transmons and weakly anharmonic circuits.
Whilst these are easy to fabricate, and have relatively large quality factors, they may not be the best platform for electro-mechanics.
Indeed, considerable phonon-sensitivity has already been achieved by coupling a GHz cooper-pair box qubit to a MHz drum~\cite{Viennot2018}.
One could also explore novel circuits rather than existing qubits which were developed for quantum computation.
An attractive avenue for research would be to tailor the qubit design to the specific task at hand: providing optimal sensitivity to a narrow frequency band of charge oscillations of a membrane.
\thumbfalse

\chapter*{Curriculum Vit\ae}
\addcontentsline{toc}{chapter}{Curriculum Vit\ae}
\setheader{Curriculum Vit\ae}

\makeatletter
\authors{\@firstname\ {\titleshape\@lastname}}
\makeatother

\noindent
\begin{tabular}{ll}
    19-08-1992 & Born in Montpellier, France.
\end{tabular}

\section*{Education}
\vspace{2em}
\begin{tabular}{ll}
    1995--2010 & Primary and secondary school \\
     & École primaire, Vendémian, France (1995-2003) \\
     & Collège Emmanuel Maffre-Baugé, Paulhan, France (2003-2006) \\
     & Collège Lo Trentanel, Gignac, France (2006-2007) \\
     & Lycée René-Gosse, Clermont l'Hérault, France (2007-2010) \\
    \vspace{2em}
    \\
    2010--2012 & Classe préparatoire MPSI/MP, Lycée Alphonse-Daudet, Nîmes, France \\
    \vspace{2em}
    \\
    2012--2014 & MSc Engineering, École Centrale de Nantes, France \\&
    \begin{minipage}{\textwidth-4\parindent-4\tabcolsep}
        \begin{tabular}{@{}p{0.2\linewidth}@{}p{0.8\linewidth-\tabcolsep}}
            \textit{Internship:} & Modelling of shape memory metal springs \\
            \textit{Company:} & Mann+Hummel
        \end{tabular}
    \end{minipage}
    \vspace{2em}
    \\
    2014--2016 & MSc\ Physics, Delft University of Technology \\
    \\&
    \begin{minipage}{\textwidth-4\parindent-4\tabcolsep}
        \begin{tabular}{@{}p{0.2\linewidth}@{}p{0.8\linewidth-\tabcolsep}}
            \textit{Internship:} & Fabrication of flexible cabling \\
            \textit{Company:} & Delft Circuits\vspace{1em}\\

            \textit{Thesis:} & Ultra-strong coupling in multi-mode circuit quantum electrodynamics \\
            \textit{Advisor:} & S.\ Bosman, Dr.\ D.\ Bothner and Prof.\ dr.\ G.\ A.\ Steele
        \end{tabular}
    \end{minipage}
\end{tabular}




\chapter*{List of Publications}
\addcontentsline{toc}{chapter}{List of Publications}
\setheader{List of Publications}
\label{publications}

Equal contribution is denoted by an asterix *

\begin{enumerate}{\small

\item F. E. Schmidt, D. Bothner, I. C. Rodrigues, \textbf{M. F. Gely}, M. D. Jenkins and G. A. Steele, \textit{Current detection using a Josephson parametric upconverter
}, {\href{https://arxiv.org/abs/2001.02521}{arXiv preprint arXiv:2001.02521 (2020)}}

\item \textbf{M. F. Gely} and G. A. Steele, \textit{QuCAT: Quantum Circuit Analyzer Tool in Python}, \href{https://iopscience.iop.org/article/10.1088/1367-2630/ab60f6/meta}{New Journal of Physics \textbf{22}, 013025 (2020)}

\item \textbf{M. F. Gely}, M. Kounalakis, C. Dickel, J. Dalle, R. Vatré, B. Baker, M. D. Jenkins and G. A. Steele, \textit{Observation and stabilization of photonic Fock states in a hot radio-frequency resonator}, {\href{https://science.sciencemag.org/content/363/6431/1072}{Science \textbf{363}, 1072 (2019)}}

\item C. F. Ockeloen-Korppi, \textbf{M. F. Gely}, E. Damskägg, M. Jenkins, G. A. Steele and M. A. Sillanpää, \textit{Sideband cooling of nearly degenerate micromechanical oscillators in a multimode optomechanical system}, {\href{https://journals.aps.org/pra/abstract/10.1103/PhysRevA.99.023826}{Phys. Rev. A \textbf{99}, 023826 (2019)}}

\item \textbf{M. F. Gely}, G. A. Steele and D. Bothner, \textit{Nature of the Lamb shift in weakly anharmonic atoms: From normal-mode splitting to quantum fluctuations}, {\href{https://journals.aps.org/pra/abstract/10.1103/PhysRevA.98.053808}{Phys. Rev. A \textbf{98}, 053808 (2018)}}

\item S. J. Bosman, \textbf{M. F. Gely}, V. Singh, A. Bruno, D. Bothner and G. A. Steele, \textit{Multi-mode ultra-strong coupling in circuit quantum electrodynamics}, {\href{https://www.nature.com/articles/s41534-017-0046-y}{npj Quantum Information \textbf{3}, 46 (2017)}}

\item S. J. Bosman*, \textbf{M. F. Gely*}, V. Singh, D. Bothner, A. Castellanos-Gomez and G. A. Steele, \textit{Approaching ultrastrong coupling in transmon circuit QED using a high-impedance resonator}, {\href{https://journals.aps.org/prb/abstract/10.1103/PhysRevB.95.224515}{Phys. Rev. B \textbf{95}, 224515 (2017)}}

\item \textbf{M. F. Gely*}, A. Parra-Rodriguez*, D. Bothner, Y. M. Blanter, S. J. Bosman, E. Solano and G. A. Steele, \textit{Convergence of the multimode quantum Rabi model of circuit quantum electrodynamics}, {\href{https://journals.aps.org/prb/abstract/10.1103/PhysRevB.95.245115}{Phys. Rev. B \textbf{95}, 245115  (2017)}}

\item S. Yanai, V. Singh, M. Yuan, \textbf{M. F. Gely}, S. J. Bosman and G. A. Steele, \textit{Mechanical dissipation in MoRe superconducting metal drums}, {\href{https://aip.scitation.org/doi/full/10.1063/1.4976831}{Appl. Phys. Lett. \textbf{110}, 083103  (2017)}}

}\end{enumerate}

\def\bibfont{\footnotesize}
\printbibliography

\end{document}